\documentclass[a4paper,12pt,times,oneside,print,scrreprt,longbibliography,index]{Classes/PhDThesisPSnPDF}

\usepackage{csquotes} 
\usepackage[utf8]{inputenc}
\usepackage{calrsfs}
\usepackage[numbers]{natbib}
\usepackage[numbers]{natbib}
\usepackage{graphicx}
\usepackage{textcomp}
\usepackage{amsmath,amssymb}
\usepackage{bbold}
\usepackage{color}
\usepackage{soul}
\usepackage{natbib}
\usepackage[numbers]{natbib}
\usepackage{natbib}
\usepackage{mathtools}
\usepackage{subcaption}
\newcommand{\newc}{\newcommand}
\usepackage{amsfonts}
\usepackage{amsbsy}
\usepackage{natbib}
\usepackage{csquotes}
\usepackage{comment}
\usepackage{amsmath}
\usepackage[bottom]{footmisc}
\usepackage{float}
\newcommand{\im}{\mathrm{i}}
\newc{\Tr}{\mbox{Tr}}
\newc{\ovl}{\overline}
\newcommand{\beqa}{\begin{eqnarray}}
\newcommand{\eeqa}{\end{eqnarray}}
\newcommand{\beq}{\begin{equation}}
\newcommand{\eeq}{\end{equation}}

\definecolor{shadecolor}{rgb}{0.8,0.8,0.8}

\input{Preamble/preamble}

\ifdefineChapter
 \fi
\ifdefineChapter
 \fi
\ifdefineChapter
 \fi
 
\ifdefineChapter
\fi

\ifdefineChapter
\fi

\ifdefineChapter
\fi
\begin{document}

\frontmatter

\begin{titlepage}
	\centering
	{\Large\bfseries  Out-of-time-order
correlation in the quantum Ising Floquet spin system and magnonic crystals\par}
	\vspace{0.2cm}
	
	\begin{figure}[htbp]
    \centering
    \includegraphics[scale=0.4]{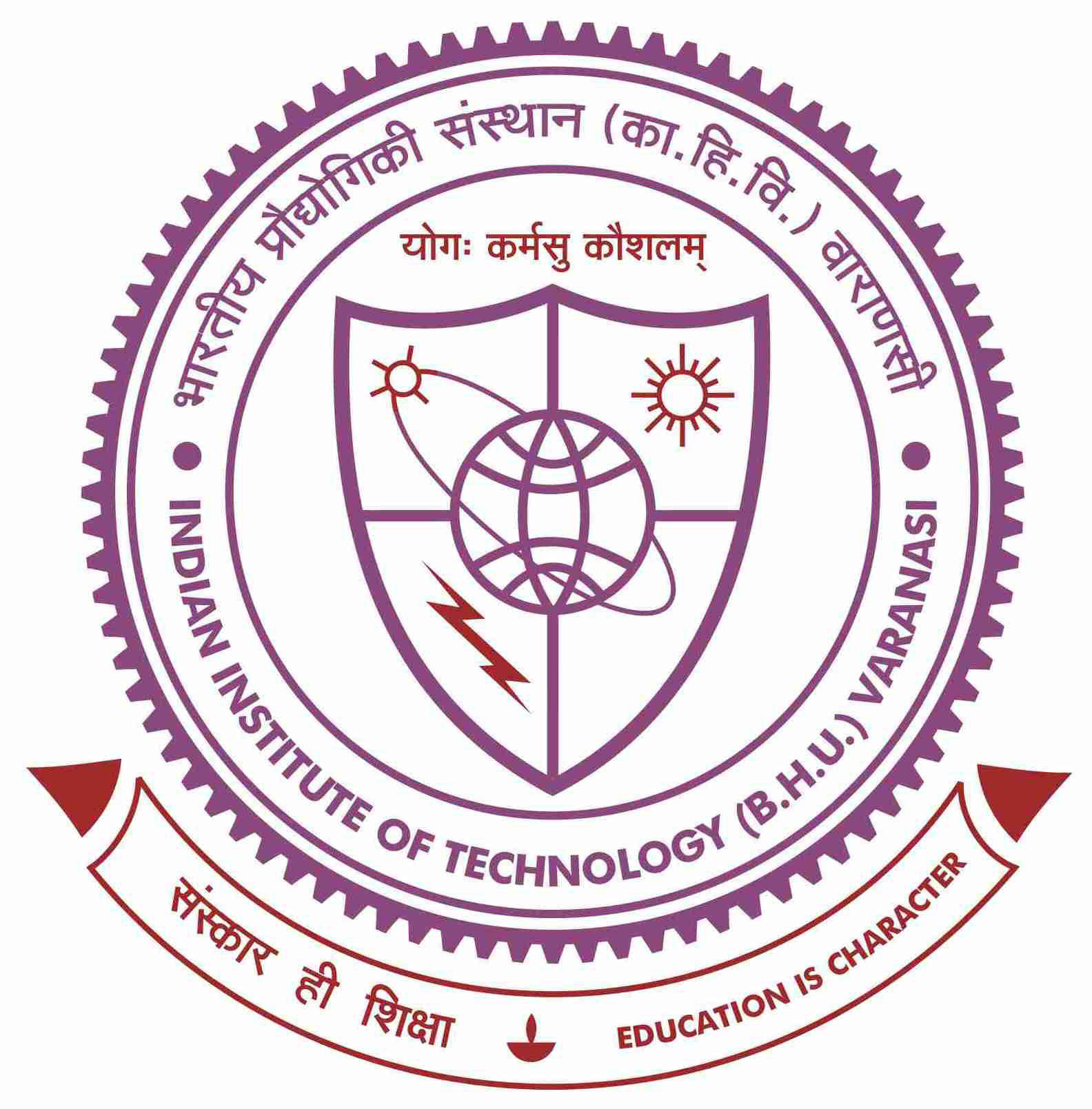}
\end{figure} 	
	
	\textbf{Thesis submitted in partial fulfillment}\\ 
	\textbf{for the Award of }\\
	{\scshape Doctor of Philosophy\par}
		{in\par}
		{\scshape \textbf{Physics}\par}

		\vfill
	{\textit{by}\par}
	{\large \scshape{Rohit Kumar Shukla}\par}
	\vfill
	{\textit{Under the supervision of}\par}
	\large {Dr. Sunil Kumar Mishra }\\

	\vfill
	{\scshape{DEPARTMENT OF PHYSICS}\par}
	\textbf{INDIAN INSTITUTE OF TECHNOLOGY}\\
	\textbf{BANARAS HINDU UNIVERSITY}\\
	\textbf{VARANASI - 221005}\\
\vspace{0.6cm}
	\begin{table*} [bthp]
		\centering
	\begin{tabular}{cc}
	ROLL NUMBER	\hspace{6cm} &  YEAR OF SUBMISSION \\ 
	17171002 \hspace{6cm}	& 2022 \\ 
	\end{tabular} 
	\end{table*}
\end{titlepage}
\doublespacing

\begin{dedication} 

I would like to dedicate this thesis to my loving parents and amazing friends, whose support and motivation kept me going through the tough times.

\end{dedication}

\begin{figure}
 \hspace{-2.5 cm}  
\includegraphics[width=1.25\textwidth]{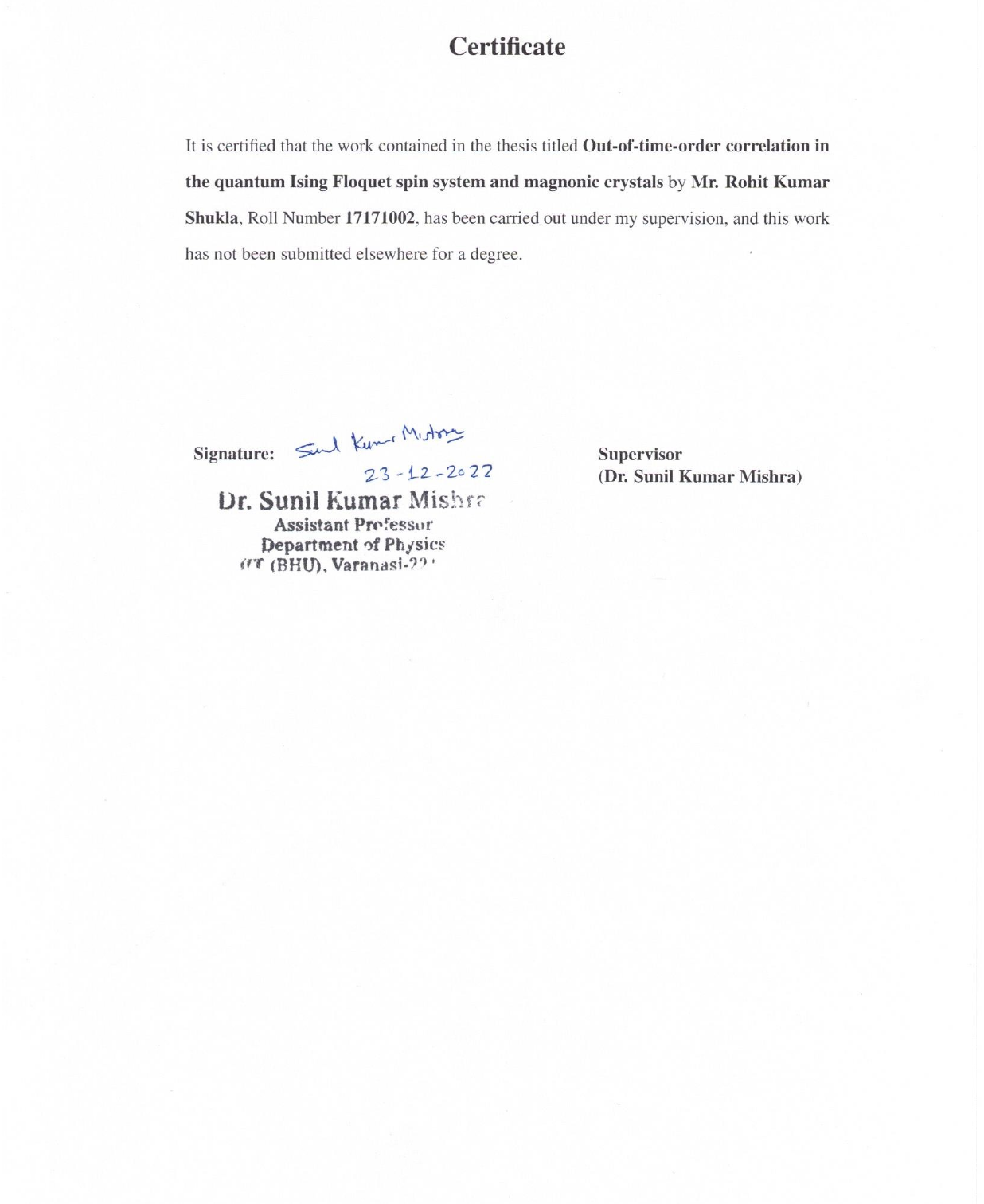}
   \end{figure}
  \begin{figure}
 \hspace{-2.5 cm}  
\includegraphics[width=1.25\textwidth]{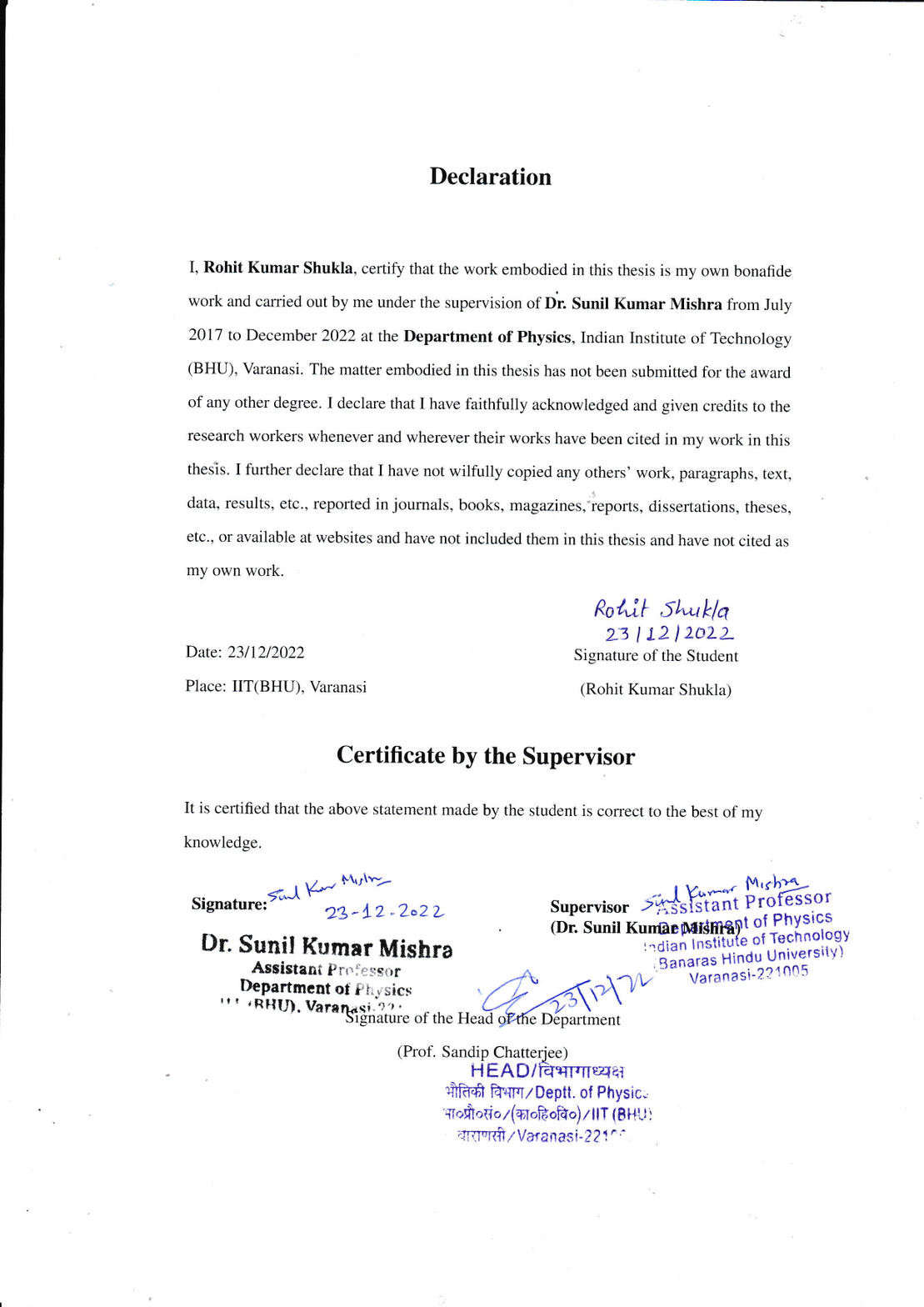}
   \end{figure}
  \begin{figure}
\hspace{-2.5 cm}  
\includegraphics[width=1.25\textwidth]{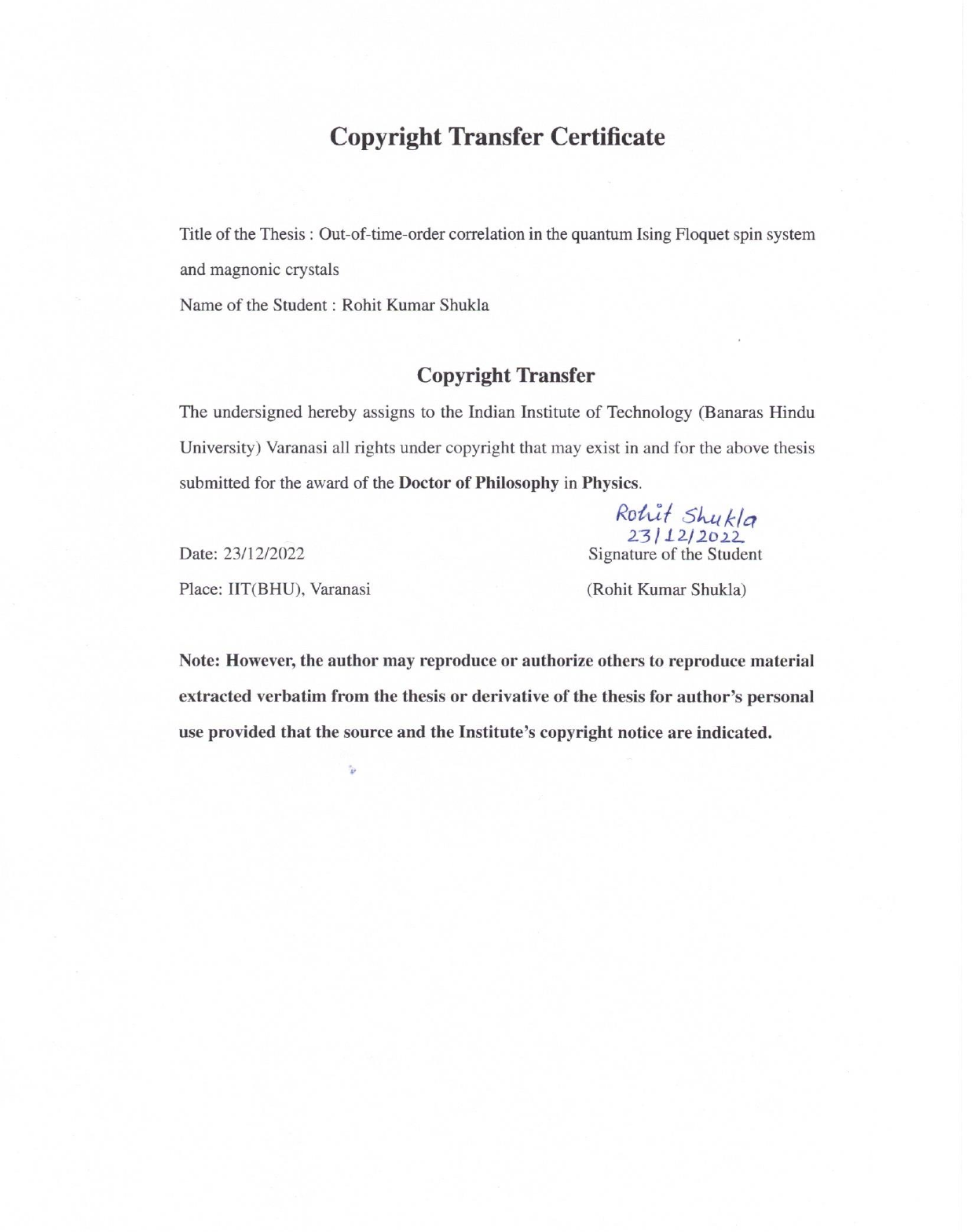}
   \end{figure}
\include{Certificates/certificate}
\include{Certificates/declaration}
\include{Certificates/copyright}

\begin{acknowledgements}
First of all, I would like to express my special thanks and
gratitude towards my supervisor \textbf{Dr. Sunil Kumar Mishra} for introducing me to the field of quantum information and theoretical condensed matter physics, and for inspiring the incredible enhancement of my interest and understanding in this particular field of study. Without his outstanding supervision, unflinching  support, unremitting discussions, and inestimable suggestions this journey would not have been possible. I thank him for giving me the opportunity to discuss each and every query related to the research work. His care and motivation for me were a strong driving force for the successful completion of my Ph. D.
\par
 I would further extend a special thanks towards my collaborators Dr. Arul Lakshminarayan, and Dr. Levan Chotorlishvili, who helped me throughout my research work. I have learned many things
from them, for which I remain greatly indebted. Special thanks to my RPEC member Dr. Rajeev Singh, and Dr. Chandan Upadhyay
for giving me valuable advice throughout this time, especially during the semester evaluation of the research progress.
\par
I am very grateful to The Department of Physics for providing me with this wonderful opportunity, support, and facilities required for completing the project. I am thankful to all the technical, non-teaching, and office staff of the Department of
Physics, IIT (BHU) for their continuous assistance and frequent co-operation at all the various stages of my Ph.D.
\par
There are number of people I would like to acknowledge, without whose support, I could have never completed my Ph.D. work. These are Prashant Dixit, Digvijay, Prashant Pandey, Balveer, Vivek, Vaibhav, Abhishek, Alam, Suraj, Raj, Vipin, Gaurav, and Upendra, Deepak.
\par
I feel short of words when expressing my thankfulness, gratitude, and indebtedness to my parents (Mr. Shankar Datta Shukla and Mrs. Baijanti Shukla), brother $\&$ sister-in-law (Mr. Rahul Shukla $\& $ Mrs. Ansu Shukla), and sister (Miss. Stuti Shukla) for their unbiased love, blessings, inspiration, and support in countless ways.
\par
Finally, I am highly obliged to the Almighty (Lord Shiva and Lord Sankat Mochan) for giving me patience and strength to make
this endeavor a success.
\\
\\
\begin{flushright}\textbf{Rohit Kumar Shukla}\end{flushright}

\end{acknowledgements}
\begin{abstract}
In recent times out-of-time-order correlators (OTOC) have been established as a tool to understand butterfly effects, quantum information scrambling, and many-body localization. They can also be useful in determining different phases of quantum critical systems. OTOCs can identify the quantum chaos within a system undergoing time evolution; and therefore, they can distinguish between chaotic and regular dynamics. This motivates us to study OTOCs in integrable and nonintegrable periodically kicked quantum spin models. A periodically kicked quantum Ising spin system, known as the quantum Ising Floquet system, is a variant of the transverse Ising model. In place of constant transverse magnetic fields in the  transverse Ising system, time-periodic fields are applied in the form of delta pulses in the quantum Ising Floquet spin system. It provides very interesting and peculiar dynamics separate from that of the transverse Ising system. 
\par
First, we explore the phase diagram of the Floquet transverse Ising model using the long-time average of OTOC as an order parameter. In the process, we present the exact analytical solution of the transverse magnetization OTOC using the Jordan-Wigner transformation.  We also calculate the speed of correlation propagation and analyze the behavior of the revival time with the separation between the observables. To get the phase structure of the Floquet transverse Ising system, we use the longitudinal magnetization OTOC. We show the phase structure numerically in the transverse Ising Floquet system by using the long-time average of the longitudinal magnetization OTOC. In both the open and the closed chain systems, we find distinct phases, out of which two are paramagnetic (0-paramagnetic and $\pi$-paramagnetic), and two are ferromagnetic (0-ferromagnetic and $\pi$-ferromagnetic) as previously defined in the literature.
\par
Next, we focus on different regimes of OTOC vs. time in the constant field transverse Floquet Ising system with and without longitudinal field. Three distinct regimes viz. characteristic, dynamic, and saturation of OTOC vs. time, are analyzed carefully. In calculating OTOC, we take local spins in longitudinal and transverse directions as observables that are respectively local and non-local in terms of Jordan-Wigner fermions. We use the exact analytical solution of OTOC for the integrable model (without longitudinal field term) with transverse direction spins as observables and provide numerical solutions for other cases.  OTOCs generated in both cases depart from unity at a kick equal to the separation between the observables when the local spins in the transverse direction and one additional kick is required when the local spins in the longitudinal direction. The number of kicks required to depart from unity depends on the separation between the observables and is independent of the Floquet period and system size. In the dynamic region, OTOCs show power-law growth in both models, the integrable (without longitudinal field) as well as the nonintegrable (with longitudinal field).  The exponent of the power-law increases with increasing separation between the observables. Near the saturation region, OTOCs grow linearly with a very small rate.
\par
Further, we calculate OTOCs using contiguous symmetric blocks of spins or random operators localized on these blocks as observables instead of localized spin observables. We find only the power-law growth of OTOC in integrable and nonintegrable regimes. In the non-integrable regime, beyond the scrambling time, there is an exponential saturation of the OTOC to values consistent with random matrix theory. This motivates the use of ``pre-scrambled" random block operators as observables. A pure exponential saturation of OTOC in both integrable and nonintegrable systems is observed without a scrambling phase. Averaging over random observables from the Gaussian unitary ensemble, the OTOC is found to be the same as the operator entanglement entropy, whose exponential saturation has been observed in previous studies of such spin chains.
\par
Finally, we utilize OTOCs as a quantifier for quantum information currents and propose a quantum information diode (QID) by exploiting the effect of nonreciprocal magnons in  a 2D Heisenberg spin system with Dzyloshinski Moriya interaction. QID is a device rectifying the amount of quantum information transmitted in opposite directions. We control the asymmetric left and right quantum information currents through an applied external electric field and quantify it through the left and right OTOC. To enhance the efficiency of the quantum information diode, we utilize a magnonic crystal. We excite magnons of different frequencies and let them propagate in opposite directions. Nonreciprocal magnons propagating in opposite directions have different dispersion relations. Magnons propagating in one direction match resonant conditions and scatter on gate magnons. Therefore, magnon flux in one direction is damped in the magnonic crystal. This fact leads to an asymmetric transport of quantum information in the quantum information diode. A quantum information diode can be fabricated from an yttrium iron garnet (YIG) film. This is an experimentally feasible concept and implies certain conditions: low temperature and small deviation from the equilibrium to exclude effects of phonons and magnon interactions. We show that rectification of the flaw of quantum information can be controlled efficiently by an external electric field and magnetoelectric effects.
\par
Overall, this thesis is focused on studying OTOC in the quantum Ising spin Floquet systems to describe the phase structure and dynamics of the systems. Additionally, it describes an application of OTOC as a quantifier of quantum information current in proposed QID based on magnonic crystals.  
\end{abstract}


\tableofcontents

\listoffigures



\printnomencl

\mainmatter

\chapter{Introduction}  

\ifpdf
    \graphicspath{{Chapter1/Figs/Raster/}{Chapter1/Figs/PDF/}{Chapter1/Figs/}}
\else
    \graphicspath{{Chapter1/Figs/Vector/}{Chapter1/Figs/}}
\fi

Correlations are commonly discussed in everyday life. In a scientific setting, particularly in the fields of statistics, classical, and quantum physics, a correlation analysis helps us in determining the degree of relationship between two or more than two variables \cite{adesso2016measures,sharma2005text}. The explanation of a significant degree of correlation of any two variables may be due to any of the following two reasons: {\it (i)} Both the variables may be mutually influencing each other. For example, the relationship between price and demand, where demand increases price and vice versa. {\it (ii)} Both variables may be influenced by other variables; for example, the production of tea is correlated with land and is affected by the amount of rainfall. The significance of the correlation is that we can estimate the change in one of the variables, given the correlation of the two related variables. 
\par
Based on the space and time dependence, we can classify the correlation as spatial and temporal. If the observable at different locations are correlated irrespective of time dependence, they are spatially correlated. For example, Sheep are highly correlated with each other in their group; however, there is no correlation with another group far away from their group. If the correlation of the observable is taken with itself at a different position (spatial variation), then it is known as spatial autocorrelation. It can be positive or negative. However, in the case of temporal correlation, the correlation of two observables takes place at different times without changing the position,  for example, GDP and life expectancy, which means that improvement in GDP improves life expectancy over time. If the correlation of the observable is taken with itself at different times, then it is known as temporal autocorrelation. In the following section, we will discuss the classical correlation of a bipartite system in the context of classical correlation theory. 
\section{Classical correlation}
Numerous measurements of correlations based on statistical analysis were proposed in Refs. \cite{Bennett1996,Bennett1995}. However, measuring the classical correlation in a bipartite system remains unclear. A measure of classical correlations between two different random variables $X$ and $Y$ is proposed in the field of classical information theory where  information in an entity is defined by the amount of data that is required to describe it completely \cite{henderson2001classical}.  It is calculated by using mutual information, which is defined as \cite{shannon1949mathematical} 
\begin{equation}
\label{BMI}
H(X:Y)=H(X)+H(Y)-H(X,Y).
\end{equation}
First and second term of Eq.~(\ref{BMI}) are known as Shannon entropy  and defined as \cite{shannon1949mathematical}
\begin{equation}
H(X) \equiv H(p)=-\sum_{i} p_i \log p_i,~~~~H(Y) \equiv H(q)=-\sum_{i} q_i \log q_i.
\end{equation}
Shannon entropy is used to find the information in a source, $X/Y$, that provides messages $x_i/y_i$ with probabilities $p_i/q_i$.  
Last term of Eq.~(\ref{BMI}) is known as joint entropy which is defined as
\begin{equation}
H(X, Y )=-\sum_{i,j} p_{ij} \log p_{ij},
\end{equation}
where, $p_{ij}$ is the probability of both outcomes $x_i$ and $y_j$. It is to be noted that correlation does not change with the change of the observables $X$ and $Y$ because it is, by definition, a property of the combined bipartite system rather than the property of either subsystem.
\par
Correlations can also be discussed in terms of a bit which is the fundamental unit of classical computation. The state of the classical bit is either $0$ or $1$. A classical bit is similar to a coin: either tails or heads up. In the next section, we will discuss quantum correlation in terms of quantum bit{\it, i.e.,} qubit, which is a fundamental concept for quantum computation.

\section{Quantum correlation}
Like a classical bit, two possible states of a quantum bit are $\vert 0 \rangle$ and $\vert 1 \rangle$. In quantum mechanics, $`\vert  ~\rangle'$ represents a state in the form of Dirac notation. Other than $\vert 0 \rangle$ or $\vert 1 \rangle$ state, a qubit can be in a superposition state. In general, it is written as
\begin{equation}
    \vert \psi \rangle=\alpha \vert  0 \rangle+\beta \vert 1 \rangle,
\end{equation}
where, $\alpha$ and $\beta$ are complex numbers. When we measure a qubit outcome will be either $0$, with probability $\vert \alpha \vert ^2$, or  $1$, with probability $\vert \beta \vert ^2$ and $\alpha$ and $\beta$ follow the condition $\vert \alpha \vert ^2+\vert \beta \vert ^2=1$. 
\par
We will discuss quantum correlation in  a composite quantum system made up of two or more distinct physical systems. For the sake of simplicity, we consider a two-qubit system.  Corresponding to this system four computational basis states denoted as $\vert 0\rangle \otimes \vert 0\rangle$, $\vert 0\rangle \otimes 1\rangle$, $\vert 1\rangle \otimes \vert 0\rangle$, and $\vert 1\rangle \otimes \vert 1\rangle$. A pair of qubits can also exist in superpositions of these four states that is given as
\begin{equation}
\vert \psi\rangle =\alpha_{00}\vert 0\rangle \otimes \vert 0\rangle+\alpha_{01}\vert 0\rangle \otimes \vert 1\rangle+\alpha_{10}\vert 1\rangle \otimes \vert 0\rangle+\alpha_{11}\vert 1\rangle \otimes \vert 1\rangle.
\end{equation}
Similar to the case for a single qubit, when we do a measurement on the state of two qubits $\vert xy\rangle=\vert x\rangle \otimes \vert y \rangle$, where $\vert x\rangle \in H_1$,  $\vert y\rangle \in H_2$, and $\vert xy\rangle \in H_1\otimes H_2$, where $H_1$ and $H_2$ are Hilbert spaces. The measurement result is $xy(= 00$, $01$, $10$ or $11$)  with probability $\vert \alpha_{xy}\vert^2$ and the probabilities add up to one, {\it i.e.,}  $\sum_{x,y \in (0,1)}\vert \alpha_{xy}\vert^2=1$. Four perfectly correlated states of two qubits are defined by Bell, named Bell states, and given as 
 \begin{equation}
   \label{bell}
    \begin{split}
& \vert \phi^+\rangle =\frac{1}{\sqrt{2}}(\vert 0\rangle_A \otimes \vert 0\rangle_B +\vert 1\rangle_A \otimes \vert 1\rangle_B),  \\
   & \vert \phi^-\rangle =\frac{1}{\sqrt{2}}(\vert 0\rangle_A \otimes \vert 0\rangle_B -\vert 1\rangle_A \otimes \vert 1\rangle_B),  \\
   & \vert \psi^+\rangle =\frac{1}{\sqrt{2}}(\vert 0\rangle_A \otimes \vert 1\rangle_B +\vert 0\rangle_A \otimes \vert 1\rangle_B),  \\
   & \vert \psi^-\rangle =\frac{1}{\sqrt{2}}(\vert 0\rangle_A \otimes \vert 1\rangle_B -\vert 0\rangle_A \otimes \vert 1\rangle_B),  
\end{split}
\end{equation}
where, ``A'' and ``B'' are acronyms of Alice and Bob. The meaning of expression  $\vert \phi^+\rangle$ in Eq.~(\ref{bell}) is that qubit held by Alice or Bob can be $0$ as well as $1$. Alice and Bob prepare a few copies of the $\vert \phi^+\rangle$ state and take a qubit each. Let us assume that Alice chooses the z-basis and measures her qubit. The measurement outcome ($0$ or $1$) would be random with  probability $1/2$. Subsequently, when Bob measures his qubit on the same z-basis, Bob's outcome would be the same result that Alice has already measured for all the copies of the $\vert \phi^+\rangle$ state prepared.   
If Alice and Bob communicate their results, it would be found that although the outcomes are seemingly random at each end, whence combined, they are perfectly correlated. 
\par
Let us discuss a generic experiment setup in which two parties, Alice and Bob, are a distance apart \cite{nielsen2002quantum}. Charlie prepares two particles for the measurement and sends one to Alice and the second to Bob. Alice (or Bob) performs measurements on one system, but there is no effect on the result of Bob's (or Alice's) measurement. Let us consider two different realities. Corresponding to these realities, Alice and Bob have two outcomes. Outcome on Alice's side: $A^{'}=\pm1$ or $A^{''}=\pm1$, and Bob's side: $B^{'}=\pm1$ or $B^{''}=\pm1$. Measurements are performed simultaneously by Alice and Bob.  As Alice receives her particle, she performs a measurement on it. Suppose she has two different apparatuses for measurement to know the reality on her side. So she has two options to perform the measurements. These measurements are label as $P_{A^{'}}$ and $P_{A^{''}}$, respectively. In advance, Alice does not make sure which measurement should perform first. She either flips the coin or uses some random technique to do a measurement. For simplicity, consider each measurement to have one of two outcomes, either $+1$ or $-1$. Let Alice’s particle has a value $A^{'}/A^{''}$ for the property $P_{A^{'}}/P_{A^{''}}$.
Now, suppose Bob also has two operations for measurements, and these are labeled as $P_{B^{'}}$  or $P_{B^{''}}$. Consider each measurement has one of two outcomes, either $+1$ or $-1$. As Bob receives his particle, he randomly selects an operator and starts to measure. Since the experiments are performed by Alice and Bob simultaneously, therefore, the results of the measurements of Alice and Bob cannot disturb one another.
\par
Let us discuss simple algebra of the quantity $S=A^{'}B^{'} + A^{''}B^{'} + A^{''}B^{''}-A^{'}B^{''}$ which includes all the correlation between possible outcomes on Alice and Bob side. It can be written as
\begin{equation}
\label{PQR}
A^{'}B^{'} + A^{''}B^{'} + A^{''}B^{''}-A^{'}B^{''} = (A^{'} + A^{''})B^{'} + (A^{''}- A^{'})B^{''}.
\end{equation}
Since, $A^{''}$ and $A^{'}=\pm 1$, we can see that either $(A^{'} + A^{''})B^{'}=0$ or $(A^{''}- A^{'})B^{''}=0$. In both cases, it is easy to see from Eq.~(\ref{PQR}) that $A^{'}B^{'} + A^{''}B^{'} + A^{''}B^{''}-A^{'}B^{''} = \pm 2.$
\par
Let $p(a^{'}, a^{''}, b^{'}, b^{''})$ is the probability that the system is in a state $A^{'} = a^{'},~A^{''} = a^{''},~B^{'} = b^{'},$ and $B^{''} = b^{''}$ before the measurements are performed and $E(S)$ is the mean value of the quantity $S$. Then we have
\begin{equation}
\begin{split}
E(A^{'}B^{'} + A^{''}B^{'} + A^{''}B^{''} -A^{'}B^{''} ) &=
p(a^{'}, a^{''}, b^{'}, b^{''})(a^{'}b^{'} + a^{''}b^{'} + a^{''}b^{''} - a^{'}b^{''}) \nonumber \\ 
& \leq p(a^{'}, a^{''}, b^{'}, b^{''}) \times 2 ~~{\rm or} ~~\geq p(a^{'}, a^{''}, b^{'}, b^{''}) \times (-2)\nonumber \\
\end{split}
\end{equation}
Also 
\begin{equation}
\begin{split}
E(A^{'}B^{'} + A^{''}B^{'} + A^{''}B^{''} -A^{'}B^{''} ) &=
p(a^{'}, a^{''}, b^{'}, b^{''})a^{'}b^{'} +p(a^{'}, a^{''}, b^{'}, b^{''})a^{''}b^{'} \nonumber \\ 
&+p(a^{'}, a^{''}, b^{'}, b^{''})a^{''}b^{''} -p(a^{'}, a^{''}, b^{'}, b^{''})a^{'}b^{''} \\ \nonumber 
&=E(A^{'}B^{'}) + E(A^{''}B^{'}) + E(A^{''}B^{''}) -E(A^{'}B^{''} ).
\end{split}
\end{equation}
Comparing the above equation, we get Bell inequalities.
\begin{equation}
\label{leq2}
  -2\leq  E(A^{'}B^{'}) + E(A^{''}B^{'}) +   E(A^{''}B^{''}) -E(A^{'}B^{''} )\leq 2
\end{equation}
\par
 Alice and Bob can determine $E(A^{'}B^{'})$, $E(A^{''}B^{'})$, $E(A^{''}B^{''})$, and $E(A^{'}B^{''})$ and repeat the experiment several times. After completing a series of tests, Alice and Bob meet to discuss and analyse their data. They examine each experiments where Alice measured $P_{A^{'}}$ and Bob measured $P_{B^{'}}$. They obtain a sample of values for $A^{'}B^{'}$ by collectively multiplying their tests' outcomes. They can estimate $E(A^{'}B^{'})$ by averaging over the sample. Likewise, they can make estimate $E(A^{'}B^{'})$, $E(A^{''}B^{'})$, $E(A^{''}B^{''})$, and $E(A^{'}B^{''})$. 
\par
All the classical experiments defined in the above manner follow Bell's inequality. Now we perform the expectation value calculations, using quantum mechanical state and the observables manifesting two different properties of the same system. For this purpose, let Charlie set up a two-qubit quantum system in the state,
\begin{equation}\label{psi_minus}
    \vert \psi^- \rangle =\frac{\alpha_{01}\vert 0\rangle \otimes \vert 1 \rangle -\alpha_{10}\vert 1\rangle \otimes \vert0 \rangle}{\sqrt{\alpha_{01}^2+\alpha_{10}^2}}.
\end{equation}
He passes the first qubit to Alice to perform measurements of the observables  
\begin{equation}
A^{'}=Z_1,~~~~~A^{''}=X_1
\end{equation}
and passes the second qubit to Bob to perform measurements of the observables:
\begin{equation}
B^{'}=\frac{-Z_2-X_2}{\sqrt{2}},~~~~~B^{''}=\frac{Z_2-X_2}{\sqrt{2}}
 \end{equation}

The expectation value of the observable $A^{'}B^{'}$ is
\begin{equation}
     \langle A^{'}B^{'}\rangle=\Bigg\langle \frac{\langle 01\vert +\langle 10\vert}{\sqrt{2}}\Bigg\vert Z_1\frac{-Z_2-X_2}{\sqrt{2}}\Bigg\vert \frac{\vert01\rangle +\vert 10\rangle}{\sqrt{2}}\Bigg\rangle =\frac{1}{\sqrt{2}}
\end{equation}
Similarly, expectation value of the observables $ A^{''}B^{'}$, $ A^{''}B^{''}$, and $ A^{'}B^{''}$ can be found to be:
\begin{equation}
   \langle A^{''}B^{'}\rangle=\frac{1}{\sqrt{2}};~~\langle A^{''}B^{''}\rangle=\frac{1}{\sqrt{2}};~~\langle A^{'}B^{''}\rangle=\frac{1}{\sqrt{2}}.
\end{equation}
Thus, quantity $S$ shall therefore be
\begin{equation}
S= \langle A^{'}B^{'}\rangle+\langle A^{''}B^{'}\rangle+\langle A^{''}B^{''}\rangle-\langle A^{'}B^{''}\rangle=2\sqrt{2}.
\end{equation}
In Eq.~(\ref{leq2}), we find that the $\langle A^{'}B^{'}\rangle+\langle A^{''}B^{'}\rangle+\langle A^{''}B^{''}\rangle-\langle A^{'}B^{''}\rangle$ can never exceed two. However, for a quantum mechanical state like Eq.~(\ref{psi_minus}), the sum of the expectation value is equal to $2\sqrt{2}$. Quantum mechanical states like Eq.~(\ref{psi_minus}) violate Bell's inequality. Therefore, the quantum mechanical states like Eq.~(\ref{psi_minus}) have a nonlocal correlation.
In the next section, we will discuss temporal correlation, defined as the correlation of the observables at different times. 
\section{Temporal correlation}
Temporal correlations are used in information-sharing processes. Depending upon the time-ordering, correlation is categorized into two parts: \\(1) Time-ordered correlation \\(2) Out-of-time-order correlation.  \\
We will discuss these two cases independently in the following subsections.

\subsection{Time-ordered correlation}
\begin{figure}[H]
    \centering
    \includegraphics[height=.5\linewidth,width=.3\linewidth]{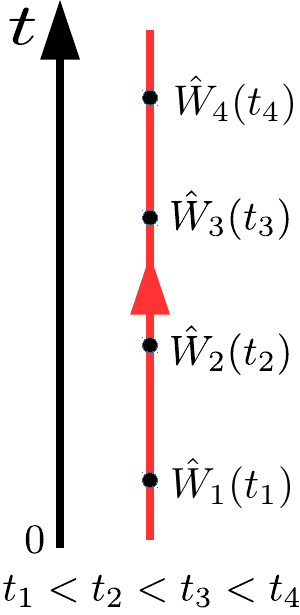}
    \caption{Contour of time folded that is showing the temporal ordered correlation of the observables $\hat W_1, \hat W_2, \hat W_3,$ and $\hat W_4$ with times  $t_1<t_2<t_3<t_3<t_4$.}
    \label{time_order}
\end{figure}

Time-ordered correlation functions play essential role in many quantum dynamical problems. A general time-order correlation function for two observables, $\hat W_1(t_1)$ and $\hat W_2(t_2)$ is defined as
$\langle \hat W_2(t_2) \hat W_1(t_1)\rangle$ at times $t_1<t_2$, where $\langle \cdots \rangle$ is the quantum mechanical averages and for four observables is given as $\langle \hat W_4(t_4) \hat W_3(t_3)\hat W_2(t_2) \hat W_1(t_1)\rangle,$ at times  $t_1<t_2<t_3<t_4$. The time-ordered correlation of four general observables at different times is shown in Fig.~\ref{time_order}.
However, a special type of behavior in some quantum systems has recently drawn considerable attention among physicists. Small changes in the initial conditions lead to drastic changes in the time-evolved state. Such problems need to overlap two states of the system prepared by a successive backward and forward evolution of the observable in time. In such cases, the correlation violates time ordering and is known as an out-of-time-order correlator (OTOC). This quantity has been found useful for determining the scrambling of information in quantum systems \cite{gu2016,bilitewski2018temperature,das2018light}, and have been used as measures of thermalization and many-body localization \cite{Fan2017, huang2017out}, chaos \cite{ray2018signature, hosur2016chaos, gu2016,ling2017},  and entanglement \cite{hosur2016chaos,yao2016interferometric}. At the same time, several experimental protocols \cite{yao2016interferometric,zhu2016measurement,swingle2016measuring} have been proposed to measure the OTOCs. A brief discussion of OTOC is given below.

\subsection{Out-of-time-order correlator}
\begin{figure}[H]
\centering
\includegraphics[height=.5\linewidth,width=.4\linewidth]{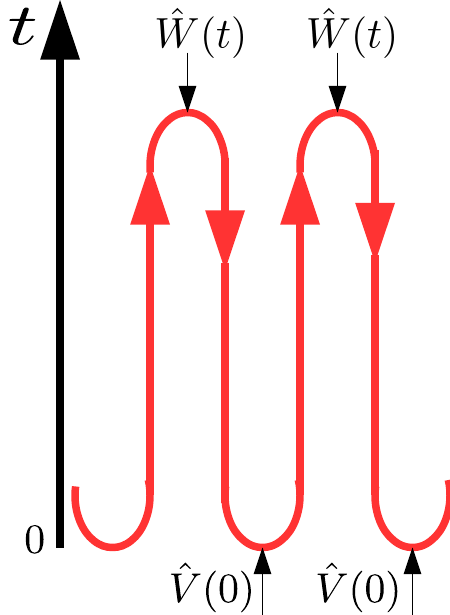}
\caption{ Illustration of observable $\hat W$ at time $t$ and observable $\hat V$ at initial time $t=0$ which are not in time ordered manner. Arrows indicate the direction of the correlation defined for the operators along the time axis.}
\label{OTOC_order}
\end{figure}

OTOC is a special type of four-point correlation that is not in time ordered. For the calculation of OTOC, we generate a correlation of two observables $\hat V(0)$ and $\hat W(t)$, where $\hat V$ at time $t=0$ and another at $t$ time and it is defined as $\langle \hat W(t)\hat V(0) \hat W(t)\hat V(0)\rangle$ in the mathematical form. 
A schematic representation of OTOC is given in Fig.~\ref{OTOC_order}.

Larkin and Ovchinnikov first introduced OTOC in 1969 \cite{larkin1969quasiclassical}. After that,  OTOC has been explored in many fields of quantum, and spin systems \cite{singh2022scrambling,kitaev2014hidden, shenker2015stringy,haake1991quantum,hosur2016chaos,rozenbaum2017lyapunov,shukla2021,garcia2018chaos,rozenbaum2020early}. In the recent years, OTOC extensively used to indicate the chaos in the quantum and semiclassical systems \cite{ray2018signature, hosur2016chaos, gu2016,ling2017}. 
\par
For two observables $\hat{W}$ and $\hat{V}$, OTOC is given as \cite{larkin1969quasiclassical}
\begin{equation}
\label{OTOC_larkin}
C\left(t\right)= \frac{1}{2}\left\langle\left[\hat{W}(t),\hat{V}(0)\right]^{\dag}\left[\hat{W}(t),\hat{V}(0)\right]\right\rangle,
\end{equation} 
where parentheses $\langle \cdots\rangle$ denotes a quantum mechanical average. Both observables $\hat W$ and $\hat V$ commute with each other at $t=0$  OTOC, defined by Eq.~(\ref{OTOC_larkin}), is  zero. At the time $t$, Heisenberg time evolution of $\hat W(0)$ is defined as $\hat W(t)=e^{\im \hat H t} \hat W e^{-\im \hat H t}$ and expansion of it is given by Baker-Campbell-Hausdorff formula that has the sum of products of many local observables given as in the mathematical form
\begin{eqnarray}
\hat W(t)=\hat W + it[\hat H, \hat W]&+& \frac{(\im t)^2}{2!}[\hat H, [\hat H, \hat W]]+ \frac{(\im t)^3}{3!}[\hat H , [\hat H, [\hat H, \hat W]]] \nonumber \\ &+& \cdots\frac{(\im t)^k}{k!}[\hat H ,  [\hat H, \cdots[\hat H, \hat W]]\cdots ]_k+\cdots 
\end{eqnarray}
$\hat V$ does not commute with the higher-order term of $\hat W(t)$, which implies the nonzero value of OTOC. This noncommutative behavior of OTOC may indicate the chaotic nature of the system.
\par
After doing some simplification in Eq.~(\ref{OTOC_larkin}), we get
\begin{equation}
\label{OTOC_general}
C(t)=\frac{1}{2}\bigg\lbrace\langle \hat V\hat W(t)^2\hat V\rangle+\langle \hat W(t)\hat V^2\hat W(t)\rangle - \langle \hat W(t)\hat V\hat W(t)\hat V\rangle-\langle\hat V\hat W(t)\hat V\hat W(t)\rangle\bigg\rbrace. 
\end{equation}
If observables $\hat W$ and $\hat V$ are Hermitian, then quantities $\langle \hat V\hat W(t)^2 \hat V\rangle$ and $\langle \hat W(t)\hat V\hat W(t)\hat V\rangle$ are complex conjugate of quanties $\langle \hat W(t)\hat V^2\hat W(t)\rangle$ and $\langle\hat V\hat W(t)\hat V\hat W(t)\rangle$, respectively. If we consider only the real part of all the expectation values, then Eq.~(\ref{OTOC_general}) simplifies in the form given as
\begin{equation}\label{otoc_Hermitian}
C(t)=\Re[\hat W^2(t)\hat V^2(0)]-\Re[\langle \hat W(t)\hat V \hat W(t) \hat V \rangle].
\end{equation}
If observables $\hat W$ and $\hat V$ are Hermitian and unitary, then the first quantity of Eq.~(\ref{otoc_Hermitian}){\it, i.e.,}  $\langle \hat W^2(t)\hat V^2(0)\rangle$ will be identity. Hence, Eq.~(\ref{otoc_Hermitian}) become
 \begin{equation}
C(t)=1-\Re[\langle \hat W(t)\hat V \hat W(t) \hat V \rangle]=1-\Re[F_z^{l,m}(t)],
\end{equation}
where, $\Re \rightarrow$ real number and $F_z^{l,m}(t)=\langle \hat W(t)\hat V \hat W(t) \hat V \rangle$.
\begin{figure}
    \centering
    \includegraphics[width=.9\textwidth,height=0.5\textwidth]{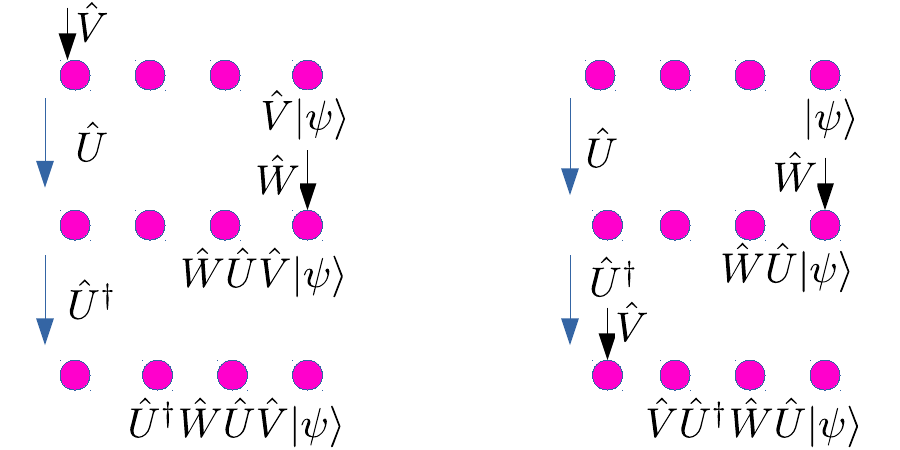}
    \caption{(Left) We consider a state $\vert\psi\rangle$ and apply operator $\hat U^{\dagger} \hat W \hat U \hat V$ and generate a state  $\vert \phi(t) \rangle=\hat U^{\dagger} \hat W \hat U \hat V\vert \psi\rangle$. (Right) We apply observable $\hat V \hat U^{\dagger} \hat W \hat U$ on state $\vert \psi\rangle$ and get state $\vert \xi(t)\rangle=\hat V \hat U^{\dagger} \hat W \hat U\vert \psi\rangle $. Inner product of $\vert \xi(t)\rangle $ and $\vert \phi(t)\rangle $ is equal to $F(t)$.}
    \label{otoc_state}
\end{figure}
\par
OTOC can be defined in terms of the inner product of differently time evolved two wave functions $\vert \phi\rangle$ and $\vert \xi\rangle$.  Let us consider an initial state wave function $|\psi\rangle$. For the generation of $\vert \phi\rangle$, the order of applied observables on state $|\psi\rangle$ is in the following manner: first, state $\vert \phi\rangle$ is perturbed  by an observable $\hat{V}$ at initial time ${t=0}$, after that, it gets evolved by unitary operator $\hat U$ till time $t$. At the time $t$, it gets perturbed by the observable $\hat{W}$ and gets evolved by $\hat{U^{\dagger}}$ in the backward direction on the time scale from $t$ to $0$. Hence, wavefuction after doing time evolution is $|\phi(t)\rangle=\hat{U^{\dagger}}\hat{W}\hat{U}\hat{V}|\psi\rangle=\hat{W}(t)\hat{V}|\psi\rangle$. Generation of the wave function $\vert \xi\rangle$ has the following steps: first, evolve forward with unitary evolution $\hat U$ till time $t$, after that perturbed with $\hat{W}$ at time $t$, evolved backward from $t$ to initial time $t=0$ and again perturbed with $\hat{V}$ at ${t=0}$. Hence, the  wave-function is $|\xi(t)\rangle=\hat{V}\hat{U^{\dagger}}\hat{W}\hat{U}|\psi\rangle=\hat{V}\hat{W}(t)|\psi\rangle$. overlapping of $\vert \xi(t)\rangle$ and $\phi(t)\rangle$ is equivalent to $1-C(t)$.  A graphical representation of the above statement is given in Fig.~\ref{otoc_state}. 

In general, OTOC is a correlation of two observables in which one observable is evolving with time by Heisenberg time evolution, and another is independent of time. Hence, OTOC shows different behavior for different observables. In this thesis, we explore different sets of observables for the study of phase structure, quantum chaos, and rectification of magnon. In the following subsection, we will discuss OTOCs taking single spin observables and block spin observables.

\subsection{OTOC using position-dependent single spin observable} 
\begin{figure}
\centering
\includegraphics[width=\linewidth, height=.30\linewidth]{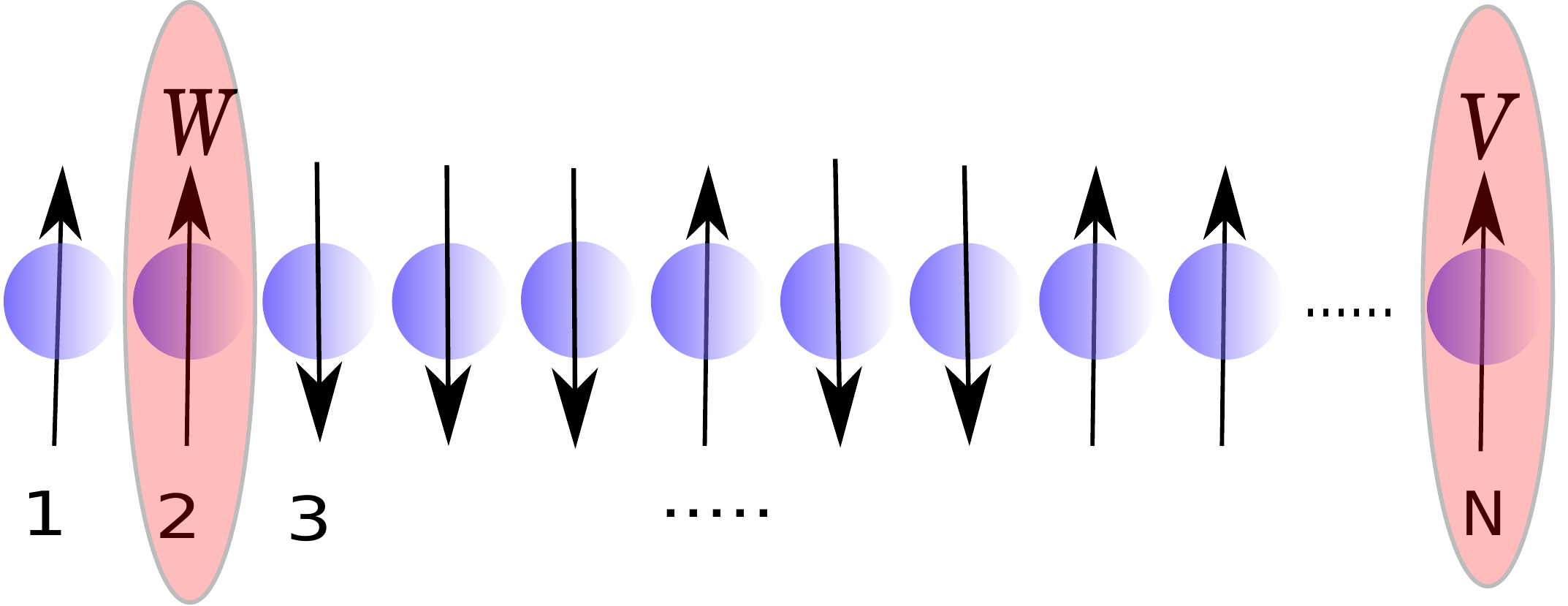}
\caption{Schematics of single spin observables. One spin is considered as observable $\hat W$ and another spin is considered as observable $\hat V$.}
\label{2d_block_operator} 
\end{figure}

Let us consider a spin chain in which spins interact in $x$ direction. Now, we consider a position-dependent pair of single spin Pauli observables at position $l$ and $m$ as $\hat W^l=\hat \sigma^l$ and $\hat V^m=\hat \sigma^m$.  OTOC [Eq.~(\ref{OTOC_larkin})] of Hermitian and unitary Pauli operator will be
\begin{equation}
C^{l,m}(t)=1-\langle \hat \sigma^l(t) \hat \sigma^m(0) \hat \sigma^l(t) \hat \sigma^m(0) \rangle=1-\Re[F^{l,m}(t)],
\end{equation}
where, $F^{l,m}(t)=\langle \hat \sigma^l(t) \hat \sigma^m(0) \hat \sigma^l(t) \hat \sigma^m(0) \rangle$. Separation between the observables, $\hat W^l$ and $\hat{V}^m$ is $\Delta l=l-m$. A graphical representation of the position-dependent observables is  given in Fig.~\ref{2d_block_operator} in which observable $\hat W$ is at position $2$ and $\hat V$ at position $N$. We can change the position of the observables. Depending upon the direction of the observables, we categorize into two parts:
\begin{enumerate}

\item Transverse magnetization OTOC (TMOTOC)
\item Longitudinal magnetization OTOC (LMOTOC)
\end{enumerate}

\begin{enumerate}
\item {\bf Transverse magnetization OTOC (TMOTOC):}\\ 
If in OTOC, two Hermitian spin observables $\hat W$ and $\hat V$ at sites $l$ and $m$ be in the direction of the z-axis. {\it, i.e.,}  $\hat W=\hat \sigma^l_z$ and $\hat V=\hat \sigma^m_z$ then we call it as TMOTOC  and defined  as: 
\begin{equation}
C_z^{l,m}(t)=1-\Re[F_z^{l,m}(t)],
\end{equation}
where, $F_z^{l,m}(t)=\langle \hat \sigma^l_z(t)\hat \sigma^m_z\hat \sigma^l_z(t)\hat \sigma^m_z\rangle$. In place of the quantum mechanical average, we consider a particular  state as $|\phi_0\rangle=| \uparrow  \uparrow  \uparrow \cdots  \uparrow \rangle$. $ \left| \uparrow \right\rangle$ denotes eigenstate of $\hat \sigma_z$ with eigenvalue $+1$. Using a special state type makes numerical and analytical calculations easier.  

\item {\bf  Longitudinal magnetization OTOC (LMOTOC) :}\\
If in OTOC, two Hermitian spin observables $\hat W$ and $\hat V$ at sites $l$ and $m$ be in the parallel direction of the Ising axis (x-axis){\it, i.e.,}  $\hat W=\hat \sigma^l_x$ and $\hat V=\hat \sigma^m_x$ then we call it as LMOTOC which is given as: 

\begin{equation}
    C_x^{l,m}(t)=1- \Re[F_x^{l,m}(t)],
\end{equation}
where,
$F_x^{l,m}(t)=\langle
 \psi_0|\hat \sigma^l_x(t)\hat \sigma^m_x\hat \sigma^l_x(t)\hat \sigma^m_x|\psi_0\rangle$. In this case, initial state is defined as  $|\psi_{0} \rangle=|\rightarrow \rightarrow \rightarrow \cdots \rightarrow \rangle$. $ \left| \rightarrow \right\rangle$ denotes the eigenstate of $\hat \sigma^x$ with eigenvalue $+1$.

In a closed chain Ising system, OTOC does not depend on $l$ and $m$ but depends on the distance between the spins {\it{i.e.}} $\Delta l=l-m$. However, in the open chain case, OTOC  depends on the $l$ and $m$ as well as the distance between the spins ({\it{i.e.}} $\Delta l=l-m$) because, in the open chain case, the last spins are connected with the environment.

\end{enumerate}
\subsection{OTOC using block observables}
Let us consider a spin chain of the length $N$ and divide it into two blocks such as 
\begin{equation}
\hat W=\frac{2}{N}\sum_{i=1}^{\frac{N}{2}}\hat \sigma_x^l ~~{\rm and}~~ \hat V=\frac{2}{N}\sum_{l=\frac{N}{2}+1}^{N}\hat \sigma_x^l. 
\end{equation}
\begin{figure}[H]
      \centering
       \includegraphics[width=\linewidth, height=.25\linewidth]{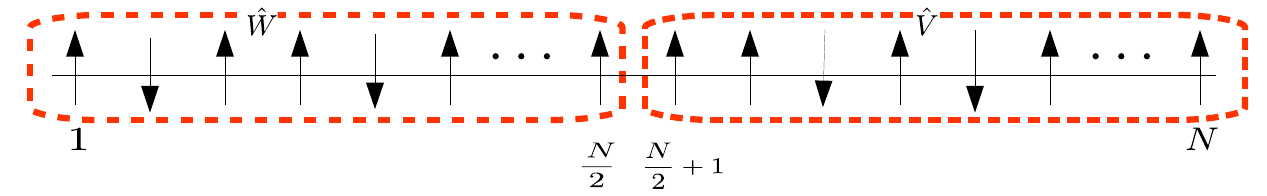}
       \caption{Illustration of SBOs $\hat W$ and $\hat V$ represented by Eq. (\ref{block}). The length of the chain is $N$ that should be even to be divided into two halved subsystems $\hat W$ and $\hat V$.}
       \label{Ch1_block_operator} 
\end{figure}
Observables $\hat W$ and $\hat V$ are defined as the first and second blocks of spins, respectively, known as spin block operators (SBOs). Graphical representations of the SBOs are given in Fig.~\ref{Ch1_block_operator}. Since, observables $\hat W$ and $\hat V$  are Hermitian but not unitary, then OTOC Eq.~(\ref{OTOC_larkin}), will be 
\begin{eqnarray}
C(t)=\langle \hat W^2(t) \hat V(0)^2\rangle-\langle \hat W(t) \hat V(0)\hat W(t) \hat V(0)\rangle=C_2(t)-C_4(t)
\label{block}
\end{eqnarray}
where, $C_2(t)$ is named as two-point correlation and $C_4(t)$ is named as  four-point
correlations, and defined as: $C_2(t)=\langle \hat W^2(t) \hat{V(0)}^2\rangle$ and $C_4(t)=\langle \hat W(t) \hat V(0)\hat W(t) \hat{V(0)}\rangle$.

In recent years, OTOCs have been used in many areas; one important use of it is to distinguish the chaotic and regular regimes in the semiclassical and quantum. The following section briefly discusses the chaos and chaotic systems.

\section{Chaos}
The word chaos comes from the Greek word ``Khaos," meaning this is a ``gaping void." Mathematicians find that it is easy to  ``recognize chaos when you see it," but no easy way exists to define it. In general, chaos is defined as a phenomenon where a small change in the input implies a large change in the output \cite{devaney1989chaos}. They find space in many disciplines such as physics,  economics, philosophy, biology, and engineering \cite{morse1967man}. The nature of the output of the chaos is highly complex, and it is not predictable \cite{robert1976simple}. Determining the output of the chaos is also a very complex process. 
\par
\cite{hilborn2000chaos}. 
Chaos is found in driven simple pendulums and double pendulums \cite{reichl1992transition}. In the study of chaos in the above system, It is found that chaos is present and have different type of behavior in different type of systems. Following, we will discuss the chaos in different systems, such as classical, quantum, and spin systems.

\section{Chaotic Systems}
A system is said to be chaotic whenever its evolution trajectory depends very strongly on the initial conditions. This property implies that even for two infinitesimally close initial conditions, the observed trajectories display large deviations that vary exponentially with time. Several natural phenomena can be recognized as chaotic and chaotic analysis can also be found in the solar system \cite{malhotra2001chaos}, meteorology, brain, and heart of living organisms \cite{walleczek2006self}, etc. The dynamic behavior of a chaotic system is very hard to predict because it involves various complicated mathematical equations. Solutions of mathematical equations of chaotic systems are complex and cannot be   easily extrapolated. Chaos in the classical and quantum systems is named classical chaotic system and quantum chaotic system, respectively. Recently, chaos has been studied in the spin system.  Following, we will discuss the chaos in the systems.

\subsection{Classical chaotic System}
Classical regular and chaotic systems are properly understood, and lots of studies have been done on them \cite{lichtenberg2013regular,ott2002chaos,strogatz2018nonlinear,thompson1990nonlinear}. There is a common technique that is used for analyzing the dynamics of the system is known as Hamilton's equations of motion.  If a system is described with $n$ degrees of freedom, then the classical dynamics of a system are described by using $2n$-dimensional phase space trajectory. Phase space is a multidimensional space in which axes are defined by position and conjugate momenta. Systems can be distinguished as integrable and nonintegrable systems by analyzing the constants of motion  \cite{reichl1992transition}. An integrable system has n independent constants of motion; however, the nonintegrable system has less than n constants of motion. Let us consider a system, a harmonic oscillator with one degree of freedom. Hamiltonian of it is given by $ H=\frac{ p^2}{2m}+\frac{1}{2} m \omega^2 x^2$. Dimension of phase space is two in which one  axis is position $(x)$, and the other is corresponding conjugate momenta $(p)$. In the harmonic oscillator,  energy remains unchanged during the entire dynamics. Thus the number of constant quantities matches the degrees of freedom making the harmonic oscillator an integrable system. Nonintegrable systems often exhibit one of the most surprising properties{\it, i.e.,} unpredictability in evolution and under-applied perturbation. This unpredictability is due to exponential variation with changing the initial conditions. This property of the dynamical system is known as chaos. 
\par
Chaos is studied in many fields of classical physics{\it, e.g.,} dynamics of fluids, stars, and some biophysical models \cite{lichtenberg2013regular,ott2002chaos,strogatz2018nonlinear,thompson1990nonlinear}). Many tools, for example, level spacing distribution, amongst several others are used to distinguish the regular and chaotic classical systems. 
 Classical chaos can be diagnosed by exponential sensitivity with the initial condition. The exponent of the exponential is defined  as the Lyapunov exponent (LE), and it is denoted by $\lambda_L$. The exponential growth of chaotic systems is known as the ``butterfly effect" \cite{gu2016,bilitewski2018temperature,das2018light}. The butterfly effect serves as a diagnostic measure of chaos which is defined as small perturbations in the initial state leading to exponential growth.
\par
Classical chaos in a system is dependent on initial conditions \cite{ott2002chaos}.
The exponential behavior of OTOC is also found in the infinitesimally small region surrounding critical points of the phase structure. The exponential growth of OTOC at a critical point is studied in the Lipkin-Meshkov-Glick (LMG) model \cite{pappalardi2018scrambling}. This is a classical system having single-degree-of-freedom in nuclear physics \cite{lipkin1965validity}, and it is realized with experiment by cold atoms \cite{gross2010nonlinear,zibold2010classical},  and nuclear magnetic resonance \cite{araujo2013classical}. 
\par
Some systems show non-chaotic behavior in classical mechanics; however, they show chaotic behavior in quantum mechanics \cite{bunimovich2019physical,rozenbaum2020early}. This is due to the instability provided by quantum mechanics in a region where classical dynamics are stable. To understand the reason for instability in quantum mechanics, we will briefly discuss quantum mechanics concepts and quantum chaos. 

\subsection{Quantum chaotic system}
 Classical physics could not be used for the explanation of a few phenomena. Explanation of these phenomena led to the advent of ideas now known as quantum physics. Quantum theory  depicts an evolving wave function in accordance with the linear Schr$\ddot{{\rm{o}}}$dinger equation, in contrast to the phase space evolution in classical physics. Here, variables of classical physics are replaced by Hermitian observables. Heisenberg’s uncertainty principle is applicable in quantum physics but it is not applied in classical physics. This principle is stated as the conjugate variables of any particle can not be determined simultaneously accurately. It fails to describe the quantum system by the phase space. In addition to this, the linear Schr$\ddot{{\rm{o}}}$dinger equation does not provide any exponential variation of the wave function by using evolution.
In quantum mechanics, the unitary property of the operator applies specific constraints under which the distance of two initial wave functions does not change under evolution. In the mathematical form, the above statement can be written as $\langle \psi_0 \vert \hat U^{\dagger}(t)\hat U(t) \vert \psi_1\rangle= \langle \psi_0 \vert\psi_1\rangle$  which is true for all $t$.
\par

The system has some specific behavior in the quantum domain, such as wave-particle duality and the uncertainty principle. Such inherent properties change the appearance of the sharp features obtained in classical dynamics, such as the sensitive accordance with initial conditions, which are implied within the butterfly effect. In the chaotic system, this effect becomes crucial because butterfly effects are destroyed. In contrast, isolated systems experience the butterfly effect after a short period of semiclassical evolution. The short period depends on the system size in  a logarithmic manner given as $t_E \equiv \log(N)$, where $t_E$ is defined as Ehrenfest time \cite{berman1978condition,toda1987quantal}. Butterfly effect is recognized by the exponential growth of OTOC after the Ehrenfest time. Exponent of exponential growth is named as Lyapunov exponent.
\par
In recent years, OTOC is also used for the discussion of the chaos and dynamics of the Ising spin systems. Researchers focused on the chaotic nature as well as the dynamics of the spin systems by OTOC. We will discuss the chaos in the spin systems.  Before the discussion of chaos and dynamics of the spin systems, it is necessary to present a brief overview of spins, spin-spin interaction, and spin chains.   

\section{Spin-1/2}
\begin{figure}
\centering
\includegraphics[
width=1\linewidth,height=.5\linewidth]{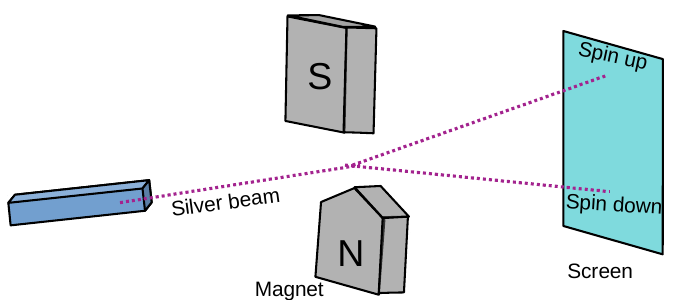}
    \caption{ Setup of Stern-Gerlach experiment.}
\label{stern}
\end{figure}

 Spin is a purely quantum mechanical concept, and it is an intrinsic quantity \cite{sakurai1995modern}. A theoretical proposal of spin is given by Goudsmit and Uhlenbech to describe the vector atom model. After that, experimental verification was given by Stern and Gerlach in their experiment known as the ``Stern-Gerlach experiment". The experimental arrangement  is illustrated in Fig~\ref{stern}. In the experiment, an oven produced a beam of the silver atom, which was passed through a nonuniform magnetic field. Two spots are observed on the screen, which is symmetric from the point of no deflection in the absence of the fields. The observation can be  explained by the spin of the electron which gives rise to the magnetic moment of an atom ${\bf S}${\it, i.e.,} ${\bf \mu} \propto {\bf S}$. The energy corresponding to the magnetic moment and the magnetic field is given by 
\begin{equation}
    E=-{\Vec{ \mu} . \vec{B}}.
\end{equation}
A nonuniform magnetic field $({\bf B }\equiv\Vec{B})$ applies force to the silver atoms. Other components of ${\bf B}$ are ignored except the z-component because the atom is very heavy. Therefore, force on the atom due to the z-component of a magnetic field is given as 
\begin{equation}
    F_z=\frac{\partial}{\partial z}({\Vec{\mu}.\Vec{B}}) \simeq \mu_z \frac{\partial B_z}{\partial z}.
\end{equation}
The direction of the force on the atom depends upon the value of the z-component of magnetic moment ($\mu_z$). If $\mu_z>0$, then a downward force is applied on the atom; however, if $\mu_z<0$, then an upward force is applied in the atom. Hence,  the beam of the silver atom  is split according to the value of $\mu_z$, and it was found that the silver atoms struck the plate only in two regions, symmetrically situated about the point of no deflection. The variation of the silver beam has only two components which dictate the magnetic moment vector of silver atoms must have only two orientations. The proportional condition of magnetic moment with spin implies that the z-component of spin also has two orientations. This confirms the theoretical proposal of spins. Now, let us discuss the interaction of two spins in the following subsection.

\subsection{Spin-spin interaction}
Let us consider two electrons with suppressed orbital degrees of freedom and having only  spin degrees of freedom. The total spin operator of this system of two electrons is written as

\begin{equation}
    {\bf S=S_1\otimes  \mathbb{1}+ \mathbb{1} \otimes S_2},
\end{equation}
where $ \mathbb{1}$ is the  identity operator of dimension $2\times 2$, and it is placed at the spin space of electron $2~(1)$ in the first (second) term. The individual spins belong to the Hilbert space of two dimensions, and the complete system of two elections is described by Hilbert space of $2\otimes 2$ {i.e.,} $4$ dimension. Commutation relations of spin operators at the same site and  different sites are given as follows:
 \begin{equation}
    [S_{1x},S_{1y}]=\im \hslash S_{1z},~ [S_{2x},S_{2y}]=\im \hslash S_{1z},~~~[S_{1x},S_{2y}]=0.
 \end{equation}
 Operators  ${\bf S^2=(S_1+S_2)^2}$, $S_z=S_{1z}+S_{2z}$,  $S_{1z}$, and $S_{2z}$ has eigenvalues                 $s(s+1)\hslash^2$, $m\hslash$, $m_1\hslash$, and $m_2\hslash$, respectively. Ket vectors corresponding to the spin state of two electrons in terms of the Eigen kets of ${\bf S^2}$ and $S_z$ can be written as $ \vert s=1,m=\pm 1,0\rangle,~\vert s=0,m=0\rangle,$ where $s=1$ represent spin-triplet ($2s+1=3$) and $s=0$ represent spin singlet  ($2s+1=1$) state. Corresponding to spin-triplet ($s=1$), there are three basis vectors $\vert s=1,m=1\rangle, \vert s=1,m=0\rangle, \vert s=1,m=-1\rangle$, and corresponding to spin-singlet ($s=0$), there is one basis vector as $\vert s=0,m=0\rangle$. The interaction of two spins is given as 
\begin{equation}
   \hat H=J_{12}{\bf S_1.S_2},
\end{equation}
where, $J_{12}$ is the interaction stength. Value of $\hat H$ depends on the ${\bf S_1.S_2}$. Let us calculate the ${\bf S_1.S_2}$ for singlet and triplet states. Since, ${\bf S^2=S_1^2+S_2^2+2S_1.S_2}$, therefore ${\bf  S_1.S_2= (S^2-S_1^2-S_2^2)}/2$.
Value of ${\bf  S_1.S_2}$ for the singlet state ($s=0$) will be,
\begin{equation}
    \begin{split}
        {\bf  S_1.S_2}&= \frac{J_{12}}{2}[s(s+1) -s_1(s_1+1)-s_2(s_2+1)], \nonumber \\
        &=\frac{J_{12}\hslash^2}{2}\Big[ 0(0+1)-\frac{1}{2}\Big(\frac{1}{2}+1\Big)-\frac{1}{2}\Big(\frac{1}{2}+1\Big)\Big], \nonumber \\
        & =-\frac{3J_{12}\hslash^2}{4}.
     \end{split}
\end{equation}
Similarly, for triplet state ($s=1$), ${\bf  S_1.S_2}=\frac{J_{12}\hslash^2}{4}$. If $J_{12}>0$, then the singlet state has lower energy than the triplet state, and the system is in a ferromagnetic state, however, if $J_{12}<0$, then the triplet state has minimum energy, and the system is represented as an antiferromagnetic state. In the next section, we will discuss a system of $N (N>2)$ interacting spins. Hamiltonian corresponding to such a system is given by Heisenberg, and  the model is named the Heisenberg model.

\section{Heisenberg model}
Consider a lattice with $N$ site ($N\rightarrow \infty$)  
and at each lattice site $i$ a spin $\vec{\bf S_i}$ is placed. The spin-spin interaction between the spins is given as $\vec{\bf S_i} \cdot \vec{\bf S_j}$. If we consider all the pairs of spin-spin interactions then 
\begin{equation}
\label{heisen}
H=\frac{1}{2}\sum_{ i\neq j} J_{ij} \vec{ \bf S_i}.\vec{\bf S_j},
\end{equation}
where, indices $i$ and $j$ are run from site $1$ to $N$ on a lattice. Coefficient $J_{ij}$ is exchange constants or interaction strength, and it is symmetric{\it, i.e.,} $J_{ij}=J_{ji}$. Interaction strength decreases as increases distance between indices $i$ and $j$.  Value of $J_{ij}$ can be either positive or negative.  Antiparallel alignment of  spin favors a positive value of $J_{ij}$. It is the case of antiferromagnetic. Parallel alignment of the spins favors a negative value of $J_{ij}$, which is the case of ferromagnetic. Factor $\frac{1}{2}$ in the Hamiltonian avoids double-counting the bonds. Spin components at the same site follow commutation relations as 
 \begin{equation}
    [S_j^{\alpha}, S_j^{\beta}]=i \epsilon_{\alpha \beta \gamma}S_j^{\gamma}~~~~~~~(\alpha, \beta, \gamma=x,y,z),
 \end{equation}
 where $\epsilon_{\alpha \beta \gamma}$ is the Levi-Civita symbol, its value will be $+1 (-1)$ if $\alpha$, $\beta$ $\gamma$ in cyclic (non-cyclic) order and $0$ when at least any two variables ($\alpha$, $\beta$ $\gamma$)  are same. However, spins at different sites commute with each other{\it, i.e.,} $[S_i^{\alpha}, S_j^{\beta}]=0$.
 
\section{Ising Model}
\begin{figure}
 \centering
\includegraphics[width=.6\linewidth,height=.65\linewidth]{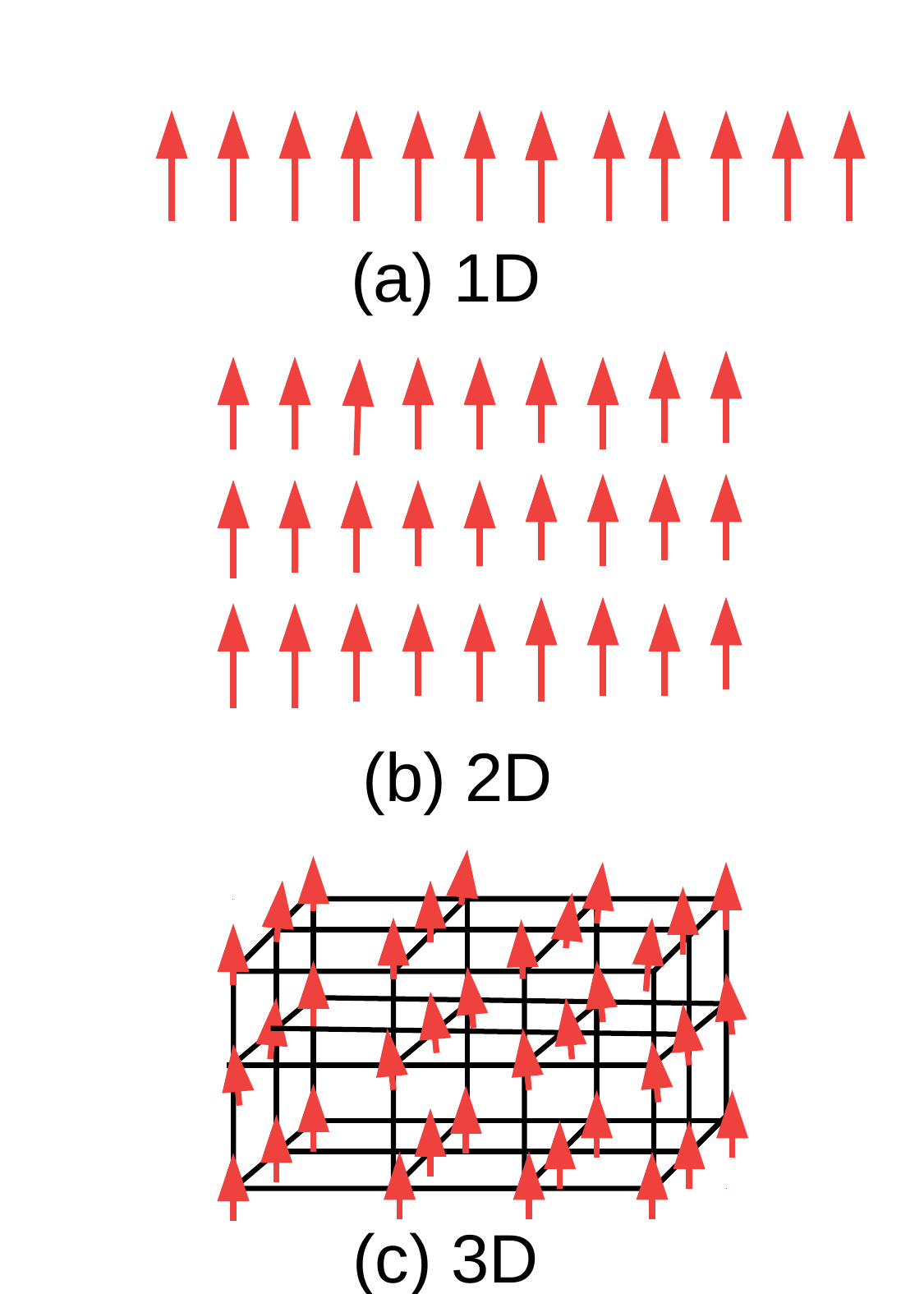}
\caption{Arrangement of spins at lattices in one, two, and three dimensions.}
     \label{ising_model}
 \end{figure}
Wilhelm Lenz first introduced the Ising model in the year 1922. He made the assumption that particles in a crystal structure can freely revolve around a given lattice point \cite{ brush1967history}.
The Hamiltonian creates the complete model, which is a combination of two pieces, one representing the energy contribution from particle-particle interaction and the other representing the energy contribution from constraints on the system. In the Ising  model, the constraint is applied from the magnetic field. The effect of the field on the quantum Ising system is to rotate the spins. To study the rotation of spins, apply anisotropic magnets to the quantum Ising chain in both transverse and longitudinal directions. 
The involvement of the field term in the Ising spin chain provides two terms in the Hamiltonian, one corresponding to the longitudinal field and another corresponding to the transverse field. Hence, the total Hamiltonian will be
\begin{equation}
\label{Hxz}
\hat H=J_{x} \hat H_{xx}+h_{x} \hat H_x +h_{z} \hat H_z,
\end{equation}
 where, $\hat H_{xx}=\sum_{l=1}^{N}\hat \sigma_l^x \sigma_{l+1}^x$, $\hat H_z=\sum_{l=1}^{N}\hat \sigma_l^z$ and $\hat H_x=\sum_{l=1}^{N}\hat \sigma_l^x$ ($\hat \sigma_l^x=\frac{\hslash}{2}\hat S_l^x$).  Replace $J_{ij}$ by $J_x$ as we consider only nearest neighbor interaction in the x-direction. $h_x$ is the strength of the continuous and constant longitudinal magnetic field, and $h_z$ is the strength of the transverse magnetic.
 The dimension of the lattice could be one, two, or three. Corresponding to the lattice's one, two, and three dimensions, the Ising model is known as the one, two, and three-dimensional Ising model. A pictorial representation of it is given in Fig.~\ref{ising_model}. The following subsection discusses the boundary condition of the spin systems.
\subsection{Boundary conditions}
We consider a one-dimensional lattice with $N$ sites and spin-1/2 particles (say electron) situated  on each site.  Spins get interact with their neighbors. The effect of a spin at a site is determined by the interactions of spin with the other spins in the model. It is commonly taken as either nearest neighbor or next nearest neighbor because the effect of interaction decreases as the distance between the spins increases. A spin chain based on the nature of boundaries can be classified into
\begin{enumerate}
    \item Periodic boundary condition 
    \item Open boundary condition 
\end{enumerate}
 
\begin{figure}
    \centering
    \includegraphics[width=1\linewidth,height=.25\linewidth]{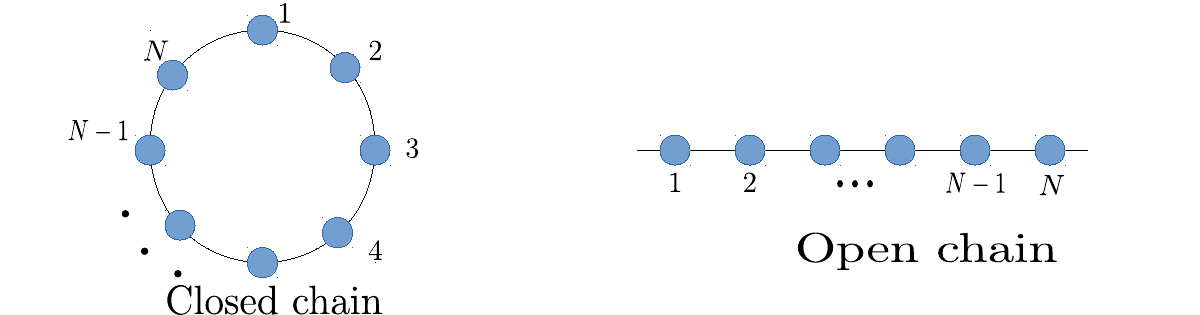}
    \caption{One-dimensional lattice configuration with (Left) periodic boundary condition  and (Right)  open boundary condition.  Solid lines denote interaction.}
    \label{closed_chain}
\end{figure}

\begin{enumerate}
\item  {\bf Periodic boundary condition:}\\
If both ends of the chain are connected, then a one-dimensional chain is called  a closed chain. Periodic boundary condition means that system repeats after $N${\rm th} spin counting {\it, i.e.,}  $\hat \sigma_{N+1} \equiv \hat \sigma_1$ as shown in Fig.~\ref{closed_chain}~(Left). The coupling term in case of periodic boundary conditions will be 
\begin{equation}
   \hat H_{xx}=J_x\sum_{l=1}^{N}\hat \sigma^x_l \hat \sigma^x_{l+1},
\end{equation}
 \item{\bf Open boundary condition:}\\
 In an open chain case, both ends of the chain are not connected with each other, as shown in Fig.~\ref{closed_chain}~(Right). In open  boundary conditions, Hamiltonian is defined as
\begin{equation}
\hat H_{xx}=J_x\sum_{l=1}^{N-1}\hat \sigma_l^x\hat \sigma_{l+1}^x.   
\end{equation}
\end{enumerate}
In the open chain case, one term $\hat \sigma_{N}^x\hat \sigma_{1}^x$ is absent from the Hamiltonian as compared to closed chain cases of the same system size. 
\section{Transverse Ising model}
The transverse Ising model is derived form the Ising model in which a constant magnetic field is applied in the coupling's transverse direction, and a longitudinal field is absent. It has long been an appreciated and well-studied model. It is dynamically interesting and can be easily implemented in quantum mechanics. The transverse Ising model has been studied in several contexts, including entanglement, state transport, and the quantum phase transition at zero temperature that separates ferromagnetic and paramagnetic phases. \cite{chakrabarti2008quantum,heyl2018detecting,su2006,sun2009}. It is integrable due to a mapping from interacting spins to a collection of noninteracting spinless fermions via the Jordan-Wigner transformation. 

\section{Floquet transverse Ising model}
\begin{figure}[H]
\includegraphics[width=.7\linewidth, height=.45\linewidth]{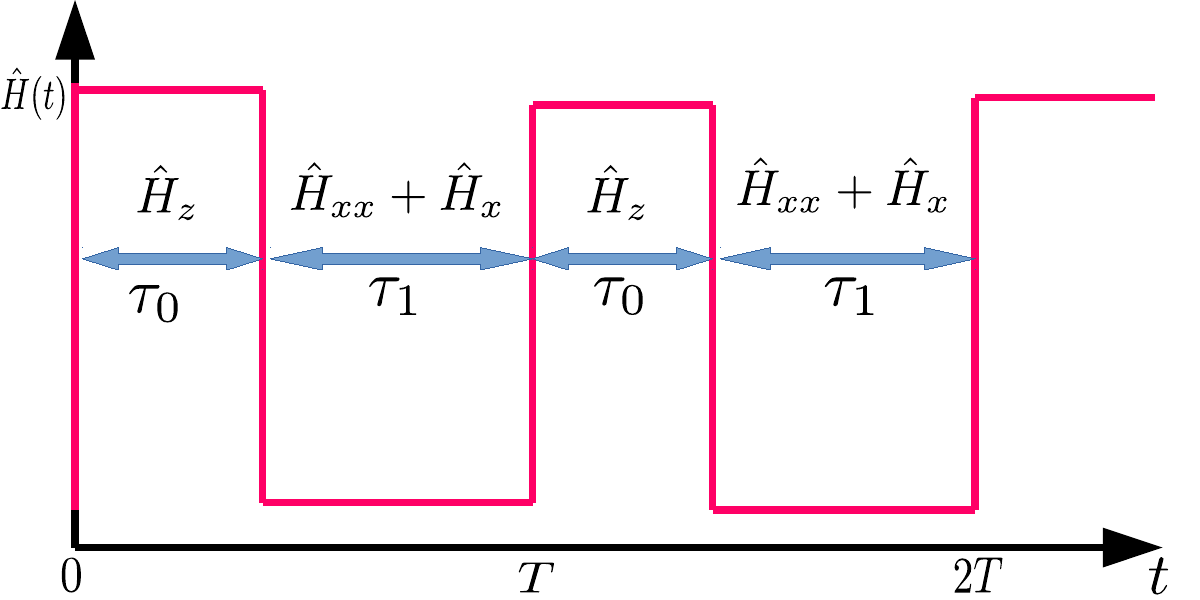}
\includegraphics[width=.20\linewidth, height=.5\linewidth,angle=0 ]{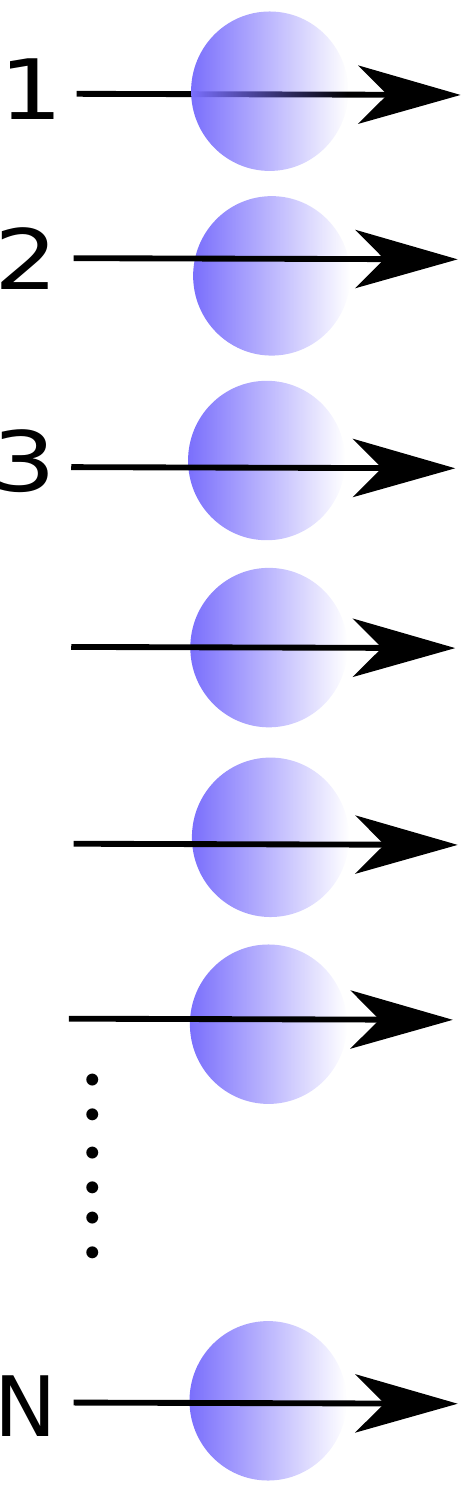}
\caption{Spin chain experience a periodic quench by non-commuting Hamiltonian functions $\hat H_z$ and $\hat H_x$ for durations $\tau_0$ and $\tau_1$, respectively. The complete  system is periodic with period $T=\tau_0+\tau_1$.}
\label{spin_chain}
\end{figure}

The Floquet spin model is a variant  of the transverse Ising model. In this model, a periodic kicked transverse field is applied for the period $\tau_0$. A longitudinal constant field is applied for a period $\tau_1$. Hence total time period of the periodic field is $T=\tau_0+\tau_1$ [Fig.~\ref{spin_chain}].  The involvement of a time-periodic kicked field displays very interesting and peculiar  behavior in the Ising system \cite{gritsev2017,lakshminarayan2005multipartite,naik2019controlled,shukla2021}. A graphical representation of a periodic field in the form of delta pulses applied on the spin chain is given in Fig.~\ref{spin_chain}. 

The Hamiltonian  defined by Eq.~(\ref{Hxz}) will take the form as
\begin{equation}
\hat H_F=J_{x} \hat H_{xx}+h_{x} \hat H_x+h_{z}\sum_{n=-\infty}^{\infty}\delta\Big(n-\frac{t}{\tau_0}\Big) \hat H_z.
\end{equation}
Here,  $\hat H_{xx}=\sum_{l=1}^{N-1}\hat \sigma_l^x\hat \sigma_{i+1}^x $ in the open chain, and  $\hat H_{xx}=\sum_{l=1}^{N}\hat \sigma_l^x\hat \sigma_{i+1}^x $ in a closed chain, $ \hat H_x=\sum_{l=1}^{N}\hat \sigma_l^x$ and $\hat H_z=\sum_{l=1}^{N}\hat \sigma_l^z$. When both longitudinal and transverse fields are present then the system will be nonintegrable; however, it will be integrable in the absence of any one field.

\subsection{Floquet map}
The wave function under time evolution  is defined as follows
\begin{equation}
\label{operator}
  \psi(x,t)=\hat U(t) \psi(x,0), 
\end{equation}
where, $\hat U(t)$ is a time evolution operator that evolve the wave-function from $t=0$ to $t$.   Time-dependent Schr$\ddot{{\rm{o}}}$dinger equation
is given as 
\begin{equation}
\label{SE}
  \im \hslash\frac{\partial  \psi}{\partial t} =\hat H \psi.
\end{equation} 
Inserting Eq.~(\ref{operator}) in Eq.~(\ref{SE}), we get
\begin{equation}
\label{SE1}
  \im \hslash \frac{\partial{\hat U(t)}}{\partial t}= \hat H \hat U(t).
\end{equation}
Initially, at time $t=0$, $\hat U(0)=1$. Since Hamiltonian is Hermitian so $\hat U(t)$ should be unitary. For the proof we take the adjoint of Eq.~(\ref{SE1}),
\begin{equation}
\label{SE2}
  -\im \hslash \frac{\partial{\hat U}^{\dagger}(t)}{\partial t}=\hat U^{\dagger}(t) \hat H ,
\end{equation}
Multiply $\hat U^{\dagger}(t)$ in both sides of Eq.~(\ref{SE1}) from the left and $\hat U(t)$ in both sides of  Eq.~(\ref{SE2}) from the right and take the difference of it.  We get
\begin{equation}
\label{diff}
    \frac{d(\hat U^{\dagger}(t) \hat U(t))}{dt}=0,
\end{equation}
From Eq.~(\ref{diff}), it is obvious $\hat U^{\dagger}(t) \hat U(t)$ is constant. From the initial condition $\hat U(0)=1$ it follows that the constant must be one,
\begin{equation}
  \hat U^{\dagger}(t) \hat U(t)=1.
\end{equation}
With time $t=\tau$ Eq.~(\ref{operator}) will take the form
\begin{equation}
  \psi(x,\tau)=\mathcal{\hat U}_x (\tau)\psi(x,0),
\end{equation}
Generalized form of the above equation is given as
\begin{equation}
  \psi(x,n\tau)=[\mathcal{\hat U}_x (\tau)]^n\psi(x,0).
\end{equation}
Hence, the time-evolution operator of the periodic system is
\begin{equation}
  [\hat U(n\tau)]=[\mathcal{\hat U}_x (\tau)]^n.
\end{equation}
Observation of a periodic system at arbitrary time $t=nT~(n=1,2\cdots)$, can be done by the knowledge of $\mathcal{\hat U}_x (\tau)$. 
In the Floquet system, steps like drive between Hamiltonians $\hat H_z$ of duration $\tau_0$ and $\hat H_x$ of duration $\tau_1$ are used. Corresponding to such type of drive, the propagator connecting states over a single time interval  $T=\tau_0+\tau_1$ is the Floquet operator, and this operator is denoted by $\mathcal{\hat U}_x$ 
\begin{equation}
\label{Uxx}
 \mathcal{\hat U}_x =\exp\left[-i\tau_1 (J_x \hat H_{xx}+h_{x}\hat H_{x})\right] \exp\left(-i \tau_0 h_{z} \hat H_{z}\right).
\end{equation}
In the absence of longitudinal fields ($h_x=0$), $\mathcal{\hat U}_x$  changes in $\mathcal{\hat U}_0$. 

\subsection{Dzyaloshinskii–Moriya interaction (DMI)}
\begin{figure}[H]
     \centering
     \includegraphics{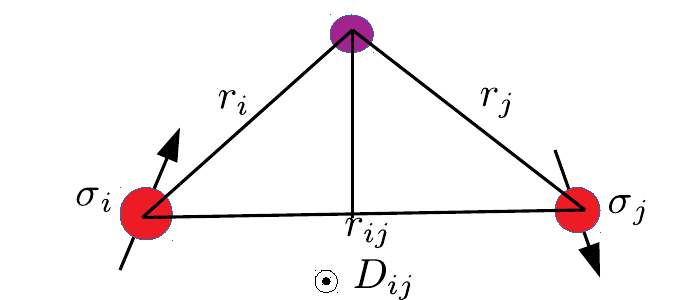}
     \caption{Local geometry to determine the DM vector's orientation.}
     \label{DMI}
\end{figure}

The DMI is a spin-spin interaction in a system that has no inversion symmetry.  The concept of DMI comes into consideration after the two proposals; first, Dzyaloshinskii proposed that the low symmetry and spin-orbit coupling lead to an antisymmetric exchange interaction \cite{dzyaloshinsky1958thermodynamic}, and second by Moriya, who explained how to use a microscopic model to determine the antisymmetric exchange interaction for localized magnetic systems.  \cite{moriya1960anisotropic}. Let us consider two magnetic spins ${\bf \sigma}_i$ and ${\bf \sigma}_{j}$. The total magnetic exchange interaction between these spins is known as DMI, and Hamiltonian corresponding to these terms can be written as
\begin{equation}
    H_{ij}^{DM}={\bf D_{ij}}.{\bf \sigma}_i\times{\bf \sigma}_{j},
\end{equation}
where ${\bf D_{ij}}$ is a DM vector. Graphical representation of the interaction of spins and orientation of the DM vector are shown in Fig.~\ref{DMI}. 
\subsection{2D square-lattice system with DMI interacation}
Let us consider a 2D square spin system and introduce ferroelectric polarization by an external electric field. Hamiltonian corresponding to the square lattice with applied electric field is given as   
\begin{equation}
    \label{Hamiltonian}
  \hat{H}_{s}=J_1\sum\limits_{\langle
  n,m\rangle}\hat\sigma_n\hat\sigma_m+
  J_2\sum\limits_{\langle\langle
  n,m\rangle\rangle}\hat\sigma_n\hat\sigma_m-{\bf
  P}\cdot{\bf E}.
\end{equation} 
Here, we replace $J_{ij}$ by $J_1$ and $J_2$ for the nearest and second nearest-neighbor interaction strength, respectively.  $\langle n,m\rangle$ and $\langle\langle n,m\rangle\rangle$) are the representation of the nearest and second nearest-neighbor  interaction, respectively. ${\bf
  P}\cdot{\bf E}$ describes a coupling of the ferroelectric polarization $\mathbf{P}=g^{\phantom{\dagger}}_{\mathrm{ME}}\mathbf{e}^x_{i,i+1}\times\left(\hat\sigma_i\times\hat\sigma_{i+1}\right)$ with an applied external electric field and mimics an effective DMI term $D=E_y g^{\phantom{\dagger}}_{\mathrm{ME}}$ breaking the left-right symmetry, where $g^{\phantom{\dagger}}_{\mathrm{ME}}$ is the magneto-electric coupling constant. This may be written as follows
\begin{equation}
-{\bf P}\cdot{\bf E}=D\sum\limits_{n}(\hat\sigma_{n}\times\hat\sigma_{n+1})_z.
\end{equation}
Here we consider only the nearest neighbor DMI and only in one direction.
Hence, Hamiltonian will be
\begin{equation}
  \hat{H}_{s}=J_1\sum\limits_{\langle
  n,m\rangle}\hat\sigma_n\hat\sigma_m+
  J_2\sum\limits_{\langle\langle
  n,m\rangle\rangle}\hat\sigma_n\hat\sigma_m-D\sum\limits_{n}(\hat\sigma_{n}\times\hat\sigma_{n+1})_z.
\end{equation}
\par
 In the next section, we will discuss the distinguisher of regular and chaotic spin systems and also discuss the dynamics of OTOC in  the systems.
\section{Chaos in spin system}
Recently, OTOCs are explored rapidly in the spin systems to describe the dynamics and saturation behavior of the systems \cite{lin2018out,xu2020accessing,xu2019locality,kukuljan2017weak,Fortes2019,craps2020lyapunov,roy2021entanglement,yan2019similar,bao2020out,dora2017out,Riddell2019,lee2019typical}. 
Some spin models such as Luttinger liquid model \cite{dora2017out}, XY model \cite{bao2020out}, XXZ model \cite{Riddell2019,lee2019typical}, Sachdev-Ye-Kitaev (SYK) model \cite{Fu2016}, integrable quantum Ising spin model with constant magnetic field \cite{lin2018out} and tilted magnetic field \cite{Fortes2019}, XXZ spin model, Heisenberg spin model with random magnetic fields \cite{Fortes2019}, and some other integrable and nonintegrable spin models  \cite{xu2020accessing,xu2019locality,kukuljan2017weak,craps2020lyapunov,roy2021entanglement,yan2019similar} are reported in the literature. In all the above studies, no exponential growth of OTOC in the dynamic region is found. Therefore, it is difficult to distinguish between the regular and chaotic systems using only OTOC. Usually, it is distinguished by spectral analysis of the systems. 
\par
In spin system, spectral properties of the systems either nearest neighbor spacing distribution (NNSD) of the energy spectrum \cite{berry1977level,bohigas1984characterization}, or the local properties of energy eigenvectors \cite{berry1977regular,deutsch1991quantum,srednicki1994chaos} are used to distinguish integrable, chaotic, near-integrable, and near-chaotic regimes. For the calculation of NNSD, initially, it is necessary to  identify the system's symmetries. After that Hamiltonian generated by removing the symmetries is block diagonalized. Different spin systems have different symmetries. Here we will discuss symmetries in the Floquet system, which is studied in this thesis. The Floquet system has only a “bit-reversal” symmetry in the open boundary conditions and let's define the bit-reversal operator by $\hat B$. The operation of this operator is given as
$\hat B \vert s_1, s_2,\cdots,s_N\rangle=\vert s_N,\cdots, s_2, s_1\rangle, $ and it follow the commutation relation with Floquet map {\it i.e.,} $[\hat U, \hat B] = 0$. $\vert s_i\rangle$ represents a basis state in the basis of $S_z$. We collect complete basis sets into two groups. The first group contains the state that will not change by after the operator as $\hat B$ as $\hat B\vert s_1, s_2,\cdots, s_N\rangle=|s_1, s_2,\cdots, s_N\rangle$. This group is called Palindrome. The second group contains the state which gets reflection by applying operator $\hat B$ as  $\hat B|s_1, s_2,\cdots, s_N\rangle=|s_N,\cdots, s_2, s_1\rangle$. This group is called  
non-palindrome. Since $\hat B^2 = 1$, the eigenvalues of bit reversal operator $\hat B$ are $\pm 1$. Corresponding to the eigenvalue $+1/-1$ of bit reversal observable $\hat B$, eigenstates can be classified as even/odd. All the palindromes come in the group of even states; however, non-palindromes contain half-even and half-odd states. The sum/difference of the non-palindrome state with its reflection provides even/odd states. The dimension of the odd subspace is equal to  $\frac{1}{2}(2^N-2^{N/2})$, while the even subspace is equal to $\frac{1}{2}(2^N+2^{N/2})$.
\par
NNSD is used as a distinguisher between chaotic and  regular systems. If NNSD displays Wigner-Dyson distribution behavior, then the system is said to be a strongly chaotic system. \cite{Luca2016,mehta1991theory,averbukh2001angular}
In mathematical form, the Wigner-Dyson distribution is defined as follows 
\begin{equation}
P_W(s)=\frac{\pi s}{2}e^{-\pi s^2/4},
\end{equation}
If NNSD displays Poisson-type distribution, then the system is said to be a regular system. In the mathematical form, it is given as
\begin{equation}
P_P(s)=e^{-s}.
\end{equation}
If the shape of the distribution lies near the Poisson type and near the Wigner-Dyson type, then it will be near-integrable and near chaotic, respectively.

 OTOC can be utilized to count magnons that flow from magnonic crystals. Before discussing the flow of magnons, we will briefly discuss magnons and magnonic crystals in the following section. 
\section{Magnons and spin wave}
In a magnetic material, the particle-like behavior of spin excitations is known as magnon; however, the wave-like behavior of spin excitation is called as  spin wave. The movement of a magnon or spin wave in the magnonic crystal is referred to as the spin dynamics phenomenon. It has attracted considerable attention among researchers in recent years. For spin excitations, deposition and nanopatterning techniques are now used in ferromagnetic materials. Other techniques also used are: localization \cite{jorzick2002spin}, spin wave quantization \cite{mathieu1998lateral},  and interference \cite{podbielski2006spin}. Spin waves have both quantum and classical properties of waves. They can tunnel through magnetic barriers and reflect when incident on magnetic potential wells. \cite{neumann2009frequency}. In the magnonic crystal, propagation of spin waves displays different behavior than in uniform media \cite{serga2010yig}. Propagation of spin waves does not display the band gap in uniform media but in the magnonic crystal. Spin waves can not propagate through the band gap.

\section{Magnonic crystal}
Magnonic crystals are synthetic magnetic materials whose magnetic characteristics exhibit regular spatial variation{\it, i.e.,} periodic variation in space. In such a periodic arrangement, the spin wave spectrum is affected by Bragg scattering, which causes band gaps. Magnonic crystals should have low-damping magnetic materials for the study of spin wave dynamics. Among all low-damping magnetic materials, mono crystalline YIG $(\rm{Y}_3\rm{Fe}_5 \rm{O}_{12})$  is the most useful material \cite{geller1957structure}. Spin waves can propagate to the centimeter distances in the YIG due to low damping. 
YIG-based MCs are characterized into two types on the basis of the characteristics of the transmission.
\begin{enumerate}
\item Simplest design of MC is one-dimensional grooved structures in which grooves are drawn on the MC to make spin-wave waveguides with periodically changing thickness \cite{chumak2008scattering}.
\item This type of magnonic crystal is controlled by current, and it has specific properties such as gradual tuning and modifying crystal characteristics quickly  \cite{chumak2009current}.
\end{enumerate}
In the transmission band, there is only one rejection band in the case of current-controlled;  however, in the case of grooved MC, which has many rejection bands. This rejection band means the region of frequency where propagation of spin waves is prohibited \cite{chumak2009current,chumak2008scattering}. The size of the rejection band can be adjusted and  \cite{tkachenko2010spectrum}. It is also possible to control the number of rejection bands \cite{chumak2009design}, which  allows the creation of microwave filters with a single or multiple bands. Micrometer and sub-micrometer size YIG-based grooved MCs with desired band gap characteristics are used for the study of spin wave dynamics. 

\subsection{Structure of magnonic crystal}
A grooved MCs can be fabricated from an YIG film. YIG poses a cubic crystal structure of dimension $12.376 \textup{~\AA}$.  Each unit cell contains eighty atoms 
Grooves were deposited on the YIG film using a lithography procedure in a few nanometer steps.  A prepared grooved thin film of YIG works as a spin wave waveguide. YIG film is deposited on a substrate. For the propagation of a spin wave, the substrate should have a similar lattice constant of YIG. The lattice constant of Gallium gadolinium garnet (GGG) is $(12.383 \textup{~\AA})$ and  is exactly matched with the lattice constant of YIG. It is used in the production of flawless films.  However, YIG can be lightly doped with gallium or lanthanum to produce the optimum matching.  
\par
A magnonic transistor has been proposed using a YIG magnonic crystal with periodic modulation thickness \cite{Chumak2}. Similar to the electronic transistor, a magnonic transistor has a source, drain, and gate antennas. A gate antenna injects magnons of a frequency $\omega_G$ into the crystal that matches the magnonic crystal band gap. The gate magnons may acquire a high density in the crystal. Magnons emitted from a source with wave vector $\textbf{k}_s$ flow in the direction of the drain. Interaction between the source magnons and the magnonic crystal magnons is a Four-magnon scattering process. Due to the scattering, the source magnon current attenuates in the magnonic crystal therefore weak signal arrives at the drain. The relaxation process is swift if the following condition holds
\cite{Chumak2,Gurevich} 
\begin{equation}
\label{Ch1_certain conditions} 
  k_s=\frac{m_0\pi}{a_0}, 
\end{equation} 
where $m_0$ is the integer, and $a_0$ is the crystal lattice constant.
\begin{figure}
    \centering
    \includegraphics[width=\linewidth, height=.45\linewidth]{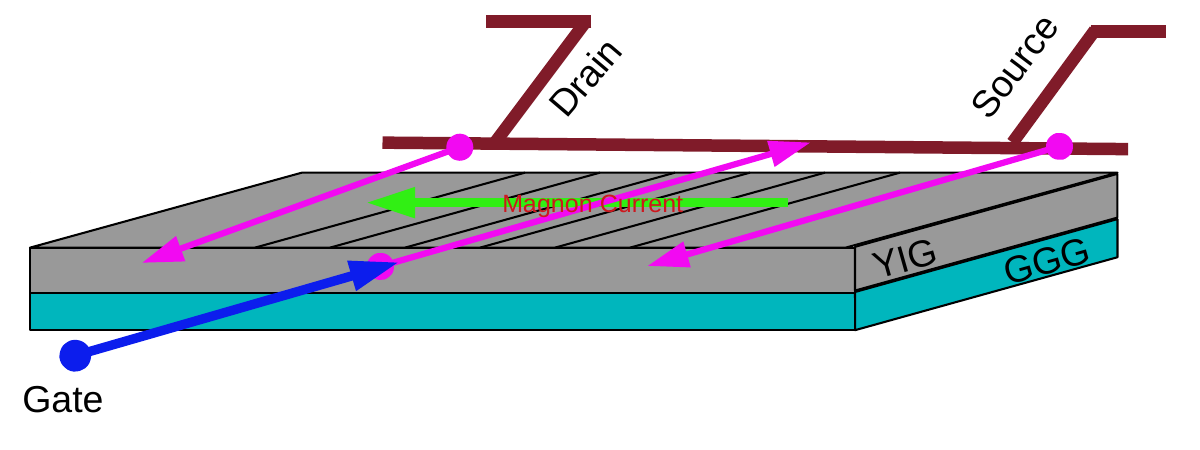}
    \caption{Pictorial representation of magnonic transistor in which YIG film with grooves is deposited on GGG substrate. Three terminals of the microstrip antenna, such as the source, drain, and gate, are added to inject and measure the magnons. }
    \label{transistor}
\end{figure}
Fig.~\ref{transistor} provides a schematic illustration of the magnonic transistor. The primary component of it is a YIG film with several parallel grooves on its surface. Microstrip antennas are used to inject magnons from a source terminal and detect them from a drain terminal. Magnon was injected from the gate terminal to control the magnon current flowing through  the source-to-drain. Each groove reflects the propagating magnons around one percent. Only those magnons will be scattered back, wavelengths of which satisfy the Bragg condition  $k_s=\frac{m_0 \pi}{a}$, leading to produce rejection bands (band gaps) in a system. 
\par
According to the transistor's operating principle, magnons are injected into its source at a frequency that falls inside the magnon transmission band. The S-magnons propagate almost distortion-free toward the drain when no magnon is in the gate of the transistor. Magnons are injected into the gate region to influence the flowing magnon through  the source to drain. To confine the magnon within the magnonic crystal, the frequency of the G-magnon should be in the center of the band gap of the magnonic crystal. The G-magnon concentration can be greatly increased because of this confinement. Injected S-magnons into the source region are scattered as they pass through the G-magnon-populated transistor gate; therefore, only partially reach the drain terminal. 

\section{Outline of the thesis}
In chapter 2, we do analytical calculations to find the formula of TMOTOC. We will do a comparative study of the revival time speed of correlation propagation in TMOTOC and LMOTOC. After that, we will verify the phase structure of the Floquet system in $\tau_0-\tau_1$ parameter space, numerically.
\par
In chapter 3, we will discuss three different regions of OTOC named characteristic (OTOC remain zero), dynamic (OTOC grow), and saturation (OTOC start to saturate) regimes of the LMOTOC and TMOTOC in the integrable and nonintegrable Floquet system. We will present a comparative study of LMOTOC and TMOTOC in all the regions.
\par
Further, in chapter 4, we use symmetric spin block observables instead of local spin observables to study OTOC in spin chains. We chose the block-spin observables to calculate OTOC in pre-scrambling and post-scrambling time regimes and analyzed the growth of OTOC, and replaced spin block observables with random block observables to analyze the saturation behavior of OTOC. We will show the averaged OTOC over random Hermitian observables is exactly the same as operator entanglement entropy.
\par
Finally, in chapter 5, we utilize OTOCs as a quantifier for quantum information currents in a 2D Heisenberg spin system with Dzyloshinski Moriya interaction. we provide a concept of quantum information diode based on magnonic crystals. 
\par
In chapter 6, we will summarize the results. We will also discuss future plans briefly.  
\chapter{Out-of-time-order correlation and detection of phase structure in Floquet transverse Ising spin system}  

\ifpdf
    \graphicspath{{Chapter1/Figs/Raster/}{Chapter1/Figs/PDF/}{Chapter1/Figs/}}
\else
    \graphicspath{{Chapter1/Figs/Vector/}{Chapter1/Figs/}}
\fi

\section{Introduction} 
In the last two decades, out-of-time-order correlation (OTOC) has gained a lot of attention among researchers in various fields. One field of interest is the butterfly effects in quantum chaotic systems
\cite{ray2018signature, hosur2016chaos, gu2016,ling2017}.  Other directions are quantum information scrambling \cite{Bohrdt2017,Yao2017,swingle2016measuring,Schleier2017,pappalardi2018scrambling,Klug2018,Khemani2018,hosur2016chaos,Alavirad2018} and many-body localization  \cite{maldacena2016bound}. The nontrivial OTOC as a holographic tool  has been instrumental in determining the interplay of scrambling, and entanglement  \cite{shenker2014black,roberts2015localized}. Many experiments have been done to measure OTOCs in various systems {\it, e.g.}, trapped-ion quantum magnets \cite{garttner2017measuring}, and nuclear magnetic resonance quantum simulator \cite{li2017measuring}.  
 \par  
In addition to the above fields of interest, the OTOCs are useful in determining phases of the quantum critical systems \cite{Shen2017,sun2020,heyl2018detecting}. The phase structures of quantum critical systems have been studied extensively in the last few decades \cite{Shen2017,Heyl2013,Pollmann2010,von2016phase,Thakurathi2013,Turner2011,Keyserlingk2016a,Feng2007,Bastidas2012,Jiang2011,sun2020,Fidkowski2011,Khemani2016,Thakurathi2014}. 
One of the simplest models to display and analyze the quantum phase transition is one dimensional transverse Ising model, which Hamiltonian is given as $H=J\sum_{i}\sigma_i^x\sigma_{i+1}^x+h\sum_i\sigma_i^z$. This system undergoes a phase transition at $J=h$ from the ferromagnetic state ($J>h$) to the paramagnetic phase ($J<h$) \cite{chakrabarti2008quantum,heyl2018detecting,su2006,sun2009}. Such phase transitions in time-independent equilibrium systems have been well-studied over the years. In the last few years, the OTOC has emerged as a tool to detect equilibrium and dynamical quantum phase transitions in the transverse field Ising (TFI) model and the Lipkin-Meshkov-Glick model (LMG)  \cite{heyl2018detecting}.
\par
\begin{figure}
\centering
\includegraphics[width=.5\linewidth, height=.45\linewidth]{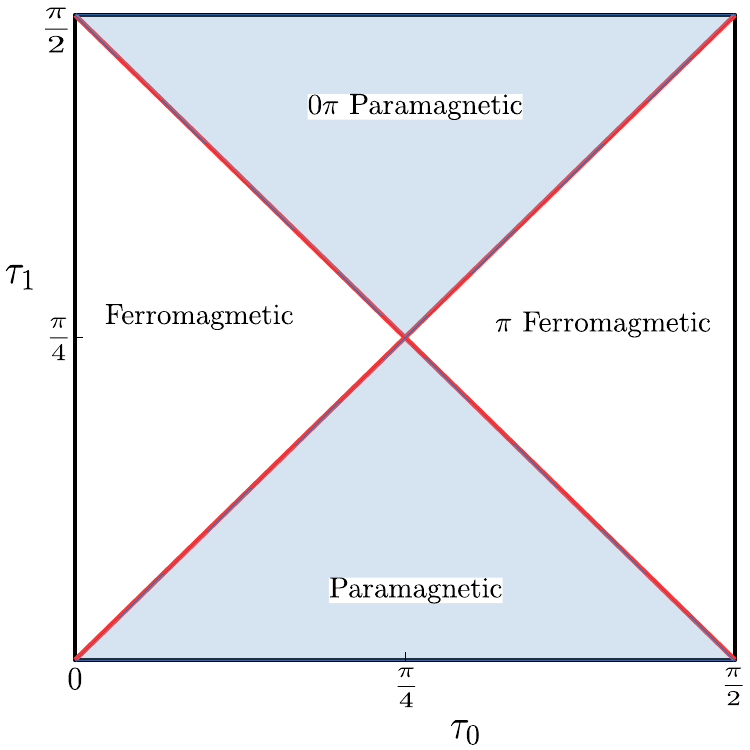}
\caption{Phase structure of the Floquet system with Floquet map given by Eq.~(\ref{U_f}). There are four distinct phases in the $\tau_0-\tau_1$ parameter space. Two of these phases, the $\pi$ ferromagnetic and the $0\pi$ paramagnetic, are phases which are unique to Floquet systems.}
\label{four_phase}
\end{figure}

  
It has been shown that the OTOC of the ground states and quenched states can diagnose the quantum phase transitions and dynamical phase transition, respectively \cite{heyl2018detecting}. The ferromagnetic ($J>h$)and paramagnetic ($J<h$) phases of the transverse Ising model can be characterized by nonzero and zero long-time averaged OTOC, respectively\cite{heyl2018detecting}. Periodically driven quantum systems, known as the Floquet systems, which have properties of the duality between time and space \cite{akila2016particle} and time-reflection symmetry \cite{iadecola2018floquet}, on the other hand, pose a different problem: one would expect generic Floquet systems to heat to infinite temperatures. However, specific cases of nonergodic phases with localization have been observed in Floquet systems \cite{Parameswaran2017,zhang2016}. In these systems, multiple nonergodic phases with differing forms of dynamics and ordering have been observed \cite{Khemani2016}. These multiple phases are characterized by broken symmetries and topological order. In the case of transverse Ising Floquet systems, Majorana modes are produced at the ends of the  chain \cite{Thakurathi2013}. The zero-energy Majorana mode corresponds to the long-range ferromagnetic order, while the nonzero-energy Majorana mode corresponds to the paramagnetic phase \cite{Thakurathi2013, von2016phase}. The sharp phase boundaries of the Floquet Ising system are explained by symmetry-protected Ising order  \cite{ Khemani2016, Keyserlingk2016a}.  For binary Floquet drive, two paramagnetic and two ferromagnetic phases can be seen in the phase diagram.  The two paramagnetic/ferromagnetic phases are distinguished by the combined eigenvalues at the edges of the Floquet drives and the parity operators. On the basis of the combined eigenvalues, the paramagnetic region is divided into two parts: $0$ and $0\pi$-paramagnetic, and ferromagnetic region is also divided into two parts: 0 and $\pi$-ferromagnetic [Fig.~\ref{four_phase}].  In the ferromagnetic region, all eigenstates have long-range Ising symmetry broken order. However, in the paramagnetic phase, all eigenstates have long-range symmetric order. 
 \par
First, we will consider transverse magnetization out-of-time-order correlation (TMOTOC) and calculate the exact solution using the Jordan Wigner transformation by mapping the spin operators onto the fermionic annihilation and creation operators. Next, we will consider the longitudinal magnetization out-of-time-order correlation (LMOTOC) and explore the various phases in the Fouquet Ising spin system. 
\par
In this chapter, we will start  discussing the model of the Floquet system. Subsequently, we will define the longitudinal and transverse magnetization  OTOCs.  We will introduce the time average of the  LMOTOC for the detection of phase structures and discuss the various phases of the Floquet Ising system using a long-time averaged LMOTOC. Later, we conclude the results.
\section{Model}
\label{model}
We consider an integrable Floquet transverse Ising system with binary Floquet drives. The Floquet map corresponding to this system is  
\begin{equation}
\label{U_f}
U=e^{-iH_{xx}\tau_{1}}e^{-iH_{z}\tau_{0}},
\end{equation} 
where $H_{xx}$ is the nearest neighbor Ising interaction given by $H_{xx}=\sum_{l=1}^{N-1}\sigma_l^x\sigma_{l+1}^x $ for open chain system and $H_{xx}=\sum_{l=1}^{N}\sigma_l^x\sigma_{l+1}^x $ for closed chain system with $\sigma_{N+1}^x=\sigma_{1}^x$.
$H_z=\sum_{l=1}^{N}\sigma_l^z$ is the transverse field in z-direction.  $\tau_0$ and $\tau_1$ are the time periods. The Hamiltonian corresponding to the above Floquet operator is:
\begin{equation}
H(t)=H_{xx}+\sum_{n=-\infty}^{\infty}\delta\Big(n-\frac{t}{\tau_1}\Big) \frac{\tau_0}{\tau_1} H_z.
\end{equation}
\section{ Out-of-time-order Correlation}
  \label{avgOTOC}
The out-of-time order correlation (OTOC) is, in general, defined as $F(t)=\langle \hat W(t) \hat V \hat W(t) \hat V \rangle $,
where $\hat V$ and $\hat W$ are two local Hermitian operators and $\hat W(t)$ is the Heisenberg evolution of the operator $\hat W$ by time $t$. We consider two different OTOCs defined as follows:
\begin{enumerate}
    \item[i)] {\it Transverse magnetization OTOC (TMOTOC) }: 
Here we consider two local spin operators $\hat W$ and $\hat V$ in the direction perpendicular to  the Ising axis (x-axis). In our generic treatment, we set the operators $\hat W=\hat \sigma_l^z$ and $\hat V=\hat \sigma_m^z$ at different sites $l$ and $m$.   The TMOTOC in our protocol is given as: 
\begin{equation}
\label{F_z}
F_z^{l,m}(n)=  \langle \phi_0|\hat \sigma_l^z(n)\hat \sigma_m^z\hat \sigma_l^z(n)\hat \sigma_m^z|\phi_0\rangle, 
\end{equation}
with the initial state as $|\phi_0\rangle=| \uparrow  \uparrow  \uparrow \cdots  \uparrow \rangle,$ 
where $ \left| \uparrow \right\rangle$ is the eigenstate of $\hat \sigma^z$ with eigenvalue $+1$.
    \item[ii)] {\it Longitudinal magnetization OTOC (LMOTOC) }:
    In this case, two local spin operators are chosen along the Ising axis{\it, i.e.} $\hat W=\hat \sigma_l^x$ and $\hat V=\hat \sigma_m^x$. The LMOTOC is given as follows:
\begin{equation}
    \label{F_x}
F_x^{l,m}(n) =  \langle
 \psi_0|\hat \sigma_l^x(n)\hat \sigma_m^x\hat \sigma_l^x(n)\hat \sigma_m^x|\psi_0\rangle.
\end{equation}
Here $|\psi_{0} \rangle=|\rightarrow \rightarrow \rightarrow \cdots \rightarrow \rangle$ is the initial state 
with, $ \left| \rightarrow \right\rangle$ is the eigenstate of $\hat \sigma^x$ with eigenvalue $+1$. 
\end{enumerate}
In what follows, $l$ and $m$ can take any value between $1$ to $N$ (even) in a closed chain system. For the open chain case, we will consider the special case with $l=m=\frac{N}{2}$. 
The time evolution of the spin operator at  the position $l$ after $n$ kicks is defined as $\hat \sigma_l^{z/x}(n)=\hat U^{\dagger n} \hat \sigma_l^{z/x} \hat U^n$.
The case $l=m$ will be treated as a special case.
\section{Analytical calculation of TMOTOC}
Considering  $t_0=2 \tau_0$, $t_1=4 \tau_1$ and $\hat \sigma_l^x=2 \hat S_l^x$ and periodic boundary condition in the  unitary operator defined in Eq.~(\ref{U_f}), we get Floquet map as:
\begin{equation}
\hat U =\exp\Big[-i t_1 \sum_{l=1}^{N}\hat S_l^x \hat S_{l+1}^x \Big] \exp\Big[-i t_0 \sum_{l=1}^{N} \hat S_l^z\Big],
\end{equation}
 We calculate the analytical expression for the TMOTOC using the Jorden-Wigner transformation (for detailed calculation, refer to the Appendix \ref{AppendixA1}):
 
\begin{eqnarray}
\label{OTOCz_gene}
 F_z^{l,m}(n) &=& 1- \Big(\frac{2}{ N}\Big)^3 \sum_{p,q,r} \Big[ e^{i(p-q)(m-l)}|\Psi_r(n)|^2 \Phi_p^{*}(n) \Phi_q(n) \nonumber \\
 &-& e^{i(-r-q)(m-l)}  \Psi_r(n)^{*} \Phi_p^{*}(n)  \Phi_q(n) \Psi_{-p}(n)  \nonumber \\
&-& e^{i(p+q)(m-l)} \Psi_{q}(n)  \Psi_{r}(n)^{*}  \Phi_{p}(n)^{*} \Phi_{-r}(n)\nonumber \\
&+& e^{i(q-r)(m-l)} \Psi_{q}(n)\Psi_r(n)^* |\Phi_p(n)|^2 \Big].  
\end{eqnarray}
Now, we take a special case in which both the local operators are at the same position i.e. $l=m$ and $\hat V=\hat W=\hat \sigma_l^z$. The expression of  TMOTOC simplifies to  
 \begin{eqnarray}
   \label{OTOCz}
 F_z^{l,l}(n) &=& 1- \Big(\frac{2}{ N}\Big)^3  \sum_{p,q,r}\Big[|\Psi_r(n)|^2 \Phi_p(n)^* \Phi_q(n)  
-\Psi_{-p}(n) \Psi_r(n)^* \Phi_p(n)^* \Phi_q(n) \nonumber \\
&& -  \Psi_q(n)  \Psi_r(n)^*  
 \Phi_{p}(n)^*  
 \Phi_{-r}(n) + \Psi_{q}(n)  
 \Psi_r(n)^* |\Phi_p(n)|^2 \Big],  
 \end{eqnarray}
where the expansion coefficients  $\Phi_q(n)$ and  $\Psi_q(n)$ are defined as 
\begin{equation}
\label{phi}
\begin{split}
&\Phi_q(n)=|\alpha_{+}(q)|^2 e^{-i n \gamma_q}+|\alpha_{-}(q)|^2 e^{i n \gamma_q},\\
&\Psi_q(n)=\alpha_{+}(q) \beta_{+}(q)e^{-i n \gamma_q}+\alpha_{-}(q)\beta_{-}(q) e^{i n \gamma_q}.
\end{split}
\end{equation}
The phase angle $\gamma_q$ and the coefficients $\alpha_{\pm}(q)$ and $\beta_{\pm}(q)$ are given by
\begin{equation}
\label{gamma}
\cos(\gamma_q)=\cos(t_0)\cos(t_1)-\cos(q)\sin(t_0)\sin \Big(\frac{t_1}{2} \Big),
\end{equation}
and
\begin{equation}
\label{apmq}
\alpha_{\pm}(q)^{-1}=\sqrt{1+\Big(\frac{\cos(\frac{t_1}{2})-cos(\gamma_q \pm t_0)}{\sin(q) \sin( t_0)\sin(\frac{t_1}{2})}\Big)^2},
\end{equation}

\begin{equation}
\label{bpmq}
\beta_{\pm}(q)=\frac{\mp\sin(\gamma_q)-\cos( t_0) \cos(q) \sin(\frac{t_1}{2})-\sin( t_0) \cos(\frac{t_1}{2})}{\sin(q)\sin(\frac{t_1}{2})}\alpha_{\pm}e^{-it_0}(q).
\end{equation}
The allowed value of $p$, $q$ and $r$ are from $\frac{-(N-1)\pi}{N}$ to $\frac{(N-1)\pi}{N} $ differing by $\frac{2 \pi}{N} $ for even number of $N_F$ ($N_F=c_l^\dagger c_l$, number of fermions).

\begin{figure}
     \centering
        \includegraphics[width=.49\linewidth, height=.30\linewidth]{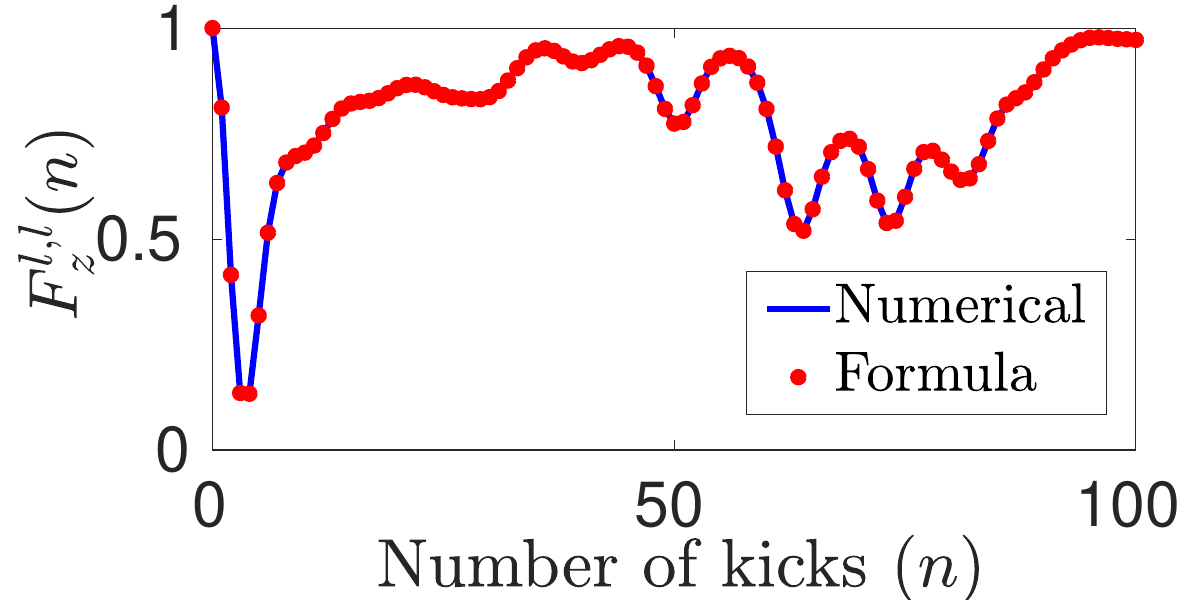}
       \caption{$F^{l,l}_z(n)$ for closed chain transverse Ising Floquet system of system size $N=12$ by using the numerical calculations (solid line) and analytical expression of Eq.~(\ref{OTOCz}) (point). Here we take $\tau_0=\tau_1=\epsilon$, where $\epsilon=\frac{\pi}{28}$.}
    \label{comp_otoc_n12}
\end{figure}

The values of $F_z^{l,l}(n)$ obtained  by the analytical expression in Eq.~(\ref{OTOCz}) exactly match with those obtained by numerical exact diagonalization as shown in Fig.~\ref{comp_otoc_n12}.

\section{Speed for correlation propagation}
\begin{figure}
\centering
  \includegraphics[width=.49\linewidth, height=.30\linewidth]{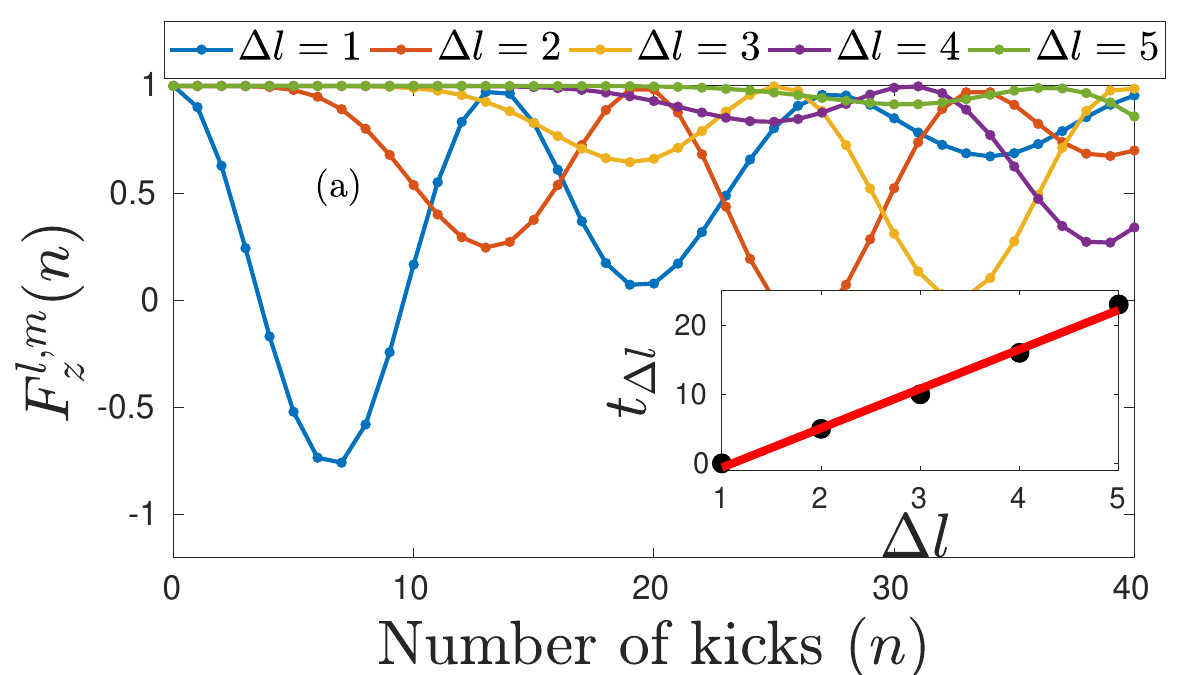}
   \includegraphics[width=.49\linewidth, height=.30\linewidth]{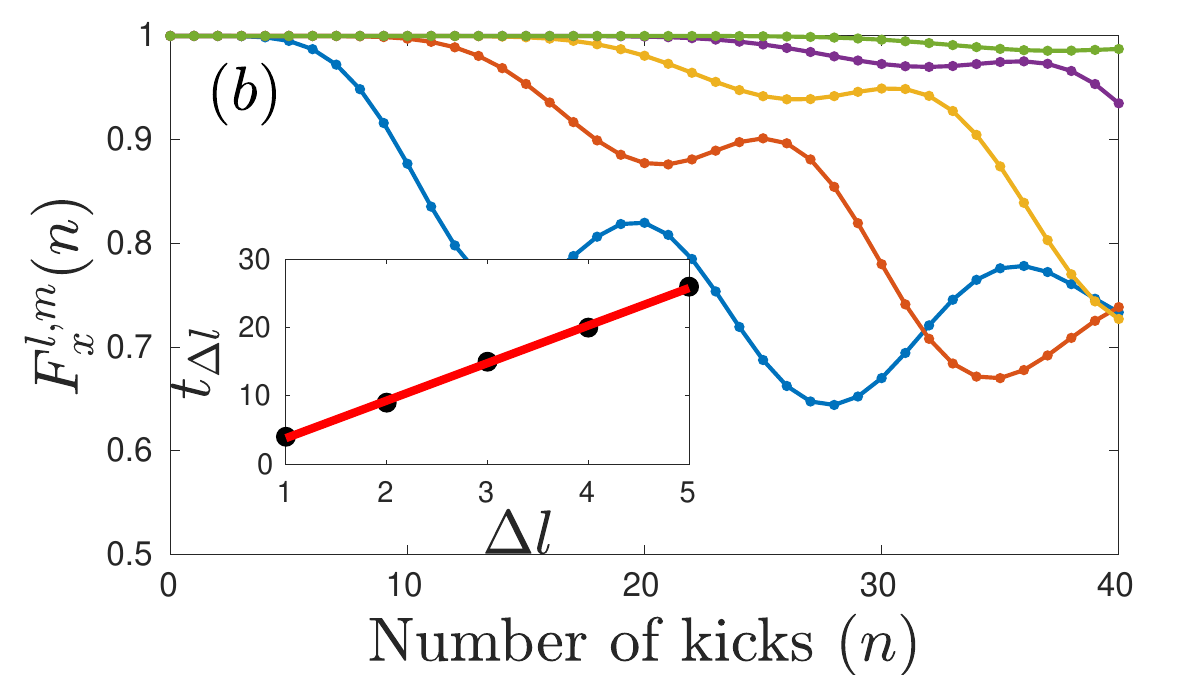}
  \caption{Behaviour of (a) $F_z^{l,m}(n)$ and (b)$F_x^{l,m}(n)$  with number of kicks for different value of $\Delta l$. Here, the parameters are: $\tau_0=\frac{\epsilon}{2}$,  $\tau_1=\epsilon$, and  $N=12$.  In both figures (a) and (b), the inset shows the behavior of time of departure from unity ($t_{\Delta l}$) as a function of separation between the observables ($\Delta l$)}.
\label{otocz_11_butterfly}
\end{figure}

Next, we use  Eq.~(\ref{OTOCz_gene}) for $l\ne m$ to calculate the speed for correlation propagation. At $t=0$, both the operators $\hat W(t=0)=\hat \sigma_l^z$ and $\hat V=\hat \sigma_m^z$, commute with each other which implies that $F_z^{l,m}(n)$ will be unity.  As time changes, the evolution of $\hat \sigma_l^z$ takes place by the Floquet operator; they no longer commute. Therefore, $F_z^{l,m}(n)$ starts to drop from the unity, which  provides us the speed of correlation propagation ($v_{cp}$). The general approach to calculate $v_{cp}$ is as follows: First, we fix $m=\frac{N}{2}$ and change $l$ from $\frac{N}{2}+1$ to  $\frac{N}{2}+5$. By using {Fig.~\ref{otocz_11_butterfly}}(a), we determine the characteristic time $t_{\Delta l}$ in which  $F_z^{l,m}(n)$ starts departing from unity and plot it as a function of the separation between the observables ($\Delta l$) [inset of {Fig.~\ref{otocz_11_butterfly}}(a)]. In the inset of {Fig.~\ref{otocz_11_butterfly}}(a), dots are the points corresponding to the given $\Delta l$ and dashed line is the best fit line.  Reciprocal of the slope of this straight line is the  speed of the correlation propagation $v_{cp}$. For comparison, we have  shown similar results for LMOTOC  in {Fig.~\ref{otocz_11_butterfly}}(b). We find that the speed of the commutator growth  of the $F_z^{l,m}(n)$ ($v_{cp}=0.175$) is nearly equal to that of the $F_x^{l,m}(n)$ ($v_{cp}=0.181$). This means that $v_{cp}$ is independent of the choice of the  observables. By comparing {Fig.~\ref{otocz_11_butterfly}}(a) and (b), we observe that the closer the operators $\hat V$ and $\hat W$ are, the smaller the characteristic time $t_{\Delta l}$ is.

\section{Revival time}
 Now we move to another interesting quantity, the revival time of $F_z^{l,m}(n)$ and $F_x^{l,m}(n)$, which is defined as the time in which OTOCs return back to their initial value. We can see from Fig.~\ref{otocz_11_butterfly}(a,b) that the early time behavior of both $F_z^{l,m}(n)$ and $F_x^{l,m}(n)$  looks very similar. For instance, both $F_z^{l,m}(n)$ and $F_x^{l,m}(n)$ start deviating from unity after a certain time. However, the long time behaviors of $F_z^{l,m}(n)$ and $F_x^{l,m}(n)$ differ widely. After decreasing from unity  to a minimum, $F_z^{l,m}(n)$ revives and recovers to its initial value{\it, i.e.,} unity, while $F_x^{l,m}(n)$ oscillates about a finite value and  never reaches to unity. Revival time depends on the distance between local operators. The larger the separation between the operators $\hat V$ and $\hat W$ is, the more the revival time is.  This can be seen from {Fig.~\ref{otocz_11_butterfly}}(a,b).
 \begin{figure}
     \centering
        \includegraphics[width=.49\linewidth,height=.30\linewidth]{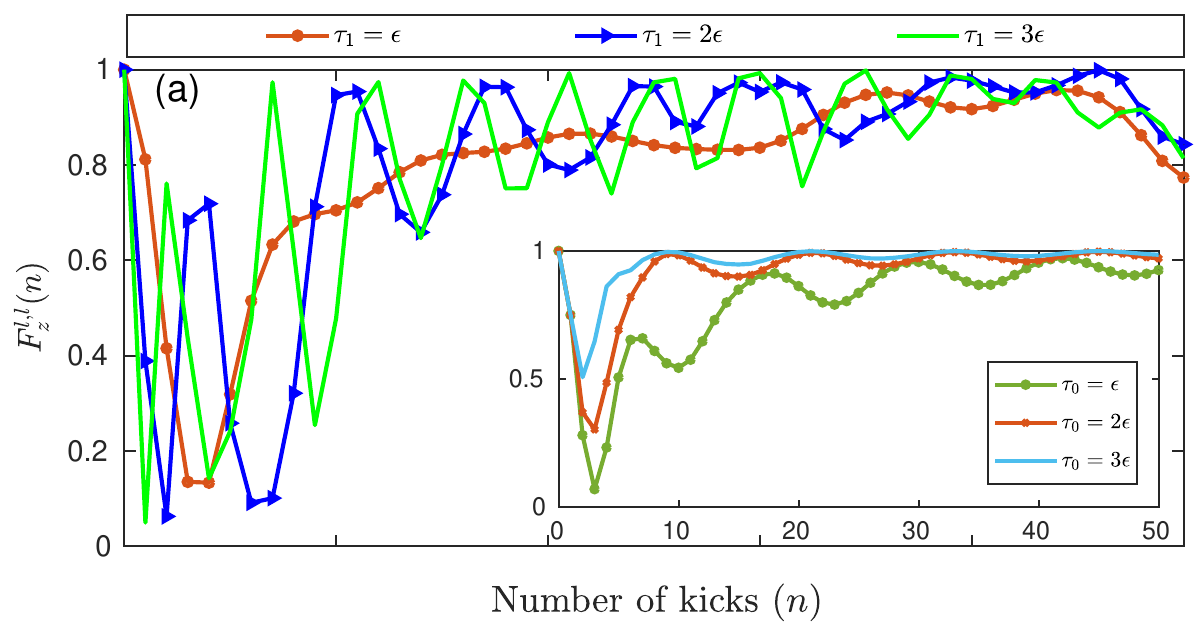}
        \includegraphics[width=.49\linewidth,height=.30\linewidth]{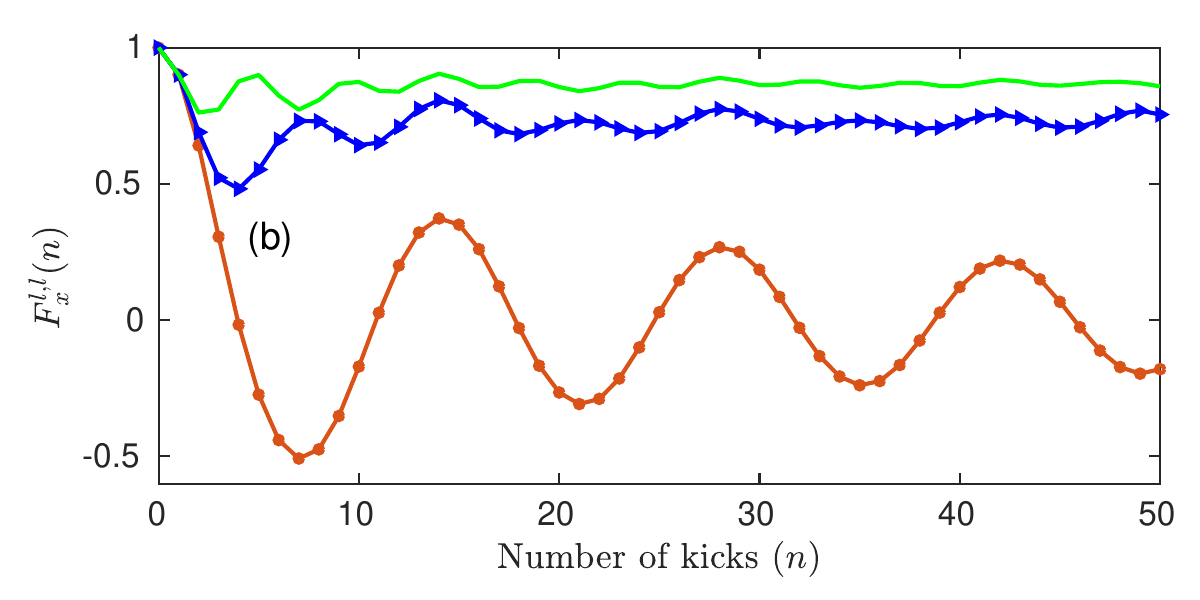}
       \caption{(a) Variation of the real part of (a) $F_x^{l,l}(n)$ and (b) $F_x^{l,l}(n)$ with number of Floquet periods for a fixed $\tau_0=\epsilon$ and $\tau_1 =\epsilon,$ $ 2 \epsilon$ and $ 3\epsilon$ in closed  chain Floquet systems with system size N=12.
Inset of the figure is the variation of $F_z^{l,l}(n)$ with the number of Floquet periods for a fixed $\tau_1=\frac{\pi}{24}$ and $\tau_0 =\epsilon,$ $ 2 \epsilon$ and $ 3\epsilon$ in closed chain Floquet systems with system size N=50.}
\label{otocz_otocx_n12_tau0_pi28_tau1_pi6}
\end{figure}
 \par
The advantage of having an easily computable formula such as Eq.~ (\ref{OTOCz}) is that we can study the TMOTOC  as a function of Floquet periods $\tau_0$ and $\tau_1$ and see the behavior at any number of kicks. The analytical expression is of $O(L^3)$, which has a significant advantage over exact diagonalization calculations of  O($2^L$) [inset of {Fig.~\ref{otocz_otocx_n12_tau0_pi28_tau1_pi6}}(a)].
\par
The LMOTOCs have been shown to be useful in detecting the phase transitions between the paramagnetic and ferromagnetic phases in spin systems \cite{heyl2018detecting}. 
However, the same cannot be said about TMOTOC using the same concept.  A comparison of the behavior of the two quantities with time is shown in Fig.~\ref{otocz_otocx_n12_tau0_pi28_tau1_pi6}. We see from Fig.~\ref{otocz_otocx_n12_tau0_pi28_tau1_pi6}(a) that  TMOTOCs always oscillate about a  positive value for all the pairs of $\tau_0$ and $\tau_1$, signaling that the long time average of  TMOTOC is always a positive quantity. However, in the case of  LMOTOCs, as shown in \ref{otocz_otocx_n12_tau0_pi28_tau1_pi6}(b), we find that the long-time average value can be either zero or a positive quantity. In order to detect the phase structure of the system, we require the order parameter characterizing the distinct phases to show a sharp contrast between the phases. We see that the  LMOTOCs qualify the criterion to be used as an order parameter, but TMOTOCs fail to do so.

Upon performing a quantum quench from a polarized state, we will use the saturation value of the  LMOTOC as the order parameter to distinguish between the two phases. It is calculated by numerical methods because a compact analytical expression is not achievable using the Jordan-Wigner transformation.
A possible approach and inability to get a compact analytical solution for LMOTOC is given in the Appendix (\ref{AppendixA2}).
\label{phase_structure}

We study the phase structure of the system [Eq.~(\ref{U_f})] by calculating the  LMOTOC [Eq.~(\ref{F_x})]. If LMOTOC  saturates to a particular value after a long period of time, this value can be determined by taking the long-time average of the LMOTOC. We define the long time average of the  LMOTOC, $\overline{F}_x^{l,l}(T)$ upto $T$ Floquet periods by: 
\begin{equation}
\overline{F}_x^{l,l}=\frac{1}{T} \sum_{n=1}^T F_x^{l,l}(n).
\end{equation}
 The long-time average of the LMOTOC has a direct link with the spectral properties of the system in consideration \cite{DeMelloKoch2019}. The averaged  LMOTOC links with the spectral form factor \cite{Dyer2017}, a well-known quantity in random matrix theory, which is a quantifier for discreteness in the spectrum.  
\section{Phase Structure}

\begin{figure}
     \centering
        \includegraphics[height=.40\linewidth,width=.99\linewidth]{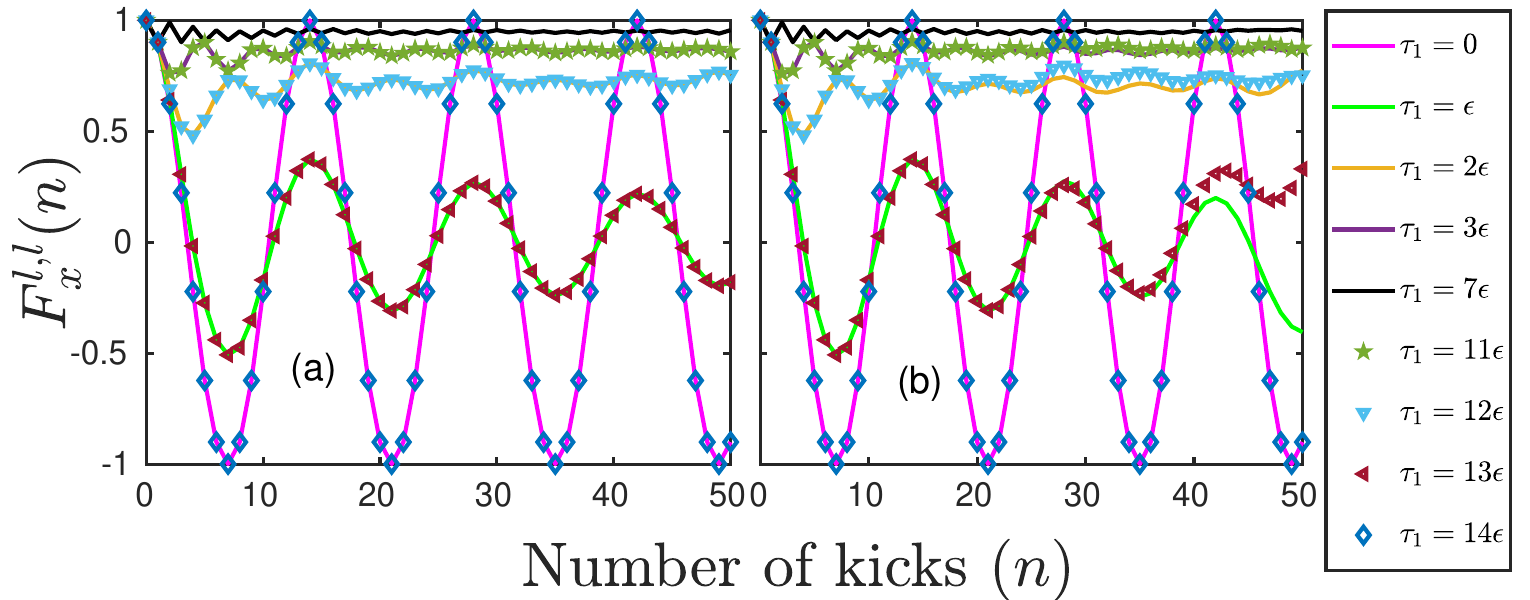}
\caption{Variation of the real part of $F_x^{l,l}(n)$) with the number of Floquet periods for a fixed $\tau_0=\epsilon=\frac{\pi}{28}$ and multiple values of $\tau_1$ in (a) closed and (b) open chain Floquet systems with system size N=12. The initial state is a direct product of the eigenstate of $\sigma^{x}$ with eigenvalue $+1$. At $\tau_1=0$, $F_x^{l,l}(n)$ shows periodic oscillations about zero. As the value of $\tau_1$ is increased, the $F_x^{l,l}(n)$ oscillates about greater non-zero values with lower amplitudes of oscillation and at $\tau_1=\frac{\pi}{4}$, it saturates to the value $1$. As increase constant value of  $\tau_1$, $F_x^{l,l}(n)$ has same and different value as $ \frac{\pi}{2}-\tau_1$ in closed  and open chain respectively.}
    \label{otocx_tau1_n12_p1_p0_new}
\end{figure}

\begin{figure}
    \centering
\includegraphics[height=.40\linewidth,width=.99\linewidth]{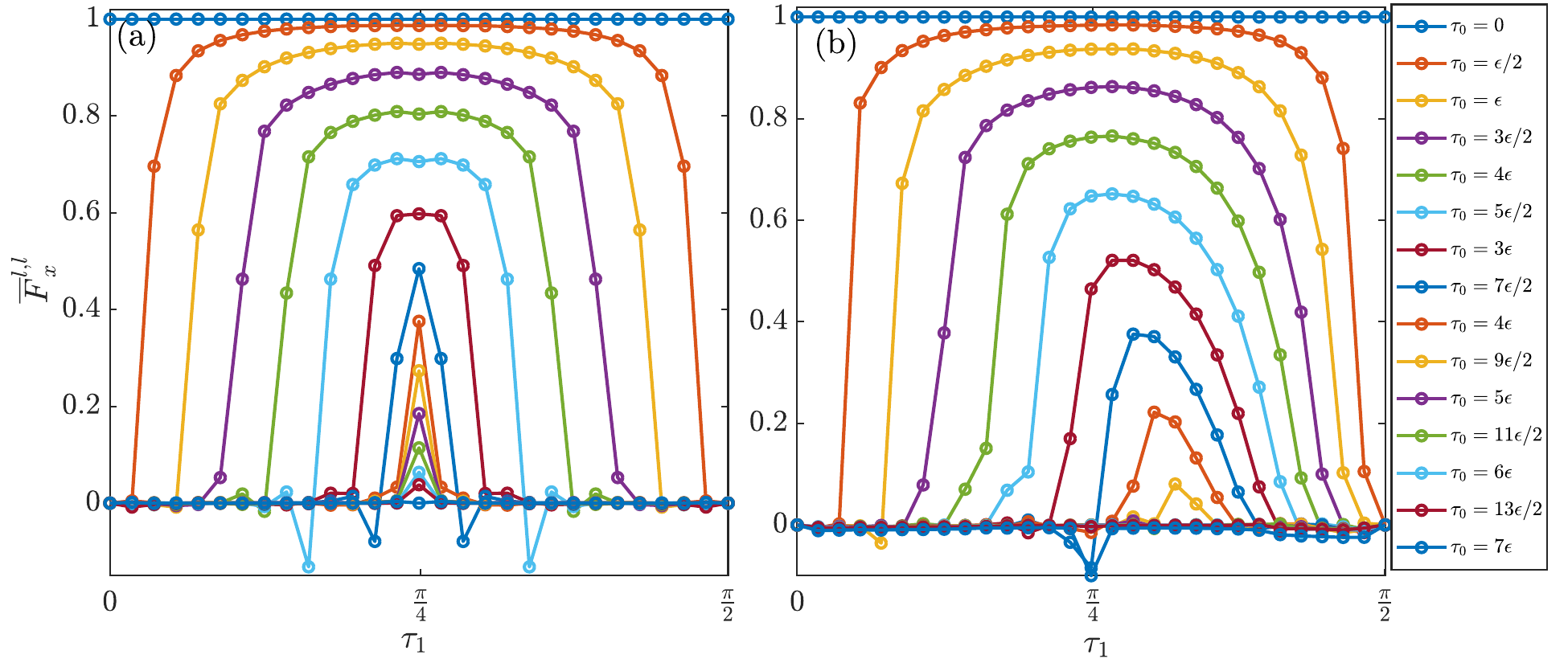}
\caption{ Plot of $\overline{F}_x^{l,l}(T)$ with $\tau_1$ for values of $\tau_{0}$ varying from $0$ to $\frac{\pi}{4}$  in intervals of $\epsilon$ in the (a) closed and (b) open chain Floquet systems of system size $N=10$ and $T=10^4$. The variation of $\overline{F}_x^{l,l}(T)$ with $\tau_1$ for $\pi/4<\tau_0<\pi/2$ is the same as that for $\frac{\pi}{2}-\tau_{0}$. This plot can be used to find the regions in the $\tau_0$ and $\tau_1$ parameter space that have $\overline{F}_x^{l,l}(T)>0$ and $\overline{F}_x^{l,l}(T)=0$ (Fig.~\ref{finite_phase}).}
    \label{otoc2_n10_tau0}
\end{figure}

\begin{figure}
\centering
\includegraphics[height=.40\linewidth,width=.99\linewidth]{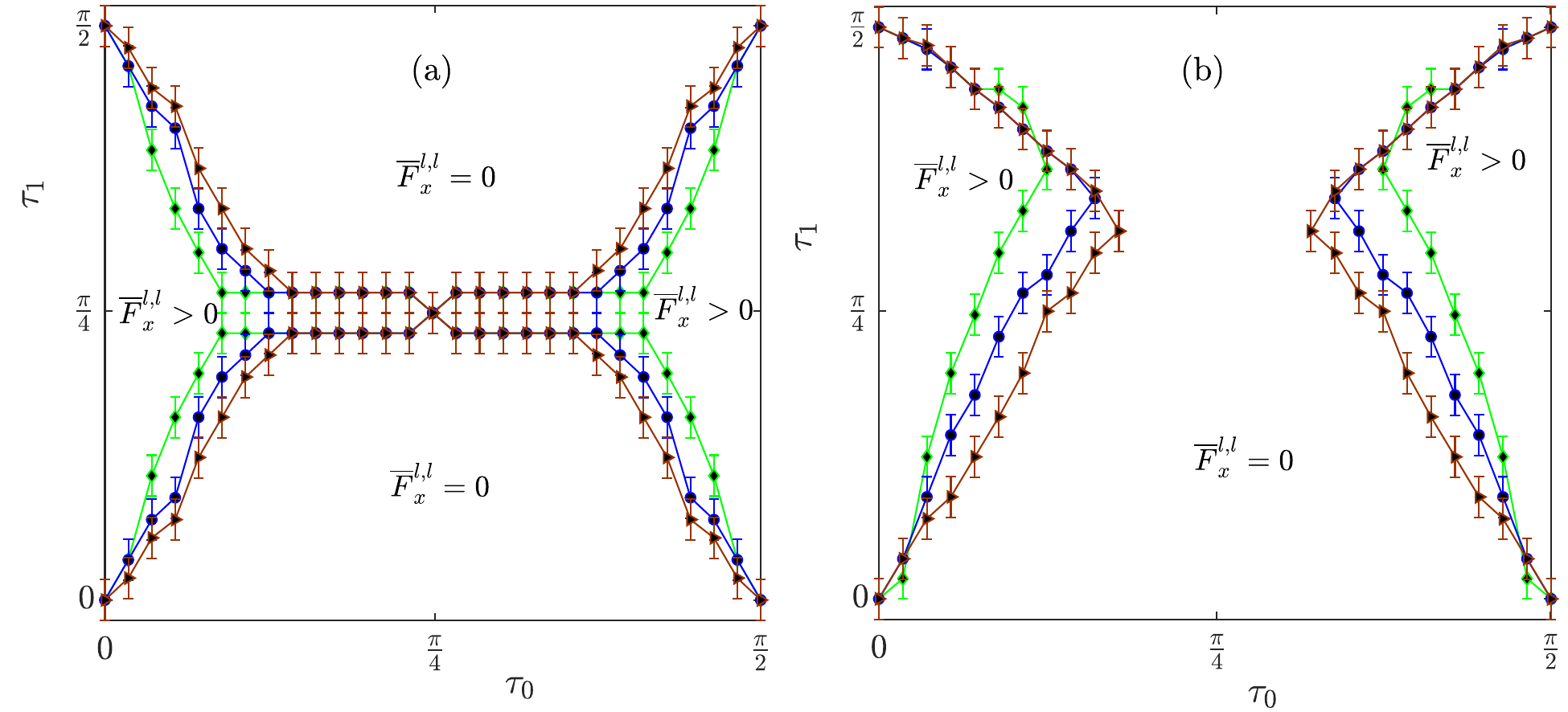}
\caption{Regions with $\overline{F}_x^{l,l}(T)>0$ and $\overline{F}_x^{l,l}(T)=0$ in the $\tau_0$ and $\tau_1$ parameter space with $T=10^4$, for the (a)closed and (b) open chain Floquet systems of system size N=6(Green), 8(Blue) and 10(Brown). As increasing system size, critical lines of  both (a) and (b) tend towards diagonal critical lines. Hence, this suggests that the phase structure of the system would contain two ferromagnetic regions (where $\overline{F}_x^{l,l}>0$) and two paramagnetic regions (where $\overline{F}_x^{l,l}=0$).}
    \label{finite_phase}
\end{figure}

The phase structure of the Floquet system given by Eq.~(\ref{U_f}) is known to have four distinct phases in the two-dimensional parameter space of $\tau_0$ and $\tau_1$. The phase diagram is shown in Fig.~\ref{four_phase} \cite{Khemani2016,von2016phase,Keyserlingk2016a}. Paramagnetic and ferromagnetic phases show behavior similar to their undriven counterparts. The other two of the phases observed, the $\pi$-ferromagnetic and $0\pi$-paramagnetic, are unique to Floquet systems and are not observed in undriven non-Floquet systems. The phase transitions between these phases in the $\tau_0$ and $\tau_1$ parameter space can be detected by calculating LMOTOCs. The  LMOTOC has been shown to saturate to non-zero values in the ferromagnetic region and zero in the paramagnetic region at long times in the undriven systems \cite{heyl2018detecting}. Hence, the long-time averaged LMOTOC serves as a good order parameter for paramagnetic and ferromagnetic regions in the undriven systems. In driven Floquet systems, LMOTOCs do not saturate to non-zero and zero values for all values of $\tau_0$ and $\tau_1$; we see a continuing oscillating behavior about a non-zero or zero mean value (Fig.~\ref{otocx_tau1_n12_p1_p0_new}). The time-averaged LMOTOC ($\overline{F}_x^{l,l}(n)$) is seen to saturate at long times in the thermodynamic limit to non-zero values in the ferromagnetic and $\pi$ ferromagnetic regions and zero in the paramagnetic and $0\pi$ paramagnetic regions of the phase space. Fig.~\ref{otoc2_n10_tau0} shows the variation of the long time-average of the real part of the LMOTOC  with $\tau_1$, for different values of $\tau_0$ in the closed and open boundary conditions for a system size $N=10$.\\ The critical points where the phase transition occurs are identified along the constant $\tau_0$ line at the points where LMOTOC  goes from zero to non-zero. These critical points, when mapped in the $\tau_0$ and $\tau_1$ parameter space for $ N=6, 8$ and $10$, give plots as shown in Fig.~\ref{finite_phase}.
\par
There exists a symmetry along $\tau_1=\frac{\pi}{4}$ in the closed chain case because the behavior of LMOTOC is the same for Floquet period $\tau_1$ and $\frac{\pi}{2}-\tau_1$ ({\it e.g.,} $\epsilon$ and $13\epsilon$, $2\epsilon$ and $12\epsilon$, $3\epsilon$ and $11\epsilon$  are same in   Fig.~\ref{otocx_tau1_n12_p1_p0_new}(a)). In the open chain case, LMOTOC for long-time is different for $\tau_1$ and $\frac{\pi}{2}-\tau_1$ (see, for example, $\epsilon$ and $13\epsilon$, $2\epsilon$ and $12\epsilon$, $3\epsilon$ and $11\epsilon$   in   Fig.~\ref{otocx_tau1_n12_p1_p0_new}(b)), therefore the symmetry along $\tau_1=\frac{\pi}{4}$ is absent in Fig.~\ref{finite_phase}(b). However, a symmetry along $\tau_0=\frac{\pi}{4}$ exists in both open and closed chain [Fig.~\ref{finite_phase}(a,b)]. We demonstrate these symmetries for the closed chain system using a toy model  of two and four spins. First, we calculate $F_x^{l,l}(n)$ for two spin system:
After the first kick ($n=1$), $F_x^{l,l}(1)=\cos(4\tau_0)$. Since the magnetization is in the direction of coupling of spins, the interaction term $H_{xx}$ (with $\tau_1$) is not involved in the state after the first kick.
After the second kick ($n=2$) LMOTOC is given by
\begin{eqnarray}
\label{L2_n2}
F_x^{l,l}(2)&=&\frac{1}{8}\Big[1-4\cos(4\tau_1) - 4\cos(4\tau_0)\Big(-1+\cos(8\tau_1)\Big) \nonumber \\
&+&3\cos(8\tau_1)+\cos(8\tau_0)\Big(3+4\cos(4\tau_1)+\cos(8\tau_1)\Big)\Big] \nonumber.
\end{eqnarray}
The symmetry along $\tau_0=\pi/4$ is evident in the above expression as $\tau_0$, and $\tau_1$ appear in the expression with a multiple of $4k$, where $k$ is an integer. Further kicking the system will also manifest the multiplicity of $4k$ with $\tau_0$ and $\tau_1$. Next, we take the toy model of four spins case.  LMOTOC after the first kick ($n=1$) will again be $F_x^{l,l}(1)=\cos(4\tau_0)$
and after the second kick ($n=2$) will be 
\begin{eqnarray}
\label{L4_n2}
F_x^{l,l}(2) &=& \frac{1}{64} \Big[\cos(4\tau_0)(21-4\cos(4\tau_1)-17\cos(8\tau_1))  \nonumber \\
&+&\cos(12\tau_0)(-5 +4\cos(4\tau_1)+\cos(8\tau_1)) \nonumber \\
&+& 2(5-12\cos(4\tau_1)  + 7\cos(8\tau_1)+  \cos(8\tau_0) \nonumber \\ 
&\times& (19+12\cos(4\tau_1)+\cos(8\tau_1)))\Big] \nonumber
\end{eqnarray}
Again we see that $F_x^{l,l}(1)$ and $F_x^{l,l}(2)$ have $\tau_0$ and $\tau_1$ are in a multiple  of $4k$, where $k$ is the integer. Therefore, LMOTOC will be same for (1) $\tau_0$ and $\frac{\pi}{2}-\tau_0$, and (2)  $\tau_1$ and $\frac{\pi}{2}-\tau_1$ and symmetric about $\tau_0=\frac{\pi}{4}$ and  $\tau_1=\frac{\pi}{4}$ in the closed chain system.

In the closed chain Floquet system, the tips of the regions with $\overline{F}_x^{l,l}=0$ can be seen to be moving closer  to each other along the line $\tau_1=\frac{\pi}{4}$, with increasing the system size [Fig.~\ref{finite_phase}(a)]. In the open boundary condition, the tips, which start out in the upper half of the parameter space, also move downwards towards the point $(\frac{\pi}{4},\frac{\pi}{4})$ with increasing system size [Fig.~\ref{finite_phase}(b)].

\subsection{Critical line with system size}
\begin{figure}
    \centering
   \includegraphics[width=.49\linewidth,height=.30\linewidth]{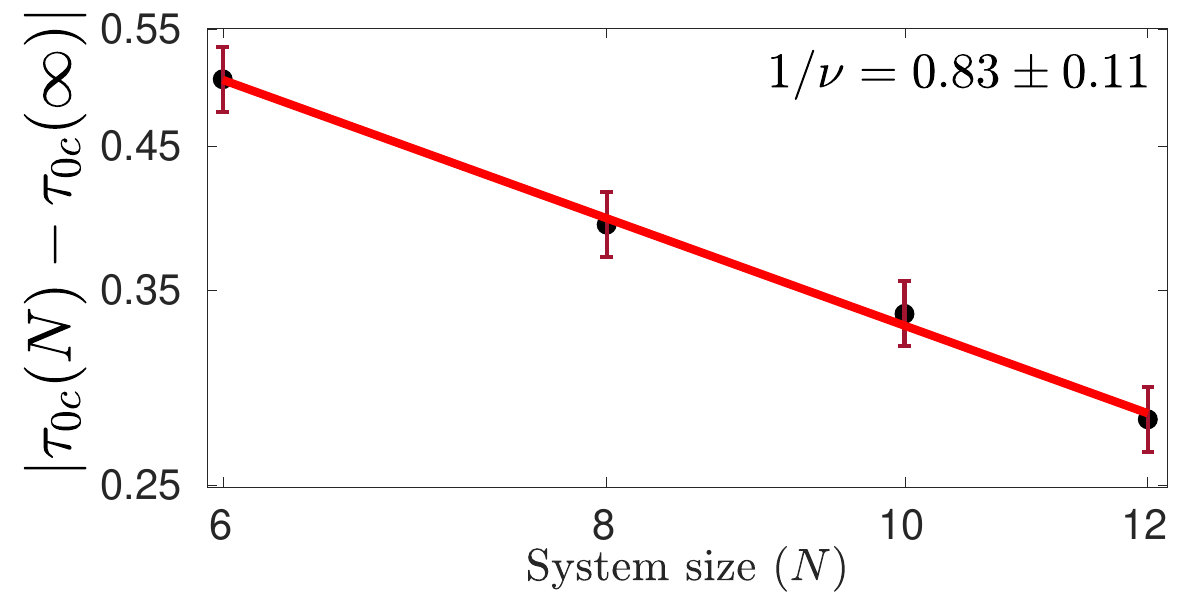}
    \caption{  Plot of the difference between finite size critical point and the infinite size critical point ($|\tau_{0c}(N)-\tau_{0c}(\infty)|$) of the phase structure of the periodic Floquet system as the function of system size (log-log plot). Black points are data points, and the red dashed line is the best fit yielding the slope  $1/\nu=0.8314\pm0.1122$.}
    \label{critical_point}
\end{figure}

Behavior of increasing the tips with increasing system size in closed chain case can be understood by the finite size effect analysis, which is given as \cite{Korniss2000} 
\begin{equation}
|\tau_{0c}(N)-\tau_{0c}(\infty)|\propto N^{-1/\nu},
\end{equation}
where $\tau_{0c}(N)$($\tau_{0c}(\infty)$) is the location of the critical point on the horizontal axis of the phase structure  of the finite system size [Fig.~\ref{finite_phase}(a)] (infinite system size  [Fig.~\ref{four_phase}]).  $\nu$ is the transverse field exponent defined as the reciprocal of the slope of the straight line drawn from $|\tau_{0c}(N)-\tau_{0c}(\infty)|$ {\it vs} system size $(N)$ (log-log plot). As evident from  Fig.~\ref{critical_point}, increasing the system size $N$ leads to closing the gap between  $\tau_{0c}(N)$ and  $\tau_{0c}(\infty)$.
In the thermodynamic limit,  we expect the tips to meet at the center, giving the diagonal lines as shown in Fig.~\ref{four_phase}. A similar argument holds true for the open chain case. Hence, the time-averaged LMOTOC [$\overline{F}_x^{l,l}(T)$] for large $T$ and $N \rightarrow \infty$ can be used as an order parameter to distinguish the phases of a driven transverse field Floquet Ising model. It must be noted that the time-averaged LMOTOC does not distinguish between the ferromagnetic and the $\pi$ ferromagnetic phase or the paramagnetic and the $0\pi$ paramagnetic phase. However, these distinct phases can be identified by observing the combined eigenvalues at the edges of the phase structures of the unitary operator which is defined in Eq.~(\ref{U_f}) and the parity operator ($P=\prod_{l}\sigma_l^z$) \cite{von2016phase,Keyserlingk2016a}. Considering the  operators $\hat U$ and $\hat P$ have eigenvalues $u$ and  $p$, respectively, the different phases can be distinguished by observing the eigenvalues along the outer edges of the phase diagram. The eigenvalues have protected multiplets of the form: $(u,p)$ in the paramagnetic, $[(u,p), (u,-p), (-u,p), (-u,-p)]$ in the  $0\pi$ paramagnetic, $[(u,p),  (u,-p)]$ in the ferromagnetic and   $[(u,p), (-u,-p)]$ in the $\pi$ ferromagnetic regions.

\begin{figure}
 \centering
   \includegraphics[width=.49\linewidth,height=.40\linewidth]{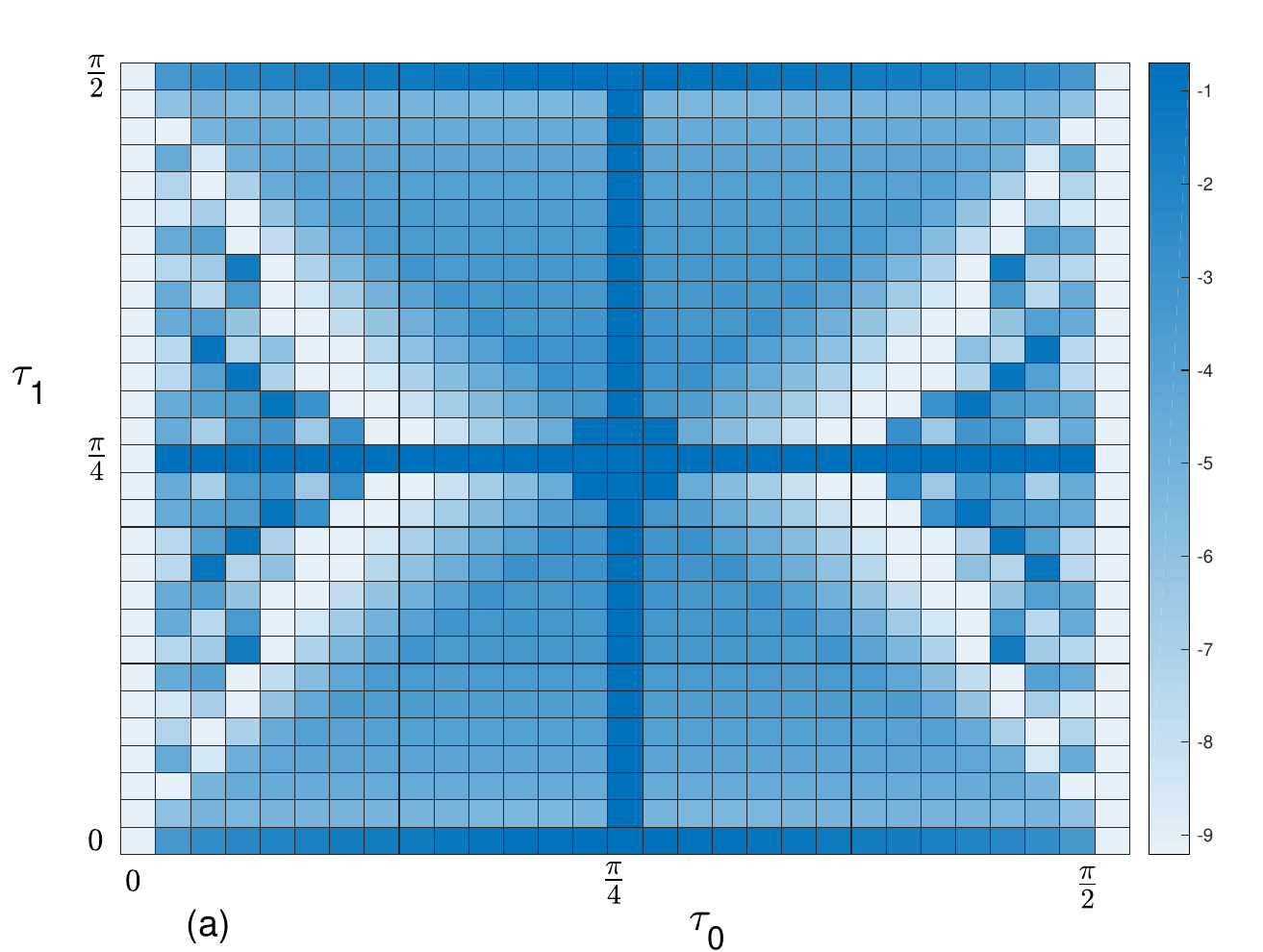}
\includegraphics[width=.49\linewidth,height=.40\linewidth]{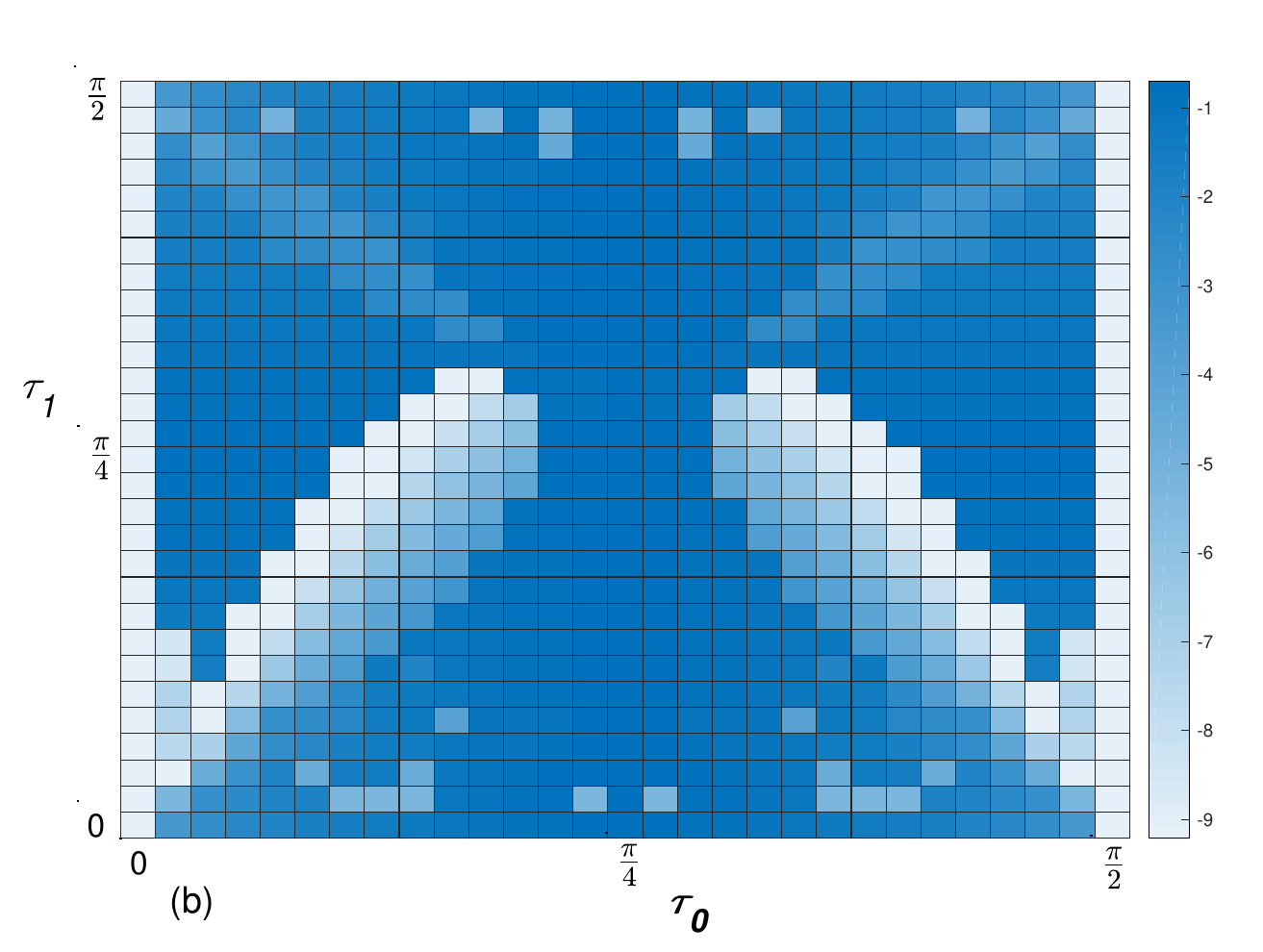}
\caption{Heat map of the logarithmic dominant frequencies of $F(n)-\overline{F}_x(T)$ in the discretized $\tau_0$-$\tau_1$ parameter space for the (Left) closed and (Right) open Floquet system of system size $N=10$ and with $T=10^4$. The heat map shows logarithmically small values close to the transition lines shown in Fig.~\ref{finite_phase}.}
    \label{heatmap_n10}
\end{figure}

\section{Phase structure by frequencies of oscillations}
The frequencies of oscillations of the LMOTOC also provide  a signature of the phase transitions.  Here, the dominant frequencies have been numerically determined by taking the argument maxima of the discrete Fourier transforms of the deviation of the LMOTOC from its mean value ($F_x(n)-\overline{F}_x(T)$). Mathematically, it can be defined as:
\begin{equation}
\mathcal{F}(\nu)=\frac{1}{T} \sum_{n=1}^T [F_x(n)-\overline{F}_x(T)]e^{-i\frac{2\pi}{T}\nu n}
\end{equation}
A heatmap of the dominant frequencies of $\mathcal{F}(\nu)$ in the logarithmic scale is shown in {Fig.~\ref{heatmap_n10}} for $N=10$ in the closed and open boundary conditions. These plots show that the dominant frequencies logarithmically drop close to the critical lines and at the edges. Comparing {Fig.~\ref{heatmap_n10}} and {Fig.~\ref{finite_phase}}, we observe that heatmap displays indications of the phase transition in the Floquet Ising system.

\section{Conclusion}
\label{conclusion}
We calculated the exact analytical expression for  TMOTOC  as a function of $\tau_0$ and $\tau_1$. With the help of the analytical formulation, we calculated the speed of commutator growth for the TMOTOC and compared it with those of the LMOTOC. We also analyzed the revival of the initial state and found that the TMOTOC revived back within a finite time while  LMOTOC did not. Further, we study the phase structure of the traverse field Floquet system given by Eq.~(\ref{U_f}) using numerical calculation of LMOTOC. We use LMOTOC defined in equation Eq.~(\ref{F_x}) to distinguish between the paramagnetic and ferromagnetic phases of the chosen Floquet system.  Ferromagnetic and $\pi$ ferromagnetic phase or paramagnetic and  $0 \pi$ paramagnetic phases are distinguished by the combined eigenvalues of unitary operator $\hat U$ and parity operator $\hat P$ along the edges of the phase structures. We numerically find the time averaged LMOTOC [$\overline{F}_x^{l,l}(T)$] for the system sizes up to $N=10$ and plot the regions of the parameter space that have $\overline{F}_x^{l,l}(T)=0$ and $\overline{F}_x^{l,l}(T)>0$ for $T=10^4$. We observe that the plot showing the critical lines of phase transition for $N \rightarrow \infty $ tends to the expected plot Fig.~\ref{four_phase}. In the limit $N \rightarrow \infty $, the regions with $\overline{F}_x^{l,l}(T)>0$ for large $T$ are ferromagnetic and those with $\overline{F}_x^{l,l}(T)=0$ for large $T$ are paramagnetic.\\ 
OTOCs can be experimentally calculated \cite{li2017measuring}, and Floquet systems can be experimentally realized \cite{Chitsazi2017}. Our study outlines the analytical calculation of the TMOTOC, its behavior with the separation between the  observables, and how LMOTOC can be a useful tool to distinguish the phases of a Floquet system.
\par
In the next chapter, we will discuss three different regimes such as  characteristic, dynamic, and saturation regions of LMOTOC and TMOTOC in both integrable and nonintegrable Ising spin Floquet systems.

%

\chapter{Characteristic, dynamic and near saturation regions of Out-of-time-order correlation in Floquet Ising models}  

\ifpdf
    \graphicspath{{Chapter1/Figs/Raster/}{Chapter1/Figs/PDF/}{Chapter1/Figs/}}
\else
    \graphicspath{{Chapter1/Figs/Vector/}{Chapter1/Figs/}}
\fi
\section{Introduction}
\label{Introduction}
 Larkin and Ovchinnikov first introduced the concept of out-of-time-order correlation (OTOC) for defining approaches from quasi-classical  to quantum systems \cite{larkin1969quasiclassical}.  In recent years OTOCs have gotten attention in various fields \cite{singh2022scrambling,kitaev2014hidden, shenker2015stringy,gutzwiller1990chaos,haake1991quantum,hosur2016chaos,rozenbaum2017lyapunov,garcia2018chaos,shukla2021,garcia2018chaos,rozenbaum2020early} such as quantum chaos and information propagation in quantum many-body systems \cite{maldacena2016bound, stanford2016many,aleiner2016microscopic,roberts2016lieb, roberts2015localized,bilitewski2018temperature,das2018light}, quantum entanglement and quantum-information delocalization  \cite{hosur2016chaos,huang2017out,wei2018exploring,lin2018out,abeling2018analysis,daug2019detection,grozdanov2018black}, static and dynamical phase transitions  \cite{heyl2018detecting,chen2020detecting, shukla2021}. Several proposals for experimental measurement of OTOC are proposed  \cite{yao2016interferometric,swingle2016measuring,zhu2016measurement,campisi2017thermodynamics,halpern2017jarzynski} using cold atoms or cavity and circuit quantum electrodynamics (QED) or trapped-ion simulations. Experimental realisations have been made using nuclear spins of molecules \cite{wei2018exploring,chen2020detecting,li2017measuring}, trapped ions \cite{landsman2019verified,joshi2020quantum}, and ultra-cold gases \cite{meier2019exploring}.
  Chaotic characteristics of OTOC are manifested if a small disturbance in the input of the system provides exponential deviation to the output of the system, which is known as butterfly effect \cite{bilitewski2018temperature,gu2016}. 
 
 Classical Hamiltonian systems, which have highest amount of randomness and chaos, are converted into the quantum domain for seeing the behavior of quantum chaos \cite{gutzwiller1990chaos,haake1991quantum}.
  OTOC finds a role in characterizing the quantum chaos in these systems. There exist a characteristic form of growth of OTOC that can distinguish different classes of information scrambling. In a chaotic case, OTOC grows very fast, which is often described by an exponential
behavior with a Lyapunov exponent. If the chaos is absent, the growth of OTOC can be much slower or even absent. In disordered systems, OTOC
distinguishes many-body localization \cite{Oganesyan2007,altman2015universal,nandkishore2015many} from the Anderson
localization \cite{anderson1958absence}.
\par

 Growth of OTOC is  also discussed in spin systems  \cite{lin2018out,xu2020accessing,xu2019locality,kukuljan2017weak,Fortes2019,craps2020lyapunov,roy2021entanglement,yan2019similar,bao2020out,dora2017out,Riddell2019,lee2019typical}. Power-law growth of OTOC is observed in the dynamic region of Luttinger liquid model \cite{dora2017out}, XY model \cite{bao2020out}, integrable quantum Ising chain \cite{lin2018out} and some systems exhibiting many-body localization \cite{Riddell2019,lee2019typical}. Similar studies have been done in the Ising model with tilted magnetic fields, perturbed XXZ model, and Heisenberg spin chain with random magnetic fields \cite{Fortes2019}. In these systems, OTOC is calculated for different types of observables. For the observables that are local, non-local or mixed in terms of the Jordon-Wigner (JW) fermions, the OTOCs grow as power-law in time \cite{lin2018out}. 
\par
The quantum systems periodically driven by external forces received considerable attention for a long time. Examples are:  kicked-rotor model in which a particle moving on a ring and field is applied in the form of kicks \cite{casati1979lecture}, Chirikov standard map \cite{chirikov1971research}, and the Kapitza pendulum \cite{kapitza1951dynamic}. These systems show a transition from  integrability to chaos, dynamical localization \cite{Rahav2003,Rahav2003t}, and dynamical stabilization \cite{kapitza1951dynamic,landau2007mechanics}. In recent years, in the quantum domain such as time crystal \cite{zhang2017observation, Russomanno2017}, the topological system with ultra-cold atoms \cite{wintersperger2020realization, Zhu2021}, a periodically-driven quantum system such as particle moving in a modulated harmonic trap
\cite{PhysRevX.4.031027}, kicked quantum rotors \cite{PhysRevLett.80.4111, PhysRevLett.87.074102,d2013many}, Floquet spin systems with constant fields \cite{gritsev2017,lakshminarayan2005multipartite,d2014long, naik2019controlled, shukla2021,Mishra2015} and quenched fields \cite{mishra2014resonance, Rossini2010, essler2016quench,Russomanno2012,Russomanno2013} got considerable attention. Periodic perturbation can be realized in experiments to understand specific properties of matter \cite{ovadyahu2012suppression,iwai2003ultrafast,kaiser2014optically, PhysRevLett.80.4111}. OTOC generated by the sum of quadratic and composite observables in terms of  Majorana fermions is studied in integrable and nonintegrable kicked quantum Ising system \cite{kukuljan2017weak} shows linear growth with time and starts to saturate at $t \simeq N/2$, where, $N$ is the system size. OTOCs using local and nonlocal observables for Floquet XY and synchronized Floquet XY
models are also studied recently  \cite{zamani2022out}. In our previous study \cite{shukla2021}, we could get the phase structure using time-averaged longitudinal magnetization OTOC (LMOTOC), but transverse magnetization OTOC (TMOTOC) failed to give us the phase diagram. While thoroughly understanding the comparison between the initial and the time-averaged behavior of integrable TMOTOC and LMOTOC, we found the different characteristic times. In this chapter, we carry out a comprehensive study of the entire region of OTOC in the integrable as well as nonintegrable Floquet spin models, not just the initial time or averaged behavior.
We will analyze whether the integrability breaking term changes the growth of OTOC.  We extract the differences and similarities of TMOTOC and LMOTOC for integrable and nonintegrable models.
\vspace{-.1 cm }\\ 
This chapter is structured as follows: In section \ref{Ch3_model}, we will discuss the Floquet transverse Ising models. Subsequently, in section \ref{Ch3_OTOC}, we will define transverse magnetization OTOC (TMOTOC) and longitudinal magnetization  OTOC (LMOTOC).  Later, we discuss results in section \ref{result} while comparing the calculations of integrable and nonintegrable Floquet transverse Ising models in both TMOTOC and LMOTOC. Finally, we conclude the results in section \ref{Ch3_conclusion}.

\section{Model}
\label{Ch3_model}
Consider a periodically driven interacting transverse Ising Floquet system. The Hamiltonian of the system is given as 
\begin{equation}
\hat H(t)=J_x\hat H_{xx}+h_{z}\sum_{n=-\infty}^{\infty}\delta\Big(n-\frac{t}{\tau}\Big) \hat H_z,
\label{int_hamiltonian}
\end{equation}
where $J_x$ is the nearest-neighbor exchange coupling strength, and $h_z$ is the external field in the transverse direction applied in the form of kicks at equal intervals of time $\tau$. $\hat H_{xx}=\sum_{l=1}^{N}\hat \sigma^l_x\hat \sigma^{l+1}_x $ is the nearest-neighbor Ising interaction term and $\hat H_z=\sum_{l=1}^{N}\hat \sigma^l_z$ is the interaction of unit magnetic field with the total transverse magnetization. 
\par 
Floquet map corresponding to the Eq.~(\ref{int_hamiltonian}) is 
\begin{equation}
\label{U0}
 \mathcal{\hat U}_0=\exp(-i \tau J_x \hat H_{xx})  \exp( -i \tau h_{z} \hat H_{z}),
\end{equation}
Since in Eq.~(\ref{int_hamiltonian}) there is only transverse field is present, and the Hamiltonian is exactly solvable using Jordan-Wigner (JW) transformation \cite{lakshminarayan2005multipartite,prosen2000exact,prosen2002general}. Now, if we introduce a longitudinal field term $  h_x \hat H_x=h_x\sum_{l=1}^{N}\hat \sigma^l_x$, the Hamiltonian can be written as  
\begin{equation}
\hat H(t)=J_{x}\hat H_{xx}+h_{x}\hat H_x+h_{z}\sum_{n=-\infty}^{\infty}\delta\Big(n-\frac{t}{\tau}\Big) \hat H_z.
\label{H0}
\end{equation}
However, the model could not be transformed into the free fermions using JW transformation because the longitudinal field term, when transformed into JW fermions, gives an interacting fermionic term \cite{prosen2000exact,prosen2002general}. 
The Floquet map corresponding to this model is
\begin{equation}
\label{Ux}
 \mathcal{\hat U}_x=\exp\big[-i \tau (J_x \hat H_{xx}+h_{x}\hat H_{x})\big]  \exp( -i \tau h_{z} \hat H_{z}).
\end{equation}
Henceforth in the chapter, we mean integrable transverse Ising Floquet model as $\mathcal{\hat U}_0$ and nonintegrable transverse Ising Floquet model as $\mathcal{\hat U}_x$.  

\section{ TMOTOC and LMOTOC}
  \label{Ch3_OTOC}
Let us consider a pair of observables $\hat W^l$ and $\hat V^m$ at $ l^{\rm th}$ and $ m^{\rm th}$ sites, respectively. OTOC of these observables is defined as
\begin{equation}
C^{l,m}(n)=-\frac{1}{2}\langle[\hat W^l(n), \hat V^m(0)]^\dagger[\hat W^l(n), \hat V^m(0)] \rangle.
\end{equation}
Observables, $\hat W^l$ and $\hat{V}^m$ are separated by distance $\Delta l=\vert l-m \vert$. Initially at $n=0$, both the observables commute to each other{\it, i.e.} $[\hat W^l(0), \hat V^m(0)]=0$. As time increases, higher order terms of the time evolution of $\hat W^l(0)$ given by the Baker-Campbell-Hausdorff formula  fail to commute with $\hat V^m$, resulting in noncommutative  $\hat W^l(n)$ and $\hat{V}^m$. By examining the noncommutativity of $\hat V^m$ at different positions, one can quantify upto some degree how $\hat W^l(n)$ spread over the space. Here $\hat W^l(n)$ is  $ (\mathcal{\hat U}_{x/0}^{\dagger})^n\hat W^l(0)  (\mathcal{\hat U}_{x/0})^n$.  If $\hat W^l$ and $\hat V^m$ are Hermitian and unitary, the OTOC simplifies in the form
\begin{equation}
\label{gene_OTOC}
C^{l,m}(n)=1-\Re[F^{l,m}(n)],
\end{equation}
where, $F^{l,m}(n)=\langle \hat W^l(n) \hat V^m(0) \hat W^l(n) \hat V^m(0) \rangle$ and $\langle\cdot \rangle$, denotes the quantum mechanical  average  over  the  initial state.
\par
OTOC is calculated with either trace over a maximally mixed state or a thermal ensemble. Trace can be replaced by employing Haar random states of $2^N$ dimensions to evaluate expectation values, that is 
\begin{equation}
\mbox{Tr}(\hat W^l(n) \hat V^m(0) \hat W^l(n) \hat V^m(0))/2^N \approx \left \langle \Psi_R\big\vert\hat W^l(n) \hat V^m(0) \hat W^l(n) \hat V^m(0)\big\vert\Psi_R\right \rangle
\end{equation}
where $|\Psi_R\rangle$ is a random state.  We replaced the random state by two fully polarized special initial states according to the observables and found that there are no remarkable differences in the characteristic, dynamic, and saturation regions of  OTOC. We observe only one difference in the saturation region, and there are comparatively small oscillations when considering a random state. Detailed discussion is mentioned in Appendix~\ref{Appendix_B1}. Moreover, the special initial states may help to get the exact analytical formula, at least for integrable OTOC cases with transverse direction spins as observables.
\par
In this chapter, we consider $\hat W^l$ and $\hat V^m$ as local Pauli operators either in the longitudinal direction $\hat \sigma^{l,m}_x$ or in the transverse direction $\hat \sigma^{l,m}_z$. For the Pauli operators in the transverse direction as local observables, the OTOC is defined as transverse magnetization OTOC (TMOTOC) and given as: 
\begin{equation}
\label{Ch3_F_z}
C_z^{l,m}(n)=1- \Re[F_z^{l,m}(n)],
\end{equation}
where, $F_z^{l,m}(n)=\langle \phi_0|\hat \sigma^l_z(n)\hat \sigma^m_z\hat \sigma^l_z(n)\hat \sigma^m_z|\phi_0\rangle$. In the fermionic representation, $\hat \sigma_z^l$ can be written as $\hat \sigma_z^{l}=-(\prod_{j<l}A^jB^j)A^l$, where, $A^l$ and $B^l$ are defined by fermionic creation ($c^{l\dagger})$ and annihilation operator ($c^l$) as, $A^l=c^{l\dagger}+c^l$  and $B^l=c^{l \dagger}-c^l$ \cite{sachdev2011}. Since $\hat \sigma_z^{l}$ contain string  operator, hence it is known as non-local operator in terms of Jordan-Wigner fermion \cite{sachdev2011,lin2018out}.

For the calculation purpose we take initial state as $|\phi_0\rangle=| \uparrow  \uparrow  \uparrow \cdots  \uparrow \rangle$, 
where $ \left| \uparrow \right\rangle$ is the eigenstate of $\hat \sigma_z$ with eigenvalue $+1$.
If the observables are taken as Pauli operators in the longitudinal direction of the Ising axis ({\it i. e.,} z-axis), then OTOC will be referred to as longitudinal magnetization OTOC (LMOTOC). The LMOTOC is given as follows:
\begin{equation}
\label{Ch3_F_x}
C_x^{l,m}(n)=1- \Re[F_x^{l,m}(n)],
\end{equation}
where,
$F_x^{l,m}(n)=\langle
 \psi_0|\hat \sigma_x^l(n)\hat \sigma_x^m\hat \sigma^l_x(n)\hat \sigma^m_x|\psi_0\rangle$. In the fermionic representation, $\hat \sigma^{l/m}_x$ can be written as $\hat \sigma_x^{l/m}=A^{l/m}B^{l/m}$. In fermionic representation $\hat \sigma_x^{l/m}$ is known as local observable \cite{sachdev2011,lin2018out}.
 In this case the initial state will be taken as  $|\psi_{0} \rangle=|\rightarrow \rightarrow \rightarrow \cdots \rightarrow \rangle$, where, $ \left| \rightarrow \right\rangle$ is the eigenstate of $\hat \sigma_x$ with eigenvalue $+1$.  
\subsection{Analytical formula of TMOTOC}
Analytical solution of the TMOTOC for the initial state $|\phi_0\rangle=| \uparrow  \uparrow  \uparrow \cdots  \uparrow \rangle$ and Floque  map defined by Eq.~(\ref{U0}) with $J_x=1$ and $h_z=1$ is derived in the Ref.~\cite{shukla2021} as
\begin{eqnarray}
\label{Ch3_OTOCz_gene}
 F_z^{l,m}(n) &=& 1- \Big(\frac{2}{ N}\Big)^3 \sum_{p,q,r} \Big[ e^{i(p-q)(m-l)}|\Psi_r(n)|^2 \Phi_p^{*}(n) \Phi_q(n) \nonumber \\
 &-& e^{i(-r-q)(m-l)}  \Psi_r(n)^{*} \Phi_p^{*}(n)  \Phi_q(n) \Psi_{-p}(n)  \nonumber \\
&-& e^{i(p+q)(m-l)} \Psi_{q}(n)  \Psi_{r}(n)^{*}  \Phi_{p}(n)^{*} \Phi_{-r}(n)\nonumber \\
&+& e^{i(q-r)(m-l)} \Psi_{q}(n)\Psi_r(n)^* |\Phi_p(n)|^2 \Big],  
\end{eqnarray}
where the expansion coefficients  $\Phi_q(n)$ and  $\Psi_q(n)$ are defined as 
\begin{equation}
\label{Ch3_phi}
\Phi_q(n)=|\alpha_{+}(q)|^2 e^{-i n \gamma_q}+|\alpha_{-}(q)|^2 e^{i n \gamma_q},
\end{equation}
\begin{equation}
\label{Ch3_psi}
\Psi_q(n)=\alpha_{+}(q) \beta_{+}(q)e^{-i n \gamma_q}+\alpha_{-}(q)\beta_{-}(q) e^{i n \gamma_q}.
\end{equation}
The phase angle $\gamma_q$ and the coefficients $\alpha_{\pm}(q)$ and $\beta_{\pm}(q)$ are given by
\begin{equation}
\label{Ch3_gamma}
\cos(\gamma_q)=\cos(2 \tau)\cos(4\tau)-\cos(q)\sin(2\tau)\sin \Big(2\tau \Big),
\end{equation}
and
\begin{equation}
\label{Ch3_apmq}
\alpha_{\pm}(q)^{-1}=\sqrt{1+\Big(\frac{\cos(2\tau)-cos(\gamma_q \pm 2 \tau)}{\sin(q) \sin(
2\tau)\sin(2\tau)}\Big)^2},
\end{equation}

\begin{eqnarray}
\label{Ch3_bpmq}
\beta_{\pm}(q)= \frac{\mp\sin(\gamma_q)-\cos( 2\tau)\sin(2\tau)\big(\cos(q) +1\big)}{\sin(q)\sin(2\tau)}   \alpha_{\pm}(q)e^{-i2\tau}.
\end{eqnarray}
The allowed value of $p$, $q$ and $r$ are from $\frac{-(N-1)\pi}{N}$ to $\frac{(N-1)\pi}{N} $ differing by $\frac{2 \pi}{N} $ for even number of $N_F$ ($N_F=\sum_l c_l^\dagger c_l$, number of fermions) and $\hslash=1$. 
 We use the above exact solution to calculate TMOTOC for integrable $\mathcal{\hat U}_0$ model. However, TMOTOC for the nonintegrable $\mathcal{\hat U}_x$ model, and LMOTOC for both integrable $\mathcal{\hat U}_0$ and nonintegrable $\mathcal{\hat U}_x$ model will be calculated numerically.

\begin{figure}[hbt!]
\centering
 \includegraphics[width=.90\linewidth, height=.50\linewidth]{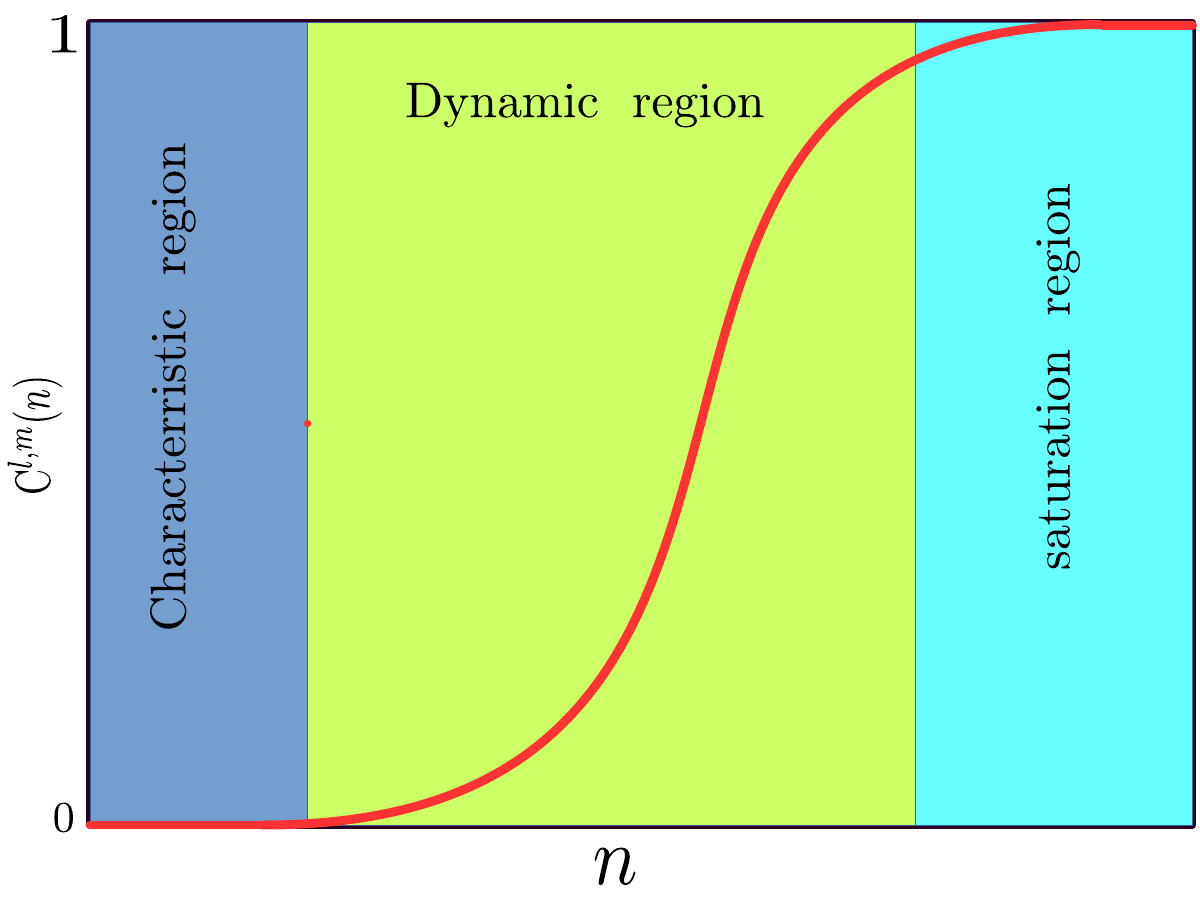}
 \caption{Schematic of the various regions of OTOC in a typical system.} 
   \label{otocfig}
  \end{figure}
 
\section{Results}
\label{result}
We analyze TMOTOC and LMOTOC for both $\mathcal{\hat{U}}_0$ and $\hat{\mathcal{U}}_x$ models in three regions as depicted in the Fig.~\ref{otocfig}. These are, namely:

\begin{enumerate}
    \item[i]) {\it Characteristic Region:} Both the observables $\hat W^l$ and $\hat V^m$ commute with each other till the characteristic time ($t_{\Delta l}$), which is defined as time after that $C^{l,m}_{z/x}(n)(F^{l,m}_{z/x}(n))$ is departed from zero (one). The Characteristic time depends upon the separation between the spins ($\Delta l=|l-m|$). As we increase the separation between the spins, the characteristic time increases, and it is independent of the  Floquet period and system size. 
    
    \item[ii]) {\it Dynamic Region:} After the characteristic time, $C^{l,m}_{z/x}(n)$ becomes nonzero.  In the dynamic region $C^{l,m}_{z/x}(n)$ increases rapidly.
    
    \item[iii]) {\it Near saturation Region:} After rapid growth, $C^{l,m}_{z/x}(n)$ starts to saturate to a finite value. However, the manner in which  $C^{l,m}_{z/x}(n)$ saturates follows some trend with an oscillating amplitude. Such trend we calculate by analysing behaviour of $\Re[F^{l,m}_{z/x}(n)]$ vs. $n$.
\end{enumerate}
\subsection{TMOTOC in the integrable Floquet system}

\begin{figure}
\centering
\begin{subfigure}{.49\textwidth} 
\includegraphics[width=.99\linewidth, height=.70\linewidth]{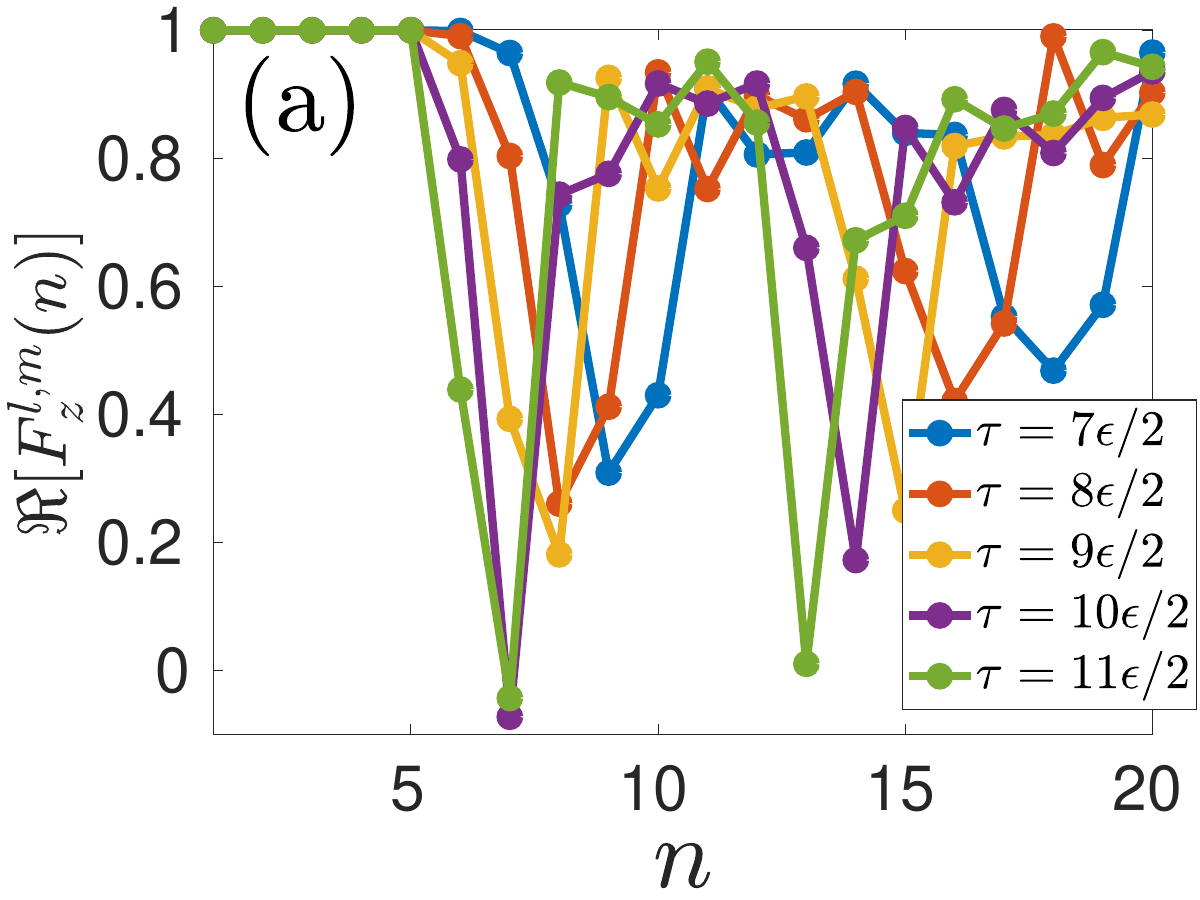}
 \end{subfigure}
 \begin{subfigure}{.49\textwidth} 
 \includegraphics[width=.99\linewidth, height=.70\linewidth]{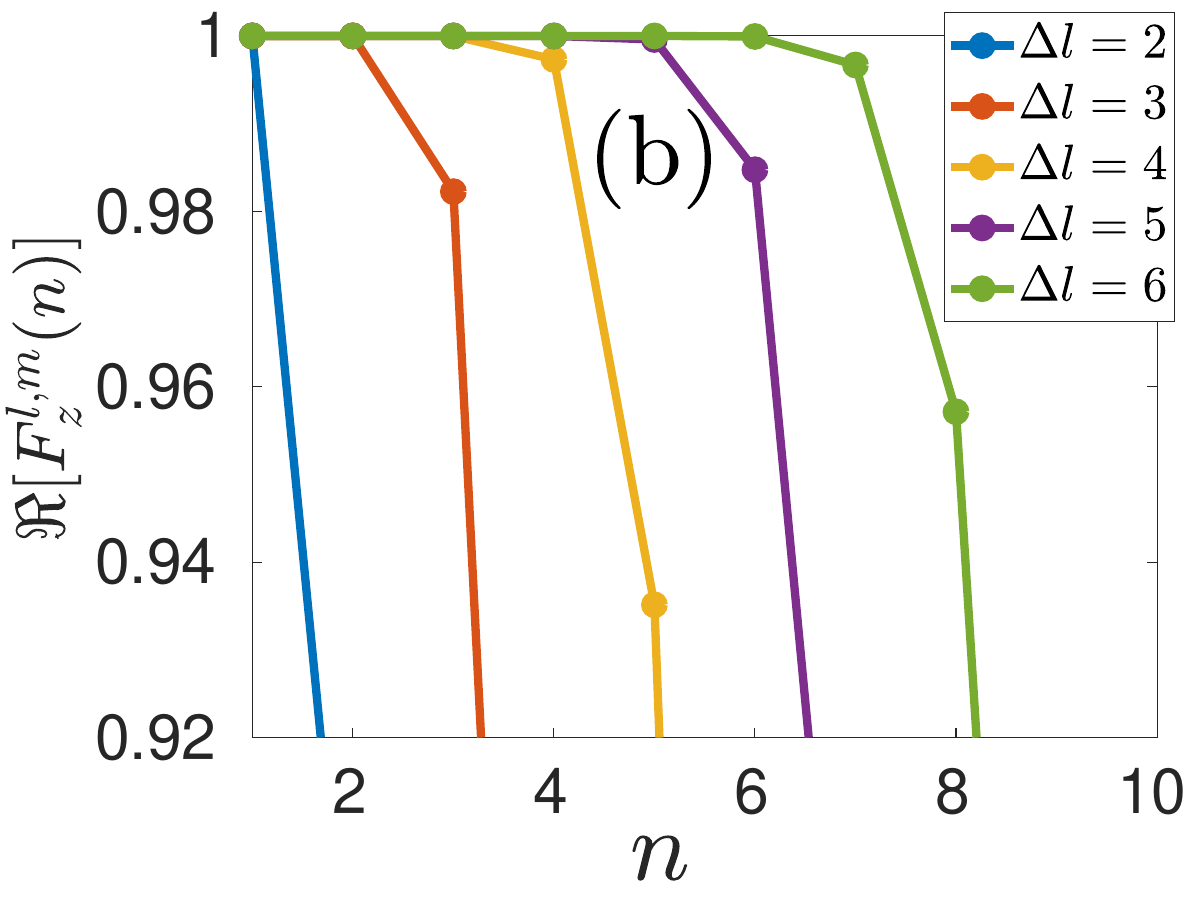}
 \end{subfigure}
\begin{subfigure}{.490\textwidth}
   \includegraphics[width=.99\linewidth, height=.70\linewidth]{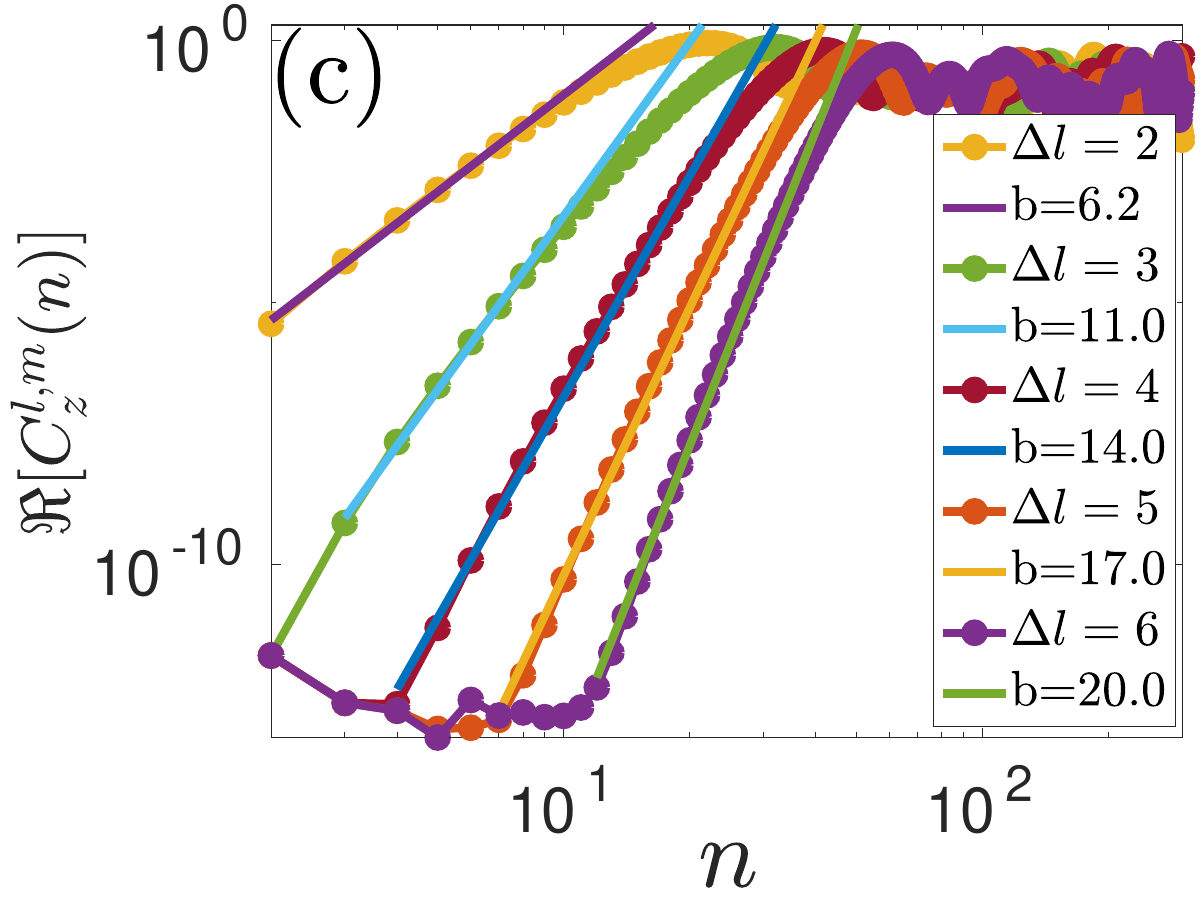}
  \end{subfigure}
     \begin{subfigure}{.490\textwidth}
   \includegraphics[width=.99\linewidth, height=.70\linewidth]{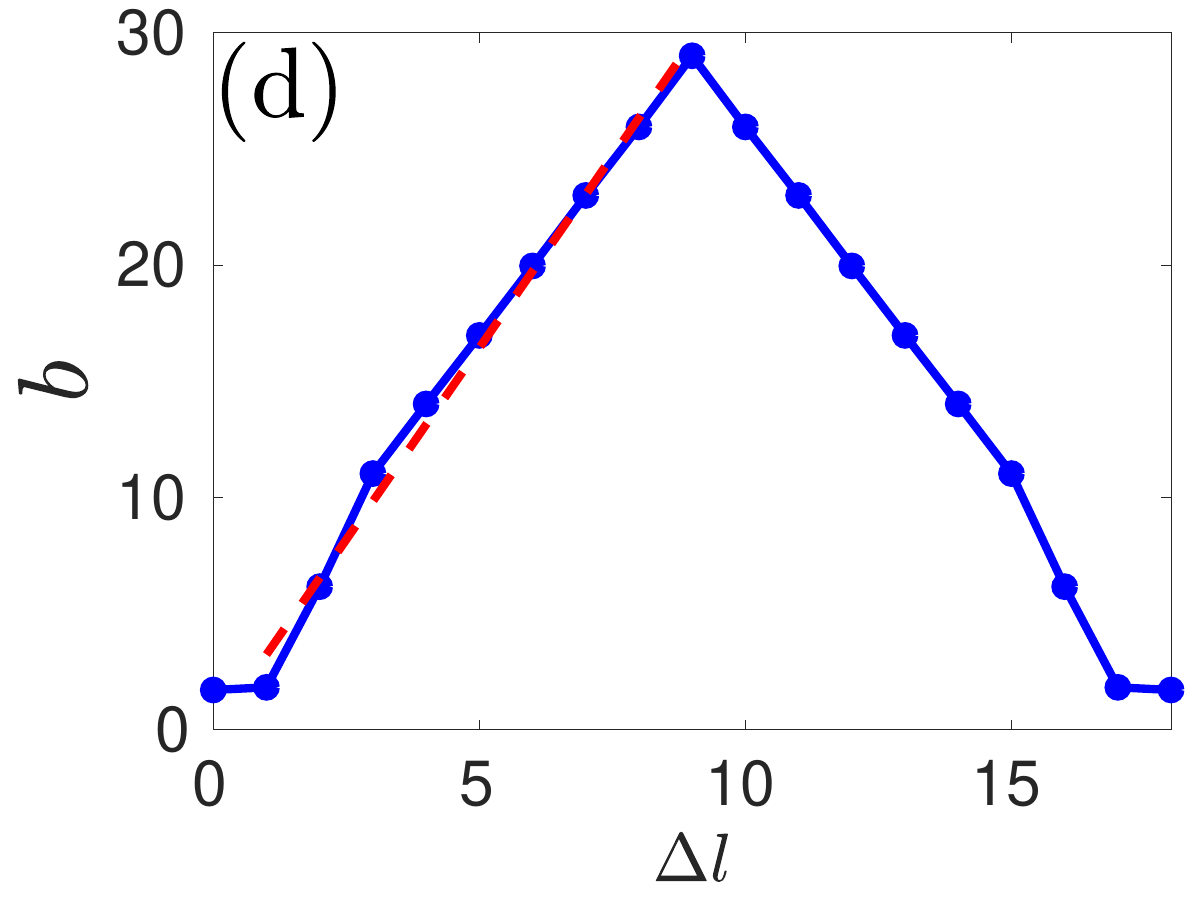}
 \end{subfigure}
  \begin{subfigure}{.490\textwidth}
   \includegraphics[width=.99\linewidth, height=.70\linewidth]{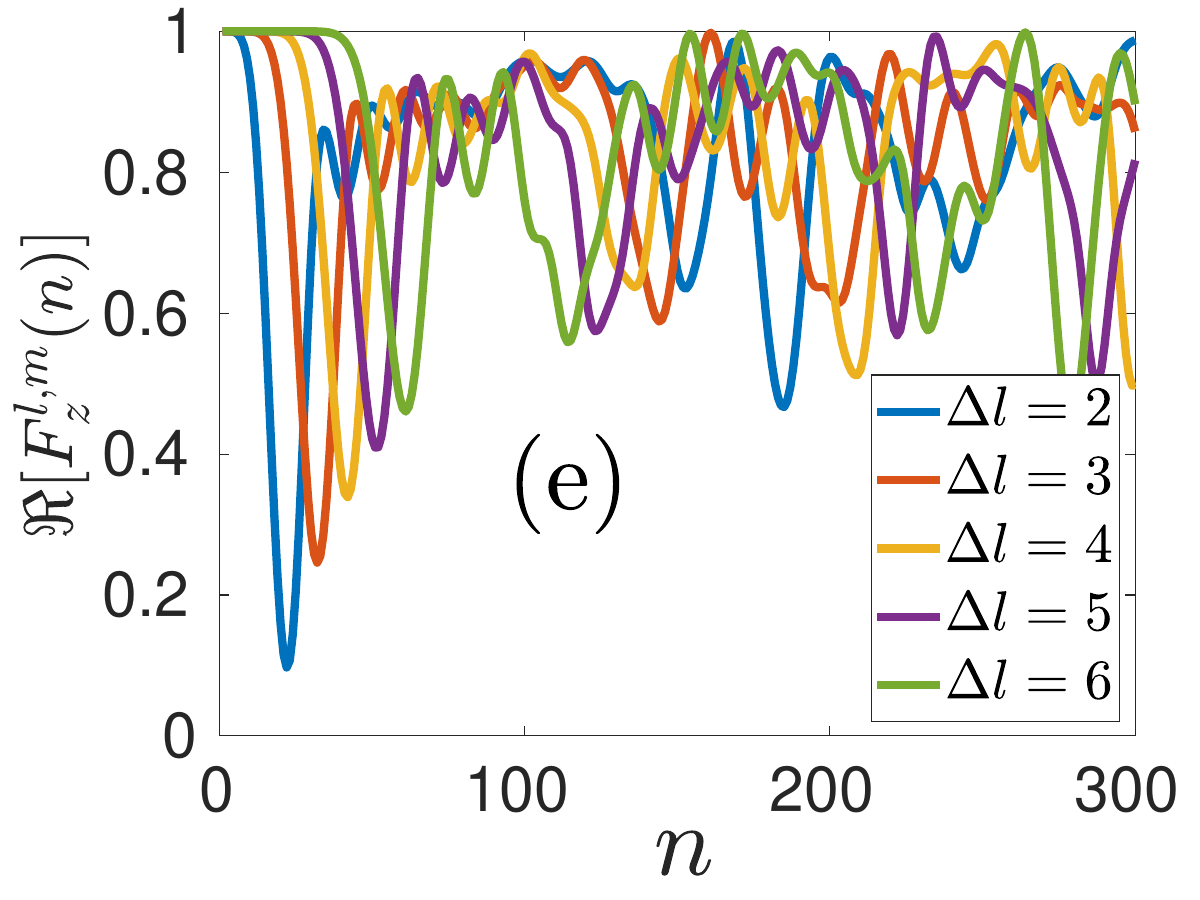}
  \end{subfigure}
 \caption{Integrable transverse Ising Floquet system  with $J_x=1$ and $h_z=1$ for $N=18$. (a) The behavior of $TMOTOC$ with the number of kicks $(n)$ by increasing the value of Floquet period from 
 $\frac{7\epsilon}{2}$ to $\frac{11 \epsilon}{2}$ differing by $\epsilon/2$ with fixed $\Delta l=6$ ($\epsilon=\frac{\pi}{28})$. (b)  $F^{l,m}_z$ with number of kicks by increasing $\Delta l$ and fixed Floquet period 
 $\tau=6\epsilon/2$. (c) $C^{l,m}_z$ with number of kicks ($\log-\log$) with increasing distances ($\Delta l$) between the spins at constant Floquet period $\tau=\frac{\epsilon}{2}$. (d) Exponent of power-law with increasing distance between the spins. (e)  $\Re[F^{l,m}_z]$ with number of kicks at different $\Delta l$.}  
 \label{cf_TMOTOC_int}
\end{figure}

Let us begin with TMOTOC for integrable $\mathcal{\hat{U}}_0$ system defined by Eq.~(\ref{U0}). First, we  focus on the characteristic region of TMOTOC with increasing Floquet period. Let us consider an operator $\hat W$ located at site $l$ initially. One can see that the considered Floquet evolution increases the size of $\hat W$ at each Floquet step. In particular, the left end of the support of $\hat W(n)$ and the right end of the support of $\hat W(n)$ will increase by one for each Floquet step. We, therefore, can see that  $\Re[F_z^{l,m}(n)]=1$ if $n < |l - m|$. However, once $n \geq |l - m|$, $\Re[F_z^{l,m}(n)]$ will start to deviate from $1$. 
\par
\begin{figure}
 \centering
\begin{subfigure}{.490\textwidth} 
 \includegraphics[width=.99\linewidth, height=.70\linewidth]{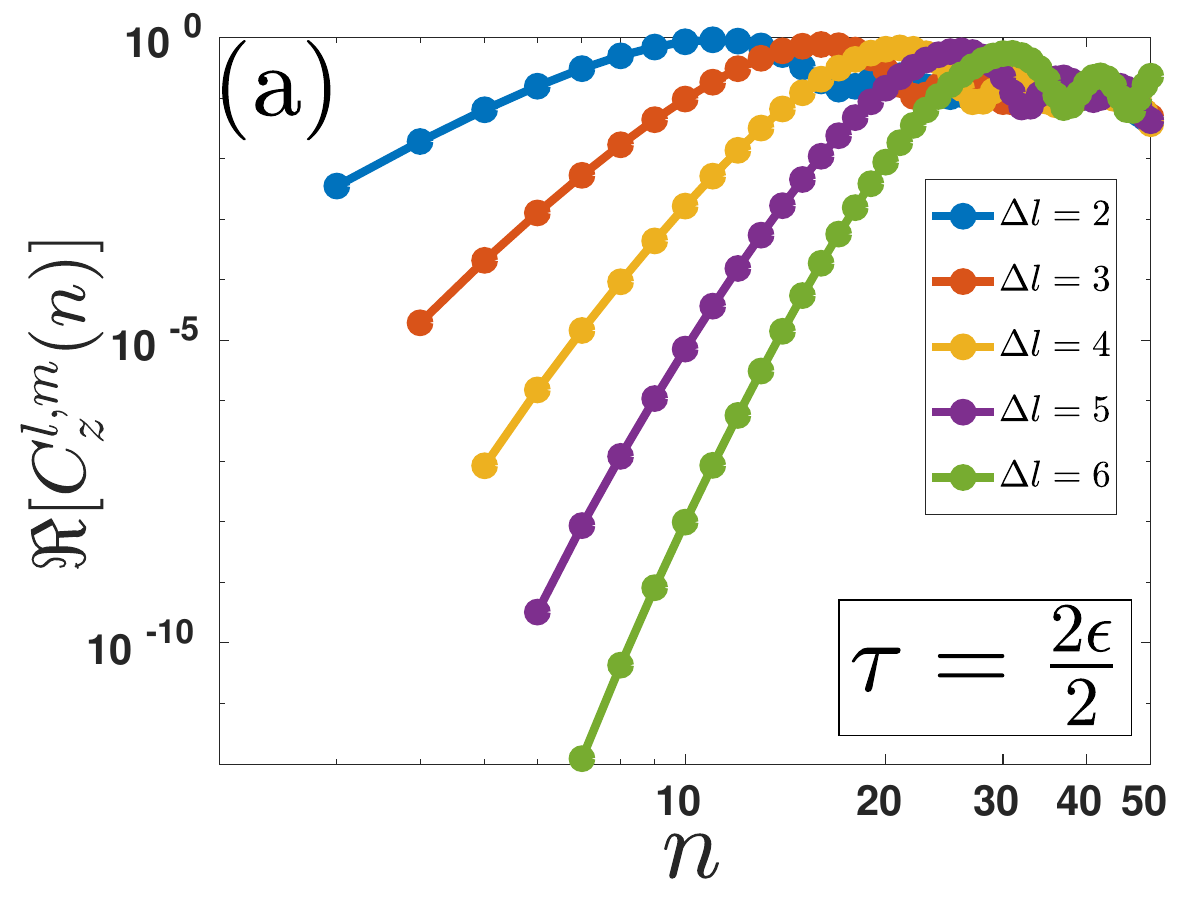}
 \end{subfigure}
 \begin{subfigure}{.490\textwidth} 
 \includegraphics[width=.99\linewidth, height=.70\linewidth]{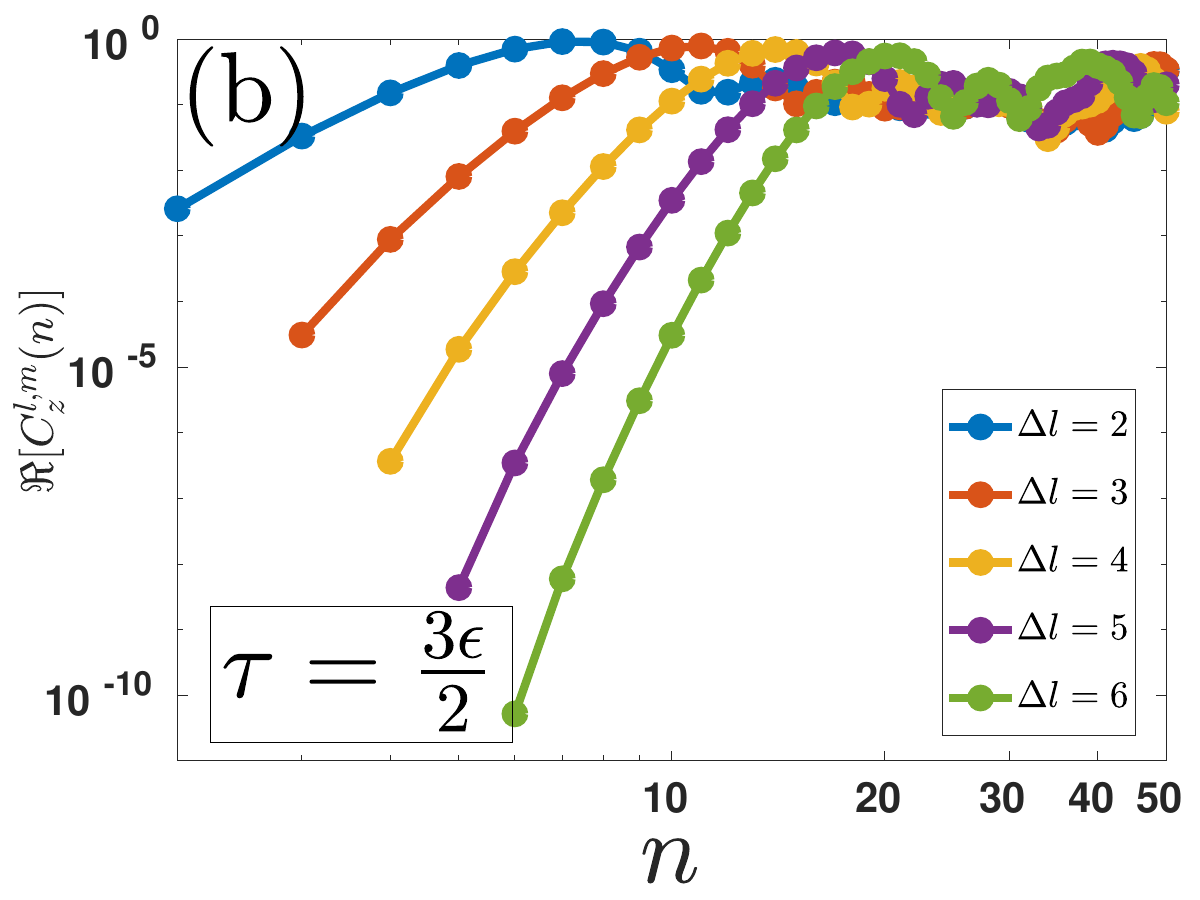}
 \end{subfigure}
 \begin{subfigure}{.490\textwidth}
   \includegraphics[width=.99\linewidth, height=.70\linewidth]{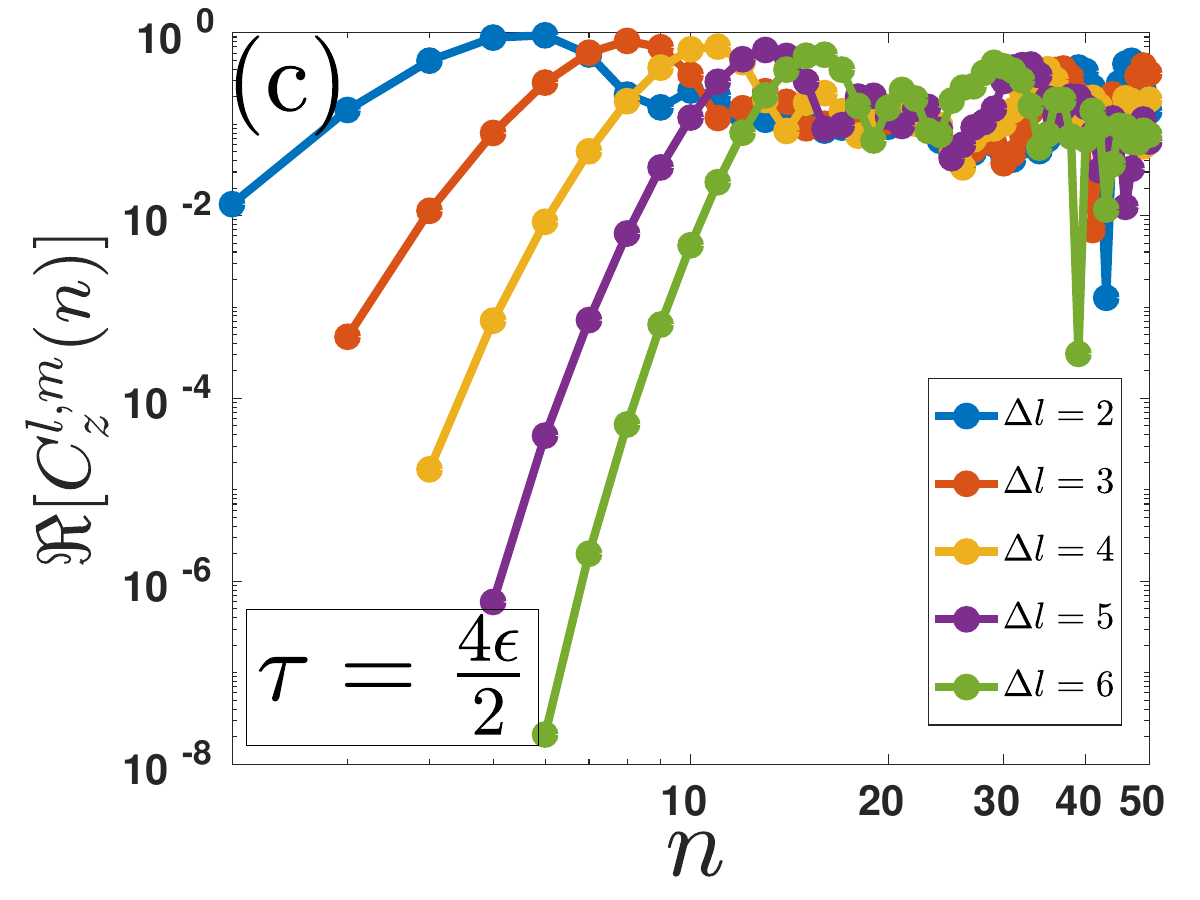}
  \end{subfigure}
     \begin{subfigure}{.490\textwidth}
   \includegraphics[width=.99\linewidth, height=.70\linewidth]{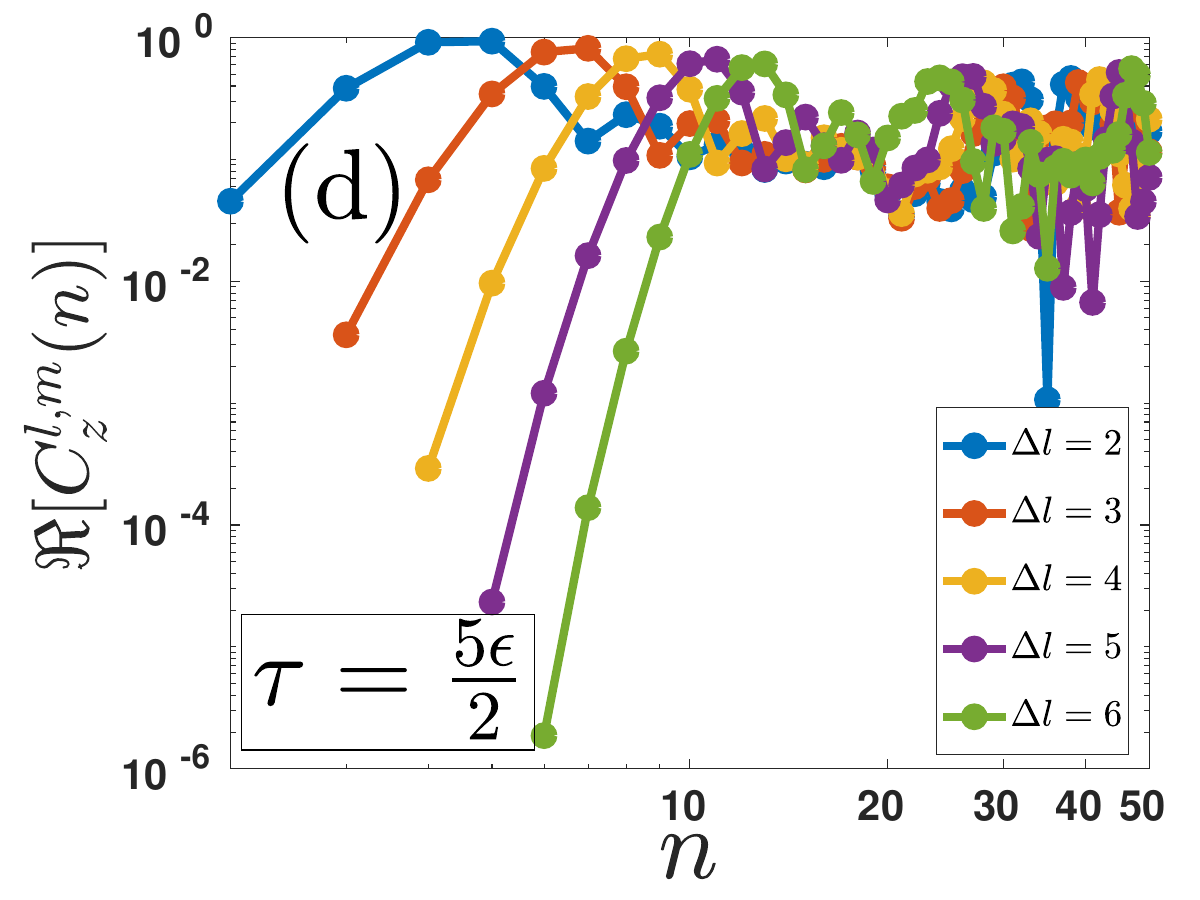}
  \end{subfigure}
  \begin{subfigure}{.490\textwidth}
   \includegraphics[width=.99\linewidth, height=.70\linewidth]{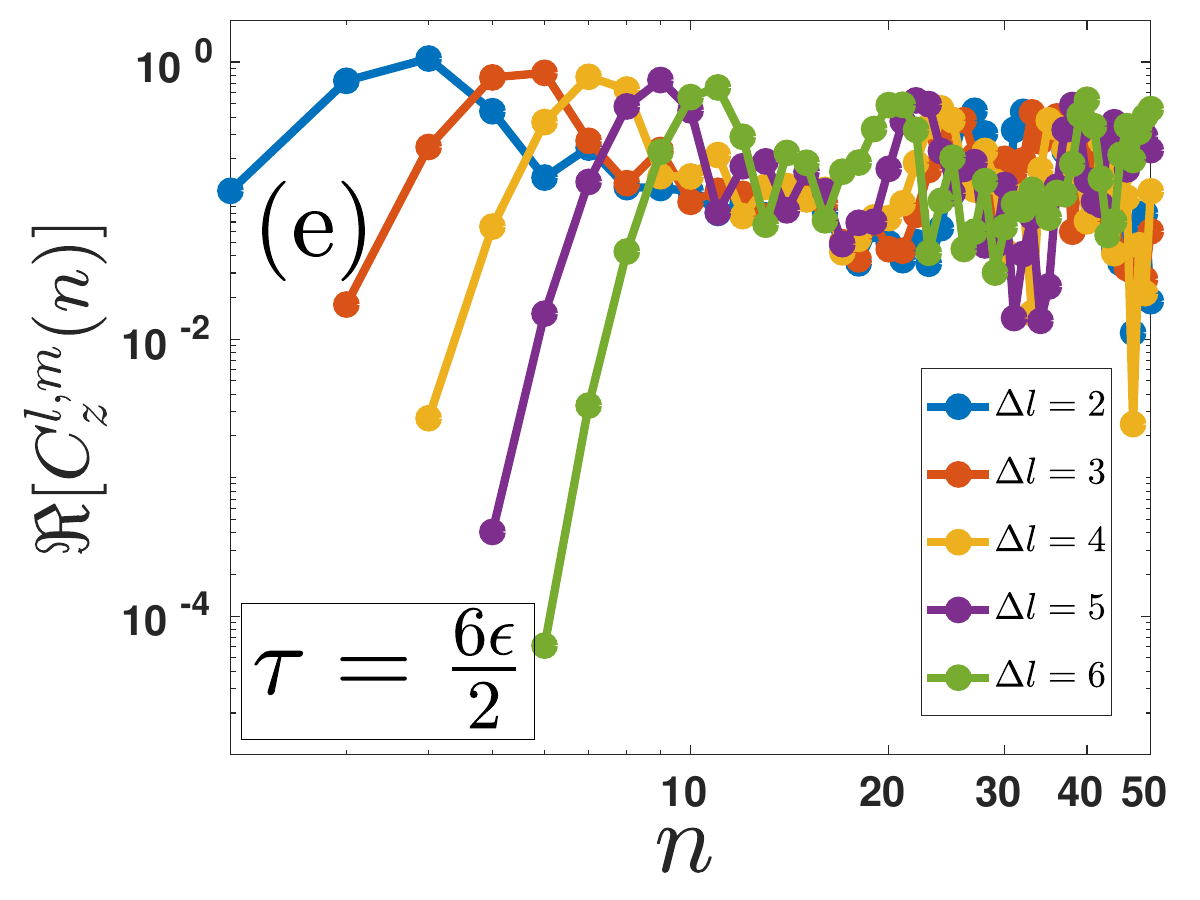}
  \end{subfigure}
  \caption{Integrable transverse Ising Floquet system with $J_x=1$ and $h_z=1$ for $N=18$.  Behaviour of $TMOTOC$ with number of kicks $(n)$  by increasing $\Delta l$ from $2$ to $6$ at different  Floquet period (a) $\tau=\frac{2\epsilon}{2}$, (b) $\tau=\frac{3\epsilon}{2}$, (c) $\tau=\frac{4\epsilon}{2}$, (d) $\tau=\frac{5\epsilon}{2}$ and (e) $\tau=\frac{6\epsilon}{2}$ ($\epsilon=\frac{\pi}{28}$)}.  
 \label{cf_TMOTOC}
\end{figure}

 Fig.~\ref{cf_TMOTOC_int}(a) is the behaviour of $\Re[F_z^{l,m}(n)]$ with increasing Floquet period and fixed $\Delta l=6$. One can see from Fig.~\ref{cf_TMOTOC_int}(a) that $\Re[F_z^{l,m}(n)]$ start to deviate at $\Delta l^{\rm th}(=6^{\rm th})$ kick for all Floquet period $(\tau)$. This characteristic time is independent
of the Floquet period and system size ($N$) (we have checked till $N=50$). For a fixed Floquet period $\tau$, we can see the behavior of $\Re[F^{l,m}_z(n)]$ with the number of kicks and see the dependence of $t_{\Delta l}$ on $\Delta l$. In Fig.~\ref{cf_TMOTOC_int}(b), for 
$\tau=\frac{6\epsilon}{2}$, we show $\Re[F^{l,m}_z(n)]$  vs. the number of kicks by changing the separation between the observables $\Delta l=|l-m|$. We see that increasing the separation between the spins, increases the characteristic time for the TMOTOC case, and number of kicks required to deviate from unity is equal to the separation between the observables ($n=\Delta l$).  The growth of TMOTOC in the dynamic region follows a power-law. The exponent of the power-law increases with increasing the separation between the local spin observables in a systematic manner [Fig.~\ref{cf_TMOTOC_int}(c)]. The exponent increases, reaches the maximum at $\Delta l=\frac{N}{2}$, and further decreases with increasing the distance between the spins [Fig.~\ref{cf_TMOTOC_int}(d)]. The exponent of the power-law can be expressed as a triangular function:
\begin{equation}
b \approx b_{\rm max}-\kappa\Big\vert\frac{N}{2}-\Delta l\Big\vert,  \quad\quad\quad 1\leq\Delta l\leq N-1.
\label{bformula}
\end{equation}
where, the constants $\kappa=3.2$, $b_{\rm max}=29$ and $b_0=1.7$.   Eq.~(\ref{bformula}) shows the dependence of the exponent of power-law with increasing the separation between the observables. It is symmetric about $\Delta l=\frac{N}{2}$ because of the periodic boundary condition of the spin chains. 
$\Re[F^{l,m}_z(n)]$ revives back to unity after a few kicks in the saturation region. Revival  time has nontrivial dependence on $n$ and $\Delta l$    [Fig.~\ref{cf_TMOTOC_int}(e)]. 
The TMOTOC extracted from the analytical expression Eq.~(\ref{OTOCz_gene}) in characteristic, dynamic, and saturation regions can be summarised as  
\begin{equation}
\label{C_int_tmotoc}
C_z^{l,m}(n) \approx \left\{
\begin{aligned}
 0, & \quad\quad\quad n\tau< t_{\Delta l},\\
 (n\tau)^{\kappa\Delta l+2}, & \quad\quad\quad  t_{\Delta l}<n\tau<t_s,\\
 {\rm revived \quad back}, & \quad\quad\quad   t_s<n\tau.
\end{aligned} \right.
\end{equation}
 In the above expression, $t_{\Delta l}$ is characteristic time,  and $t_s$ is the time at which TMOTOC starts saturating.  Dynamic region of TMOTOC decreases with increasing the Floquet period $\tau$ as shown in Fig~\ref{cf_TMOTOC}(a-e). In general the dependence on $\tau$ is such that we can define $C_z^{l,m}\propto (n\tau)^{\kappa\Delta l+2}$ in dynamic region. 

\subsection{TMOTOC in the nonintegrable Floquet system}
\begin{figure}
\centering
\begin{subfigure}{.490\textwidth} 
\includegraphics[width=.99\linewidth, height=.70\linewidth]{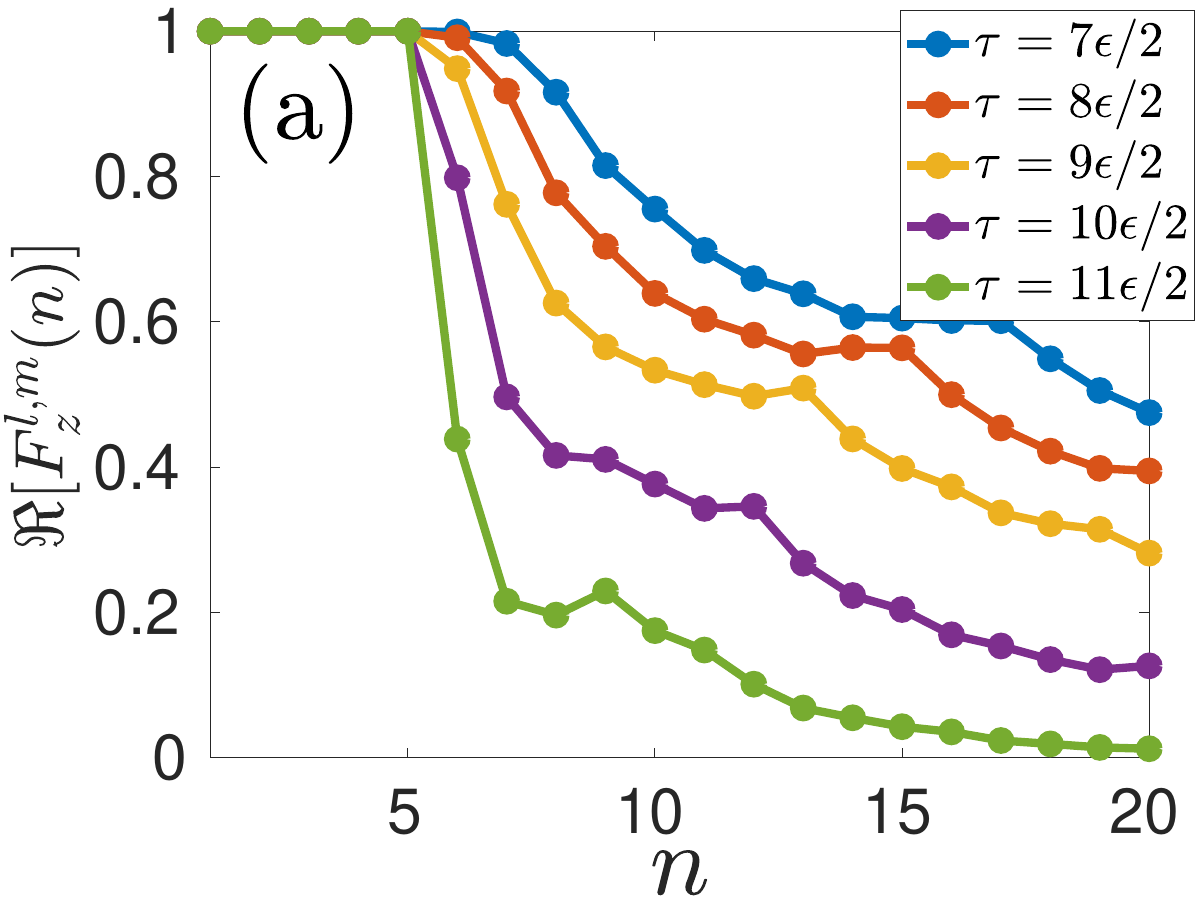}
 \end{subfigure}
 \begin{subfigure}{.490\textwidth} 
 \includegraphics[width=.99\linewidth, height=.70\linewidth]{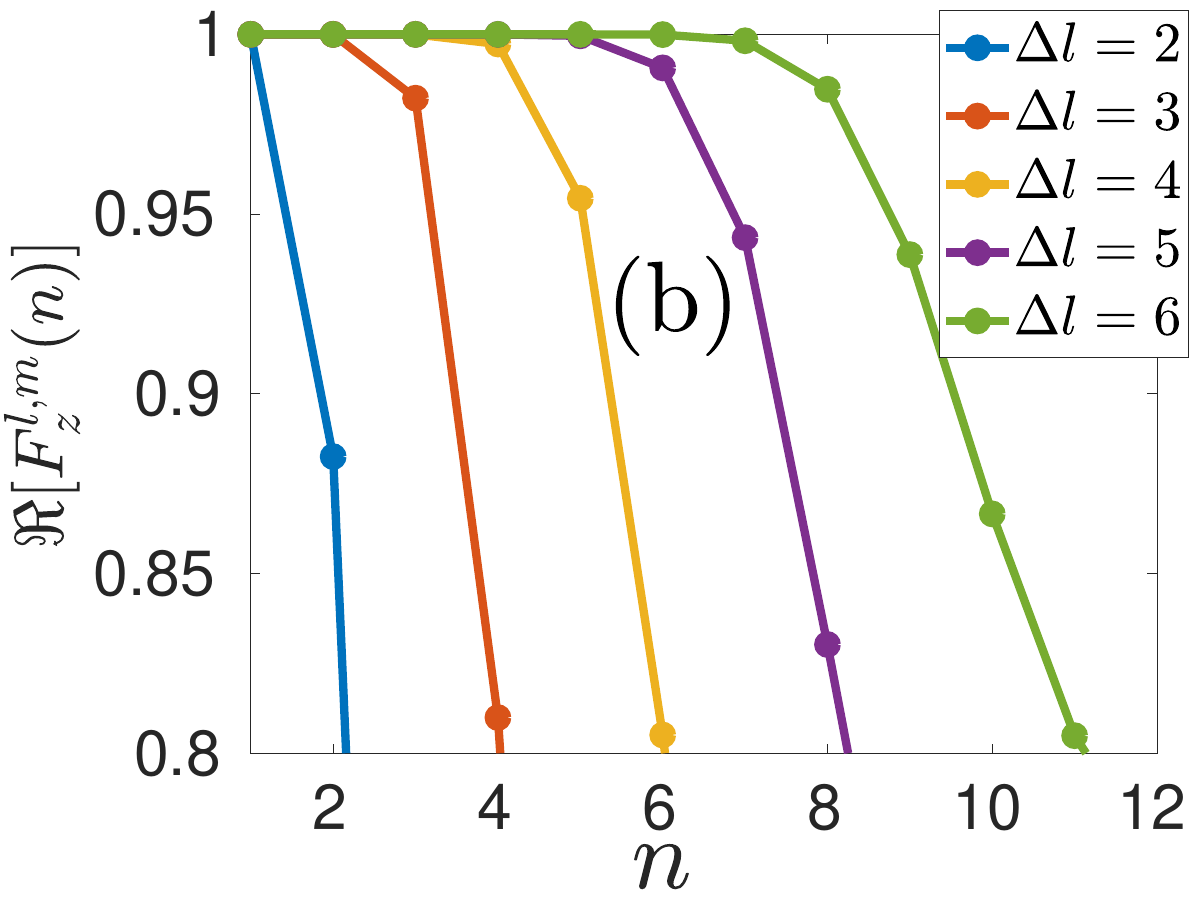}
 \end{subfigure}
 \begin{subfigure}{.490\textwidth}
  \includegraphics[width=.99\linewidth, height=.70\linewidth]{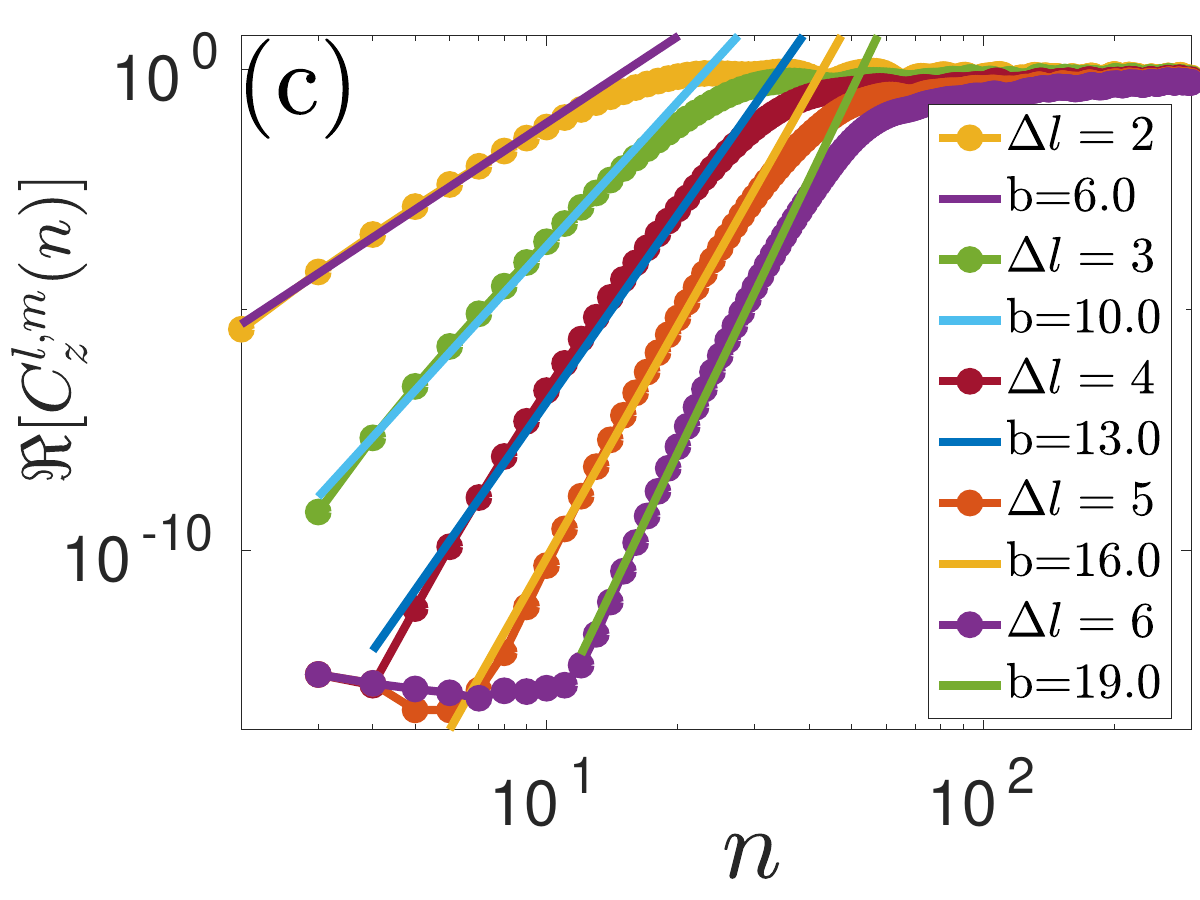}
  \end{subfigure}
      \begin{subfigure}{.490\textwidth}
  \includegraphics[width=.99\linewidth, height=.70\linewidth]{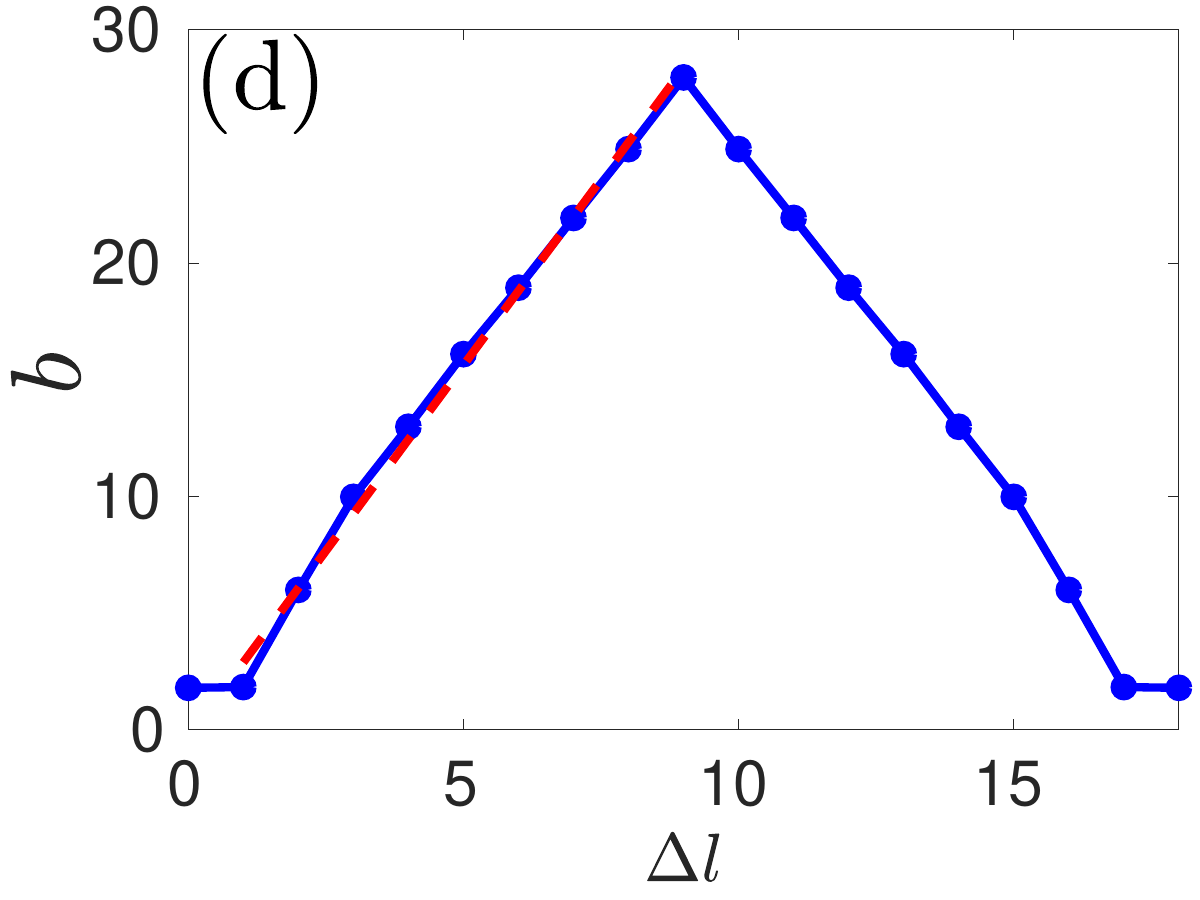}
  \end{subfigure}
  \begin{subfigure}{.490\textwidth}
  \includegraphics[width=.99\linewidth, height=.70\linewidth]{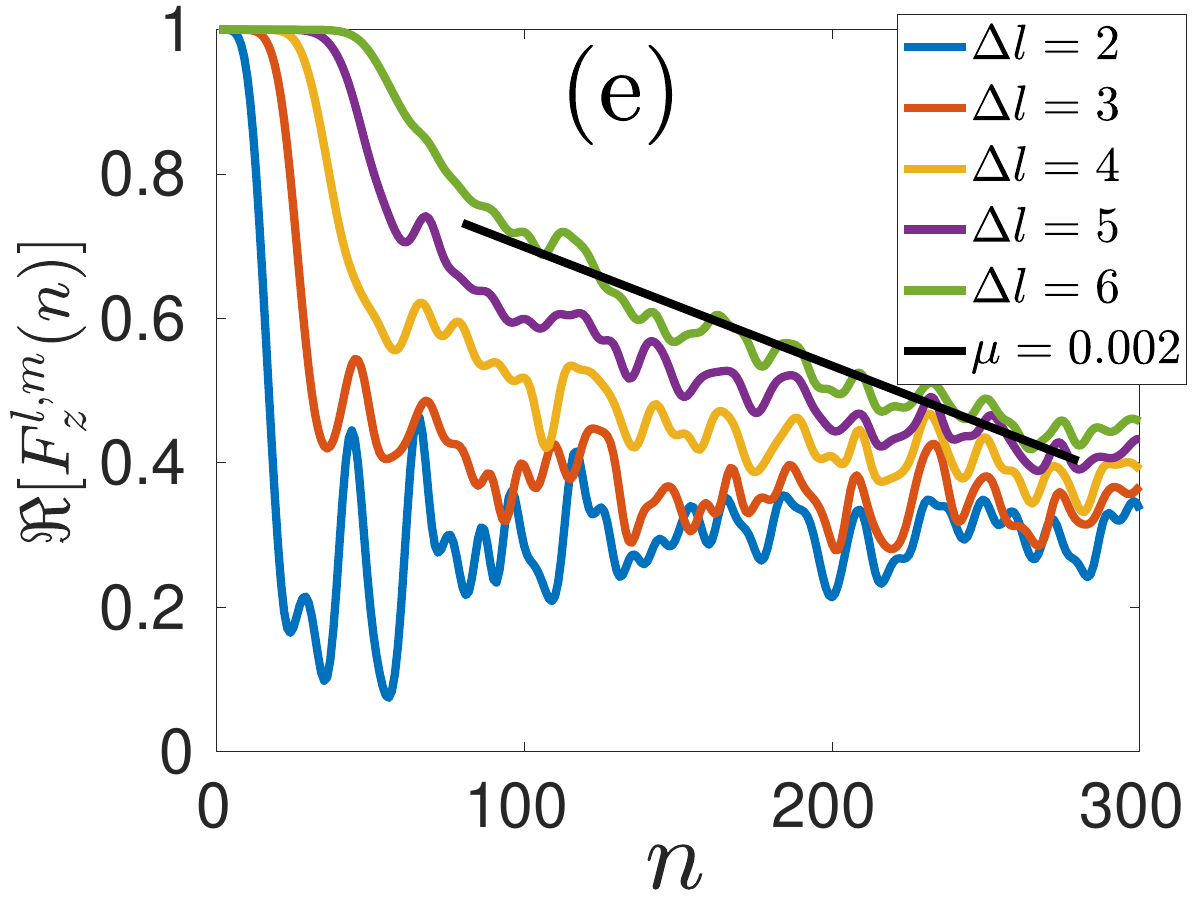}
  \end{subfigure}
 \caption{Non-integrable closed chain transverse Ising Floquet system with $J_x=1$, $h_z=1$, and $h_x=1$ of size $N=18$. (a) Behavior of $TMOTOC$ with number of kicks $(n)$ by increasing value of Floquet period from 
 $\frac{7\epsilon}{2}$ to $\frac{11\epsilon}{2}$ differing by  $\frac{\epsilon}{2}$ with fixed $\Delta l=6$ ($\epsilon=\frac{\pi}{28})$.  (b)  Initial region of $F^{l,m}_z$ with number of kicks with increasing distances between the spins ($\Delta l$) and fixed Floquet period  
 $\tau=6\epsilon/2$.   
 (c)   $C^{l,m}_z$ with number of kicks ($\log-\log$) with increasing ($\Delta l$) at fixed $\tau=\frac{\epsilon}{2}$.  (d) Changing of power with $\Delta l$. (e) Saturation of $F^{l,m}_z$ with the number of kicks.}
\label{cf_TMOTOC_nint}
\end{figure}

Now, we use the nonintegrable $\hat{\mathcal{U}}_x$ model given by Eq.~(\ref{Ux}) and analyze the TMOTOC. Fig.~\ref{cf_TMOTOC_nint}(a) shows the behavior of $\Re[F^{l,m}_z(n)]$ for varying $\tau$ and fixed $\Delta l=6$. From Fig.~\ref{cf_TMOTOC_nint}(a), one can see that number of kicks required for $\Re[F_z^{l,m}(n)]$ depart from unity is equal to the separation between the observables ($n=\vert l-m\vert$). Hence characteristic time does not depend on the Floquet periods. Let us explore the behavior of TMOTOC with the distance between the spins for a fixed $\tau$ (say $\tau=\frac{6\epsilon}{2}$) and increase the separation between the spins $\Delta l$. As $\Delta l$ increases, the characteristic time $(t_{\Delta l})$ increases  in such a way that $n=\Delta l$ [Fig.~\ref{cf_TMOTOC_nint}(b)]. Dynamic region of TMOTOC for the nonintegrable is again showing power-law, and the exponent of the power-law $(b)$ depends on $\Delta l$ [Fig.~\ref{cf_TMOTOC_nint}(c)]. $b$ increases with increasing $\Delta l$ and reaches a maximum $(b_{\rm max})$  at $\Delta l=\frac{N}{2}$ and afterwards decreases symmetrically with increasing $\Delta l$ before coming down to $b_1$ at $\Delta l=N-1$. 
Since we consider the periodic boundary condition, the exponent of the power-law is symmetric about $\Delta l=\frac{N}{2}$ [Fig.~\ref{cf_TMOTOC_nint}(d)]. In a mathematical form we can express $b$, approximately, by Eq.~(\ref{bformula}) with $\kappa =3.2$, $b_{\rm max}=28$ and $b_{\rm min}=1.78$.
Saturation of $\Re[F_z^{l,m}(n)]$ in this nonintegrable model is following a linear decaying behavior with a very small slope  for all $\Delta l$ [Fig.~\ref{cf_TMOTOC_nint}(e)].
 TMOTOC for $\hat{\mathcal{U}}_x$ model in all the regions is summed up as
 \begin{equation}
 \label{C_tmotoc_nint}
C_z^{l,m}(n) \approx \left\{
\begin{aligned}
 0, & \quad\quad\quad n\tau< t_{\Delta l},\\
 (n\tau)^{\kappa\Delta l+1}, & \quad\quad\quad  t_{\Delta l}<n\tau<t_s,\\
 1- \mu n, & \quad\quad\quad   t_s<n\tau.
\end{aligned} \right.
\end{equation}
where $\mu=0.002$ and $\kappa=3.2$.  We calculate the exponent of the power-law by using the HBC formula for $\Delta l=1,2$ and find approximate matches with the exponent of the power-law in the dynamic region of Eq.~(\ref{C_tmotoc_nint}).  Detailed calculation is given in the Appendix \ref{HBC_TMOTOC}. 

\subsection{LMOTOC in the integrable Floquet system}

\begin{figure}
\centering
\begin{subfigure}{.490\textwidth} 
 \includegraphics[width=.99\linewidth, height=.70\linewidth]{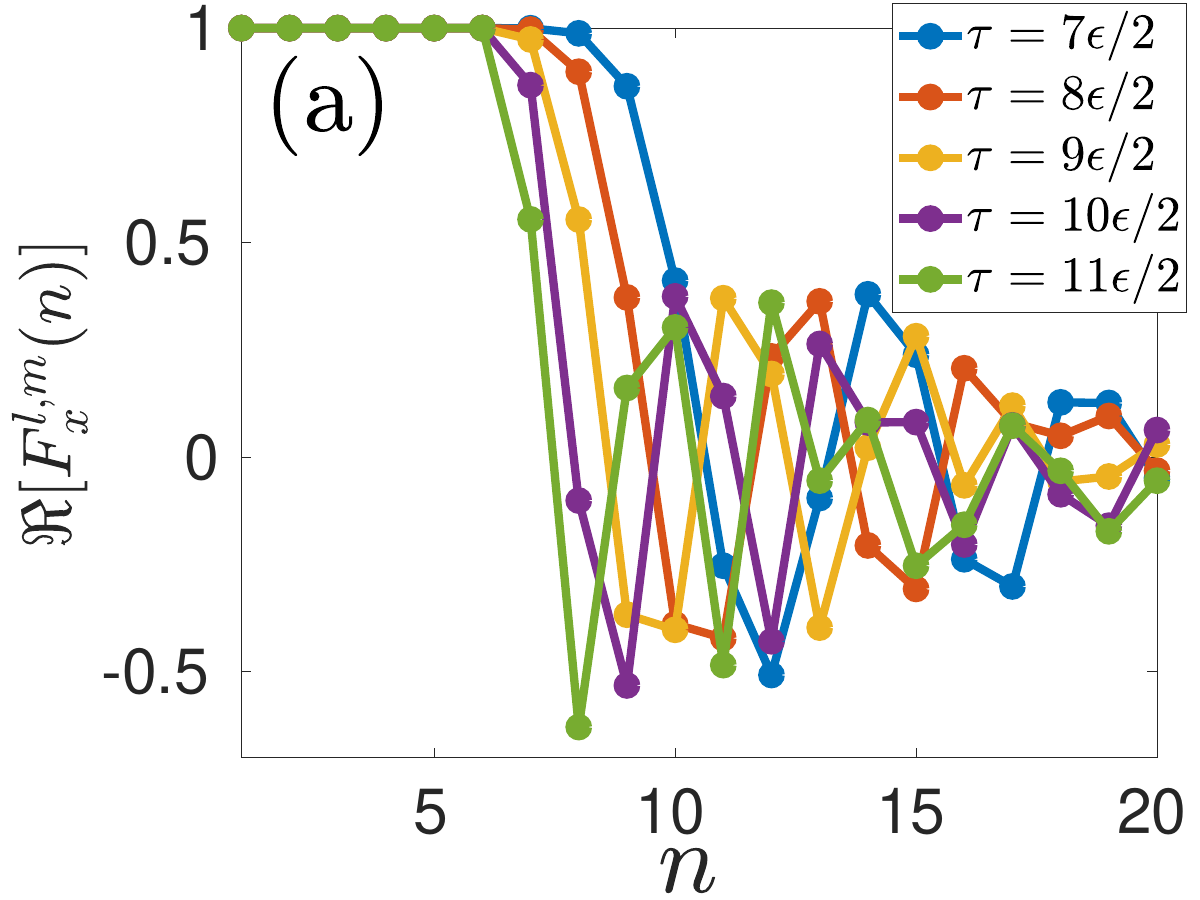}
 \end{subfigure}
 \begin{subfigure}{.490\textwidth} 
 \includegraphics[width=.99\linewidth, height=.70\linewidth]{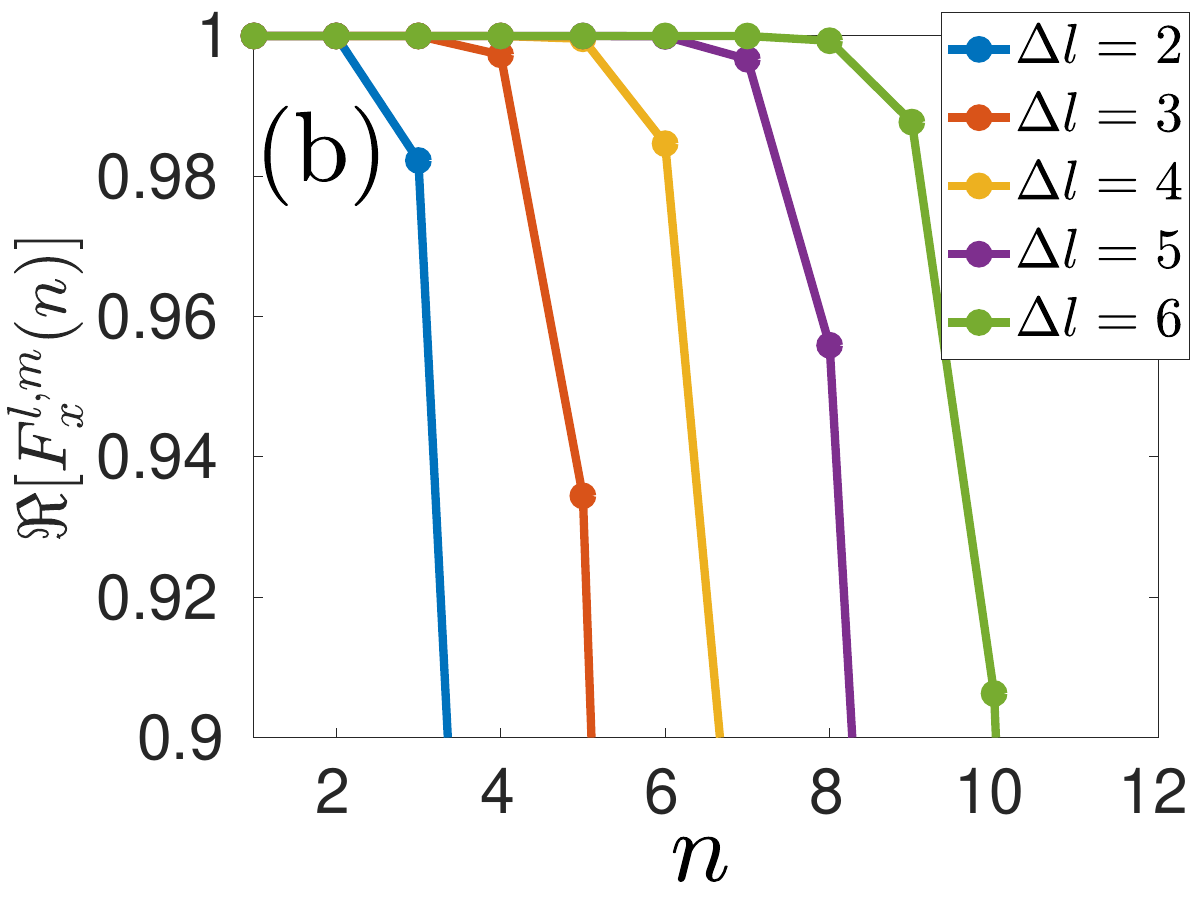}
   \end{subfigure}
 \begin{subfigure}{.490\textwidth}
  \includegraphics[width=.99\linewidth, height=.70\linewidth]{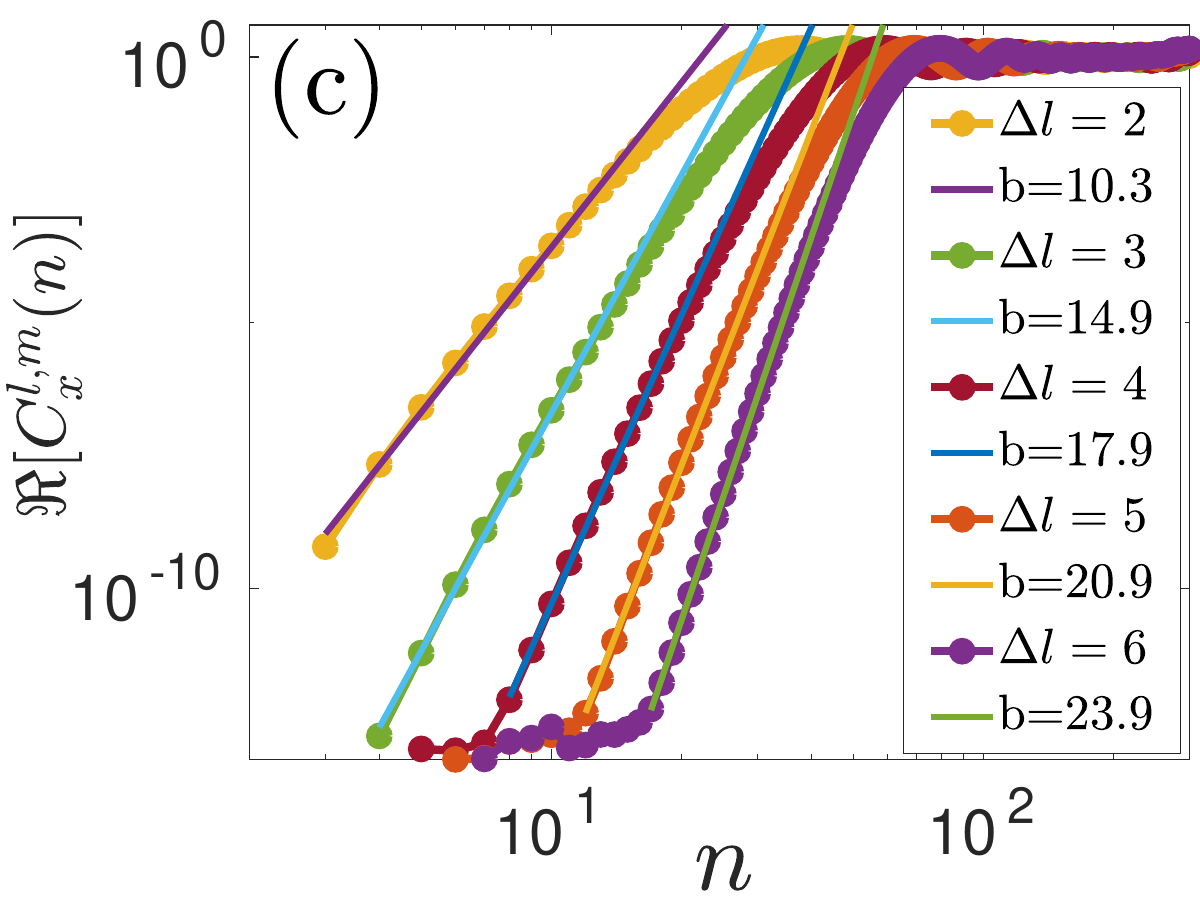}
  \end{subfigure}
      \begin{subfigure}{.490\textwidth}
  \includegraphics[width=.99\linewidth, height=.70\linewidth]{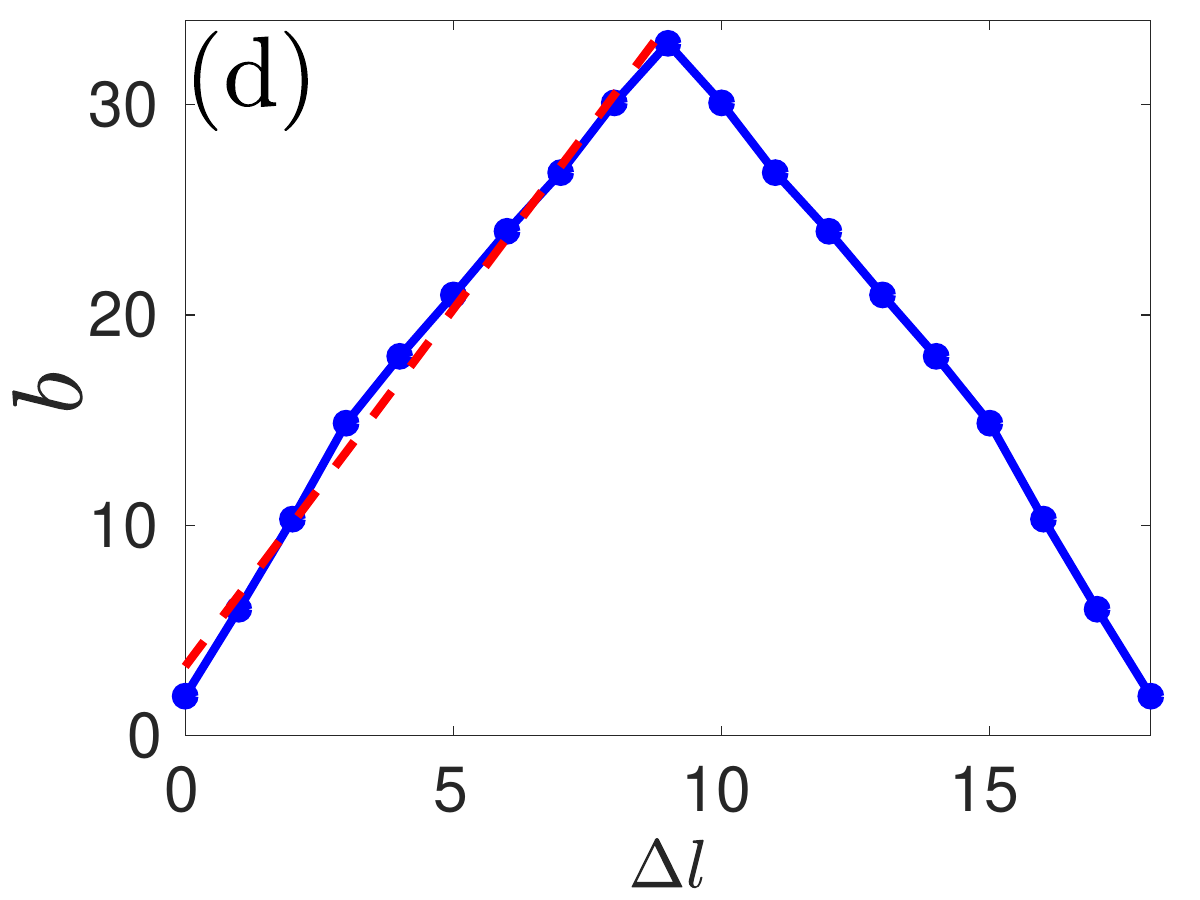}
  \end{subfigure}
  \begin{subfigure}{.490\textwidth} 
 \includegraphics[width=.99\linewidth, height=.70\linewidth]{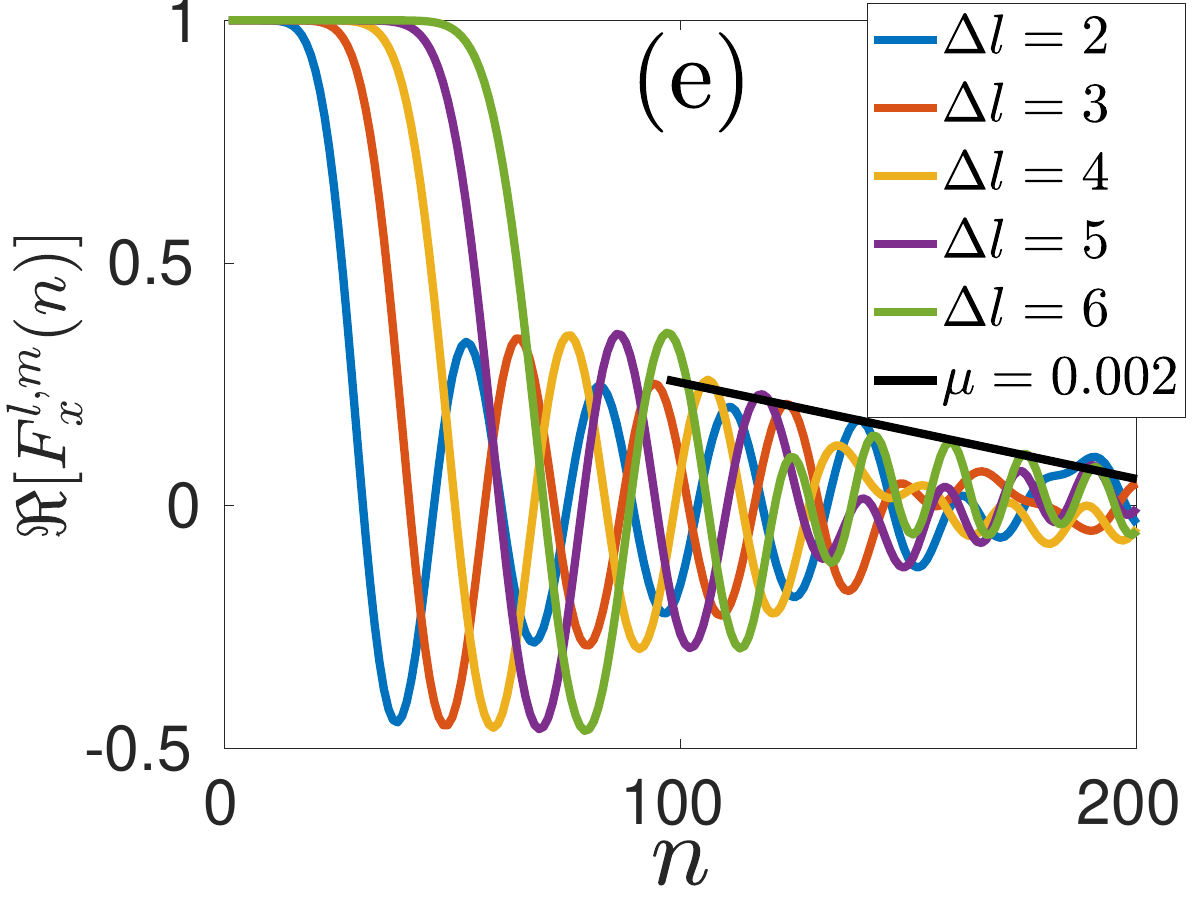}
 \end{subfigure}
 \caption{Integrable closed chain transverse Ising Floquet system  with $J_x=1$ and $h_z=1$ of size $N=18$. (a) Behaviour of $LMOTOC$ with number of kicks $(n)$ with increasing value of Floquet periods from  $\frac{7\epsilon}{2}$ to $\frac{11\epsilon}{2}$ differing by $\frac{\epsilon}{2}$ and  $\Delta l=6$ ($\epsilon=\frac{\pi}{28})$. (b) $F^{l,m}_x(n)$ with number of kicks with increasing ($\Delta l$) and fixed Floquet period $\tau=6\epsilon/2$. (c)  $C^{l,m}_x(n)$ with number of kicks ($\log-\log$) with increasing $\Delta l$  at fixed $\frac{\epsilon}{2}$. (d) Changing of power with $\Delta l$. (e)  $F^{l,m}_x(n)$ with number of kicks at different  $\Delta l$. Black line represents the exponential decreasing of maxima of saturation amplitude.}
 \label{cf_LMOTOC_int}
\end{figure}

 Now we focus on the LMOTOC for the integrable $\hat{\mathcal{U}}_0$ model which shows a similarity with TMOTOC for the same model. Fig.~\ref{cf_LMOTOC_int}(a)  is the behavior of LMOTOC at different Floqeut periods and fixed $\Delta l=6$. In the LMOTOC, number of kicks required to deviate from unity is $n=\Delta l+1$. In comparison with TMOTOC, LMOTOC required one more kick to deviate $\Re[F_x^{l,m}(n)]$  from unity because $\hat \sigma_x^l$ (using Baker–Campbell–Hausdorff formula) provides spreading terms after the first kick. 
Hence, characteristic time does not depend on the Floquet period, however, it depends on the separation between the observables in such a way that characteristic time increases linearly with increasing the separation between the observables ($n=\Delta l+1)$ [Fig.~\ref{cf_LMOTOC_int}(b)]. In the dynamic region of LMOTOC at a small Floquet period, we get a power-law behavior similar to the TMOTOC case. The exponent of the power-law increase with $\Delta l$ in the same manner as in the TMOTOC case [Fig.~\ref{cf_LMOTOC_int}(c)]. We can approximate the exponent with $\Delta l$ by Eq.~(\ref{bformula}) with $\kappa=3.4$, $b_{\rm max}=32.9$ and $b_{0}=1.9$ [Fig.~\ref{cf_LMOTOC_int}(d)].   
 Saturation region of LMOTOC for  $\hat{\mathcal{U}}_0$ shows oscillating behavior. The envelope of the oscillation decays linearly with a constant slope for all $\Delta l$ [Fig.~\ref{cf_LMOTOC_int}(e)]. This behavior is a contrast to the saturation region of TMOTOC for $\hat{\mathcal{U}}_0$ which displays a revival to early-time behavior.  All the regions of LMOTOC  for  $\hat{\mathcal{U}}_0$ can be encapsulated as
\begin{equation}
\label{C_lmotoc_int}
C_x^{l,m}(n) \approx \left\{
\begin{aligned}
 0, & \quad\quad\quad n\tau< t_{\Delta l},\\
 (n\tau)^{\kappa\Delta l+6}, & \quad\quad\quad   t_{\Delta l}<n\tau<t_s,\\
 1- \mu n, & \quad\quad\quad   t_s<n\tau.
\end{aligned} \right.
\end{equation}

\newpage
\subsection{LMOTOC in the nonintegrable Floquet system}
\begin{figure}
\centering
\begin{subfigure}{.490\textwidth} 
\includegraphics[width=.99\linewidth, height=.70\linewidth]{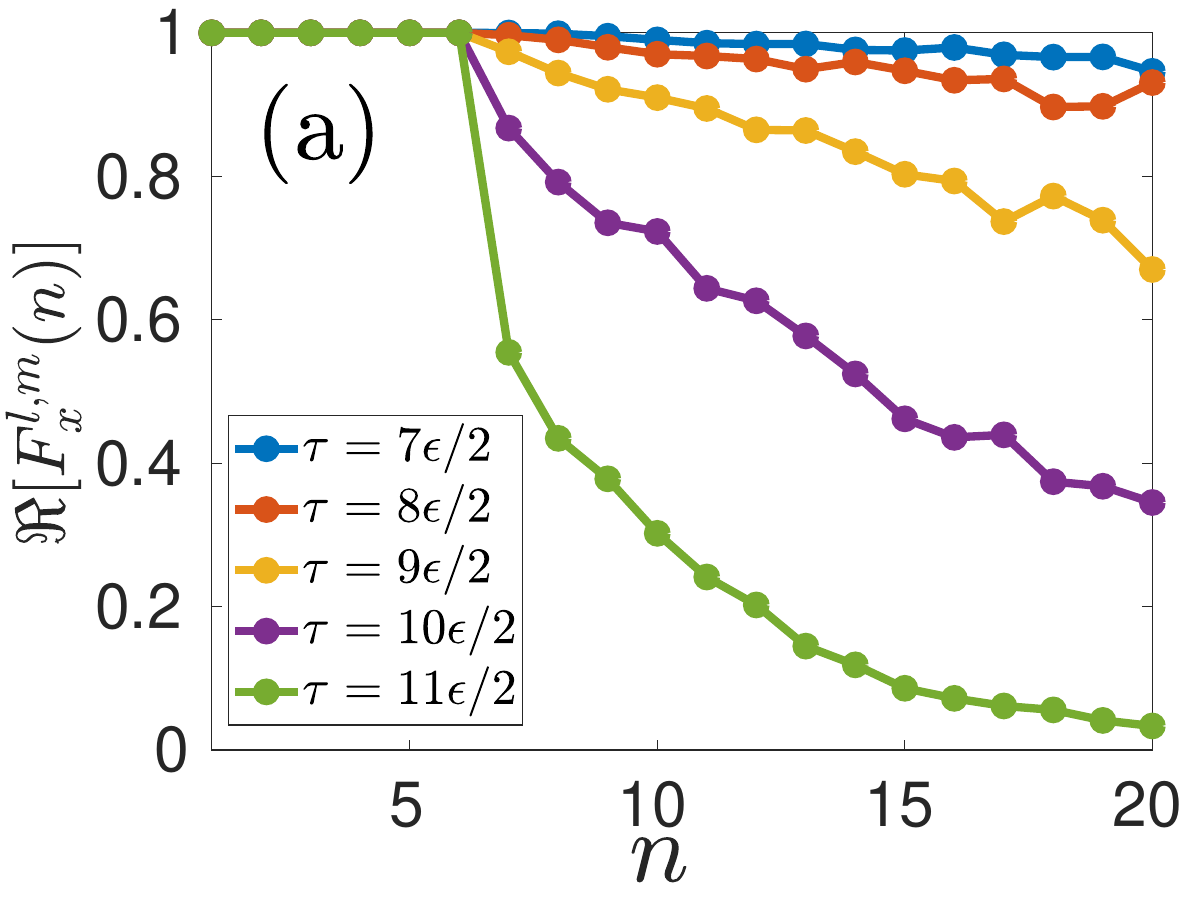}
\end{subfigure}
\begin{subfigure}{.490\textwidth} 
\includegraphics[width=.99\linewidth,height=.70\linewidth]{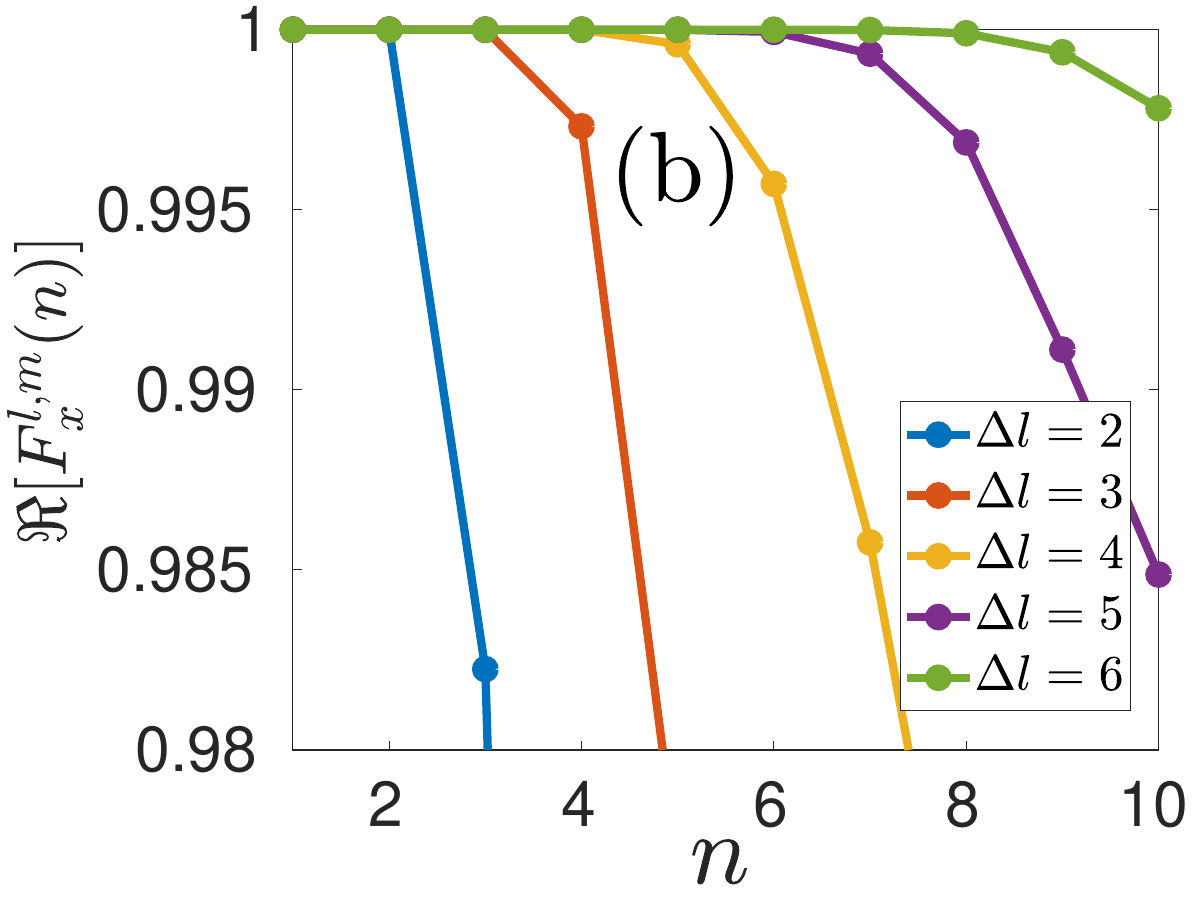}
\end{subfigure}
\begin{subfigure}{.490\textwidth}
\includegraphics[width=.99\linewidth, height=.70\linewidth]{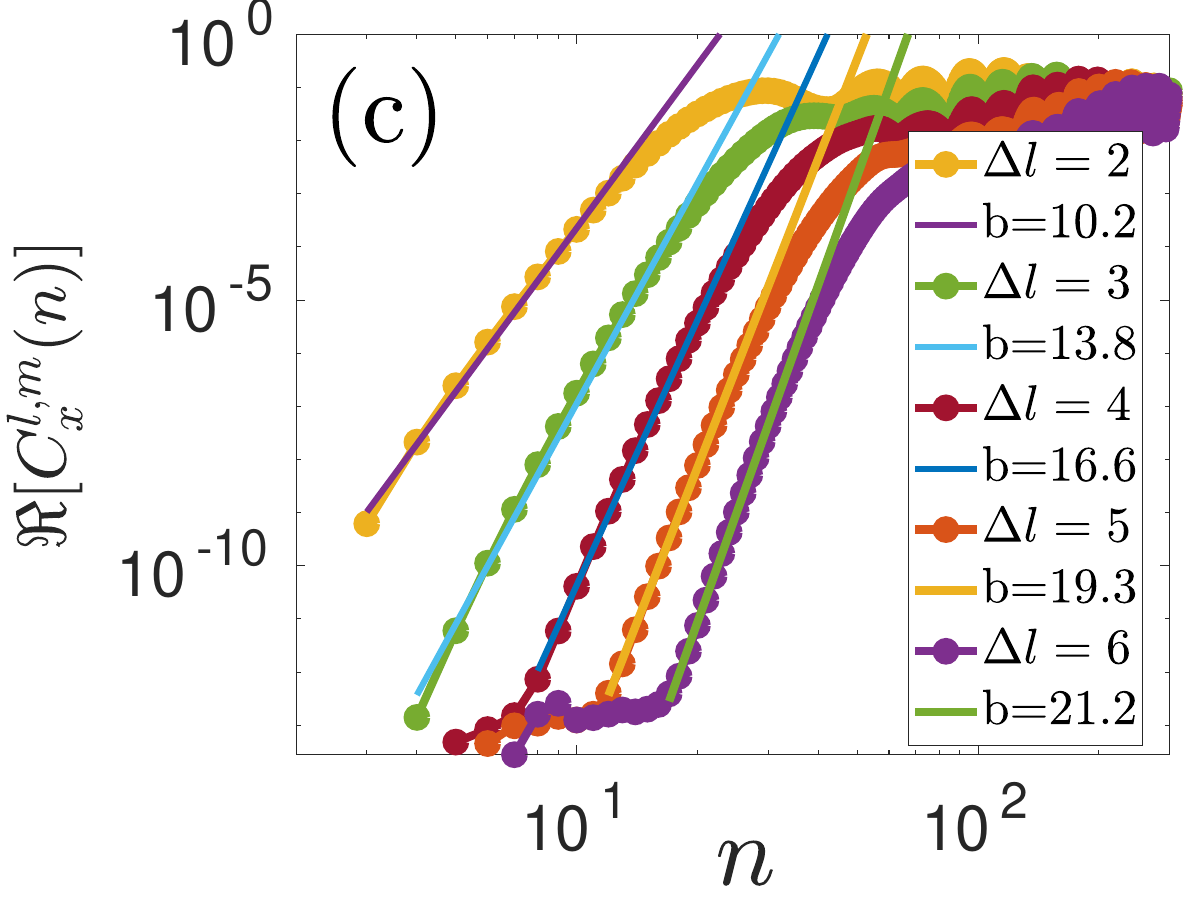}
\end{subfigure}
\begin{subfigure}{.490\textwidth}
\includegraphics[width=.99\linewidth, height=.70\linewidth]{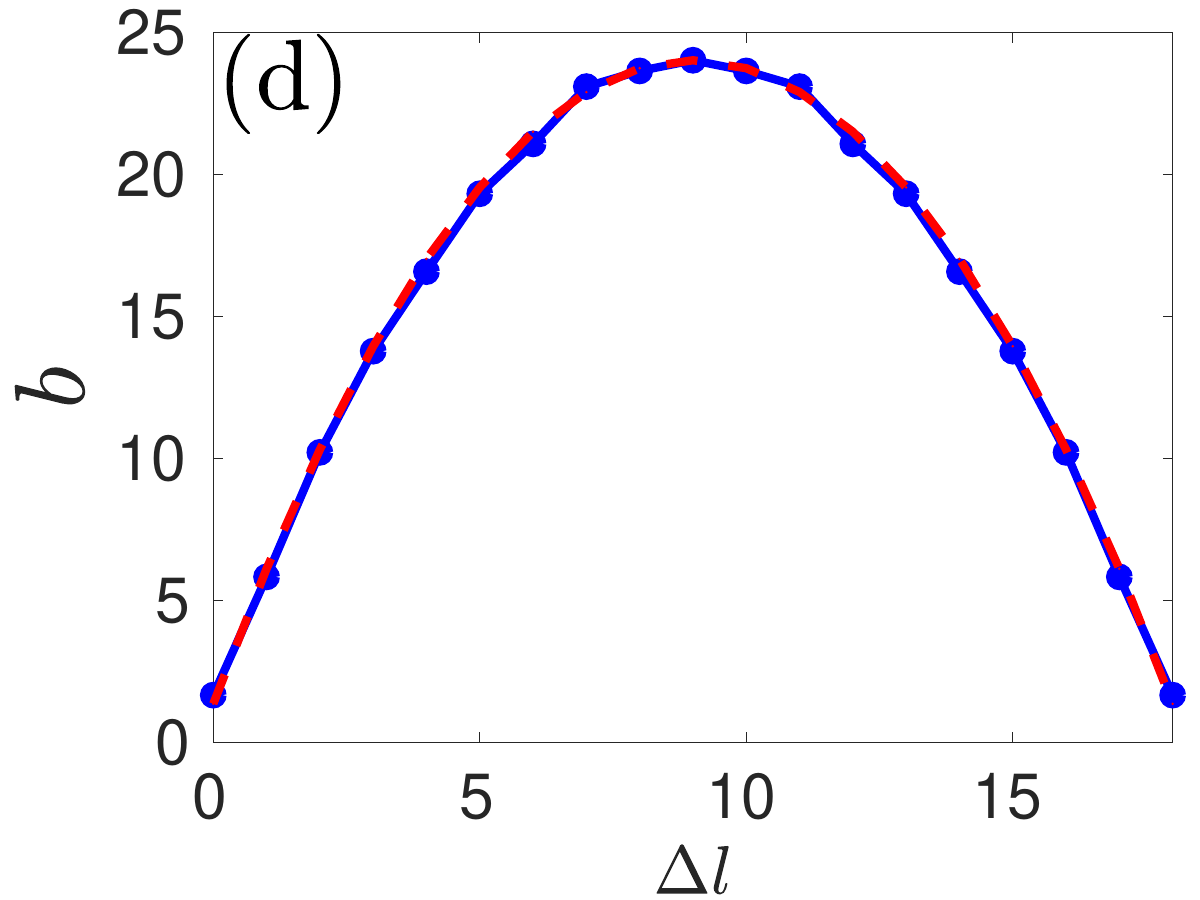}
\end{subfigure}
\begin{subfigure}{.490\textwidth}
\includegraphics[width=.99\linewidth, height=.70\linewidth]{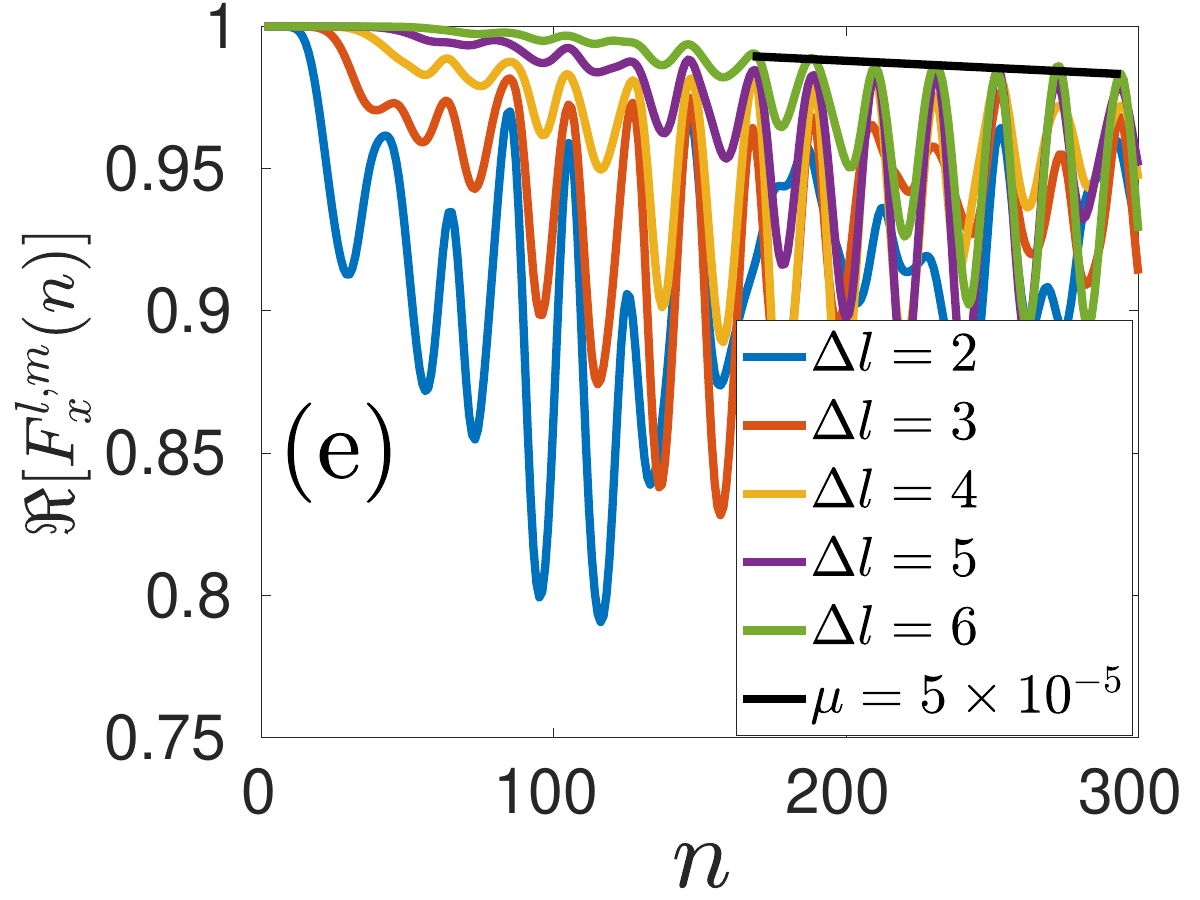}
\end{subfigure}
\caption{Non-integrable closed chain transverse Ising Floquet system with $J_x=1$, $h_z=1$, and $h_x=1$ for $N=18$. (a) $LMOTOC$ with number of kicks $(n)$ by increasing value of Floquet period from $\frac{7\epsilon}{2}$  to $\frac{11\epsilon}{2}$ differing by $\frac{\epsilon}{2}$ with fixed $\Delta l=6$ ($\epsilon=\frac{\pi}{28}$). (b)  $F^{l,m}_x(n)$ with increasing $\Delta l$ from $6$ to $6$ and fixed period $\tau=\frac{6\epsilon}{2}$. (c)  $C^{l,m}_x(n)$ with number of kicks ($\log-\log$) by increasing $\Delta l$ from $2$  to $6$ and fixed Floquet period $\tau=\frac{\epsilon}{2}$. (d) Changing of power with $\Delta l$. (e) $F^{l,m}_x(n)$ with number of kicks at different $\Delta l$. Black line represents the exponential decrease of local maxima of saturating amplitude. }
 \label{cf_LMOTOC_nint}
\end{figure}

Finally, we consider nonintegrable $\hat{\mathcal{U}}_x$ model for LMOTOC calculations. We get similar behavior in the characteristic regime as that for LMOTOC with $\hat{\mathcal{U}}_0$ model [Fig.~\ref{cf_LMOTOC_nint}(a) and (b)]. In the dynamic region, the growth is again a power-law, and the exponent increase with $\Delta l$ [Fig.~\ref{cf_LMOTOC_nint}(c)] but the trend is a bit different than the previous cases. Unlike the previous cases, we see a quadratic increase of the exponent by increasing $\Delta l$, till a maximum is reached. After the maximum $b_{\rm max}$ at $\Delta l=\frac{N}{2}$, we see a symmetric decrease in the exponent till $\Delta l=N$ [Fig.~\ref{cf_LMOTOC_nint}(d)]. We approximate $b$ as follows:
\begin{equation}
\label{bformula_lmotoc}
b \approx \Big(b_{\rm max}-\lambda\Big\vert\frac{N}{2}-\Delta l\Big\vert^2\Big).\quad\quad\quad 0\leq \Delta l \leq N
\end{equation}
Where $\lambda=2.8$, $b_{\rm max}=24.0$ and $b_{0}=1.7$.  Eq.~(\ref{bformula_lmotoc}) describes the variation of exponent of power-law with increasing the separation between the observables. It is a parabolic form with vertex at $\frac{N}{2}$ and also symmetric about $\Delta l=\frac{N}{2}$ because of closed chain consideration. We calculate the exponent of the power-law by using the HBC formula for $\Delta l=1$ and find that exponent approximately matches the Eq.~(\ref{bformula_lmotoc}). Detailed calculation is mentioned in Appendix \ref{HBC_LMOTOC}.  
Saturation of LMOTOC for a nonintegrable case is  oscillating, and the maxima of the oscillation decrease linearly (with a very small slope $ \mu=10^{-5}$, and same for all $\Delta l$ ) with the number of kicks [Fig.~\ref{cf_LMOTOC_nint}(e)]. 
The complete region of LMOTOC for $\hat{\mathcal{U}}_x$ system is given as 
\begin{equation}
\label{C_lmotoc_nint}
C_x^{l,m}(n) \approx \left\{
\begin{aligned}
 0, & \quad\quad\quad n\tau< t_{\Delta l},\\
 (n\tau)^{\lambda(\Delta l)^2}, & \quad\quad\quad   t_{\Delta l}<n\tau<t_s,\\
 1- \mu n, & \quad\quad\quad   t_s<n\tau.
\end{aligned} \right.
\end{equation}

In a nutshell, we see that the characteristic regions of LMOTOCs have similar behavior for $\hat{\mathcal{U}}_0$ and $\hat{\mathcal{U}}_x$ systems. In both  cases, the commutator propagation varies with $\tau$ in a similar way. But the dynamic region displays a contrast between $\hat{\mathcal{U}}_0$ and $\hat{\mathcal{U}}_x$. In the integrable case, the exponent of the power-law increases linearly with $\Delta l$, but in the nonintegrable exponent, we see a quadratic growth of power-law with $\Delta l$. In the saturation region, both are oscillating, and the envelope decreases with different rates.
\par
In this chapter, we considered single spins as observables in our calculation of OTOCs. The experimental procedure of calculating OTOC using single spin observables and  initial product state has been done in Ref.~\cite{joshi2020quantum}.  Implementation of the unitary operator on observable $\hat W^l$  \big[$\hat W^l(n)=(\mathcal{\hat U}_x^{\dagger})^{n}\hat W^l(0)(\mathcal{\hat U}_x)^n$\big] followed by perturbation of observable $\hat V^m$ is discussed in the Ref.~\cite{garttner2017measuring}. The OTOC is obtained by measuring the expectation value of the observable $(\mathcal{\hat U}_x^{\dagger})^n \hat W^l(0)(\mathcal{\hat U}_x)^n \hat V^m (\mathcal{\hat U}_x^{\dagger })^n \hat W^l(0)(\mathcal{\hat U}_x)^n \hat V^m$ \cite{joshi2020quantum}. Therefore, LMOTOCs and TOMOTOCs can be calculated experimentally.

\section{Conclusion}
\label{Ch3_conclusion}
We studied the behavior of TMOTOC and LMOTOC comprehensively using  $\mathcal{\hat{U}}_0$ and $\mathcal{\hat{U}}_x$ systems. We divided LMOTOC and TMOTOC into three distinct regimes: characteristic-time, dynamic-time, and saturation-time regimes. 
\par
Characteristic times of TMOTOC and LMOTOC are independent of the integrability of the system. It is also independent of the Floquet period and system size; however, it depends on the separation between the observables. The number of kicks required for the deviation of $\Re[F^{l,m}]$ from unity is equal to the numerical value of the separation between the observables in the case of TMOTOC; however, one extra kick is required in the case of LMOTOC. Behavior of the dynamic region is also independent of the integrability of the system. In both systems, $\mathcal{\hat U}_0$ and $ \mathcal{\hat U}_x$, LMOTOC, and TMOTOC show the power-law growth. There is no signature of Lyapunov exponent. This power-law growth depends on the separation between the spins and the Floquet period. The rate of change of exponent with respect to the separation between the spins is independent of the integrability of the system in the TMOTOC; however, we see a dependence in the case of LMOTOC. In TMOTOC for both the systems $\mathcal{\hat U}_0$, and $\mathcal{\hat U}_x$, the exponent varies as a  triangular function.  In the case of LMOTOC, we see a triangular function with linear increase/decrease for $\mathcal{\hat U}_0$ system but a quadratic increase/decrease for $\mathcal{\hat U}_x$ system. Saturation region of TMOTOC is different in both systems:  $\mathcal{\hat U}_0$ system revives back, but $\mathcal{\hat U}_x$ system decays linearly. Saturation behavior of LMOTOC shows the oscillating decay with envelop decaying linearly in both systems. Saturation of TMOTOC and LMOTOC are independent of $\Delta l$. 
\par
In the next chapter, we will calculate OTOCs using contiguous symmetric blocks of spins or random operators localized on these blocks as observables instead of localized spin observables. In the calculation of OTOC, we consider both integrable and nonintegrable Ising spin Floquet systems.

%

\chapter{Out-of-time-order correlation of the nonlocal block observables in Floquet Ising spin chain}  

\ifpdf
    \graphicspath{{Chapter3/Figs/Raster/}{Chapter3/Figs/PDF/}{Chapter3/Figs/}}
\else
    \graphicspath{{Chapter3/Figs/Vector/}{Chapter3/Figs/}}
\fi

\section{Introduction} 
Periodically driven Floquet systems have been extensively studied
in the recent past in both classical and quantum systems. A popular set of models are driven by fields applied in the form of kicks \cite{d2014long, naik2019controlled, shukla2021,Mishra2015}, as analytical forms of the time evolution operator are easy to find.
One textbook example is the kicked-rotor model of a particle moving on a ring \cite{reichl1992transition}. These models show interesting behavior displaying transition from integrability to chaos, dynamical Anderson localization \cite{chirikov1981dynamical,fishman1982chaos,reichl1992transition}, and dynamical stabilization \cite{kapitza1965dynamical,broer2004parametrically}. These systems are of interest in  both classical
as well as quantum systems. Such periodic forcing has been realized in experiments to study various phenomena \cite{wintersperger2020realization,Franca2021simulating,zhang2017observation,choi2017observation,Santhanam2022}.
\par
In contrast to the kicked rotor,  the Ising model with time-periodic transverse and longitudinal magnetic fields is an example of a many-body Floquet system of current interest \cite{gritsev2017,lakshminarayan2005multipartite,naik2019controlled,shukla2021}.
 Absence of a transverse component renders the system trivially integrable. Presence of both a longitudinal and transverse magnetic component makes this system nonintegrable. However, in the absence of longitudinal field, the system is rendered integrable as a system of noninteracting fermions.   
These systems have been studied using sudden quenches \cite{Polkovnikov2011}, and slow annealing \cite{Santoro2002}. In the quenched case, the system is out of equilibrium and leads to interesting dynamics of the observables,  and has drawn considerable attention in the last decade with significant theoretical and experimental observations \cite{Russomanno2012,Russomanno2013,mishra2014resonance}. 

\par
A typical way to distinguish between integrable, non-integrable, and near-integrable regimes has been to use spectral properties and random matrix theory. This mostly leaves aside the question of dynamics. However, a quantity that has been extensively used recently to distinguish the chaotic and integrable dynamics is the out-of-time-order correlator (OTOC) \cite{ garcia2018chaos,rozenbaum2020early,yan2019similar,rozenbaum2017lyapunov,lee2019typical,rozenbaum2019Universal}. In classical physics, one hallmark of chaos is that a small difference in the initial condition results in the exponential deviation of the trajectory, which is responsible for the so-called ``butterfly effect" \cite{gu2016,bilitewski2018temperature,das2018light}. Classical Hamiltonian systems can have such pure deterministic chaos, which is used in the quantum domain for the study of quantum chaos \cite{gutzwiller1990chaos,haake1991quantum}. It has been proposed that quantum chaos be characterized by the growth rate of OTOC \cite{maldacena2016bound}, an  exponential growth defining a quantum Lyapunov exponent. 
\par
Spin systems have been a playground for understanding  many-body physics in general, and the growth of OTOCs in particular \cite{lin2018out,xu2020accessing,xu2019locality,kukuljan2017weak,Fortes2019,craps2020lyapunov,roy2021entanglement,yan2019similar,bao2020out,dora2017out,Riddell2019,lee2019typical}.  Growth of OTOC is discussed in systems such as Luttinger liquids \cite{dora2017out}, XY model \cite{bao2020out}, Sachdev-Ye-Kitaev (SYK) model \cite{Fu2016}, Heisenberg XXZ model and  Aubry–Andr\'e–Harper model \cite{Riddell2019,lee2019typical}.
Lin and Motrunich \cite{lin2018out} calculated OTOC for single spin observables in the integrable transverse field Ising model and observed power-law growth, with the power varying with the separation between the localized spins.

Fortes {\it et al.} \cite{Fortes2019} studied OTOCs in the time-independent Ising model with tilted magnetic fields, perturbed XXZ model, and Heisenberg spin model with random magnetic fields. In all these models with single-spin  observables, only power-law growth has been reported despite the presence of quantum chaos. 
OTOCs in integrable and nonintegrable Floquet Ising  models were studied by Kukuljan {\it et al.} \cite{kukuljan2017weak} using  extensive observables. In one dimension case, the growth of OTOC density was still found to be linear in time.

The cases where exponential growth has been definitely reported involve semiclassical models such as the quantum kicked rotor \cite{rozenbaum2017lyapunov}, coupled kicked rotors \cite{prakash2020scrambling,Santhanam2022}, the kicked top, which may be considered to be a transverse field kicked Ising model but with the interactions being all-to-all \cite{yin2021quantum,sreeram2021out}, the bakers map \cite{lakshminarayan2019out}, and so on. Our motivation herein is to allow for a large Hilbert space for the observables, which are restricted to blocks of spins. We may consider the spin chain as a bipartite chaotic system, each consisting of $N/2$ spins, to explore the possibility of exponential growth. We will see that such spin-1/2 nonintegrable models, even for block operators, have only power-law OTOC growth, implying that their quantum Lyapunov exponents are $0$. 

\par
In nonintegrable systems including spin chains such as studied here, the long-time  saturation value of the OTOC is consistent with an estimate from random matrix theory. The approach of the OTOC to the saturation value was found to be at an exponential rate in a weakly interacting bipartite chaotic system \cite{prakash2020scrambling}. Exponential approach to saturation was also found in a semiclassical theory of OTOC \cite{rammensee2018many}.
We find such an exponential approach to the random matrix value in spin chains with block observables for the nonintegrable cases. 

To understand the exponential approach, we consider the case when the block operators are random. Averaging over random unitary operators in a bipartite system, the OTOC has been shown to be exactly the operator entanglement of the propagator \cite{anand2021brotocs}. We show this is also the case with random Hermitian observables drawn from the Gaussian Unitary Ensemble (GUE). 

Thus the exponential saturation of the OTOC is qualitatively consistent with the behavior previously observed for the operator entanglement growth of the propagator \cite{Pal2018}.  

According to the BGS conjecture \cite{bohigas1984characterization}, the spectral properties of the quantum analogue of a chaotic classical system will follow Wigner-Dyson statistics,  unlike the quantum analogue of an integrable classical system following Poisson distribution.  Thus, the spectral statistics of spacing between the consecutive energy levels of a quantum system works as a tool to differentiate a chaotic system from an integrable one    \cite{craps2020lyapunov,Pal2018,karthik2007entanglement,ray2018signature,chen2018measuring,ray2018signature,mehta1991theory,averbukh2001angular}.
\par
This chapter is organised as follows. In subsection \ref{Ch4_model}, we will discuss the Floquet map with and without longitudinal fields. In subsection \ref{OTOC}, we will define the OTOC for the block spin operators. In subsection \ref{avg}, we will discuss the relation of OTOC with operator entanglement entropy (OPEE).
In subsection \ref{LSD}, we will elaborate the nearest-neighbor spacing distribution (NNSD) and its behavior in the integrable and nonintegrable cases. We will elaborate the behavior of OTOC and NNSD in section \ref{CF} for the constant-field Flqouet system, and in section \ref{special_CF}, a special case of constant-field Flquet system. Finally, in section \ref{Ch4_conclusion}, we will conclude the results of this chapter.

\section{The spin model and background}
\subsection{The spin model}
\label{Ch4_model}

Consider a periodically driven Ising spin system with the Hamiltonian
\begin{equation}
\label{Ch4_Hxz}
\hat H(t)=J_{x} \hat H_{xx}+h_{x} \hat H_x+h_{z}\sum_{n=-\infty}^{\infty}\delta\Big(n-\frac{t}{\tau}\Big) \hat H_z.
\end{equation}
Here $\hat H_{xx}=\sum_{l=1}^{N-1}\hat \sigma_l^x\hat \sigma_{i+1}^x $ is the nearest-neighbor Ising interaction term,  $ \hat H_x=\sum_{l=1}^{N}\hat \sigma_l^x$ and $\hat H_z=\sum_{l=1}^{N}\hat \sigma_l^z$. The interaction strength is $J_x$, the continuous and constant longitudinal magnetic field in $x$-direction is given by $h_{x}$ and the transverse magnetic field in the $z$-direction, which is applied in the form of delta pulses at regular interval $\tau$ is $h_{z}$.
 
The Floquet operator is the propagator connecting states across one time period $\tau$. Denoting this as $\mathcal{\hat U}_x$, we have (with $\hslash=1$)
\begin{equation}
\label{Ch4_Ux}
 \mathcal{\hat U}_x =\exp\left[-i\tau(J_x \hat H_{xx}+h_{x}\hat H_{x})\right] \exp\left(-i \tau h_{z} \hat H_{z}\right), 
\end{equation}
and will be referred to as $``\mathcal{\hat U}_x$ systems" below,  
when the longitudinal field is absent, the model is solvable by the Jordan–Wigner transformation and renders the system as one of noninteracting fermions. In the presence of the longitudinal field, these fermions are interacting, and there is evidence that there is a transition to quantum chaos \cite{prosen2004ruelle,prosen2000exact,prosen2002general,lakshminarayan2005multipartite,else2016floquet}. The Floquet map of the integrable model is a special case of Eq.~(\ref{Ux}) with $h_{x}=0$ will be referred to as the $\mathcal{\hat U}_0$ system below.

\subsection{Out-of-time-order correlation and block operators}
\label{OTOC}

Dynamics of quantum systems lead to the spreading of initially localized operators under the unitary time evolution.  Let the discrete time evolution of operator $\hat W \equiv \hat W(0)$ be $\hat W(n)=\hat U(n)^{\dagger}\hat W(0)\hat U(n)$, where $\hat U(n)$ is time$-n$ propagator. For example, if the time evolution is governed by Eq.~(\ref{Ux}), $\hat U(n)= \mathcal{\hat U}_x^n$.
If $\hat V$ and $\hat W$ are two Hermitian operators that are localized on different sets of spins (say $A$ and $B$), we consider as the out-of-time-order correlation (OTOC)  \cite{larkin1969quasiclassical,shenker2014black,shenker2014multiple,almheiri2013apologia,shenker2015stringy,roberts2015diagnosing,maldacena2016bound,stanford2016many}:
\begin{equation}
C(n)=-\frac{1}{2 \, d_A d_B}\mbox{Tr} \left([\hat W(n), \hat V]^2\right),
\label{Ch4_Cn1}
\end{equation}
where $d_A$ and $d_B$ are dimensions of the subspaces, and  $d_A=d_B=2^{N/2}$ as we consider only the case of equal blocks. The OTOC $C(n)$ is clearly a measure of the non-commutativity of these two operators via its norm.

This separates as $C(n)$ =$C_2(n)-C_4(n)$, where $C_2(n)$ and $C_4(n)$ are two-point and four-point correlations respectively:
\begin{equation}
    \begin{split}
&C_2(n)=\frac{1}{d_A d_B}\mbox{Tr}(\hat W^2(n) \hat{V}^2),\\ 
&  C_4(n)=\frac{1}{d_A d_B}\mbox{Tr}(\hat W(n)\hat V\hat W(n)\hat V).
\end{split}
\end{equation}
These are infinite temperature quantities and involve the entire spectrum of $2^{N}$ states. We will use the trick of evaluating this by employing 
Haar random states of $2^N$ dimensions to evaluate expectation values, that is 
$\mbox{Tr}(\hat A)/2^N \approx \left \langle \Psi_R|\hat A|\Psi_R\right \rangle$
were $|\Psi_R\rangle$ is such a state. Averages over a few random states are used.

\begin{figure}[hbt!]
      \centering
       \includegraphics[width=\linewidth, height=.20\linewidth]{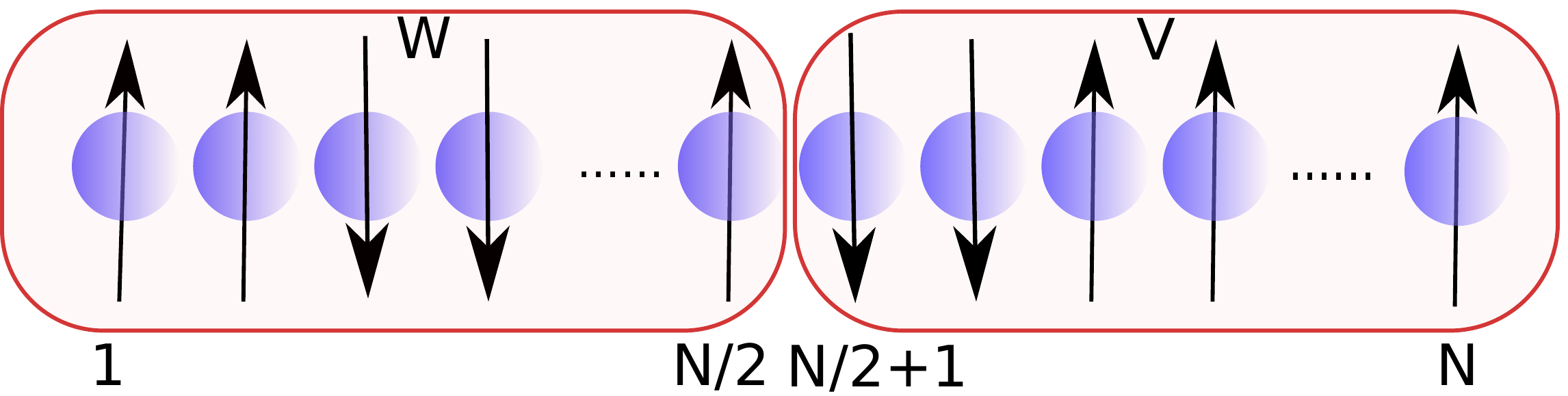}
       \caption{Schematics of SBOs defined in Eq.~(\ref{Block}).  Even $N$ is considered and halved into subsystems $W$ and $V$.}
       \label{block_operator} 
\end{figure}

Almost all studies of OTOC in such spin models thus far concentrate on operators that are localized on single spins, in contrast, we consider operators $\hat V$ and $\hat W$ initially isolated on the first and second block of spins, see Fig. \ref{block_operator}, referred to here as spin-block-operators (SBOs): 
\begin{equation}
\label{Block}
\hat W=\frac{2}{N} \sum_{l=1}^{\frac{N}{2}}\hat \sigma_l^x \quad {\rm and} \quad \hat V=\frac{2}{N} \sum_{l=\frac{N}{2}+1}^{N}\hat \sigma_l^x.
 \end{equation}
Note that the behavior of these OTOCs is genuinely different and does not follow from a knowledge of the single site OTOCs involving correlations such as $\langle \hat{\sigma}^x_{l_1} \, \hat{\sigma}^x_{l_2}(n) \,\hat{\sigma}^x_{l_3}\, \hat{\sigma}^x_{l_4}(n) \rangle$ for general values of $l_i$.

For $n>0$, $\hat W(n)$ is no longer confined to the first $N/2$ spins, and the OTOC becomes nonzero.  Previous studies with single-site localized observables show no exponential growth of OTOC even for nonintegrable cases in such spin models. The cases where exponential growth has been definitely reported involve semiclassical models such as the quantum kicked rotor, coupled kicked rotors, the kicked top wherein the interactions are all-to-all, the bakers map, and so on. Our motivation herein is to allow for a large Hilbert space for the local operators. We may consider the spin chain as a bipartite chaotic system, each consisting of $N/2$ spins, to explore the possibility of exponential growth.

If short-time growth is exponential \cite{hosur2016chaos,garcia2018chaos,chen2018measuring,rozenbaum2017lyapunov} then it is related to quantum chaos and quantum Lyapunov exponents.
Can the OTOC help define a quantum Lyapunov exponent for spin models with short-range interactions such as in Eq.~(\ref{Ux})? Integrable system show power-law growth of OTOC before the scrambling time \cite{chen2018measuring,rozenbaum2017lyapunov,hosur2016chaos,garcia2018chaos,Fortes2019,lin2018out}.
However, it is unclear under what circumstances OTOC of nonintegrable systems with other signatures of quantum chaos can fail to grow exponentially. We will see that such spin-1/2 nonintegrable models continue even for block operators to not have an unambiguous exponential OTOC growth.

\subsection{Average and asymptotic OTOC values}
\label{avg}
 As $\hat{V}$ and $\hat{W}$ are block restricted sums of spin operators, $\hat{V}+\hat{W}$ is the total spin in the ${x}$ direction and appears as a term in the Hamiltonian. Thus these are special operators with dynamical significance, as would be natural to assume. In contrast, if they are random operators on the space of 
$N/2$ spins, the OTOC behaves quite differently till possibly the
scrambling time. Beyond the scrambling time, we may expect that the local operators have largely become random if there is nonintegrability and quantum chaos. Thus, it is of interest to compare the behavior of random operator OTOC with non-random ones: to separate the effects of dynamics and scrambling. In a semiclassical model of weakly coupled chaotic systems, it was noted that the post-scrambling time OTOC of non-random operators did behave as that of ``pre-scrambled" random operators \cite{prakash2020scrambling}. We find some similarities in the case of spin chains but also interesting differences. 

In the case of random operators for $\hat V$ and $\hat W$, ergodicity may be expected and hence an average over them is done. It has been observed \cite{styliaris2021information} that if these operators are random unitaries chosen uniformly (Haar measure, circular unitary ensemble, CUE), the average OTOC is remarkably related to the operator entanglement. As we are using Hermitian operators, we average over random Hermitian ensembles for which we naturally choose the GUE, and the result is identical.

Let there be a bipartite space $\mathcal{\hat H}_{A}\otimes \mathcal{\hat H}_B$, such as the space of the first and second $N/2$ spins in the chain. The Schmidt decomposition of the unitary propagator on this bipartition is of the form 
\begin{equation}
 \hat U(n)=2^{N/2}\sum_{i=1}^{2^N}\sqrt{\lambda_i(n)} \hat A_i(n) \otimes \hat B_i(n).
 \end{equation}
Here $\hat A_i(n)$ and $\hat B_i(n)$ are orthonormal operators on individual spaces $\mathcal{\hat H}_{A,B}$,  satisfying, $\Tr (\hat A_i(n)^{\dagger} \hat A_j(n))= \Tr(\hat B_i(n)^{\dagger} \hat B_j(n))=\delta_{ij}$. The numbers $\lambda_i(n)>0$ satisfy the condition $\sum_i\lambda_i(n)=1$ which is a consequence of the unitarity of $\hat U(n)$.
     
 Operator entanglement entropy (OPEE) is used for the measure of entanglement \cite{Pal2018,styliaris2021information,wang2002quantum,wang2004entanglement} and defined via the linear entropy as
 \begin{equation}
 E_l[\hat U(n)]=1-\sum_{i=1}^{2^{N}}\lambda^2_i(n).
  \end{equation}
This vanishes if and only if $\hat U(n)$ is of product form and is maximum when all $\lambda_i(n)=2^{-N}$ and the OPEE is equal to $1-2^{-N}$.

Let an element of the GUE be $\hat W=(\hat M+\hat M^{\dagger})/2$, where $\hat M$ is a $d$ dimensional matrix whose entries are such that its real and imaginary parts are zero centered, unit variance,  independent normal random numbers, the Ginibre ensemble. It is straightforward to see that $\ovl{\hat W^2}=d\, \hat I_d$, where $\hat I_d$ is the $d$ dimensional identity matrix, and the overline indicates the GUE average. The average of $C_2(n)$ is then 
\begin{equation}
\ovl{C_2(n)}^{\hat W,\hat V}=\frac{1}{d^2}\ovl{\Tr \left(\hat U(n)^{\dagger} \hat W^2\hat U(n)\hat \hat V^2\right) }^{\hat W,\hat V}=d^2,
\end{equation}
where $\hat V$ is also a GUE realization independent of $\hat W$. 

To evaluate the 4-point function $C_4(n)$, we need to use the standard ploy of doubling the space: $\Tr(\hat A^2)=\Tr((\hat A \otimes \hat A)\;\hat S)$ where $\hat S$ swaps the original and ancilla spaces. With $\hat A=\hat W(n)\hat V$. The only relevant average needed is 
\beq
\ovl{\hat W \otimes \hat W}^{\hat W}= \hat S,
\eeq 
and it follows using identities known for the operator entanglement \cite{anand2021brotocs,styliaris2021information} that $\ovl{C_4(n)}^{\hat W,\hat V}= d^2[1-E_l(\hat U(n))]$ and hence the OTOC averaged over the observables is 
\beq
\label{OPEE_OTOC}
\ovl{C(n)}^{\hat W,\hat V}=d^2 E_l[\hat U(n)].
\eeq
Thus the observable averaged OTOC is identical to the OPEE. Based on ergodicity, the case of a single random realization may then be expected to be represented by this average.

In the asymptotic limit of large times, if the dynamics are chaotic, we may expect that $\hat U(n)$ is a complex operator on the whole Hilbert space and treat it as being sampled according to the random CUE of size $2^{N}$ while keeping the $\hat W$ and $\hat V$ as fixed or non-random operators. The averaged quantities for traceless operators $\hat V$ and $\hat W$ are (see Appendix \ref{appendix_C} for details) 
\begin{subequations}
\begin{align}
\overline{C_2(n)}^{U}&= \frac{1}{d^2} \Tr (\hat W^2) \Tr (\hat V^2)\\
\overline{C_4(n)}^{U}&=\frac{-1}{d^2(d^2-1)}\Tr (\hat W^2) \Tr (\hat V^2)\\
\overline{C(n)}^{U}&=\frac{1}{d^2-1} \Tr (\hat W^2) \Tr (\hat V^2).
\end{align}
\end{subequations}
For the $\hat W$ and $\hat V$ in Eq.~(\ref{Block}), the asymptotic value of the OTOC, ignoring the $C_4$ value, which is of lower order in the Hilbert space dimension, 
is this average and denoted below as
\beq
C(\infty)=4/N^2.
\eeq
For the GUE random $\hat V$ and $\hat W$ used above $\overline{\Tr \hat W^2}=d^2$ and hence in this case $C(\infty)=d^2=2^{2N}$ for large $d$. We will always study scaled OTOC, dividing by the relevant $C(\infty)$; thus for the random operator case, the averaged and scaled OTOC is exactly the OPEE $E_l[\hat U(n)]$.

\subsection{Nearest-neighbour spacing distribution}
 \label{LSD}
\par
 Spectral statistics of the spacing between consecutive energy levels are used to differentiate the chaotic and integrable regimes. In order to calculate the NNSD, first, we need to identify the symmetries of the Hamiltonian. Next, the Hamiltonian is block diagonalized in the symmetry sectors. Our system with open boundary conditions has a “bit-reversal” symmetry at all the Floquet periods. This bit-reversal symmetry is due to the fact that the field and interaction do not distinguish the spins by interchanging the spins at the sites $i$ and $N-i+1$ for all $i=1,\cdots, N$. Let us consider $\hat B$ a bit-reversal operator given by
\begin{equation}
\hat B|\mathtt{s}_1,\mathtt{s}_2,\cdots,\mathtt{s}_N\rangle=|\mathtt{s}_N,\cdots,\mathtt{s}_2,\mathtt{s}_1\rangle,\hspace{.5cm} [\hat U,\hat B]=0,\nonumber\\
\end{equation}
where $|\mathtt{s}_i\rangle$ is any single-particle basis state in standard $(\mathtt{s}_z)$ basis. We divide whole basis sets into two groups of basis states, one with the palindrome in which there is no change in the state after applying the operator $\hat B$ {\it{i.e.}}, $\hat B|\mathtt{s}_1,\mathtt{s}_2,\cdots,\mathtt{s}_N\rangle=|\mathtt{s}_1,\mathtt{s}_2,\cdots,\mathtt{s}_N\rangle$. The other one with the non-palindrome in which states get reflected after applying the operator $\hat B$ {\it{i.e.}} $\hat B|\mathtt{s}_1,\mathtt{s}_2,\cdots,\mathtt{s}_N\rangle=|\mathtt{s}_N,\cdots,\mathtt{s}_2,\mathtt{s}_1\rangle$.  Since $\hat B^2=1$, the eigenvalues of $\hat B$ are $\pm1$. The eigenstates can be classified as odd or even states under bit reversal. All the palindromes are even states; however, all the non-palindromes have one even and one odd state. Sum and difference of the non-palindrome and its reflection generate even and odd states. 
The dimension of the odd subspace is equal to half the number of
non-palindromic binary words of length $N${\it, i.e.,} $\frac{1}{2}\Big(2^N-2^\frac{N}{2}\Big)$, while the even
subspace is equal to the sum of half the number of palindromic bit sequences and half the number of total space of the
same length {\it i.e.} $\frac{1}{2}\Big(2^N+2^\frac{N}{2}\Big)$.
\par
In the NNSD, it is necessary to concentrate on the fluctuations properties of the spectrum, which display universal effects. For this, one should do the unfolding of the spectrum in order to get rid of the non-universal properties (level density). Unfolding is usually done by parameterizing numerically obtained level densities in terms of a smooth function, typically a polynomial, followed by mapping the energies to unfolded ones such that the mean energy spacing is unity.
\par
We consider the ensemble of differences between the consecutive energy levels. The average spacing between the consecutive eigenvalues is controlled by the local mean density of states. If within a region $\delta E$ of the spectrum, there are $D(E)\delta E$ eigenvalues, then the average spacing between the consecutive eigenvalues will be $1/D(E)$. If we re-scale the differences between the consecutive eigenvalues by the local mean density of states, the average difference will be one.  We study the shape of distribution by using the NNSD, which may be used as an indicator of quantum chaos and nontrivial integrable models.
\par In NNSD, strongly chaotic points are those where the unfolded level-spacings are well described by the Wigner distribution \cite{Luca2016,mehta1991theory,averbukh2001angular} which is given as
\begin{equation}
P_W(s)=\frac{\pi s}{2}e^{-\pi s^2/4},
\end{equation}
 where $s$ is drawn from the ensemble of consecutive energy level separation. On the other hand, nontrivial integrable models are those where the unfolded NNSD follows Poisson statistics,
\begin{equation}
P_P(s)=e^{-s}.
\end{equation}




\section{Constant field Floquet system}
\label{CF}
\begin{figure}[H]
\centering
\begin{subfigure}{.49\textwidth} 
\includegraphics[width=.99\linewidth, height=.70\linewidth]{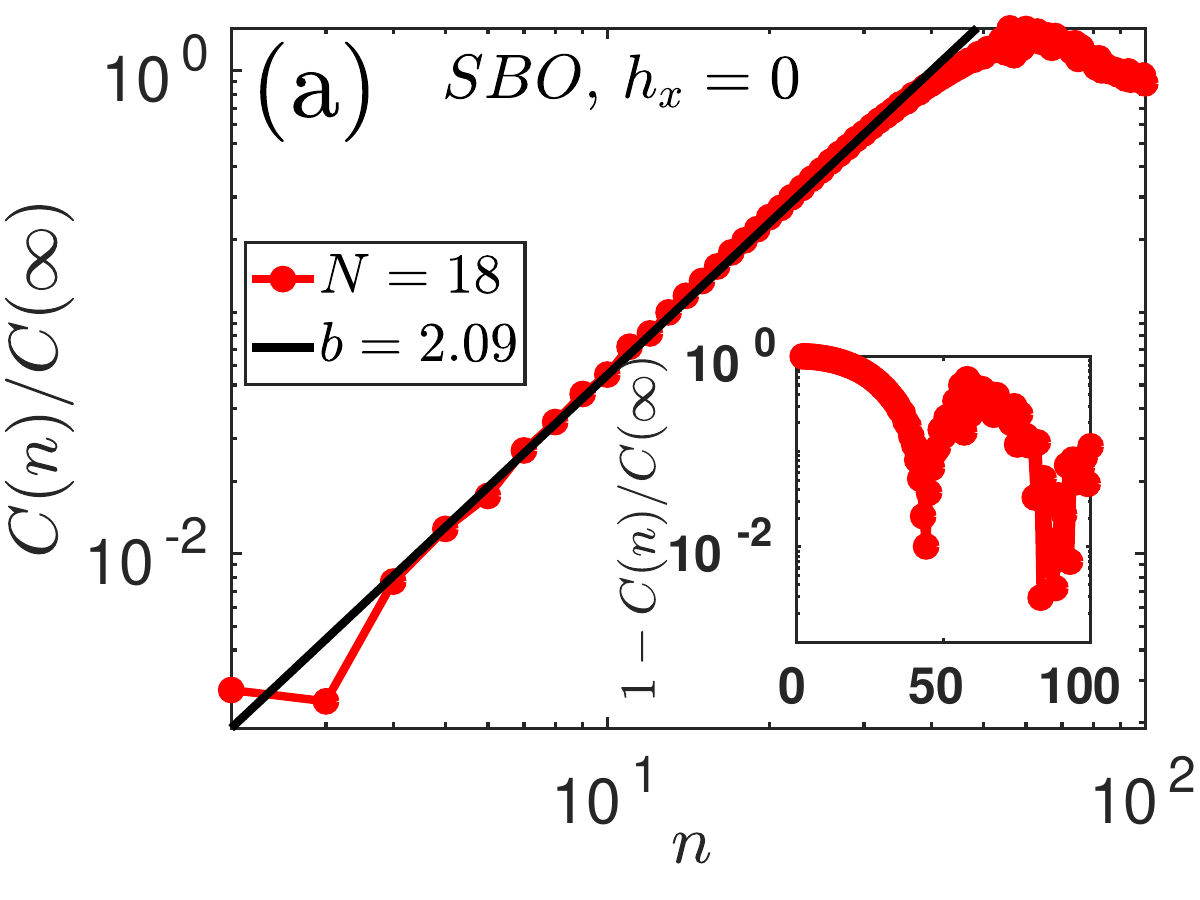}
\end{subfigure}
\begin{subfigure}{.49\textwidth} 
\includegraphics[width=.99\linewidth,height=.70\linewidth]{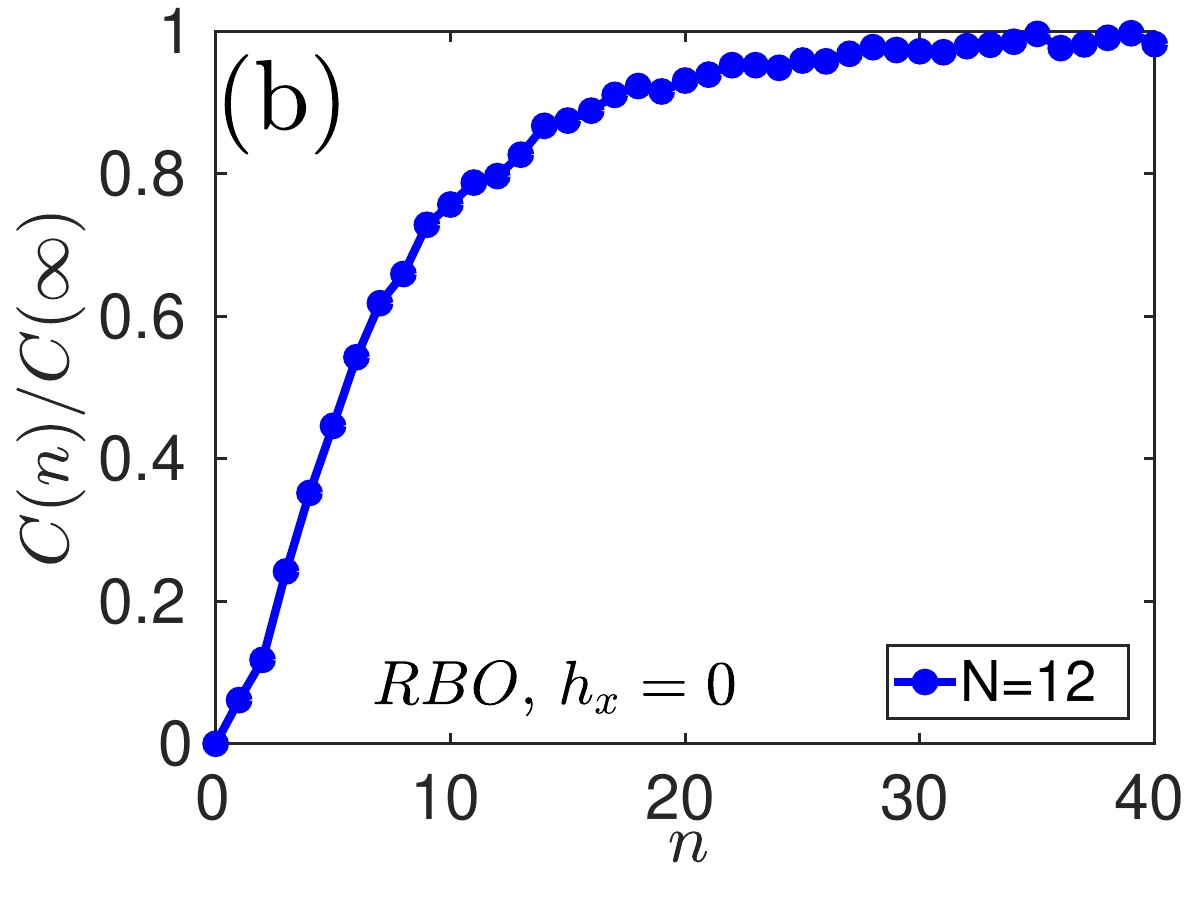}
\end{subfigure}
\begin{subfigure}{.49\textwidth} 
\includegraphics[width=.99\linewidth,height=.70\linewidth]{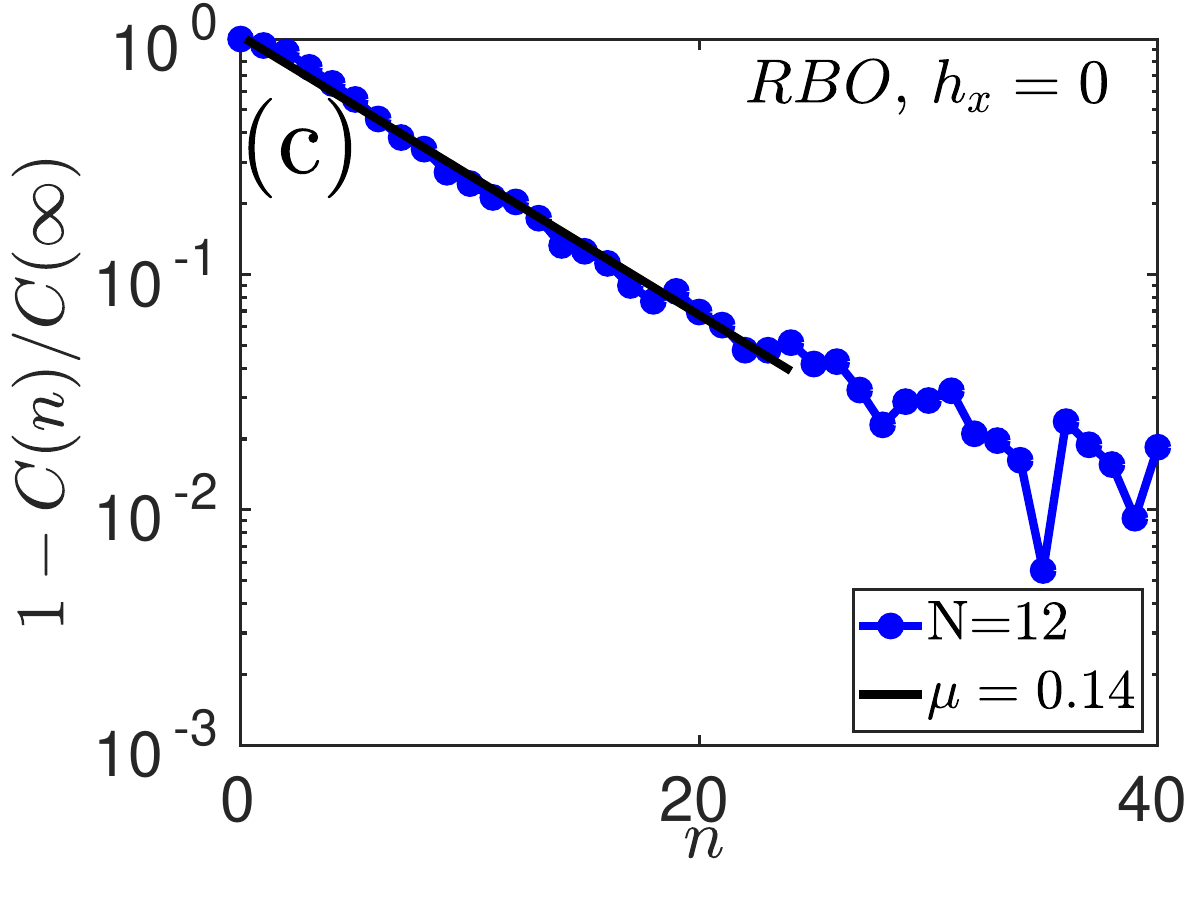}
\end{subfigure}
\begin{subfigure}{.49\textwidth}
\includegraphics[width=.99\linewidth, height=.70\linewidth]{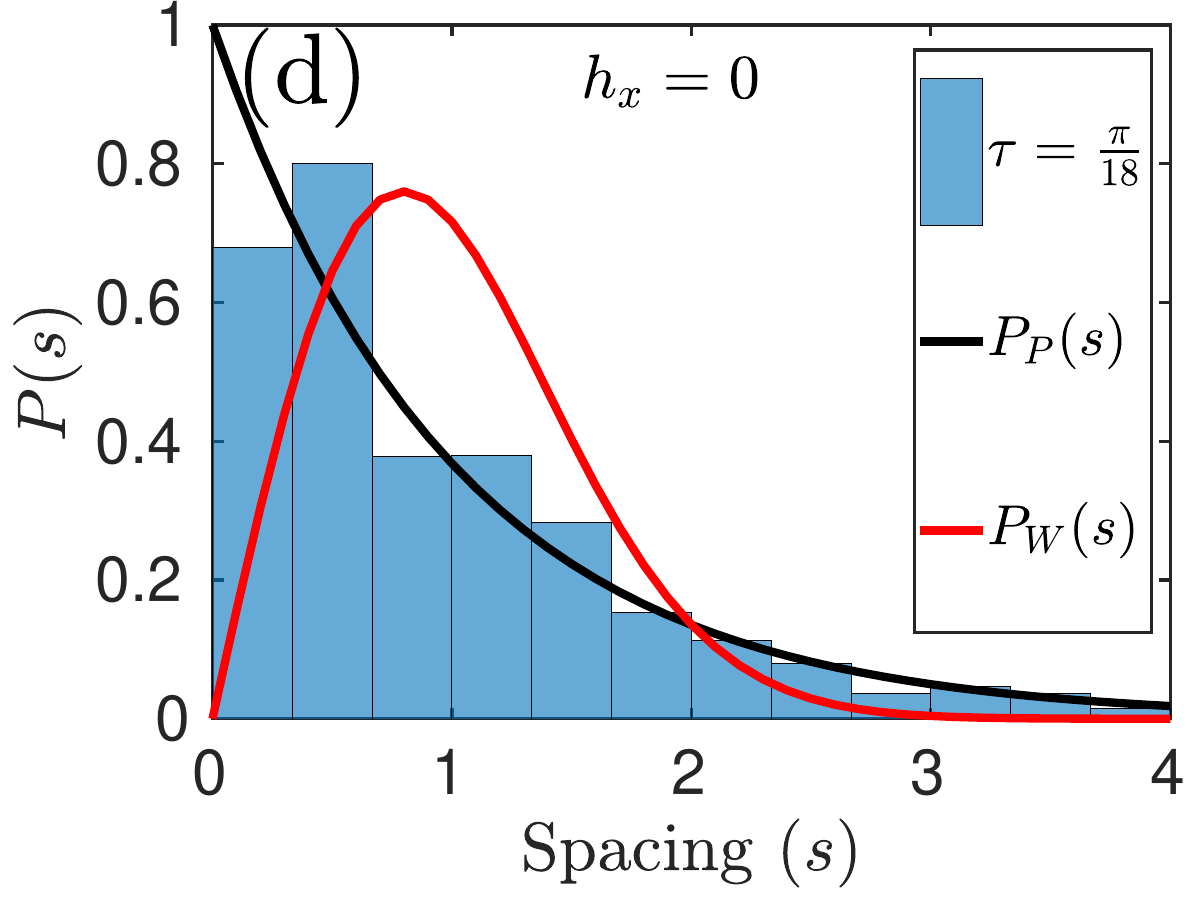}
\end{subfigure}
\caption{ Integrable $\mathcal{\hat U}_0$ system with parameters: $\tau=\frac{\pi}{18}$, $J_x=1$, $h_{x}=0$ and $h_{z}=4$.  (a) $C(n)/C(\infty)$ generated by SBOs vs. $n$ for $N=18$  ($\log-\log$). Line with points represents data from the numerical calculation, and the solid line is the polynomial fitting. Inset shows  $1-C(n)/C(\infty)$ vs. $n$ ($\log-$linear).   (b) $C(n)/C(\infty)$ vs. $n$ for $N=12$ and RBOs as observables.  (c) $1-C(n)/C(\infty)$ vs. $n$ for $N=12$ and RBOs as observables. Line with points is data generated numerically, and a solid line is the exponential fitting. (d) NNSD for $N=12$. In all the cases, open boundary condition is considered.}
       \label{pi18_hx0_hz4_int_N_p0} 
\end{figure}

 We analyze the OTOC given by Eq.~(\ref{Ch4_Cn1}) for integrable $\mathcal{\hat{U}}_0$ and nonintegrable $\mathcal{\hat{U}}_x$ systems defined in section \ref{Ch4_model}. The Floquet period $\tau$ acts as a parameter to drive the system into interesting dynamical regimes. In this chapter, we will discuss the dynamic (pre-Ehrenfest time) and saturation (post-Ehrenfest time) regions of OTOC generated by SBOs defined in Eq.~(\ref{Block}). We will only focus on the behavior of OTOC in the range of Floquet periods from $0$ to $\frac{\pi}{4}$ as OTOCs for our Floquet systems $\mathcal{\hat{U}}_0$ and $\mathcal{\hat{U}}_x$ are exactly the same for $\tau$ and $\frac{\pi}{2}-\tau$. This peculiarity of OTOC  can be shown below by taking $J_x=1$, $h_{x}=h_{z}=4$ and  replacing $\tau$ by $\frac{\pi}{2}-\tau$ in Eq.~(\ref{Ux}).
\begin{eqnarray}
\mathcal{\hat U}_x\Big(\frac{\pi}{2}-\tau\Big) &=&e^{-i(\hat H_{xx}+4\hat H_{x})(\frac{\pi}{2}-\tau)} e^{-i 4 \hat H_{z} (\frac{\pi}{2}-\tau)}, \nonumber \\
&=&e^{i(\hat H_{xx}+4\hat H_{x})\tau} e^{4i\hat H_{z}\tau}e^{-i(\hat H_{xx}+4\hat H_{x})\frac{\pi}{2}} e^{-4i \hat H_{z}\frac{\pi}{2}}, \nonumber \\
&=&e^{i(\hat H_{xx}+4\hat H_{x})\tau} e^{4i\hat H_{z}\tau}=\mathcal{\hat U}^{\dagger}_x(\tau).
\end{eqnarray}
We see that the Floquet map at $(\frac{\pi}{2}-\tau)$ is a complex conjugate of the Flqouet map at  $\tau$. Therefore, OTOC behavior will be exactly the same at both $\tau$ and $(\frac{\pi}{2}-\tau)$. 
In the integrable $\mathcal{\hat{U}}_0$ case, at $\tau=\frac{\pi}{18}$, dynamic region of the OTOC shows power-law growth with the exponent of the power-law $b=2.03$ (approximately quadratic) [Fig. \ref{pi18_hx0_hz4_int_N_p0}(a)]. In the range $\tau=0$ to $\tau=\frac{\pi}{2}$, for any $\tau$, OTOC shows power-law growth except at $\tau=\frac{\pi}{4}$. While approaching the saturation value, OTOC is not showing exponential behavior, as seen in the inset of  Fig. \ref{pi18_hx0_hz4_int_N_p0}(a). Let us check the behavior of OTOC by replacing the SBOs with random block operators (RBOs). With random block observables, OTOC thermalizes quickly as compared to SBOs. This led to the disappearance of power-law growth in the dynamic region for $\tau=\frac{\pi}{18}$. [Fig. \ref{pi18_hx0_hz4_int_N_p0}(b)]. However, after the Ehrenfest time, the OTOC saturates exponentially with a rate $\mu=0.14$ [Fig. \ref{pi18_hx0_hz4_int_N_p0}(c)].
Fig.~\ref{pi18_hx0_hz4_int_N_p0}(d) shows the NNSD of the  $\mathcal{\hat U}_0$ system at 
$\tau=\frac{\pi}{18}$  is Poisson type rather than Wigner-Dyson type \cite{Fortes2019,lin2018out}. 
The system displays Poisson statistics at all the Floquet periods between $0$ to $\frac{\pi}{2}$ except at $\frac{\pi}{4}$. At $\tau=\frac{\pi}{4}$, multiplication of the Floquet period and amplitude of the transverse magnetic field ($h_{z}\tau$) is equal to $\pi$ resulting in a constant contribution of the field term in the Floquet map. Hence, for $\tau=\pi/4$, only the coupling term is present in the Floquet map, which provides degenerate eigenvalues. Due to this fact, NNSD is unable to specify the behavior of either the Poisson or  Wigner-Dyson type.
Mathematically, it can be given as:
\begin{equation}
\mathcal{\hat U}_x=e^{-i(\hat H_{xx}+4\hat H_{x})\frac{\pi}{4}} e^{-4i \hat H_{z}\frac{\pi}{4}},
=e^{-i\hat H_{xx}\frac{\pi}{4}}.
\end{equation}
\begin{figure}
\centering
    \begin{subfigure}{.32\textwidth}
     \includegraphics[width=.99\linewidth, height=.90\linewidth]{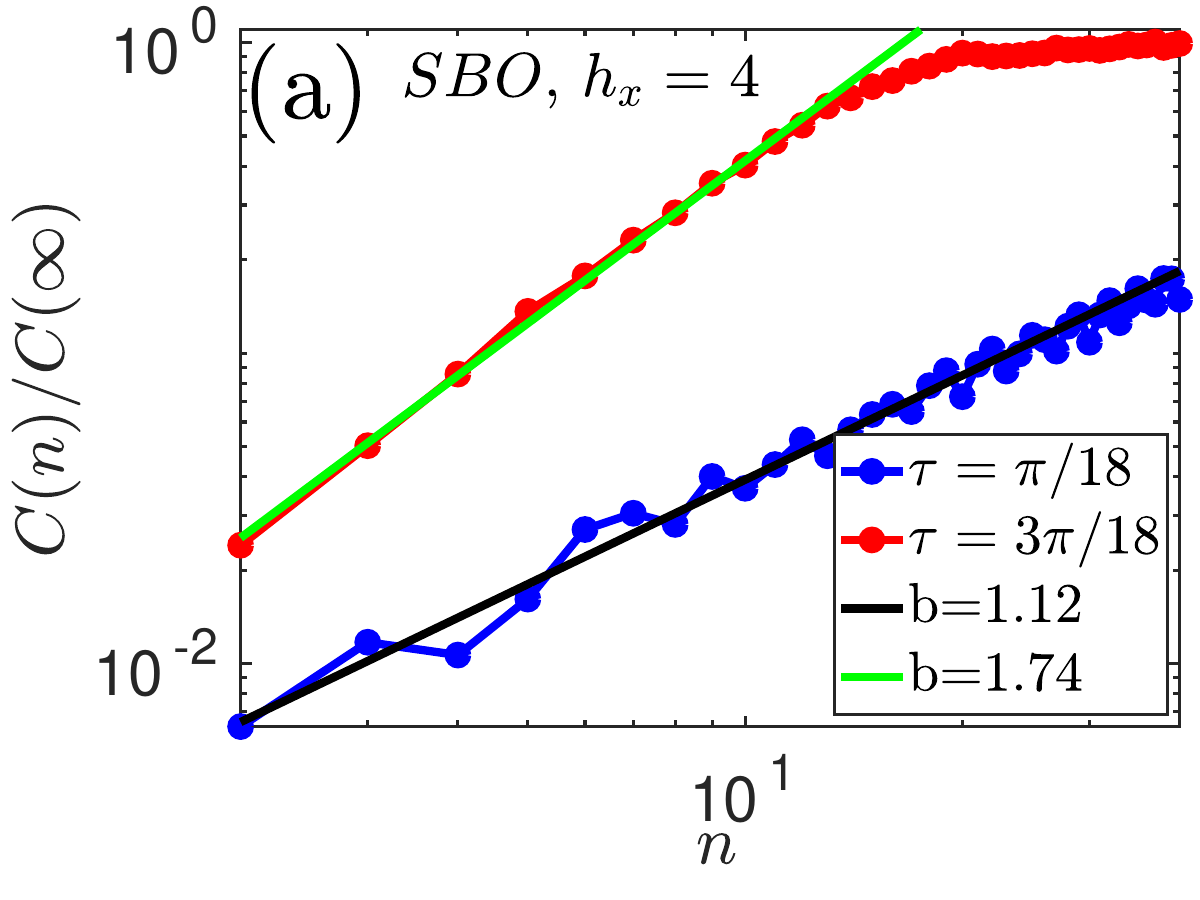}
       \end{subfigure}
       \begin{subfigure}{.33\textwidth}
\includegraphics[width=.99\linewidth, height=.90\linewidth]{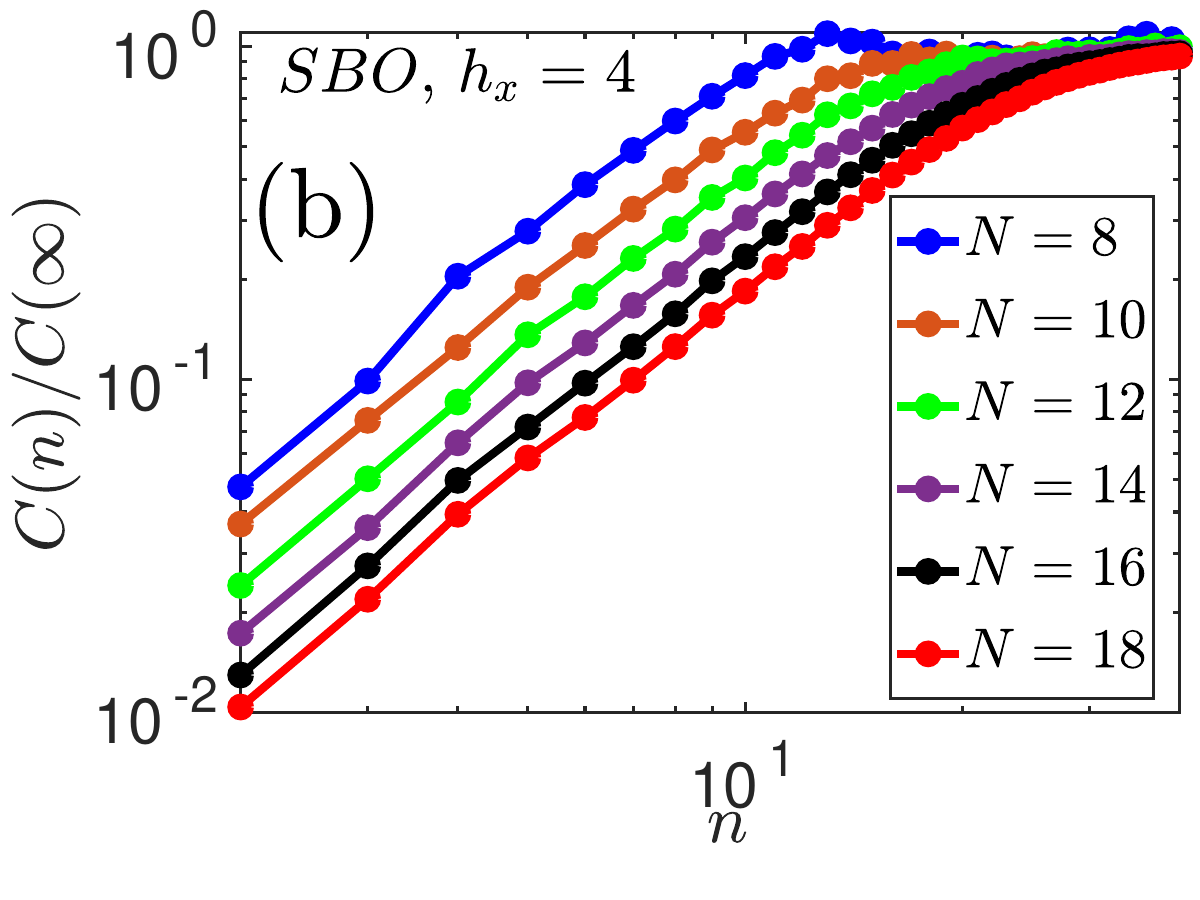}
\end{subfigure}
       \begin{subfigure}{.33\textwidth}
\includegraphics[width=.99\linewidth,height=.90\linewidth]{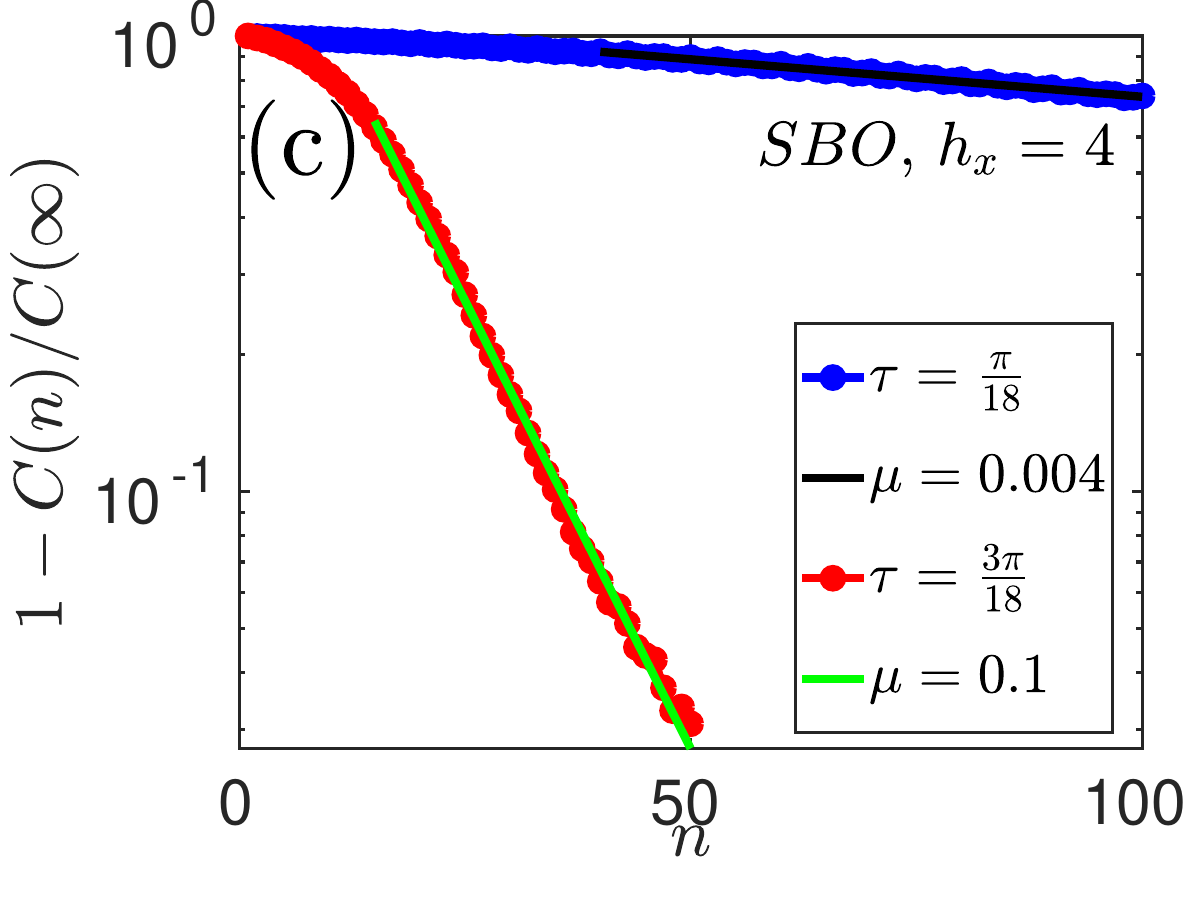}
\end{subfigure}
\begin{subfigure}{.49\textwidth}
\includegraphics[width=.99\linewidth, height=.70\linewidth]{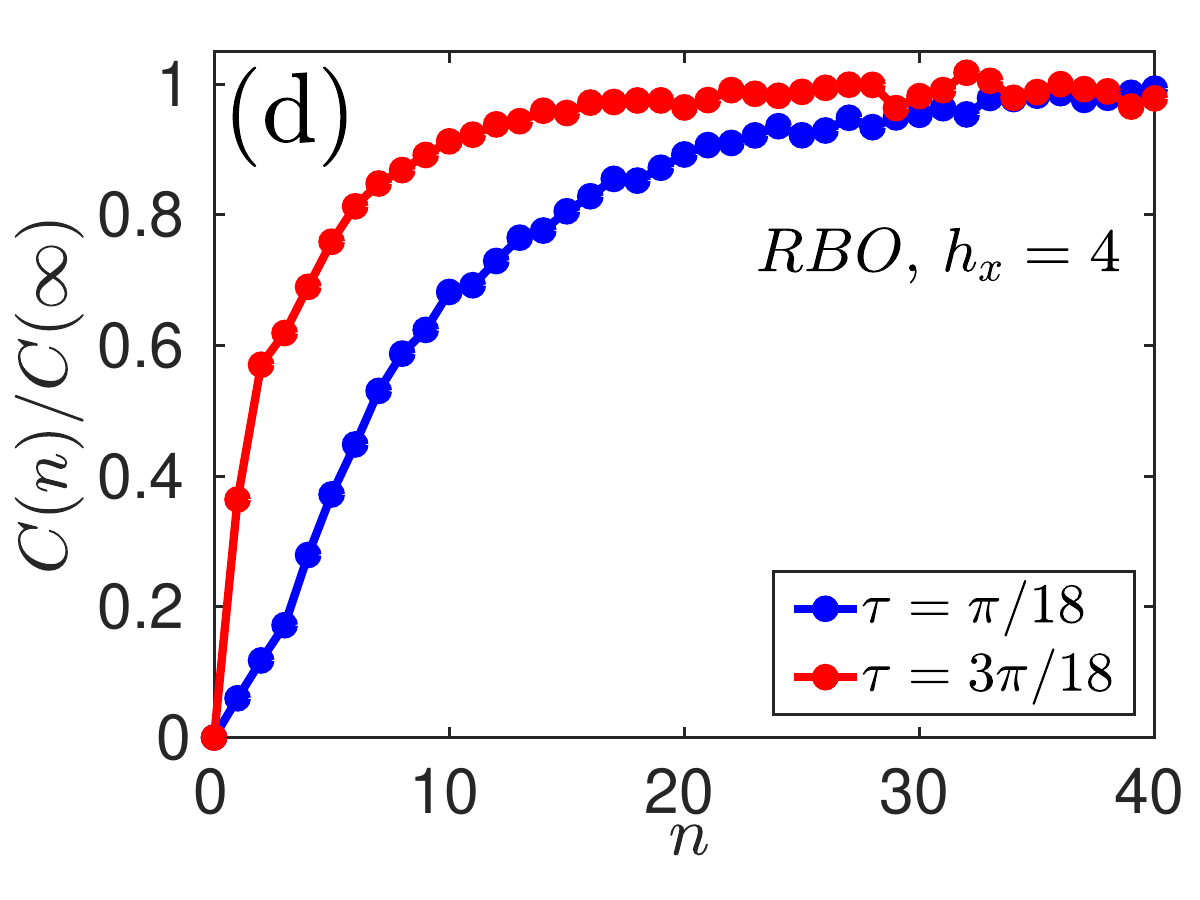}
\end{subfigure}
       \begin{subfigure}{.49\textwidth}
\includegraphics[width=.99\linewidth, height=.70\linewidth]{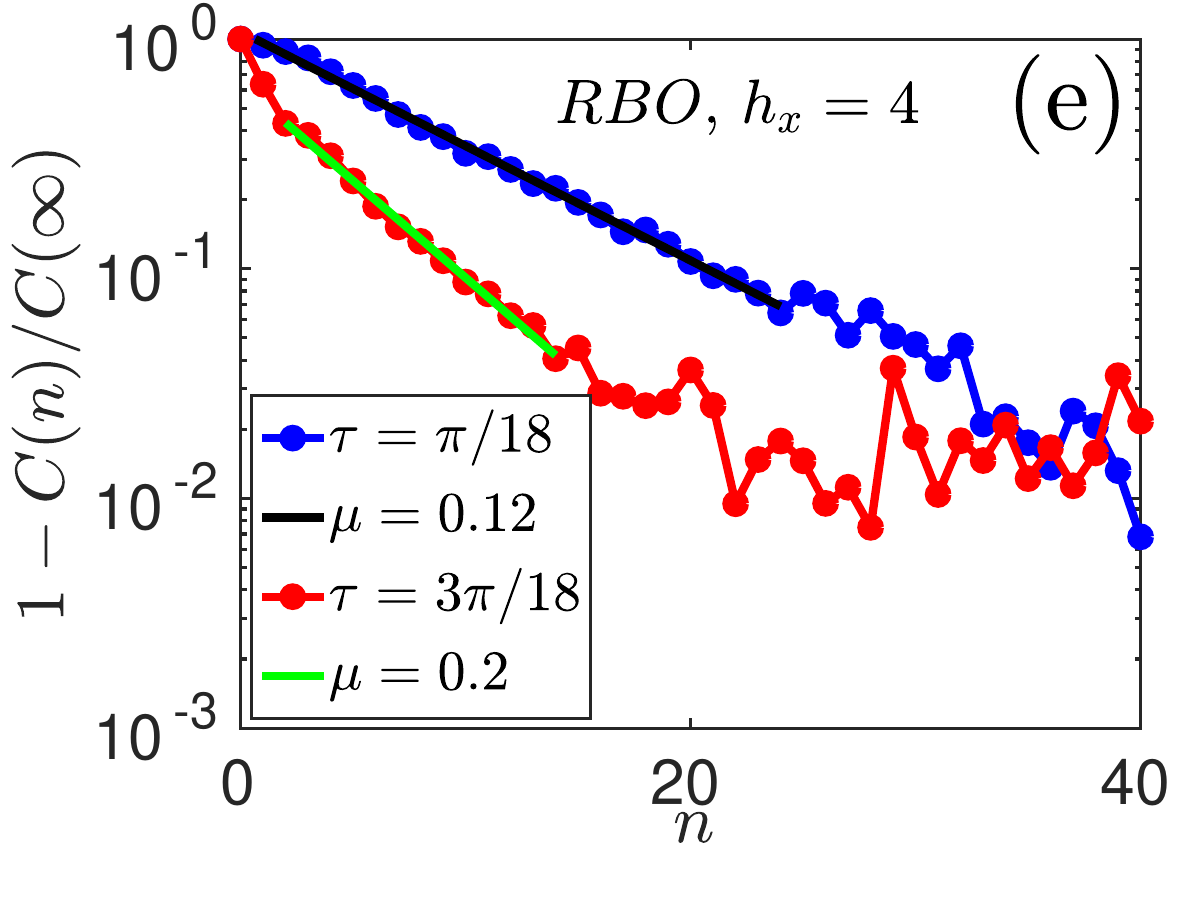}
\end{subfigure}
       \begin{subfigure}{.49\textwidth}
\includegraphics[width=.99\linewidth,height=.70\linewidth]{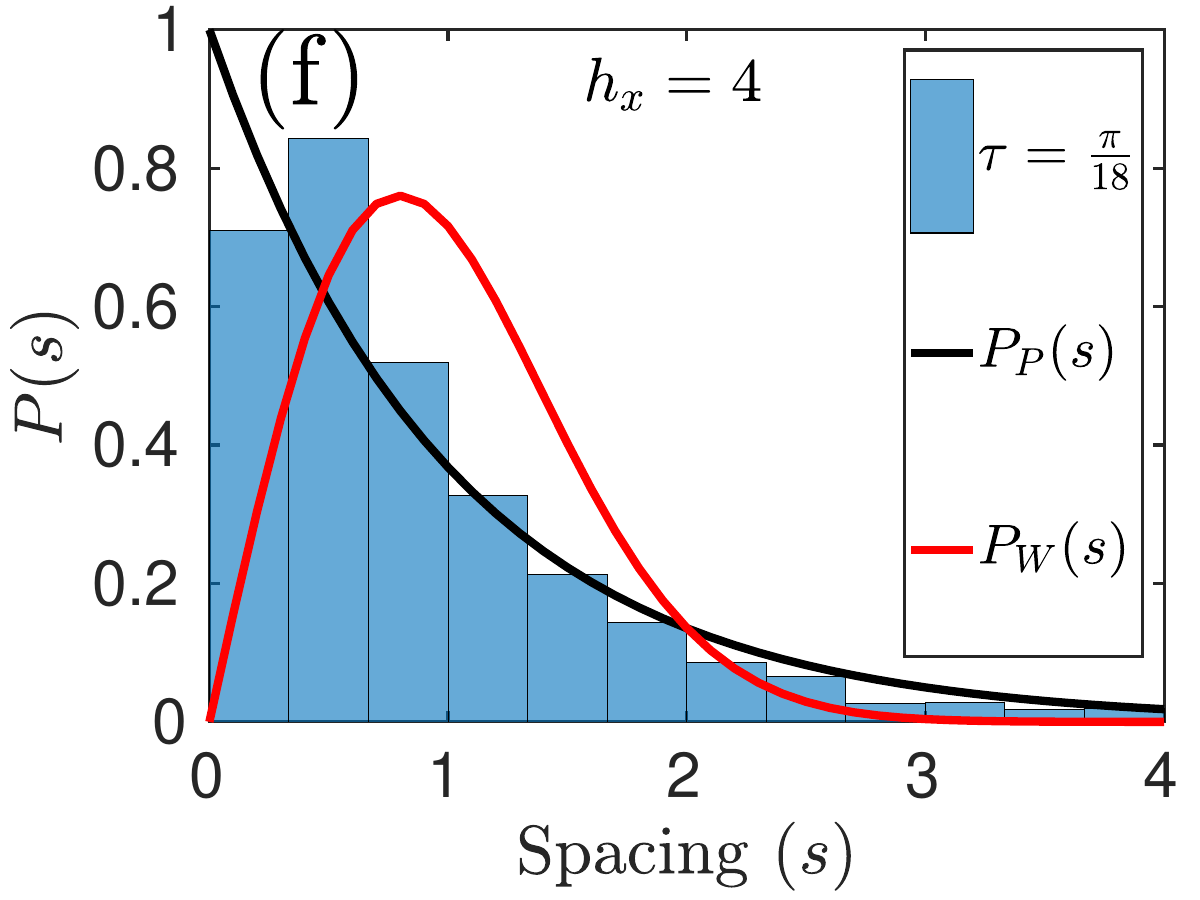}
\end{subfigure}
\begin{subfigure}{.49\textwidth}
\includegraphics[width=.99\linewidth,height=.70\linewidth]{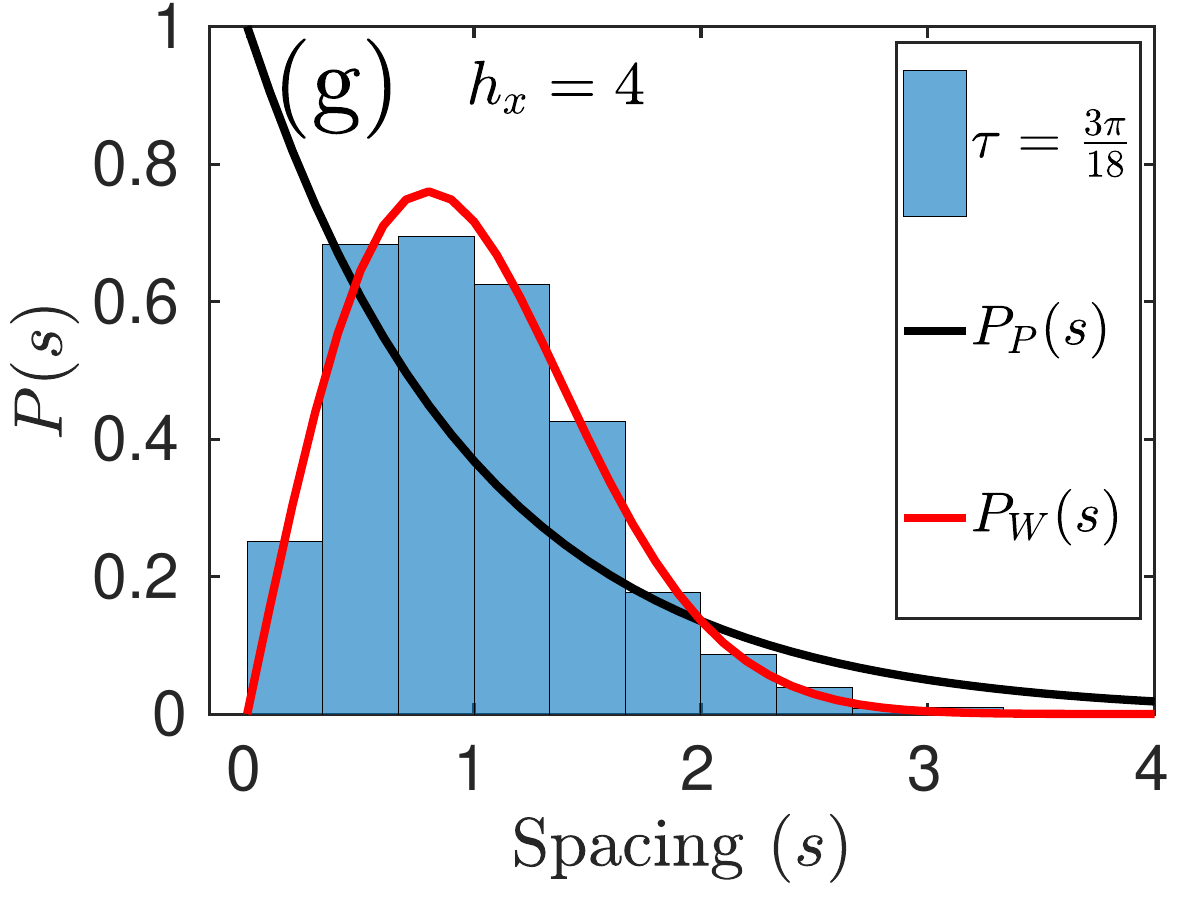}
\end{subfigure}
\caption{Nonitegrable $\mathcal{\hat U}_x$ system with parameters:  $J_x=1$, $h_{x}=4$, $h_{z}=4$ and $\tau=\frac{\pi}{18}$, $\frac{3\pi}{18}$.  (a) Illustrates the $C(n)/C(\infty)$ by using the SBOs  vs. $n$ for $N=18$ ($\log-\log$). Lines with points represent data from the numerical calculation, and solid lines are the polynomial fitting with exponent $b=1.12$ at $\tau=\frac{\pi}{18}$  and $b=1.74$ at $\tau=\frac{3\pi}{18}$. (b) $C(n)/C(\infty)$ by using the SBOs  vs. $n$ at different $N$ for $\tau=\frac{3\pi}{18}$. (c) $1-C(n)/C(\infty)$ vs. $n$ ($\log-$linear). Lines with points are data generated numerically, and solid lines are the exponential fitting. (d) Illustrates the OTOCs of RBOs vs. $n$ for $N=12$ (g) $1-C(n)/C(\infty)$ vs. $n$ ($\log-$linear). Lines with points are data generated numerically, and solid lines are the exponential fitting. NNSD of the $\mathcal{\hat U}_x$ system at  (f) $\tau=\frac{\pi}{18}$ and (g) $\tau=\frac{3\pi}{18}$ with $N=12$. In all cases, an open boundary chain is considered.}
\label{pi18_hx4_hz4_cf_nint_p0_N} 
\end{figure}
OTOC in the nonintegrable $\mathcal{\hat U}_x$ system shows a power-law growth similar to that in the integrable case. However, in the $\mathcal{\hat U}_x$ case, the exponent of the power-law is smaller as compared to the integrable case. The exponent increases with increasing $\tau$. At  $\tau=\frac{\pi}{18}$ and $\frac{3\pi}{18}$ exponent of the power-law is $1.12$ and $1.74$, respectively. Hence, at $\tau=\frac{3\pi}{18}$, the exponent is nearly quadratic in a power-law growth [Fig. \ref{pi18_hx4_hz4_cf_nint_p0_N}(a)]. The exponent of the power-law is independent of the system size, but the saturation of the OTOC depends on the system size. Longer the size, longer time it takes for saturation. Hence, the saturation value of OTOC exhibits the finite-size effect [Fig. \ref{pi18_hx4_hz4_cf_nint_p0_N}(b)]. As $N\rightarrow\infty$, saturation will occur after the infinite number of kicks. OTOC reached to saturation exponentially at all the Floquet periods. As the Floquet period increases, the rate of saturation increases [Fig. \ref{pi18_hx4_hz4_cf_nint_p0_N}(c)].
\par
Now, if we  replace the observables $\hat V$  and $\hat W$ to random matrices, growth of OTOC does not show Lyapunov or power-law type at any $\tau$ [Fig. \ref{pi18_hx4_hz4_cf_nint_p0_N}(d)]. OTOC saturates exponentially, and the exponent of the exponential increases with increasing $\tau$, which can be seen in Fig.~\ref{pi18_hx4_hz4_cf_nint_p0_N}(e).
\par 
NNSD of the nonintegrable Floquet system displays the Wigner-Dyson distribution at Floquet period $\frac{\pi}{3}$ and crossover to the Poisson distribution as the Floquet periods changes away from the $\frac{\pi}{3}$. This point is the most chaotic point in the Floquet system [Fig.~\ref{pi18_hx4_hz4_cf_nint_p0_N} (f, g)]. 
\par
Floquet system at $\tau=\frac{\pi}{4}$ is a special case, which was reported in different contexts earlier as well \cite{naik2019controlled,Pal2018}. For the choice of parameters in this section {\it i. e.}, $h_{x}=4$ and $h_{z}=4$, and $\tau=\pi/4$ we get $h_{x/z}\tau= \pi$. 
This results in a constant contribution from the magnetic field terms in the Floquet map defined in subsection \ref{Ch4_model} and only the spin-spin interaction term ${\hat H}_{xx}$ term evolves the SBO in the OTOC calculation. Since the SBOs are also in the direction of spin-spin interaction {\it{i.e.}}, along the longitudinal direction,  the SBO will be stationary at all times. Therefore, OTOC remains constant (equal to one) at all the kicks.
\section{Special case}
\label{special_CF}
\begin{figure}
\centering
\begin{subfigure}{.49\textwidth}
\includegraphics[width=.99\linewidth,height=.70\linewidth]{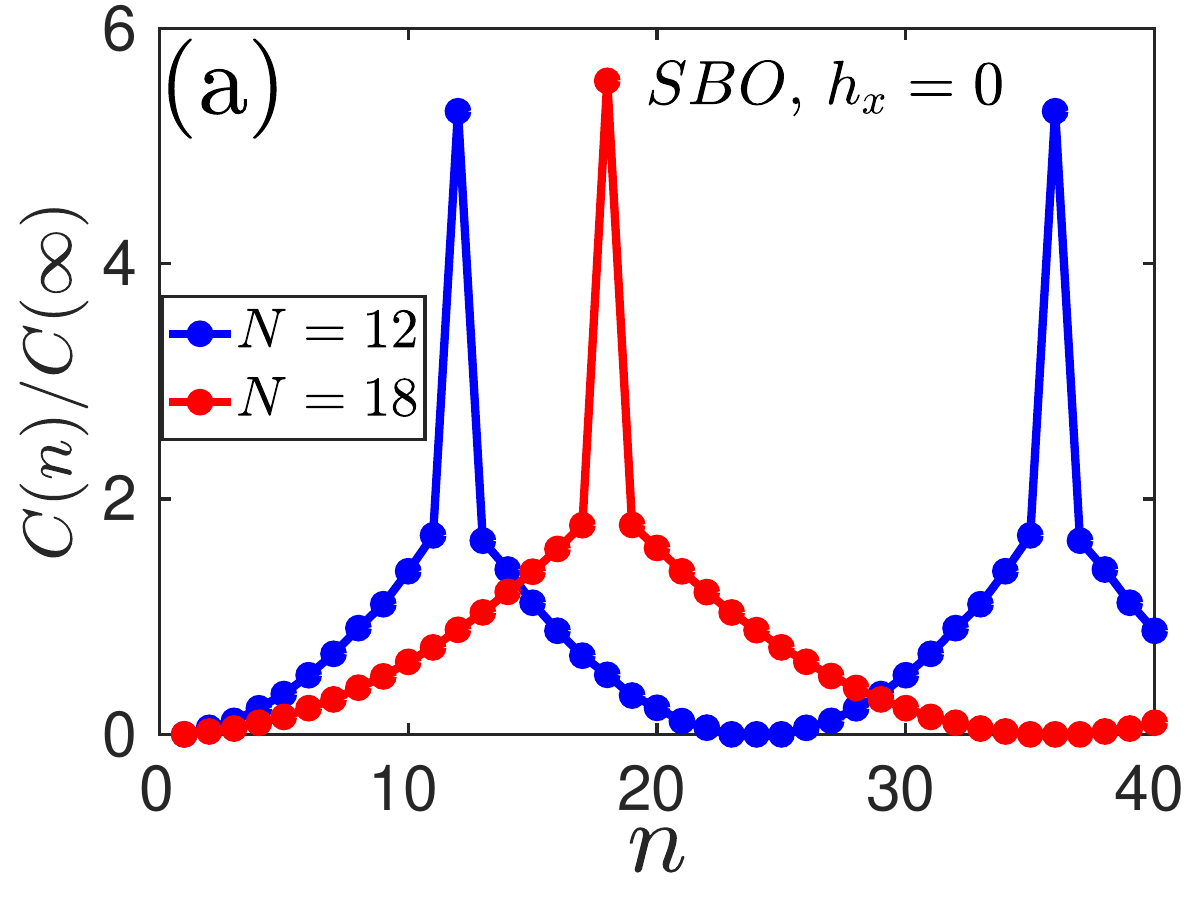}
\end{subfigure} 
\begin{subfigure}{.49\textwidth}
\includegraphics[width=.99\linewidth,height=.70\linewidth]{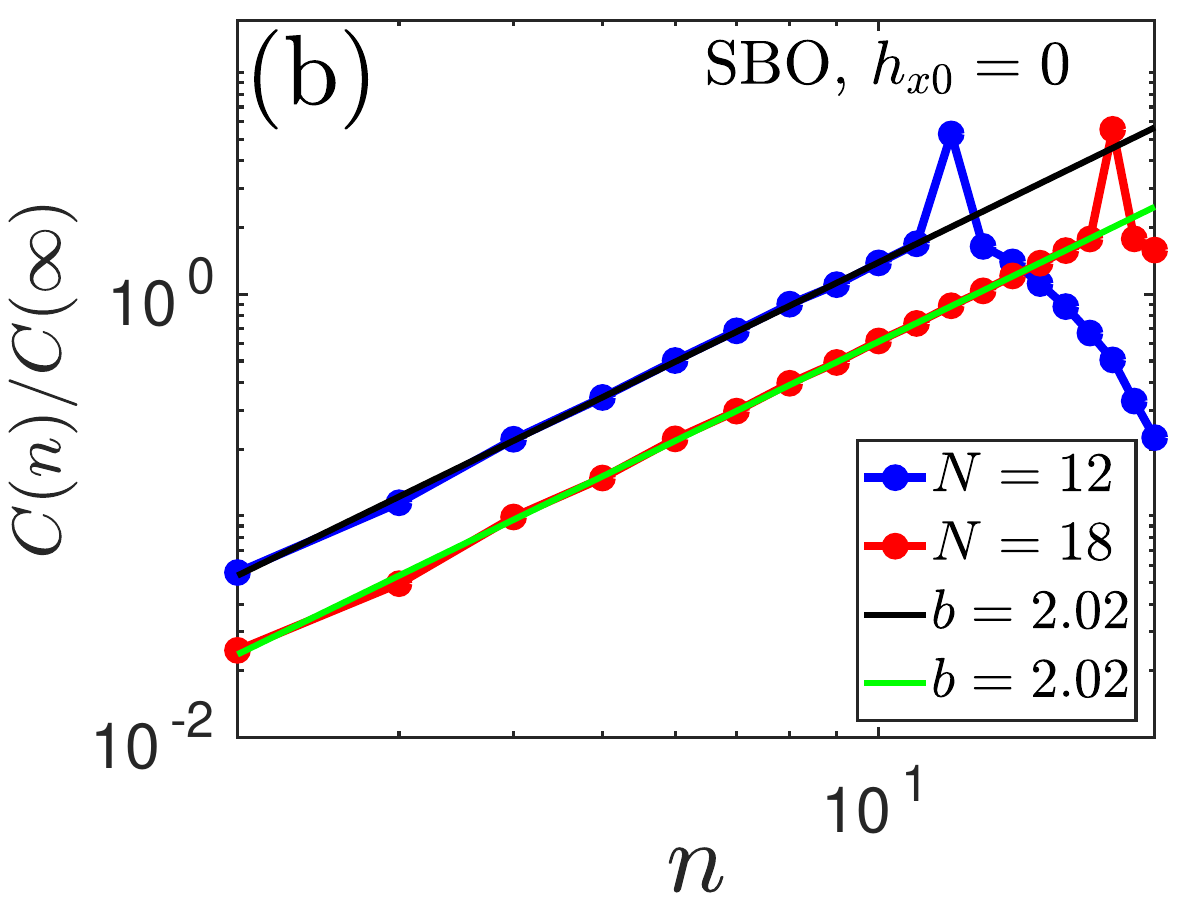}
\end{subfigure} 
\begin{subfigure}{.49\textwidth}
\includegraphics[width=.99\linewidth, height=.70\linewidth]{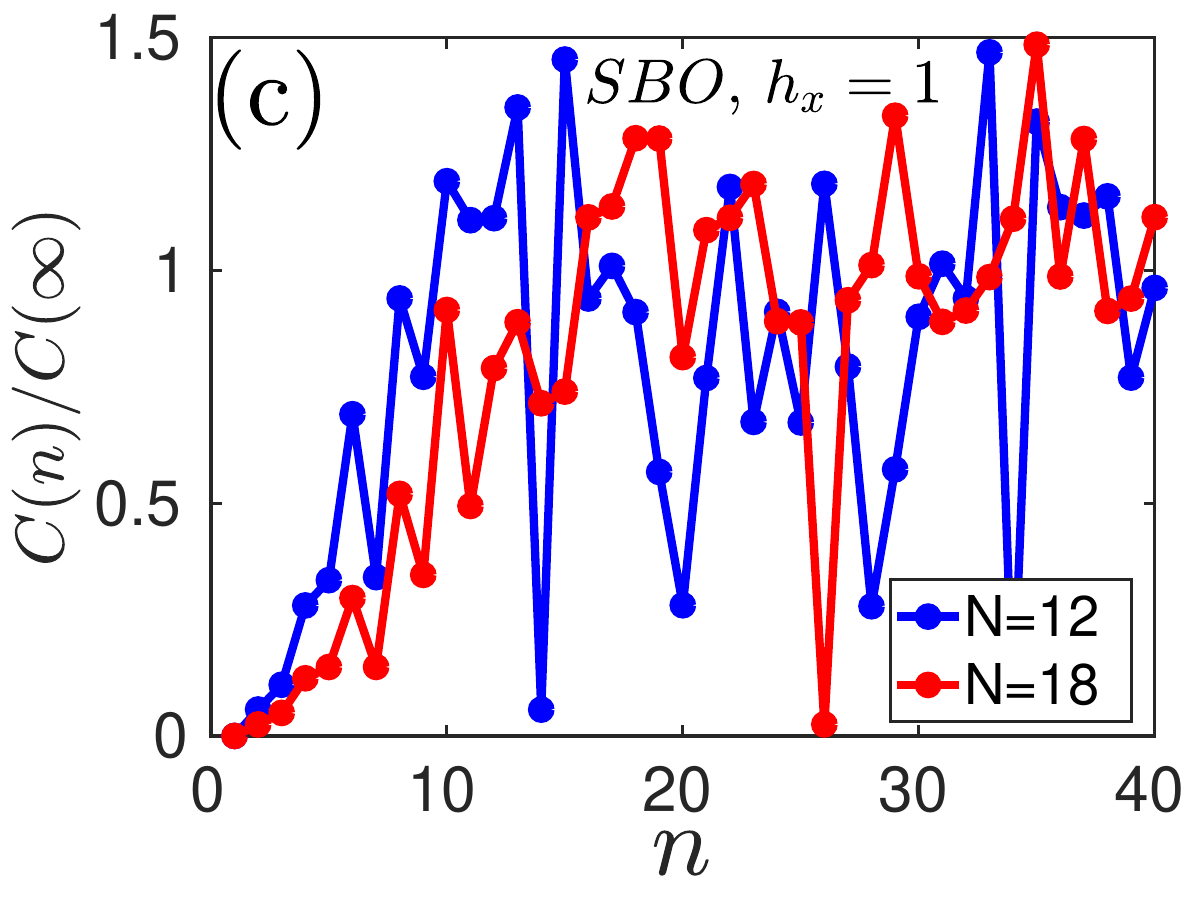}
\end{subfigure}
\begin{subfigure}{.49\textwidth}
\includegraphics[width=.99\linewidth, height=.70\linewidth]{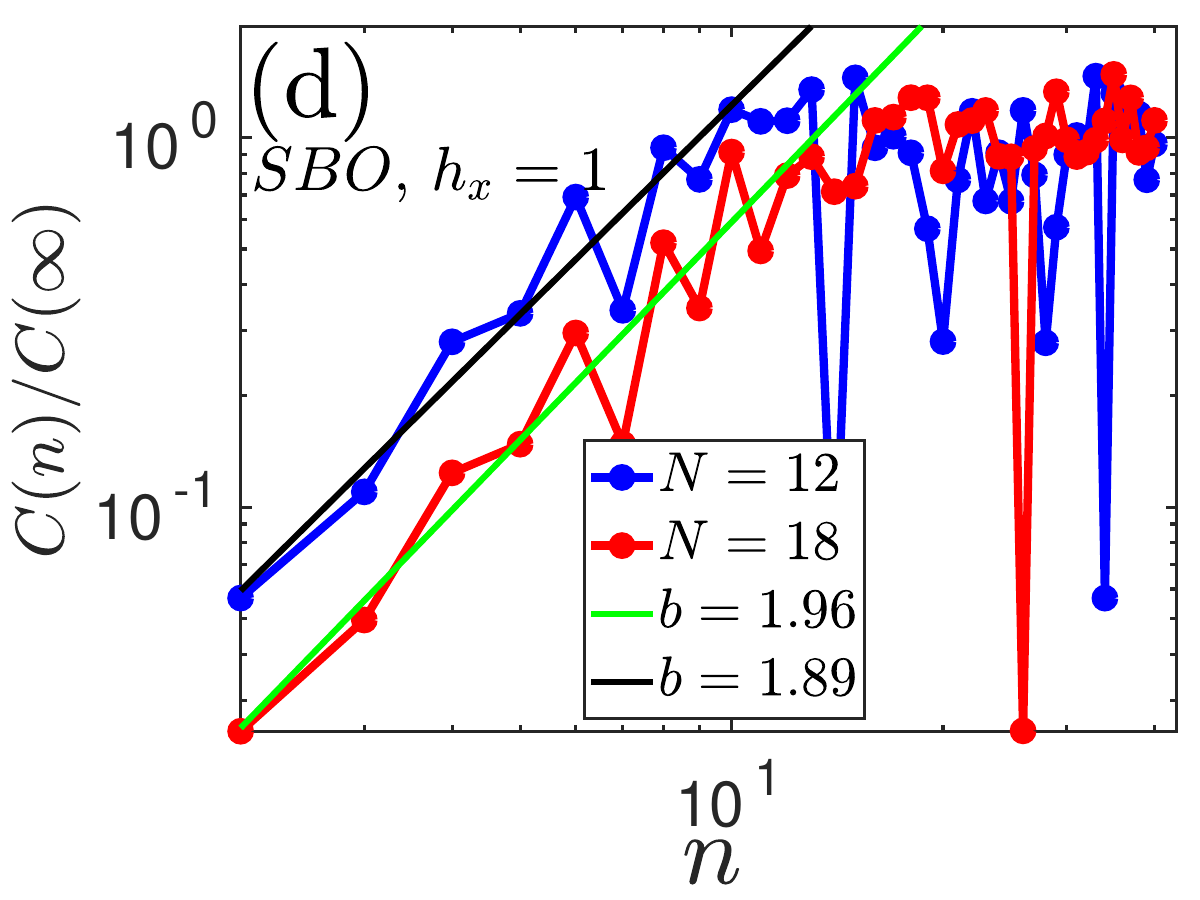}
\end{subfigure} 
\begin{subfigure}{.49\textwidth}
\includegraphics[width=.99\linewidth, height=.70\linewidth]{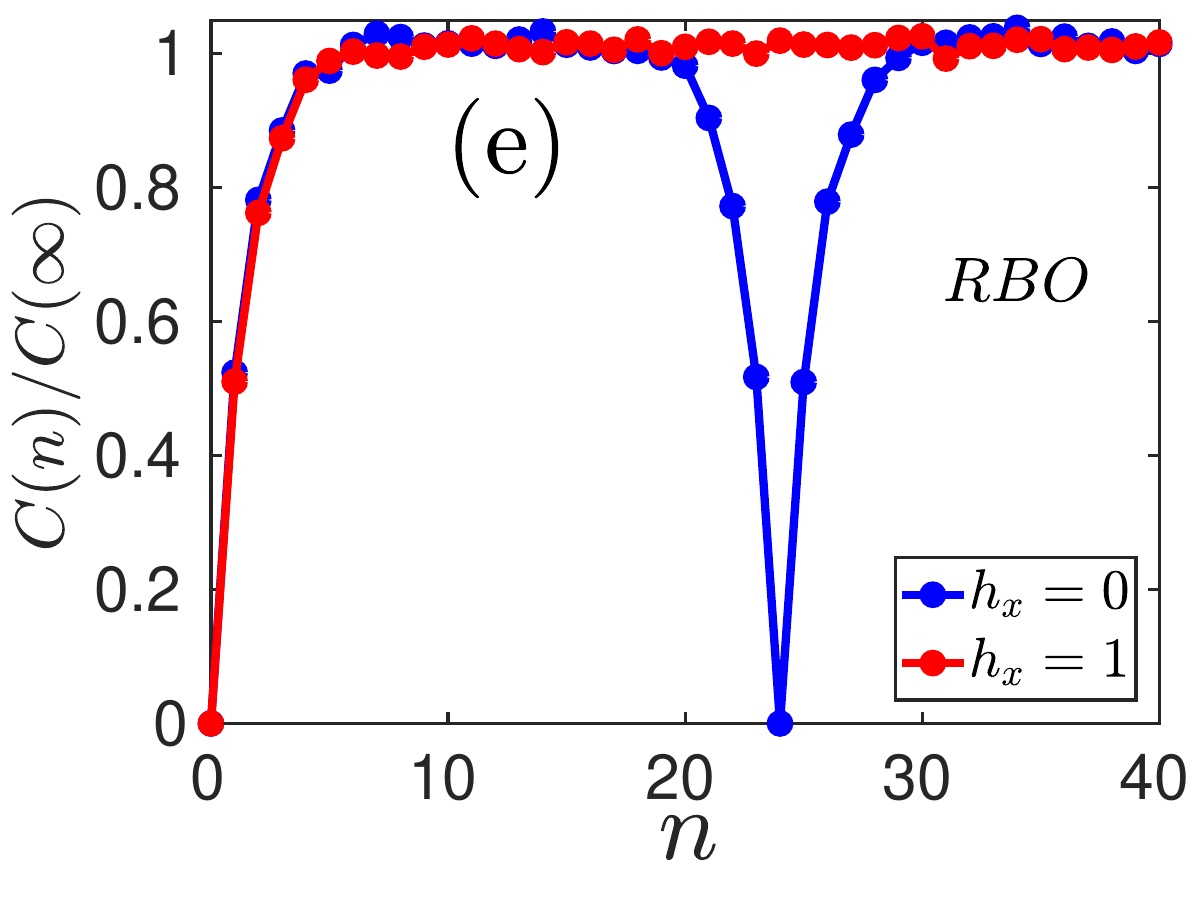}
\end{subfigure} 
\begin{subfigure}{.49\textwidth}
\includegraphics[width=.99\linewidth, height=.70\linewidth]{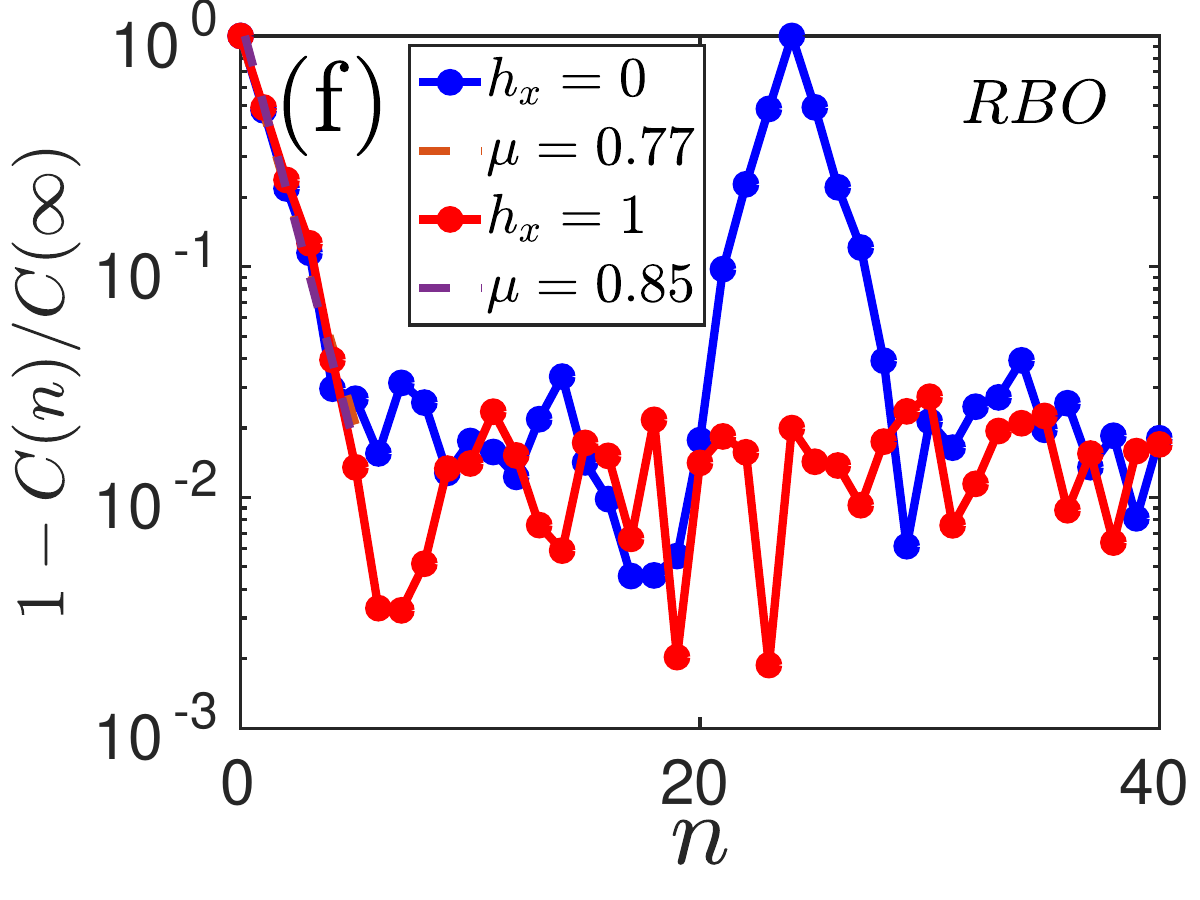}
\end{subfigure} 
\label{pi4_hx1_hz1_cf_log_rand5_N_nint_p0} 
\caption{ (a) $C(n)/C(\infty)$ of SBOs vs. $n$ in the $\mathcal{\hat U}_0$ system for $N=18$. (b) $\log-\log$ behavior of ``a" in which lines with points represent data from the numerical calculation, and solid lines are the polynomial fitting. (c) $C(n)/C(\infty)$ of SBOs with $n$ in the $\mathcal{\hat U}_x$ system for $N=18$. (d) $\log-\log$ behavior of ``c" in which lines with points represent data from the numerical calculation, and solid lines are the polynomial fitting. (e) $C(n)/C(\infty)$ of RBOs vs. $n$ in the $\mathcal{\hat U}_0$ and $\mathcal{\hat U}_x$ system for $N=12$. (f) $1-C(n)/C(\infty)$ vs. $n$ for $N=12$ ($\log-$linear). Lines with points are data generated numerically, and solid lines are the exponential fitting. Other parameters: $J_x=1$, $h_{0x}=0/1$, $h_{0z}=1$ and  $\tau=\frac{\pi}{4}$. In all cases, an open boundary chain is considered.}
  \label{pi4_hx1_hz1_cf_nint_p0_N} 
\end{figure}
In the transverse Ising Floquet system, there is a peculiar set of parameters {\it{viz.} $h_{x}=0/1,$ $h_{z}=1$ \rm{and} $\tau=\frac{\pi}{4}$} in both $\mathcal{\hat U}_0$ and $\mathcal{\hat U}_x$ systems. At this particular set of parameters, OTOC shows periodic oscillation in both integrable, as well as  nonintegrable systems. In the integrable case, OTOC shows periodic behavior with a time period equal to $2N$. 
 It  jumps to a maximum value at $n=(2m+1)N$ and goes to zero at $n=2mN$, where $m$ is the positive integer and $N$ is the system size [Fig. \ref{pi4_hx1_hz1_cf_nint_p0_N}(a)].
OTOC shows quadratic growth till $N-1$ kicks [Fig. \ref{pi4_hx1_hz1_cf_nint_p0_N}(b)]. 
Similar to the  $\mathcal{\hat U}_0$ case, in the $\mathcal{\hat U}_x$ case, also OTOC  shows a periodic behavior, but periodicity is not related to the system size [Fig. \ref{pi4_hx1_hz1_cf_nint_p0_N}(c)]. Again, the OTOC grows quadratic [Fig. \ref{pi4_hx1_hz1_cf_nint_p0_N}(d)]. Taking $\hat V$ and $\hat W$ as random matrices drawn from GUE, we do not see power-law growth of OTOC because spins in both the blocks are already thermalized before the time evolution starts. OTOC  saturates exponentially in both $\mathcal{\hat U}_0$ and  $\mathcal{\hat U}_x$ systems and for a given $\tau$ the exponent is nearly equal in both the cases ($\mu=0.77$ for $\tau=\frac{\pi}{18}$ and $\mu=0.85$ for $\tau=\frac{3\pi}{18}$) as shown in Fig. \ref{pi4_hx1_hz1_cf_nint_p0_N}(f). 
 OTOC shows identical behavior to that of OPEE with time in both $\mathcal{\hat U}_0$ and $\mathcal{\hat U}_x$ systems (The equivalence is mathematically shown in Eq.~(\ref{OPEE_OTOC}). For the $\mathcal{\hat U}_0$ system, we see a periodic behavior with a time period equal to $2N\tau$. During this periodic behavior, OTOC starts from zero, goes to a maximum at $N^{\rm th}$ kick, and returns to zero at $t=2N\tau$ and repeats the pattern thereafter. It should be noted that the entanglement entropy for the $\mathcal{\hat U}_0$ model with open boundary condition \cite{Mishra2015} and the entangling power of the $\mathcal{\hat U}_0$  model \cite{Pal2018} is 
maximum at these points where OTOC is maximum. The reason lies to the fact that the OTOC at the infinite temperature is related to second Renyi entropy $S_V^2$ as $ C(n)\sim e^{-S_V^2}$\cite{hosur2016chaos,Fan2017}, where $S_V^2=-\log Tr_V(\rho_V^2)$, behaves like von Neumann entropy \cite{Fan2017,Bergamasco}. $\rho_V=Tr_W[\rho]$ is the reduced density matrix for the partition scheme for the block operators defined in Fig.~\ref{block_operator}.
At $n=2mN$, where $m$ is positive integer, due to quantum resonance \cite{mishra2014resonance}, $\hat W(n=2mN)=\hat W$ and the commutator $[\hat W(n=2mN),V]$ becomes zero, therefore OTOC vanishes.  
\par 
 For this special set of parameters, the spectrum of the Floquet systems $\mathcal{\hat U}_0$ and  $\mathcal{\hat U}_x$  are highly degenerate, and we could not conclude the nature of distribution from the shape of NNSD. We observe that a small shift in $\tau$ from $\pi/4$ lifts this degeneracy. Therefore, it is useful to explore the proximity of $\tau=\pi/4$ by defining a small parameter (let's say, $\epsilon=\frac{\pi}{50}$) such that the natural behavior of NNSD and OTOC does not change by adding/subtracting $\epsilon$ to $\tau=\pi/4$. We explore not only NNSD but also OTOC at the proximity of $\tau=\frac{\pi}{4}$.

\begin{figure}
\centering
\begin{subfigure}{.32\textwidth}
\includegraphics[width=.99\linewidth,height=.90\linewidth]{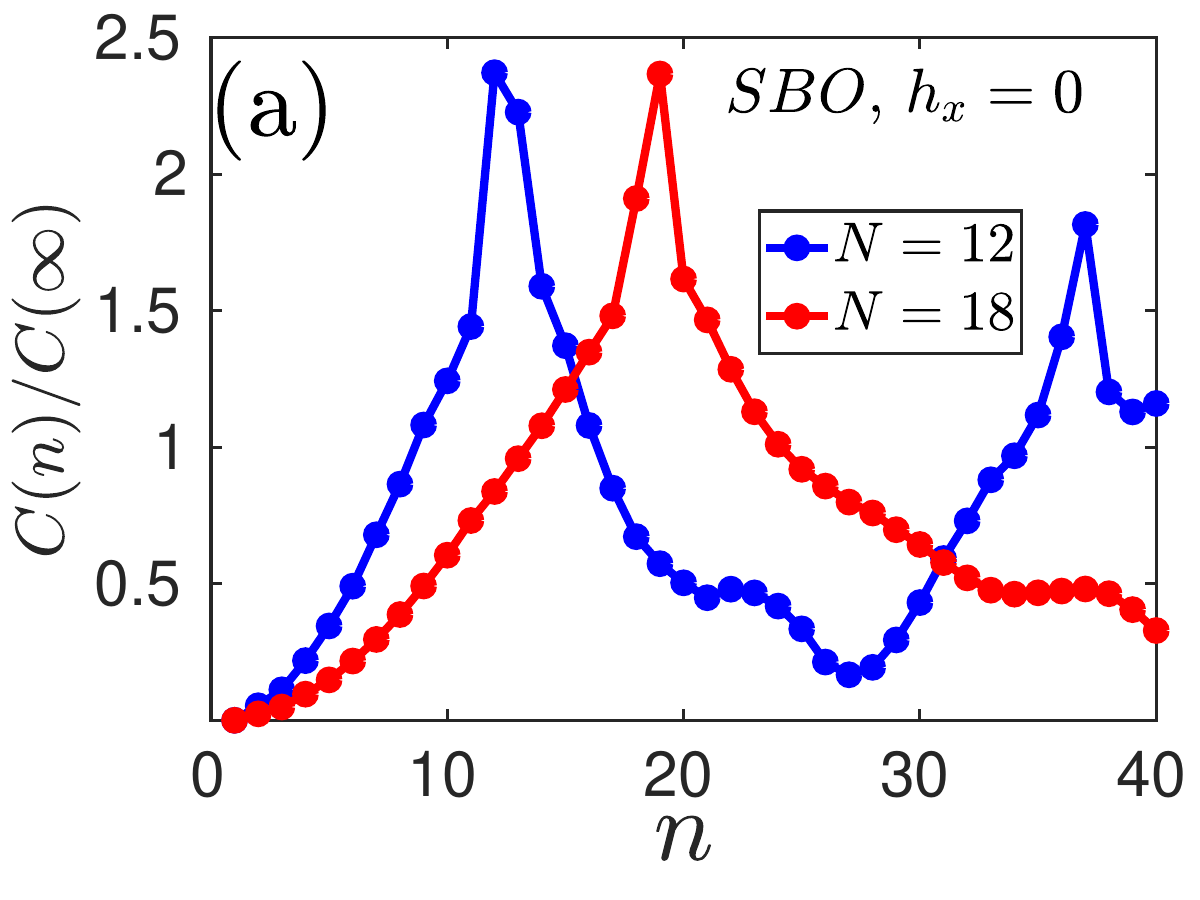}
 \end{subfigure} 
\begin{subfigure}{.33\textwidth}
  \includegraphics[width=.99\linewidth,height=.90\linewidth]{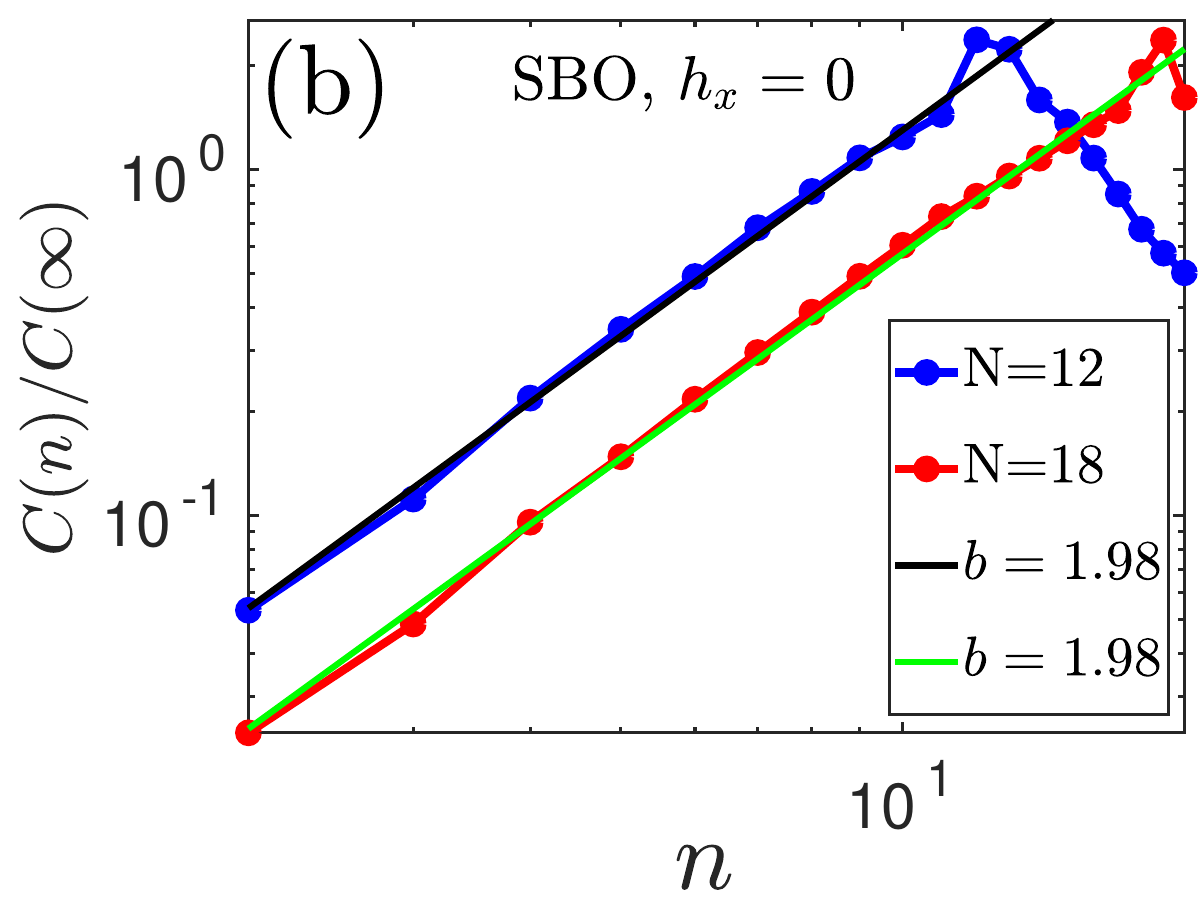}
 \end{subfigure} 
\begin{subfigure}{.33\textwidth}
  \includegraphics[width=.99\linewidth,height=.90\linewidth]{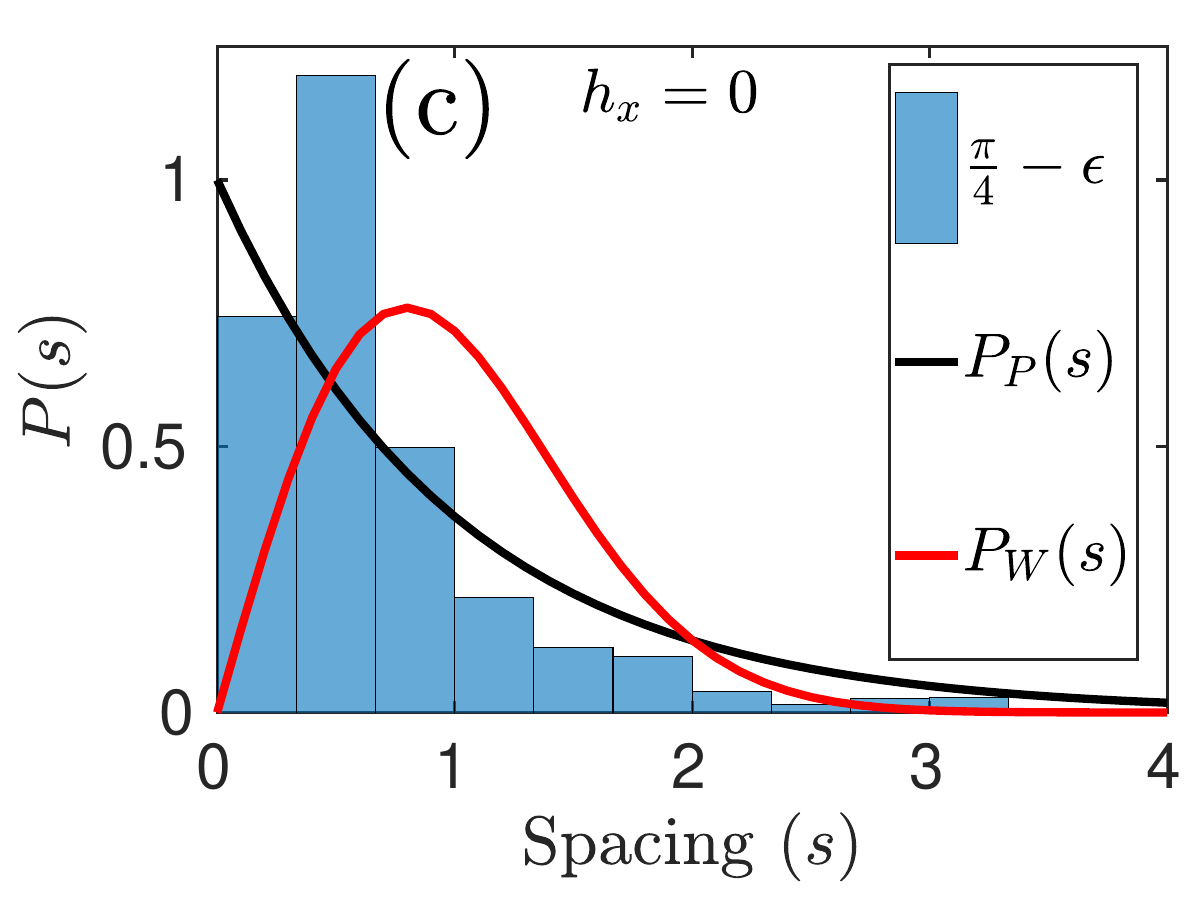}
  \end{subfigure} 
\caption{Integrable $\mathcal{\hat U}_0$ system with parameters: $\tau=\frac{\pi}{4}-\epsilon(=\frac{\pi}{50})$, $J_x=1$, $h_{x}=0$ and $h_{z}=1$. (a) $C(n)/C(\infty)$ of SBOs vs. $n$ in the $\mathcal{\hat U}_0$ system for $N=18$. (b) $\log-\log$ behavior of ``a" in which lines with points represent data from the numerical calculation, and solid lines are the polynomial fitting. (c) NNSD of the $\mathcal{\hat U}_0$ system with $N=12$.}
\label{pi4e_hx0_hz1_cf_int_p0_N} 
\end{figure}

\begin{figure}
\begin{subfigure}{.49\textwidth}
\includegraphics[width=.99\linewidth,height=.70\linewidth]{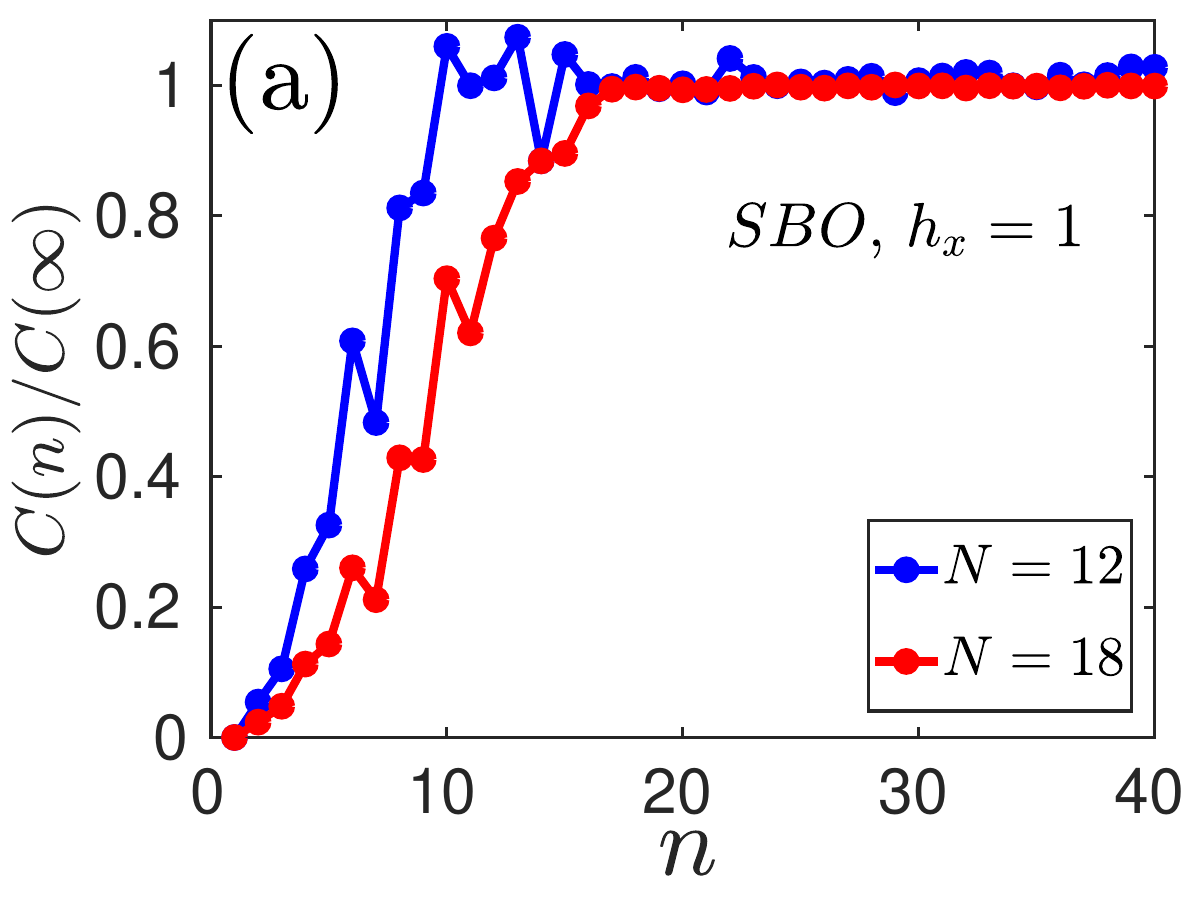}
   \end{subfigure} 
   \begin{subfigure}{.49\textwidth}
  \includegraphics[width=.99\linewidth,height=.70\linewidth]{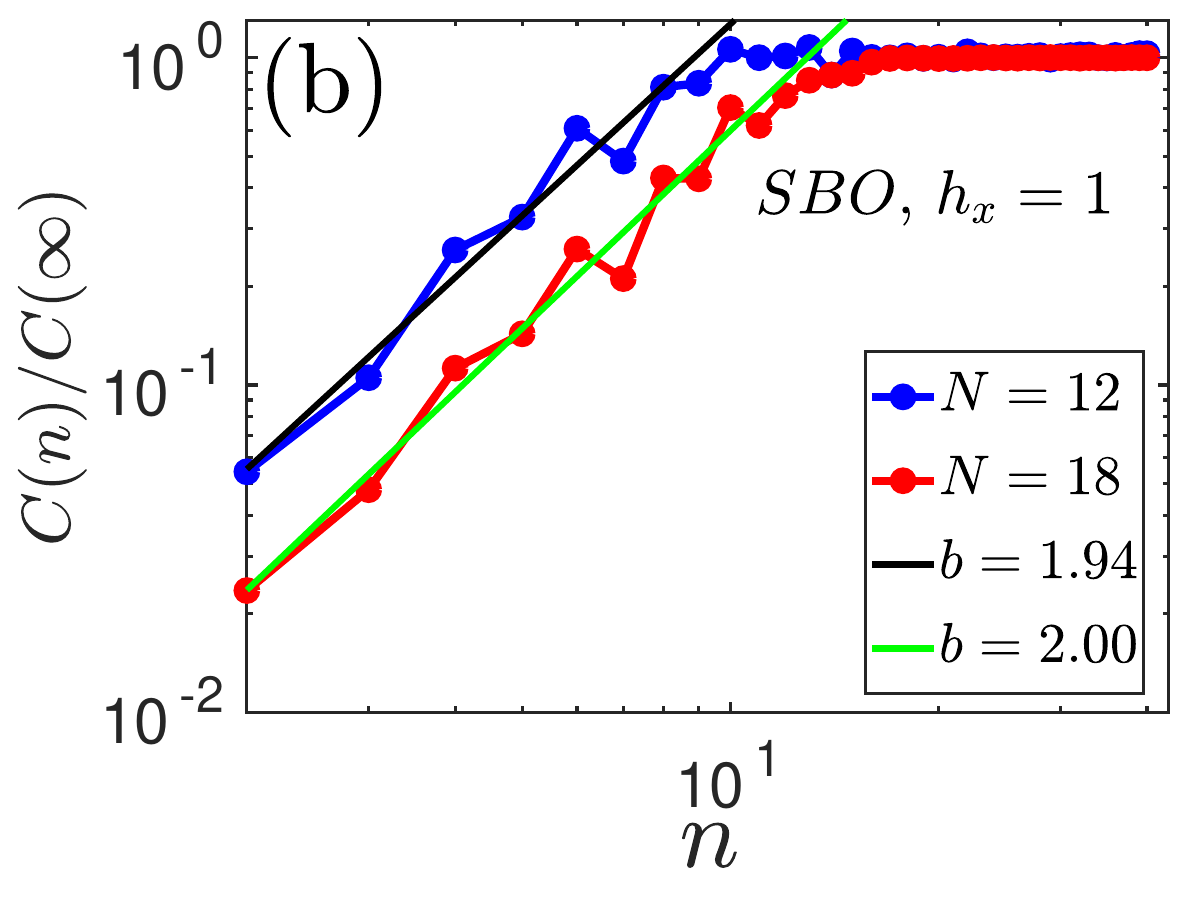}
  \end{subfigure} 
   \begin{subfigure}{.49\textwidth}
  \includegraphics[width=.99\linewidth, height=.70\linewidth]{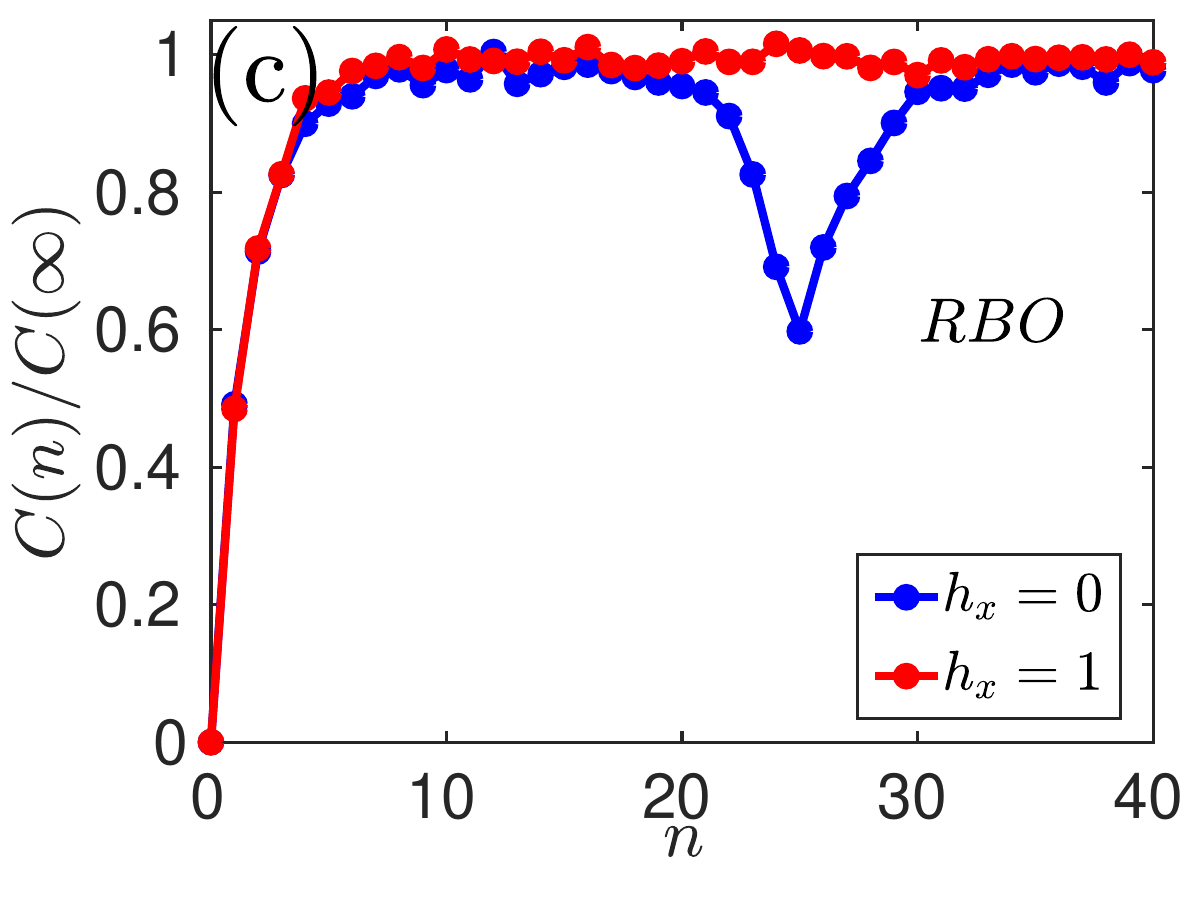}
\end{subfigure} 
\begin{subfigure}{.49\textwidth}
\includegraphics[width=.99\linewidth, height=.70\linewidth]{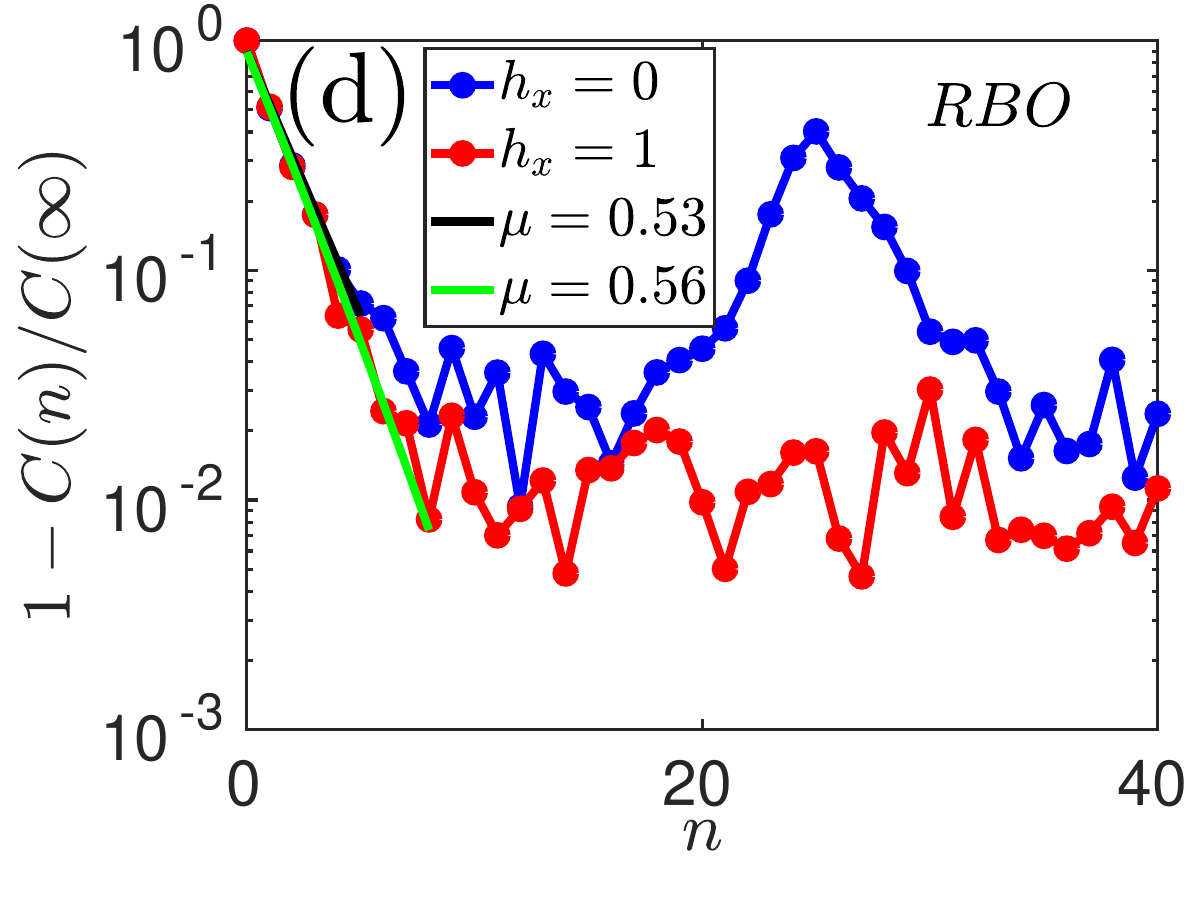}
\end{subfigure} 
   \begin{subfigure}{.49\textwidth}
\includegraphics[width=.99\linewidth,height=.70\linewidth]{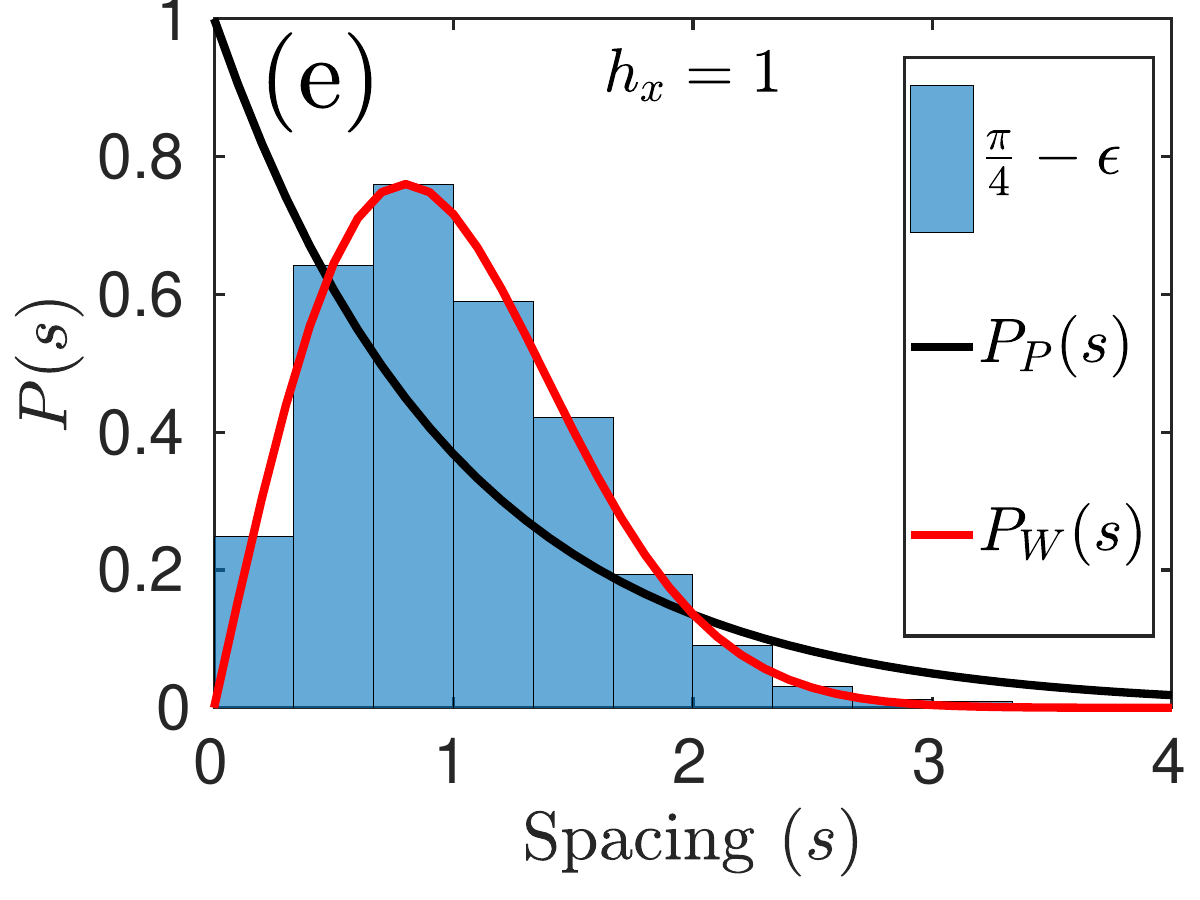}
\end{subfigure} 
   \caption{(a) $C(n)/C(\infty)$ of SBOs vs. $n$ in the $\mathcal{\hat U}_x$ system for $N=12$ and $18$. (b) $\log-\log$ behavior of ``a" in which lines with points represent data from the numerical calculation, and solid lines are the polynomial fitting. (c) $C(n)/C(\infty)$ of RBOs vs. $n$ in the $\mathcal{\hat U}_0$ and $\mathcal{\hat U}_x$ system for $N=12$. (d) $1-C(n)/C(\infty)$ vs. $n$ for $N=12$ ($\log-$linear). Lines with points are data generated numerically, and solid lines are the exponential fitting. (e) NNSD of the $\mathcal{\hat U}_x$ system for $N=12$. Other parameters: $J_x=1$, $h_{x}=0/1$, $h_{z}=1$ and  $\tau=\frac{\pi}{4}-\epsilon(=\frac{\pi}{50})$.}
\label{pi4e_hx1_hz1_cf_nint_p0_N} 
\end{figure}
In the $\mathcal{\hat U}_0$ system with $\tau=\frac{\pi}{4}-\epsilon$, we see OTOC deviates from the periodic behaviour at $\tau=\pi/4$. Though we still see maxima and minima of OTOC  near $t=(2m+1)N\tau$ and $2mN\tau$  for positive integer $m$, respectively. We observe that smaller the $\epsilon$, sharper the maxima/minima approaching to $t=(2m+1)N\frac{\pi}{4}/2mN\frac{\pi}{4}$ 
[Fig. \ref{pi4e_hx0_hz1_cf_int_p0_N}(a)]. We again get a quadratic power-law growth at $\tau=\frac{\pi}{4}-\epsilon$ [Fig.~\ref{pi4e_hx0_hz1_cf_int_p0_N}(b)]. Corresponding NNSD displays  nearly Poisson statistics in the $\mathcal{\hat U}_0$ system [Fig.~\ref{pi4e_hx0_hz1_cf_int_p0_N}(c)].\\
\vspace{-.15 cm}\hspace{.8 cm}On the other hand, OTOC in the $\mathcal{\hat U}_x$ system at $\tau=\frac{\pi}{4}-\epsilon$ has different behaviour than that at $\frac{\pi}{4}$. At this period, OTOC grows till $N$ kicks after that saturates at a value of $1$  [Fig. \ref{pi4e_hx1_hz1_cf_nint_p0_N}(a)]. Although the growth of OTOC is again quadratic power-law as shown in Fig. \ref{pi4e_hx1_hz1_cf_nint_p0_N}(b). 
Replacing the observable $\hat V$ and $\hat W$ by random matrices, we see the behavior of OTOC in both $\mathcal{\hat U}_0$ and $\mathcal{\hat U}_x$ system. We get a similar behavior of OTOC as that at $\tau=\frac{\pi}{4}$, in the $\mathcal{\hat U}_x$ system; however, in the $\mathcal{\hat U}_0$ system, OTOC does not reach to the value zero at $n=2mN$. This is due to the parameter $\epsilon$, which, if tending towards zero, leads to a coinciding $\tau=\pi/4-\epsilon$ case with $\tau=\pi/4$. Ideally, OTOC for RBOs should also vanish at $t=2mN\pi/4$ due to the same reason that $\hat W(t=2mN\pi/4)=\hat W$  but with $\tau=\pi/4-\epsilon$, we skip the moment of vanishing OTOC at $2mN$ kicks and get a dip only [Fig.~\ref{pi4e_hx1_hz1_cf_nint_p0_N}(c)]. OTOCs with RBOs are identical to OPEE [Eq.~\ref{OPEE_OTOC}]. Fig.~\ref{pi4e_hx1_hz1_cf_nint_p0_N}(d) displays the exponential saturation of OTOC with nearly equal exponent in both $\mathcal{\hat U}_0$ and $\mathcal{\hat U}_x$ system. 
 At period $\frac{\pi}{4}-\epsilon$, there is no degeneracy in the spectrum. Therefore,  NNSD shows Wigner-Dyson distribution [Fig.~\ref{pi4e_hx1_hz1_cf_nint_p0_N}(e)].
\section{Conclusion} 
\label{Ch4_conclusion}
In this chapter, we study the growth and saturation behavior of OTOC in both $\mathcal{\hat U}_0$ and $\mathcal{\hat U}_x$ systems. Initially, we calculated OTOC by using the SBOs for various time periods and analyzed the early time behavior and saturation behavior. Later, we used analytically solvable RBOs to learn about the saturation region of the system.  
\par
Growth of OTOC in both $\mathcal{\hat U}_0$ and $\mathcal{\hat U}_x$ system shows a quadratic power-law for all Floquet periods in between $0$ to $\frac{\pi}{2}$ except $\frac{\pi}{4}$. At kick interval $\tau=\frac{\pi}{4}$, the field terms do not change the state; therefore, OTOC remains constant.
\par
Later we take special parameters ($J_x=1$, $h_{z}=1$, and $h_{z}=0/1$ and $\tau=\frac{\pi}{4}$) and calculate the OTOC. In the integrable system, we see a periodic trend, and the period of oscillation is twice the system size. We also observe that the maxima/minima are those points where von Neumann entropy is also maxima/minima.  In the nonintegrable case, periodic behavior does not  show a trivial dependence on the system size. For $\tau=\pi/4$, OTOC shows a quadratic power-law growth in the integrable system till $n=N-1$ kicks. We see a quadratic power-law for the nonintegrable system as well. Large degeneracy at $\tau=\frac{\pi}{4}$ makes NNSD inconclusive whether it is Poisson or Wigner-Dyson type. In order to study the behavior approaching to this Floquet period,  we define a very small quantity (say $\epsilon=\frac{\pi}{50}$) and take a slightly lesser Floquet period, $\tau=\frac{\pi}{4}-\epsilon$. At this $\tau$, NNSD is Poisson type in the $\mathcal{\hat U}_0$ system and Wigner-Dyson type in the $\mathcal{\hat U}_x$ system. We also studied the near-saturation behavior of OTOC. Near saturation behavior can not be exactly defined by using the SBOs; therefore, we calculate OTOCs by RBOs. For the observables in consideration, the OTOC with RBOs is exactly the same as the operator entanglement entropy. We are getting an exponential increase of OTOC near the saturation region in all the cases.
\par
In the next chapter, we will utilize OTOCs as a quantifier for quantum information currents and propose a quantum information diode (QID) by exploiting the effect of nonreciprocal magnons in  a 2D Heisenberg spin system with Dzyloshinski Moriya interaction.

\chapter{Quantum information diode based on the magnonic crystal}  

\ifpdf
    \graphicspath{{Chapter5/Figs/Raster/}{Chapter5/Figs/PDF/}{Chapter5/Figs/}}
\else
    \graphicspath{{Chapter5/Figs/Vector/}{Chapter5/Figs/}}
\fi

\section{Introduction} 
A diode is a device designated to support asymmetric transport. Nowadays, household electric appliances or advanced experimental scientific equipment are all inconceivable without extensive use of diodes. Diodes with a perfect rectification effect permit electrical current to flow in one direction only. The progress in nanotechnology and material science passes new demands to a new generation of diodes; futuristic nano-devices that can rectify either acoustic (sound waves), thermal phononic, or magnonic spin current transport. Nevertheless, we note that at the nano-scale, the rectification effect is never perfect{\it, i.e.,} backflow is permitted, but amplitudes of the front and backflows are different \cite{Liang,ZhangCheng,Chen,Maldovan,Ren,Lepri,Komatsu,Majumdar,Terasaki,Wang,Casati,li2004thermal,LiRen, Etesami}. In the present work, we propose an entirely new type of diode designed to rectify the quantum information current. We do believe that in the foreseeable future the quantum information diode (QID) has a perspective to become a benchmark of quantum information technologies.  
\par
The functionality of a QID relies on the use of magnonic crystals, {\it i.e.,} artificial media with a characteristic periodic lateral variation of magnetic properties. Similar to photonic crystals, magnonic crystals possess a band gap in the magnonic excitation spectrum. Therefore, spin waves with frequencies matching the band gap are not allowed to propagate through the magnonic crystals \cite{chumak2008scattering,Chumak2,Nikitov,Kruglyak,ZKWang,Gubbiotti,ABUstinov}.
\par 
The essence of a magnonic transistor is an YIG strip with a periodic modulation of its thickness (magnonic crystal). The transistor is complemented by a source, a drain, and gate antennas. A gate antenna injects magnonic crystal magnons with a frequency $\omega_G$ matching the magnonic crystal band gap. Therefore, the gate magnons cannot leave the crystal and may reach a high density. Magnons emitted from a source with a wave vector $ \textbf{k}_s$ flowing towards the drain run into the magnonic crystal. The interaction between the source magnons and the magnonic crystal magnons is a four-magnon scattering process. The magnonic current emitted from the source attenuates in the magnonic crystal, and the weak signal reaches the drain due to the scattering. The relaxation process is swift if the following condition holds
\cite{Chumak2,Gurevich} 
\begin{equation}
\label{certain conditions} 
  k_s=\frac{m_0\pi}{a_0}, 
\end{equation}
 where $m_0$ is the integer,
and $a_0$ is the crystal lattice constant. The magnons with wave vectors satisfying the Bragg conditions Eq.~(\ref{certain conditions}) will be resonantly scattered back, resulting in the generation of rejection bands in a spin-wave spectrum over which magnon propagation is entirely prohibited. Experimental verification of this effect is given in Ref.~\cite{Chumak2}.
\par
This chapter is organized as follows. In subsection \ref{Set_up_of_QID}, we briefly describe the proposed set-up for QID. In subsection \ref{model_2d}, we will discuss a model of a 2D square lattice spin system. OTOC is defined in subsection \ref{OTOC_ch5}, and rectification is defined in subsection \ref{rectification}. At last, we conclude the results in section \ref{conclusion_ch5}

\section{Result}
\subsection{Proposed set-up for QID}
\label{Set_up_of_QID}
\begin{figure} 
\includegraphics[width=.99\linewidth,height=.450\linewidth]{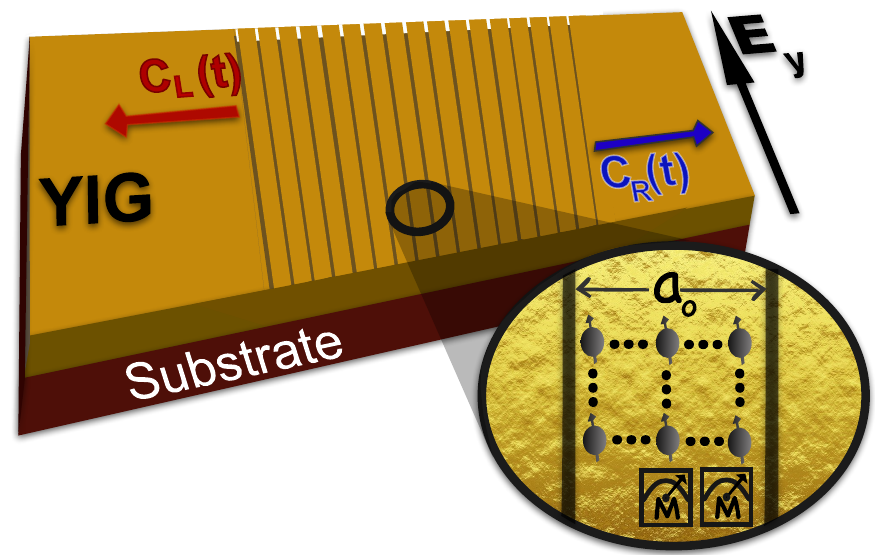}
\caption{Illustration of a quantum information diode: A plane of an YIG film with grooves orthogonal to the direction of the propagation of quantum information. In the middle of the QID, we pump extra magnons to excite the system. A quantum excitation propagates toward the left, and the right ends asymmetrically. To describe the propagation process of quantum information, we introduce the left and right OTOC $C_L(t)$ and $C_R(t)$. Because the left-right inversion is equivalent to $D\rightarrow-D$ meaning $E_y\rightarrow-E_y$, we can invert the left and right OTOCs by switching the applied external electric field.}
\label{QI_DIODE}
\end{figure}

A pictorial representation of a QID is shown in Fig.~\ref{QI_DIODE}. A magnonic crystal can be fabricated from an YIG film. Grooves can be deposited using a lithography procedure in a few nanometer steps, and for our purpose, we consider parallel lines in width of 1$\mu$m spaced with 10$\mu$m from each other. Therefore, the lattice constant, approximately $a_0=11\mu$m{\it, i. e.,} is much larger than the unit cell size $a=10$nm used in our coarse-graining approach. Due to the capacity of our analytical calculations, we consider quantum spin chains of length about $N=1000$ spins and the maximal distance between the spins $r_{ij}=d$ (in the units of $a$), $d=i-j=40$. In what follows, we take $k(\omega)a\ll1$. The mechanism of the QID is based on the effect of direction dependence of nonreciprocal magnons \cite{Takashima,Matsumoto,Shiomi}. In the chiral spin systems, the absence of inversion symmetry causes a difference in dispersion relations of the left and right propagating magnons, {\it i. e.,} $\omega_{s,L}(\textbf{k})\neq\omega_{s,R}(-\textbf{k})$. Due to the Dzyaloshinskii–Moriya interaction (DMI), magnons of the same frequency $\omega_s$ propagating in opposite directions have different wave vectors \cite{LevanWang}: $a\left(k^+_s-k^-_s\right)=D/J$, where $J$ is the exchange constant, and $D$ is the DMI constant. Therefore, if the condition Eq.~(\ref{certain conditions}) holds for the left propagating magnons, it is violated for the right propagating magnons and vice versa. These magnons propagating in different directions decay differently in the magnonic crystal. Without loss of generality, we assume that the right propagating magnons with $k^+_s$ satisfy the condition Eq.~(\ref{certain conditions}), and the current attenuates due to the scattering of source magnons by the gate magnons. The left propagating magnons $k^-_s$ violate the condition Eq.~(\ref{certain conditions}), and the current flows without scattering. Thus, reversing the source and drain antenna's positions rectifies the current. Following ref.~\cite{Chumak2}, we introduce a suppression rate of the source to drain the magnonic current $\xi(D)=1-n_D^+/n_D^-$, where $n_D^+<n_D^-$ are densities of the drain magnons with and without scattering. The parameter $\xi(D)$ is experimentally accessible, and it depends on a particular setup. Therefore, we take $\xi(D)$ as a free theory parameter. Multiferroic (MF) materials are considered as a good example of a system with broken inversion symmetry (see Refs.\cite{Nagaosa,Mostovoy,Ramesh,Bibes,Fiebig,Hemberger,Meyerheim,Cheong,Vedmedenko}) and references therein. MF properties of YIG are studied in ref.~\cite{Vignale}. Moreover, in accordance with scanning tunneling microscopy experiments, a change of the spin direction at one edge of a chiral chain was experimentally probed by tens of nanometers away from the second edge \cite{Vedmedenko}.

\subsection{Model}
\label{model_2d}
We consider a 2D square-lattice spin system with nearest-neighbor $J_1$ and the next nearest-neighbor $J_2$ coupling constants: 
\begin{equation}
\label{Hamiltonian_ch5}
\hat{H}=J_1\sum\limits_{\langle n,m\rangle}\hat\sigma_n\hat\sigma_m+ J_2\sum\limits_{\langle\langle n,m\rangle\rangle}\hat\sigma_n\hat\sigma_m-{\bf P}\cdot{\bf E}, 
\end{equation}
where $\langle n,m\rangle$, and $\langle\langle n,m\rangle\rangle$ indicate all the pairs with nearest-neighbor and next nearest-neighbor interactions, respectively. The last term in Eq.~(\ref{Hamiltonian_ch5}) describes a coupling of the ferroelectric polarization $\mathbf{P}=g^{\phantom{\dagger}}_{\mathrm{ME}}\mathbf{e}^x_{i,i+1}\times\left(\hat\sigma_i\times\hat\sigma_{i+1}\right)$ with an applied external electric field and mimics an effective Dzyaloshinskii–Moriya interaction term $D=E_y g^{\phantom{\dagger}}_{\mathrm{ME}}$ breaking the left-right symmetry, where $g^{\phantom{\dagger}}_{\mathrm{ME}}$ is the magneto-electric coupling constant. This can be written as 
\begin{equation}
-{\bf P}\cdot{\bf E}=D\sum\limits_{n}(\hat\sigma_{n}\times\hat\sigma_{n+1})_z.
\end{equation} 
Here we consider only the nearest neighbor DMI and only in one direction. As a consequence, the left-right inversion is equivalent to $D\rightarrow-D$, or $E_y\rightarrow-E_y$. The broken left-right inversion symmetry can be exploited in rectifying the information current by an electric field. More importantly, the procedure is experimentally feasible. We can diagonalize the Hamiltonian in Eq.~(\ref{Hamiltonian_ch5}) by using  the Holstein-Primakoff transformation \cite{FNori,Udvardi,Zheng,Stagraczynski}~[See Appendix \ref{diag_2d_Ham} for detailed derivation] as:
  \begin{eqnarray}
\label{2Hamiltonian}
  &&  \hat{H} = \sum\limits_{\vec{k}}\omega(\pm D,\textbf{k})\hat{a}^{\dagger}_{\vec{k}}\hat{a}_{\vec{k}},\,\,\omega(\pm D,\textbf{k})=\big(\omega(\vec{k})\pm\omega_{DM}(\vec{k})\big),  ~~\omega_{DM}(\vec{k})=D\sin(k_xa),\nonumber\\
 &&\omega_k=2J_1(1-\gamma_{1,\textbf{k}})+2J_2(1-\gamma_{2,\textbf{k}}),\,\,\gamma_{1,\textbf{k}}=1/2(\cos k_x+\cos k_y),\nonumber\\
 && \gamma_{2,\textbf{k}}=1/2(\cos (k_x+k_y)+\cos (k_x-k_y)).
  \end{eqnarray}
Here $\pm D$ corresponds to the magnons propagating in opposite directions, and the sign change is equivalent to the electric field direction change.  We note that a 1D character of the DM term is ensured by the magnetoelectric effect \cite{Nagaosa} and to the electric field applied along the $\textbf{y}$ axis. 
\par
The speed limit of information propagation is usually given in terms of Lieb-Robinson (LR) bound, defined for the Hamiltonians that are locally bounded and short-range interacting \cite{kuwahara2021lieb,hastings2006spectral,nachtergaele2006propagation}. Since the Hamiltonian in  Eq.~(\ref{Hamiltonian_ch5}) satisfies both conditions, the LR bound can be defined for the spin model. However, when we transform the Hamiltonian using Holstein-Primakoff bosons, we have to take extra care as the bosons are not locally bounded. To define LR bound, we take only a few noninteracting magnons and exclude the magnon-magnon interaction to truncate terms beyond quadratic operators. In a realistic experimental setting, low density of propagating magnons in YIG can easily be achieved by properly controlling the microwave antenna. In the case of low magnon density, the role of the magnon-magnon interaction between propagating magnons in YIG is negligible. Therefore, for YIG, we have a quadratic Hamiltonian, which is a precise approach in a low magnon density limit. Our discussion is valid for the experimental physical system \cite{Chumak2}, where magnons of YIG do not interact with each other, implying that there is no term in the Hamiltonian beyond quadratic. We can estimate LR bounds \cite{LiebRobinson} defining the maximum group velocities of the left-right propagating magnons $v_{g}^{\pm}(\vec{k})=\frac{\partial (\omega(\vec{k})\pm\omega_{DM}(\vec{k}))}{\partial k}$. Taking into account the explicit form of the dispersion relations, we see that the maximal asymmetry is approximately equal to the DM constant{\it, i. e.,} $v_{g}^{+}(0)-v_{g}^{-}(0)\approx 2D$. We note that the effect of nonreciprocal magnons is already observed experimentally {\cite{KZakeri, IguchiUemura, ShibataKubota, TaguchiArima, Gitgeatpong}} but up to date, never discussed in the context of the quantum information
theory.
\par 
We formulate the central interest question as follows: At $t=0$, we act upon the spin $\hat\sigma_{n}$ to see how swiftly changes in the spin direction  can be probed tens of sites away $d=n-m\gg 1$ and whether the forward and backward processes ({\it, i.e.,} probing for $\hat\sigma_{m}$ the outcome of the measurement done on $\hat\sigma_{n}$) are asymmetric or not.  Due to the left-right asymmetry, the chiral spin channel may sustain a diode rectification effect when transferring the quantum information from left to right and in the opposite direction. We note that our discussion about the left-right asymmetry of the quantum information flow is valid until the current reaches boundaries. Thus the upper limit of the time reads $t_{\rm max}=Na/v_{g}^{\pm}(\vec{k})$, where $N$ is the size of the system.

\subsection{Out-of-time-order correlator}
\label{OTOC_ch5}
Larkin and Ovchinnikov \cite{larkin1969quasiclassical} introduced the concept of the out-of-time-ordered correlator (OTOC), and since then, OTOC has been seen as a diagnostic tool of quantum chaos. The concern of delocalizations in the quantum information theory ({\it i.e.,} the scrambling of quantum entanglement) was renewed only recently, see Refs.~ \cite{maldacena2016bound,roberts2015localized, Iyoda2018, Chapman2018, Swingle, Klug2018, Campo2017, campisi2017thermodynamics,hosur2016chaos, halpern2017jarzynski} and references therein. We utilize the OTOC to characterize the left-right asymmetry of the quantum information flow and thus infer the rectification effect of a diode. 
\par
Let us consider two unitary operators $\hat{V}$ and $\hat{W}$
describing local perturbations to the chiral spin system
Eq.~(\ref{Hamiltonian_ch5}), and the unitary time evolution of one of the
operators
${\hat{W}\left(t\right)=\exp(i\hat{H}t)\hat{W}(0)\exp(-i\hat{H}t)}$.
Then the OTOC is defined as \begin{equation}\label{OTOC1}
  C\left(t\right)=
  \frac{1}{2}\left\langle\left[\hat{W}(t),\hat{V}(0)\right]^{\dag}\left[\hat{W}(t),\hat{V}(0)\right]\right\rangle,
\end{equation} where parentheses $\langle \cdots\rangle$ denotes a quantum mechanical average over the propagated quantum state. Following the definition, the OTOC at the initial moment of time is  zero ${C(0)=0}$, provided that ${[\hat W(0),\hat V(0)]=0}$. In particular, for the local unitary and Hermitian operators of our choice $\hat W_m^{\dagger}\left(t\right)\equiv\hat \sigma_m^z(t)=\hat \eta_m(t)=\exp(i\hat Ht) \hat \eta_m\exp(-i\hat Ht)$, and $\hat V_n^\dagger=\hat \sigma_n^z=\hat \eta_n$, where $\hat \eta_n=\hat 2a^\dagger_n\hat{a}_n-1$. The bosonic operators are related to the spin operators via $\sigma_n^-=2a_n^{\dagger}, \ \ \sigma_n^+=2a_n, \ \ \sigma_n^z=2a_n^{\dagger}a_n-1$. In terms of the occupation number operators, the OTOC is given as
\begin{equation}
\label{occupation number}
C(t)=\frac{1}{2}\bigg\lbrace\langle\eta_n\eta_m(t)\eta_m(t)\eta_n\rangle+\langle \eta_m(t)\eta_n\eta_n\eta_m(t)\rangle-\langle \eta_m(t)\eta_n\eta_m(t)\eta_n\rangle-\langle \eta_n\eta_m(t)\eta_n\eta_m(t)\bigg\rbrace. 
\end{equation}
Indeed, the OTOC can be interpreted as the overlap of two wave functions, which are time evolved in two different ways for the same initial state $|\psi(0)\rangle$. The first wave function is obtained by perturbing the initial state at ${t=0}$ with a local unitary operator $\hat{V}$, then evolved further under the unitary evolution operator ${\hat{U}=\exp (-\im\hat{H}t)}$ until time $t$. It is then perturbed at time $t$ with a local unitary operator $\hat{W}$, and evolved backwards from $t$ to ${t=0}$ under $\hat{U^{\dagger}}$. Hence, the time evolved wave function is $|\psi(t)\rangle =\hat{U^{\dagger}}\hat{W}\hat{U}\hat{V}|\psi(0)\rangle=\hat{W}(t)\hat{V}|\psi(0)\rangle$. To get the second wave function, the order of the applied perturbations is permuted{\it, i. e.,} first $\hat{W}$ at $t$ and then $\hat{V}$ at ${t=0}$. Therefore, the second wave-function is $|\phi(t)\rangle=\hat{V}\hat{U^{\dagger}}\hat{W}\hat{U}|\psi(0)\rangle=\hat{V}\hat{W}(t)|\psi(0)\rangle$ and their overlap is equivalent to $F(t)=\langle\phi(t)|\psi(t)\rangle$. The OTOC is calculated from this overlap using $C(t)=1-\Re[{F(t)}]$. What breaks the time inversion symmetry for the OTOC is the permuted sequence of operators $\hat{W}$ and $\hat{V}$. However, in spin-lattice models with a preserved spatial inversion symmetry $\mathcal{\hat{P}}\hat{H}=\hat{H}$, the spatial inversion ${\mathcal{\hat{P}}d(\hat{W}, \hat{V})}={-d(\hat{W}, \hat{V})}={d(\hat{V}, \hat{W})}$ can restore the permuted order between $\hat{V}$ and $\hat{W}$,  where $d(\hat{W}, \hat{V})$ denotes the distance between observables $\hat{W}$ and  $\hat{V} $. Permuting just a single wave-function one finds $C(t)=1-\Re(\langle\phi(t)|\mathcal{\hat{P}}\mathcal{\hat{T}}|\psi(t)\rangle)=C(0)$. Thus, a scrambled quantum entanglement formally can be unscrambled by a spatial inversion. However, in chiral systems ${\mathcal{\hat{P}}\hat{H}\neq\hat{H}}$ and the unscrambling procedure fails.
\begin{figure}
\centering
\begin{subfigure}{.32\textwidth}
\includegraphics[width=\linewidth,height=.80\linewidth]{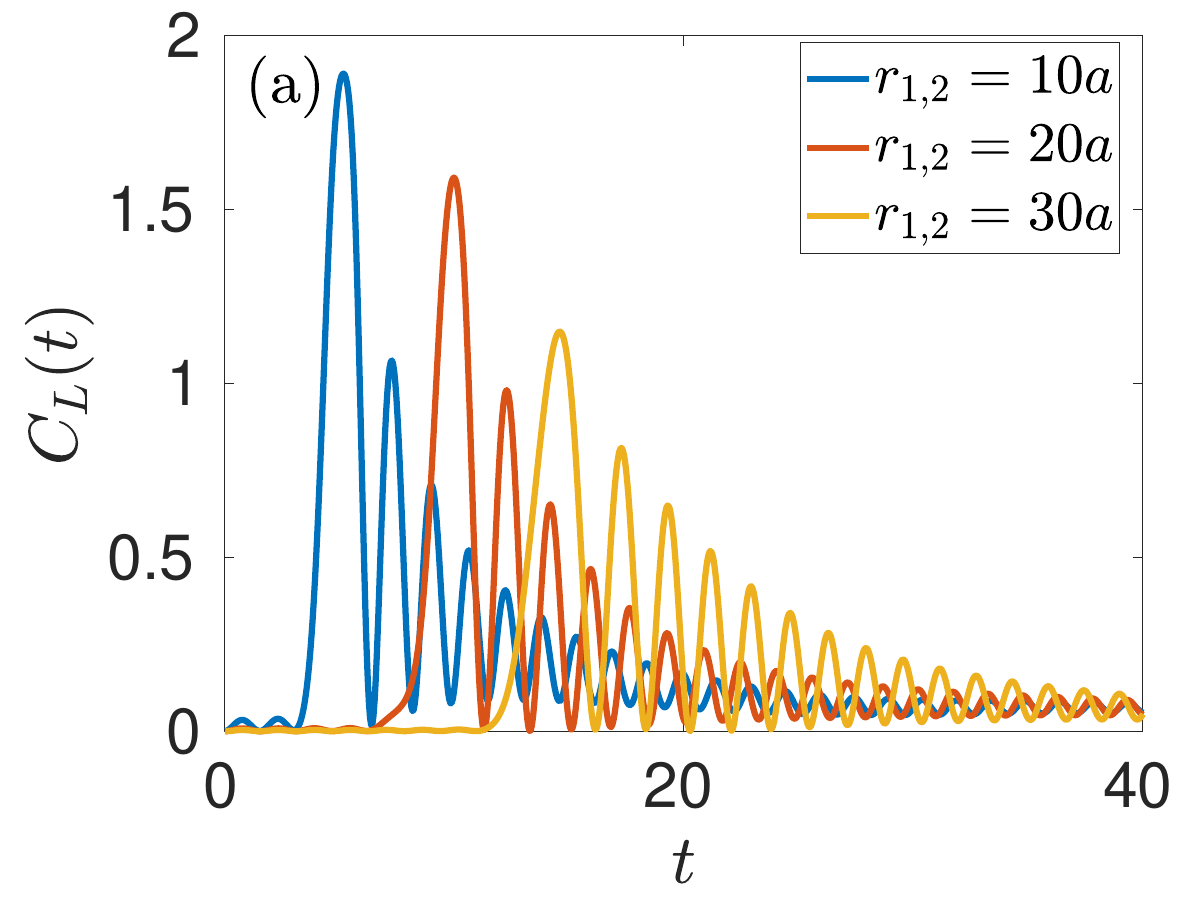}
\end{subfigure}
\begin{subfigure}{.33\textwidth}
\includegraphics[width=\linewidth,height=.80\linewidth]{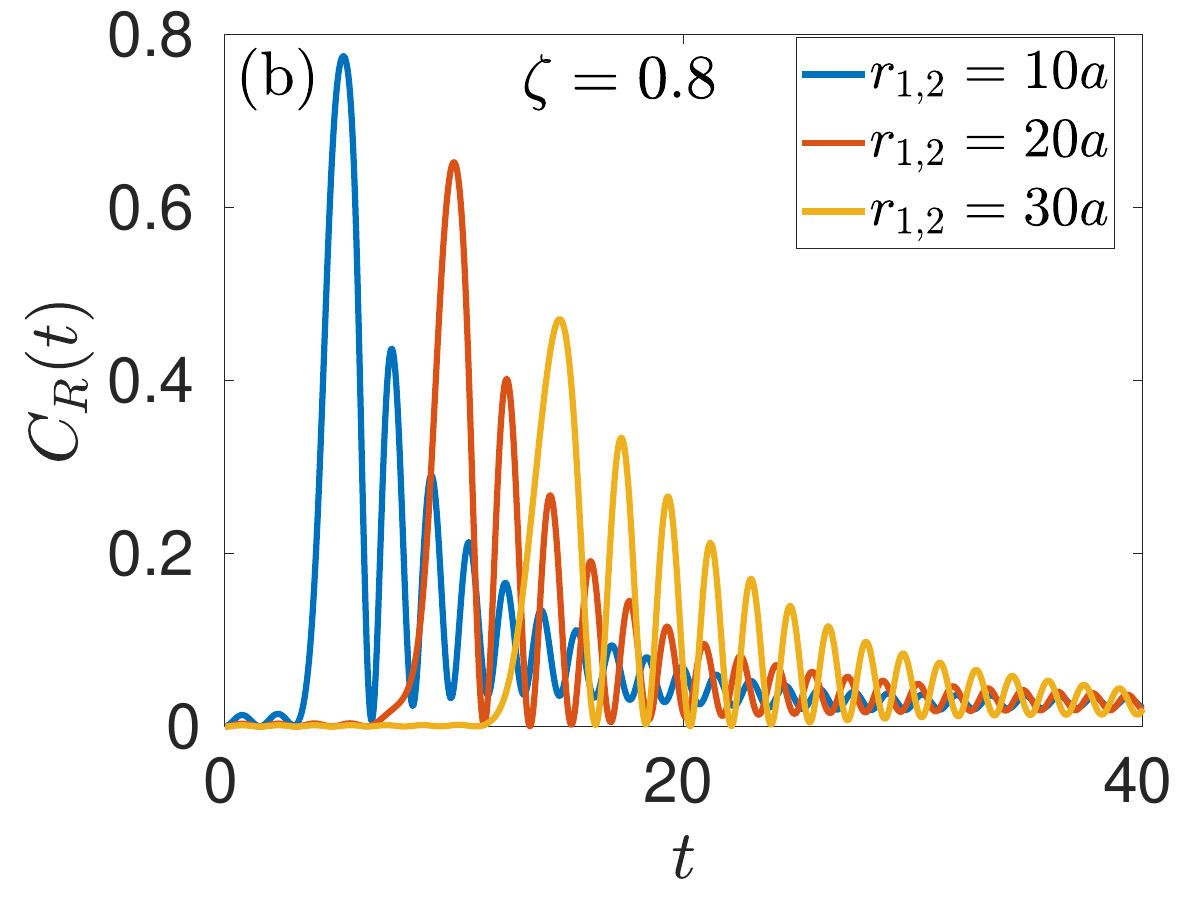}
  \end{subfigure}
\begin{subfigure}{.33\textwidth}
 \includegraphics[width=\linewidth,height=.80\linewidth]{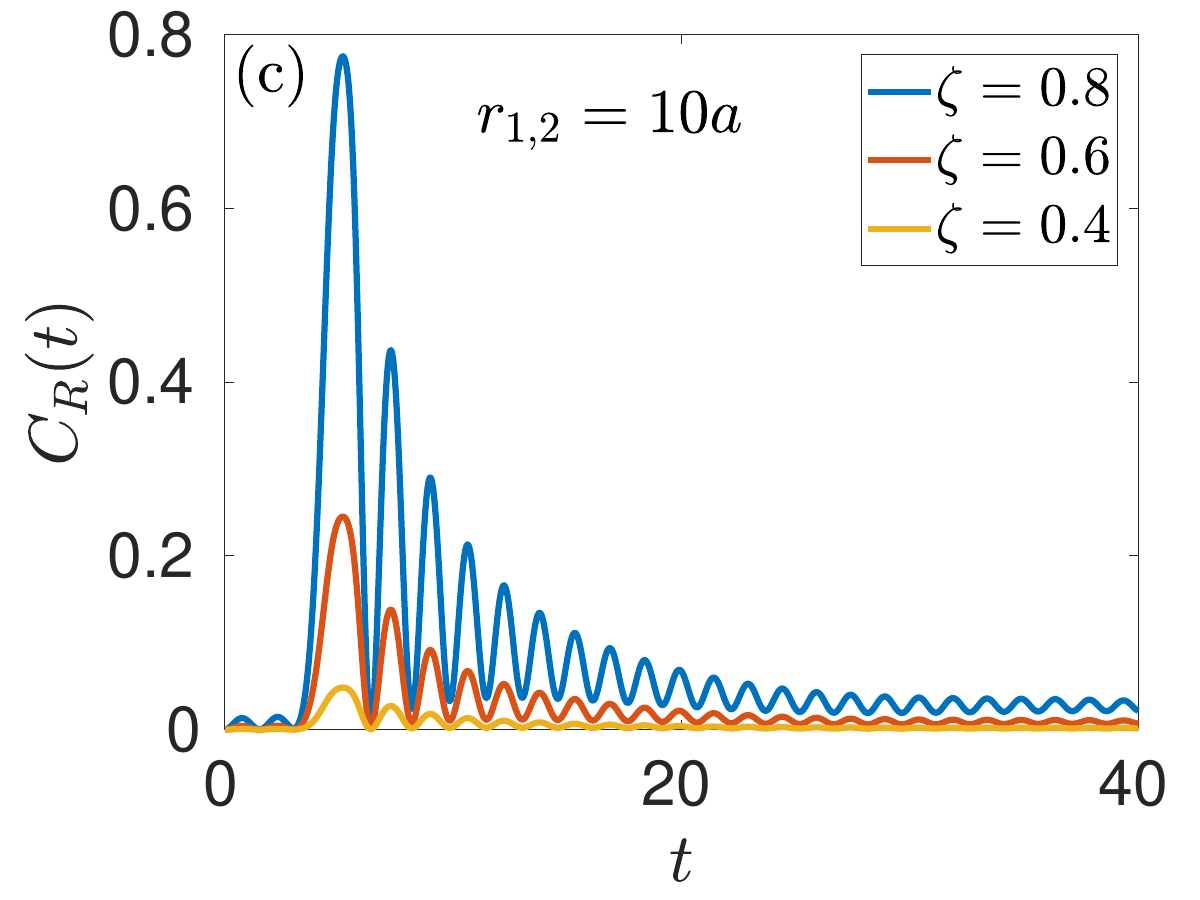}
  \end{subfigure}
\caption{\textbf{(a)} Left-OTOC and \textbf{(b)} Right-OTOC with time $t$ (in the units of $1/J$) for different distances $r_{1,2}=10a$, $20a$ and $30a$. \textbf{(c)} Right-OTOC with time for $r_{1,2}=10a$ and suppression rates of the magnon current $\zeta=0.8$, $0.6$ and $0.4$. Parameters are $N=1000$, $D=J_1=2J_2=1$. Periodic boundary conditions are considered. The values of the parameters: $m_0=1$ to $N$, $a=10^{-3}$ and $a_0=1$.}
\label{QI_DIODE_OTOC}
\end{figure}

 Taking into account Eq.~(\ref{2Hamiltonian}), we analyze quantum information scrambling along the $\textbf{x}$ axis{\it, i. e.,}  $\omega(\pm D,\textbf{k})=\omega(\pm D, k_x, 0)$ and along the $\textbf{y}$ axis, $\omega(0,\textbf{k})=\omega(0, 0, k_y)$. It is easy to see that the quantum information flow along the $\textbf{y}$ axis is symmetric, while along the $\textbf{x}$ axis, it is asymmetric and depends on the sign of the DM constant, {\it, i.e.,} the flow along the $\textbf{x}$ is different from $-\textbf{x}$. Let us assume that Eq.~(\ref{QI_DIODE}) holds for right-moving magnons  and is violated for left-moving magnons. Excited magnons with the same frequency and propagating into different directions have different wave vectors  $\omega_s\left(D,k_s^+\right)=\omega_s\left(-D,k_s^-\right)$ where: 
\begin{eqnarray}\label{propagated wave modes}
\omega_s\left(\pm D,k_s^{\pm}\right)&=&2J_1(1-1/2\cos k^{\pm}_xa)+2J_2(1-\cos k^{\pm}_xa) \pm D\sin k^{\pm}_xa,
\end{eqnarray}
$k^+_{m_0x}=\frac{m_0\pi}{a_0}$, $m_0=\mathbb N$ and $k^-_{m_0x}$ we find from the condition $\omega_s\left(D,k_s^+\right)=\omega_s\left(-D,k_s^-\right)$ leading to $k^-_{m_0x}=k^+_{m_0x}+\frac{2}{a}\tan^{-1}\left(\frac{D}{J_1+2J_2}\right)$. Here we use shortened notations $\omega_{m_0}=\omega_s\left(D,k_s^+\right)=\omega_s\left(-D,k_s^-\right)$ and set dimensionless units $J_1=2J_2\equiv J=1$. We excite in the diode magnons of different frequencies   $m_0=[1,\, N]$.  Considering Eq.~(\ref{occupation number}), Eq.~(\ref{propagated wave modes}) and following Ref.~\cite{Zheng} we obtain expressions for the left and right OTOCs $C_L(t)$ and $C_R(t)$. Those expressions and details of involved derivations are presented in Appendix \ref{LROTOC_calculation}. In Fig.~\ref{QI_DIODE_OTOC}, $C_L(t)$ and $C_R(t)$ is shown for $|n^+-m|$ and $|n^--m|$ distant spins, respectively. $C_R(t)$ is independent of the separation between the spins; however, the decay amplitude varies due to the suppression coefficient $\zeta$. In the case of the dominant attenuation by the gate magnons, the OTOC decreases significantly. The difference in $C_L$ and $C_R$ originated due to the asymmetry arising from the DMI term.
\par
 A high density of magnons can invalidate the assumption of a pure state or spin-wave approximation that works only for a low density of magnons. However, the key point in our case is that one has to distinguish between two sorts of magnons, gate magnons and propagating nonreciprocal magnons. The density of the propagating magnons can be regulated in the experiment through a microwave antenna, and one can always ensure that their density is low enough. It is easy to regulate the density of the gate magnons, and an experimentally accessible method is discussed in Ref.~\cite{Chumak2}.
\par
We proposed a novel theoretical concept that can be directly realized with the experimentally feasible setup and particular material. There are several experimentally feasible protocols for measuring OTOC in the spin systems \cite{nie2020experimental,li2017measuring}. According to these protocols, one needs to initialize the system into the fully polarized state, then apply quench and measure the expectation value of the first spin. All these steps are directly applicable to our setup from YIG. The fully polarized initial state can be obtained by switching on and off a strong magnetic field at a time moment $t = 0$. Quench, in our case, is performed by a microwave antenna which is an experimentally accessible device. Polarization of the initial spin can be measured through the STM tip. Overall our setup is the experimentally feasible setup studied in Ref. \cite{Chumak2}.

\subsection{Rectification}
\label{rectification}
\begin{figure}
\centering
\includegraphics[width=.70\linewidth,height=.40\linewidth]{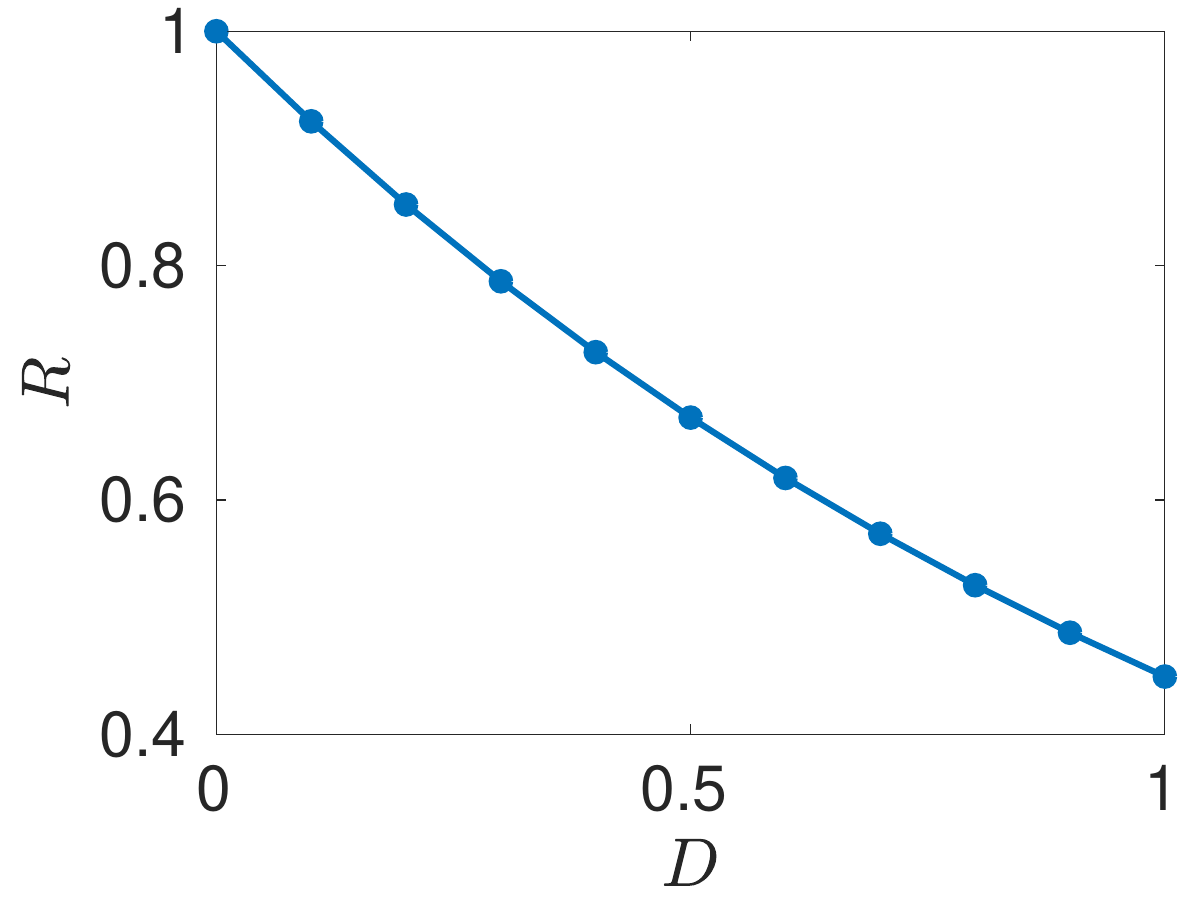}
  \caption{Rectification coefficient $R$ is plotted against DMI coefficient ($D$) for suppression rate  $\zeta(D)\approx e^{-D/5}$. The parameters are $J_1=2J_2=1$, $N=1000$, $r_{12}=10a$, $a_0=1$ and $m_0=1$ to $N$.}
  \label{R}
\end{figure}

Let us calculate the total amount of correlations transferred in opposite directions followed by the rectification coefficient, a function of the external electric field as $R=\frac{\int\limits_0^\infty C_R(t) dt}{\int\limits_0^\infty C_L(t) dt}$. We interpolate the suppression rate as a function of the DMI coefficient in the form $\zeta(D)\approx e^{-D/5}$. The coefficient $\zeta(D)$ mimics a scattering process of the drain magnons on the gate magnons \cite{Chumak2}. In Fig.~\ref{R}, we see the variation of the rectification coefficient as a function of $D$. The electric field has a direct and important role in rectification. In particular, DMI constant $D$ depends on the electric field $E_y$ as $D=E_yg_{ME}$, where $g_{ME}$ is the magneto-electric coupling constant. In the case of zero electric fields, $D$ will be zero, implying the absence of rectification effect $R=1$. 
As the electric field increases, $D$ also increases linearly, and rectification decreases exponentially. A detailed study of the role of the electric field in DM has been done in  Ref. \cite{Vignale}.

\section{Conclusions}
\label{conclusion_ch5}
We studied a quantum information flow in a  spin quantum system. In particular, we proposed a quantum magnon diode based on YIG and magnonic crystal properties. The flow of magnons with wavelengths satisfying the Bragg conditions $k=m_0\pi/a_o$ is reflected from the grooves. Due to the absence of  inversion symmetry in the system, left and right-propagating magnons have different dispersion relations and wave vectors. While for the right propagating magnons, the Bragg conditions hold, left magnons violate them, leading to an asymmetric flow of the quantum information.
\par
We found that the strength of quantum correlations depends on the distance between spins and time. The OTOC for the spins separated by longer distance shows an inevitable delay in time, meaning that the quantum information flow has a finite "butterfly velocity." On the other hand, the OTOC amplitude becomes smaller at longer distances between spins. The reason is that the initial amount of quantum information spreads among more spins. After the quantum information spreads over the whole system, which is pretty large ( $N=1000$
sites), the OTOC again becomes zero. 
\par
In the next chapter, we will summarise our complete results and discuss future plans that could be done on the basis of our previous work.

%

\chapter{Summary and Future Plans}  

\ifpdf
    \graphicspath{{Chapter6/Figs/Raster/}{Chapter6/Figs/PDF/}{Chapter6/Figs/}}
\else
    \graphicspath{{Chapter6/Figs/Vector/}{Chapter6/Figs/}}
\fi
\section{Summary}
In this thesis, we calculated OTOC by using different observables. For the calculation of OTOC, we consider    integrable and nonintegrable periodically kicked quantum Ising spin Floquet models. In this model, we see the dynamic and saturation behavior of OTOC using single-spin, block-spin, and random observables. We also studied the forward and backward flow of the magnons by using the left and right OTOC. The flow of the magnon is considered as quantum information current and using this we proposed a quantum information diode based on magnonic crystals. The results are concluded below. The ideas of future work that is related to these results are also discussed in section \ref{future_scope}.
\par 
we defined longitudinal magnetization OTOC (LMOTOC) and transverse magnetization OTOC (TMOTOC) by considering position-dependent observables in  the longitudinal and transverse directions of coupling of spins, respectively. We calculate the analytical formula of TMOTOC by using the Jorden-Wigner transformation and present the exact analytical solution of  TMOTOC. We do comparative study of the revival time, and speed of correlation propagation in TMOTOC and LMOTOC. After that, we will verify the phase structure of the transverse Ising Floquet system in $\tau_0-\tau_1$ parameter space, numerically. We use the long-time average of LMOTOC as an order parameter to distinguish the ferromagnetic and paramagnetic phases.  
\par
Subsequently, we discussed the characteristic, dynamic, and saturation regimes of the LMOTOC and TMOTOC in the Floquet system with and without a longitudinal field. we present a comparative study of LMOTOC and TMOTOC in all the regions: (1) characteristic regime when it is about to grow, (2) dynamic region when it is sharply growing,  and (3) saturation region when it starts to saturate. We focused on the role of integrability in all the regions of OTOC. 
\par

we used symmetric blocks of spins or random operators localized on these blocks are used as observables  to study OTOC in spin chains. we choose the nonlocal block-spin observables and random observables to see the possibility of exponential growth of OTOC and exponential saturation of OTOC. We calculated OTOC in pre-scrambling and post-scrambling time regimes and analyzed the growth and saturation, respectively. OTOC with spin-block observables shows a power-law growth in both integrable and nonintegrable systems.  The exponential saturation of OTOC is analyzed using the pre-scrambled random-block observables.  We have explicitly derived a connection between OTOC averaged over random observables drawn from the Gaussian unitary ensemble and the operator entanglement entropy.
\par
Later,  we provide the concept of quantum information diode, {\it i.e.,} a device rectifying the amount of quantum information transmitted in the opposite directions. we control the asymmetric left and right quantum information currents through an applied external electric field and quantify it through the left and right OTOC. To enhance the efficiency of the quantum information diode, we utilize a magnonic crystal. A quantum information diode can be fabricated from an YIG film. This is an experimentally feasible concept and implies certain conditions: low temperature and small deviation from the equilibrium to exclude effects of phonons and magnon interactions. We find that rectification of the flaw of quantum information can be controlled efficiently by an external electric field and magnetoelectric effects.

\section{Future plans}
\label{future_scope}
The OTOC, as described so far in the thesis, can serve as a reliable metric to determine if the system's dynamic behavior is chaotic or not. In particular, it allows us to study and potentially use the exotic behavior of chaotic dynamics within quantum systems. For example, quantum processors modeled over such chaotic systems shall be particularly suited for advanced searching and optimization problems in the near future.
\par
We have calculated the dynamics and saturation of OTOC by using single spin and block spin observables in the constant field Floquet system. Growth and saturation of OTOC can also be discussed in the Floquet system with time-dependent fields. Time-dependent fields can be either periodic or linear fields. In the case of a periodic field, one can take a longitudinal field in sine form and a transverse field in the form of kicks whose amplitude of kicks varies as cosine form. In the calculation of OTOC, OTOC can consider, Single spin observables, block spin observables, and half-body observables. We expect that with time-dependent fields, there will be exponential growth of OTOC in the chaotic Ising spin systems.


\begin{spacing}{0.9}

\bibliographystyle{naturemag}
\cleardoublepage
\bibliography{References/references} 

\end{spacing}


\begin{appendices} 



\renewcommand{\thesection}{A-\Roman{section}}

\renewcommand{\citenumfont}[1]{S#1}
\renewcommand{\bibnumfmt}[1]{[A#1]}

\chapter{Out-of-time-order correlation and detection of phase structure in  Floquet transverse Ising spin system}

\section{Calculation of transverse magnetization OTOC}
\label{AppendixA1}
For the calculation of the transverse magnetization OTOC due to the local operators placed at different sites, we consider  $\hat V=\hat \sigma^m_z$ and $\hat W=\hat \sigma^l_z$. Hence TMOTOC is defined as:
 \begin{equation}
 \label{F_z_g}
F_x^{l,m}(n) =  \langle \phi_0|\hat \sigma^l_z(n)\hat \sigma^m_z \hat \sigma^l_z(n) \hat \sigma^m_z|\phi_0\rangle, 
 \end{equation}
 We transform the spin variables to fermionic creation $c^{l\dagger}$ and annihilation $c^l$ operators at site $l$ by using the Jordan-Wigner transformation \cite{Jordan1928}
 \begin{equation}
S^l_x= -  \frac{1}{2}\prod_{j=1}^{l-1}(2 c^{j\dagger} c^j-1)(c^{l\dagger}+c^l)   \quad {\rm and}  \quad
S^l_z= c^{l\dagger} c^l -  \frac{1}{2}.
\end{equation}
The operators $ c^l$ and $c^{l\dagger}$ obey the the usual fermion  anticommutation rules.  The unitary operator for the closed chain is given as 
\begin{eqnarray}
\hat U &=& \exp \Big[\frac{-i  t_1}{4} \Big( \sum_{l=1}^{N-1}(c^{l\dagger} - c^l)(c^{{l+1}\dagger} - c^{l+1}) -(-1)^{N_F}(c_N^\dagger - c_N) (c^{{l+1}\dagger} - c_{N+1})\Big) \Big]  \nonumber \\ 
&&\times\exp\Big[-i t_0\sum_{l=1}^{N}\big(c^{l\dagger} c^l-\frac{1}{2}\big)\Big],
\end{eqnarray}
where $N_F=\sum_{l=1}^{N}c^{l\dagger} c^l$ is the total number of fermions. 
We move in the momentum space using the Fourier transform of $c^l$  which is defined as
\begin{equation}
c_{q}=\frac{exp(i \frac{\pi}{4})}{\sqrt{N}}\sum_{l=1}^{N} e^{-i q l}c^l.
\end{equation}

Hence U can be written as \cite{lakshminarayan2005multipartite}
\begin{equation}
\hat U=e^{(-i t_0 \frac{N}{2})} \prod_{q>0} \mathcal{V}^q.
\end{equation}
The operator $\mathcal{V}_q$ in the above expression has the form
\begin{eqnarray}
\mathcal{V}_q &=& \exp\Big(- i \frac{t_1}{2}[\cos(q)(c^{q\dagger} c_q+c_{-q}^\dagger c_{-q})+\sin(q)(c_q c_{-q} c_{-q}^\dagger c_{q}^\dagger)] \Big)  \nonumber \\ 
&&\times\exp\Big(-2 i t_0 (c^{q\dagger} c_q +c_{-q}^\dagger c_{-q})\Big).
\end{eqnarray}

For $\mathcal{V}_q$, the four basis state are $|0\rangle$ , $|\pm q\rangle=c_{\pm q}^\dagger |0\rangle$,   $|-q q\rangle=c_{-q}^\dagger c_{q}^\dagger |0\rangle$. The eigenstates of $\mathcal{V}_q$ are given by
\begin{eqnarray}
\mathcal{V}_q|\pm q\rangle=e^{\big(-\frac{t_1}{2}cos(q)- it_0 \big)}|\pm q\rangle, \quad and \quad \mathcal{V}_q|\pm \rangle=e^{\big(-\frac{t_1}{2} cos(q)- it_0} \big))e^{\pm i \gamma_q}|\pm \rangle, \nonumber
\end{eqnarray}
where 
\begin{equation}
|\pm \rangle = \alpha_{\pm q}|0 \rangle+ \beta_{\pm q}|-q q \rangle.
\end{equation}
 In the above equation $\alpha_{\pm q}$ and $ \beta_{\pm q}$ are given by eq. (11) and eq. (12) of the manuscript, respectively.
The initial unentangled state is $|\psi_N(0)\rangle=|0\rangle^{\otimes N}$. In a Fock space, it is treated as vacuum. Time evolution operator of the fermionic annihilation operator in the momentum space is given as
\begin{equation}
\label{cqt}
c_{q}(n)=\mathcal{V}_q^{\dagger n} c_{q} \mathcal{V}_q^n=\Phi_q(n)^{*} c_q-\Psi_q(n) c_{-q}^\dagger.
\end{equation}
The expansion coefficients  $\Phi_q(n)$ and  $\Psi_q(n)$ are defined in eq. (8) and eq. (9) of the manuscript, respectively, and phase angle ($\gamma_q$) is defined in eq. (10) of the manuscript.
Let us apply the first spin operator on the initial state, we get
\begin{equation}
  S_m^z(0)|0\rangle=\Big(c_m^\dagger c_m-\frac{1}{2}\Big)|0\rangle=- \frac{1}{2}|0\rangle.
\end{equation}
In the above, $c_m^\dagger c_m$ is a number operator. The operation of the number operator on the vaccum gives zero eigenvalue. Time evolution of the spin operator at position $l$ is
\begin{equation}
S_l^z(n) =  \frac{1}{N}\sum_{a,b} e^{i(a-b)l} c_a^\dagger(n) c_b(n)-\frac{1}{2},\nonumber \end{equation}
where $a$ and $b$ are  indices in momentum space. By using eq. (\ref{cqt}), we can write 
\begin{eqnarray}
S_l^z(n)  &=& \frac{1}{N}\sum_{a,b} e^{i(a-b)l} \Big[\Phi_a(n) c_a^{\dagger}-\Psi_a(n)^{*} c_{-a} \Big] \Big[ \Phi_b(n)^{*} c_b- \Psi_b(n) c_{-b}^\dagger \Big] - \frac{1}{2},\nonumber \\
  &=& \frac{1}{N}\sum_{a,b} e^{i(a-b)l} \Big[ \Phi_a(n)  \Phi_b(n)^{*} c_a^{\dagger} c_b -  \Phi_a(n)  \Psi_b(n) c_a^{\dagger} c_{-b}^\dagger \nonumber \\
  &&-  \Psi_a(n) \Phi_b(n)^{*}   c_{-a} c_b +\Psi_a(n)^{*}  \Psi_b(n) c_{-a} c_{-b}^\dagger \Big] - \frac{1}{2}. \nonumber
 \end{eqnarray}
Application of time evolved spin operator on the vacuum gives
 \begin{eqnarray}
 S_l^z(n)|0\rangle = \Big[ -\frac{1}{N}  \sum_{a,b} e^{i(a-b)l} \Phi_a(n) \Psi_b(n)  c_a^\dagger c_{-b}^\dagger 
 + \frac{1}{N} \sum_{a}|\Psi_a(n)|^2 -\frac{1}{2}\Big]|0\rangle,
\end{eqnarray} 
and the Hermitian conjugate of the above equation is
\begin{eqnarray} \label{A12}
\langle 0 | S_l^z(n)=  \langle 0 |\Big[-\frac{1}{N}\sum_{p,r} e^{-i(p-r)l} \Phi_p(n)^* \Psi_r(n)^* c_{-r} c_p 
+ \frac{1}{N} \sum_{p}|\Psi_p(n)|^2 -\frac{1}{2}\Big],
\end{eqnarray}
where $p$ and $r$ are indices in the momentum space. We can calculate $S_l^z(n)S_m^z(0)|0\rangle $ as 
\begin{eqnarray}
S_l^z(n)S_m^z(0)|0\rangle = - \frac{1}{2}\Big[ -\frac{1}{N}  \sum_{a,b} e^{i(a-b)l} \Phi_a(n) \Psi_b(n)  c_a^\dagger c_{-b}^\dagger 
 +  \frac{1}{N} \sum_{a}|\Psi_a(n)|^2 -\frac{1}{2}\Big]|0\rangle. \nonumber
\end{eqnarray}
Applying the third spin operator $S_m^z(0)$ on the state 
$S_l^z(n)S_m^z(0)|0\rangle$  we get
\begin{eqnarray}\label{A14}
S_m^z(0)S_l^z(n)S_m^z(0)|0\rangle   &=& -\frac{1}{2} \Big[\frac{1}{N}\sum_{x,y} e^{i(x-y)m}c_x^\dagger c_y - \frac{1}{2} \Big]  
    \nonumber \\ 
&\times& \Big[ -\frac{1}{N}  \sum_{a,b} e^{i(a-b)l} \Phi_a(n) \Psi_b(n)  c_a^\dagger c_{-b}^\dagger+  \frac{1}{N} \sum_{a}|\Psi_a(n)|^2 -\frac{1}{2}\Big]|0\rangle,\nonumber \\
&=& -\frac{1}{2} \Big[-\frac{1}{N^2}\sum_{x,y,a,b} e^{i(x-y)l}  e^{i(a-b)l}
  \Phi_a(n) \Psi_b(n)  \Big(c_x^\dagger c_{-b}^\dagger \delta(a,y)  \nonumber \\ 
&&-c_x^\dagger c_{a}^\dagger \delta (-b,y)\Big)+  \frac{1}{2N}\sum_{a,b} e^{i(a-b)l} \Phi_a(n) \Psi_b(n)  c_a^\dagger c_{-b}^\dagger \nonumber \\
&&- \frac{1}{2} \Big(\frac{1}{N} \sum_{a}|\Psi_a(n)|^2 - \frac{1}{2} \Big)\Big]  |0\rangle,
\end{eqnarray}
where $x$ and $y$ are the indices in the momentum space. Now we take the scalar product of the states given by eq. (\ref{A12}) and eq. (\ref{A14}) and get TMOTOC as

\begin{eqnarray}
F_x^{l,m}(n) &=& 2^4 \langle 0|S_l^z(n) S_m^z(0)S_l^z(n)S_m^z(0)|0\rangle, \nonumber \\
&=&-2^3\langle 0| \Big[ -\frac{1}{N}\sum_{p,r} e^{-i(p-r)l} \Phi_p(n)^{*} \Psi_r(n)^* c_{-r} c_p 
+ \frac{1}{N} \sum_{p}|\Psi_p(n)|^2 -\frac{1}{2} \Big] \nonumber \\
&&\Big[-\frac{1}{N^2}\sum_{x,y,a,b} e^{i(x-y)m}  e^{i(a-b)l}\Phi_a(n)\Psi_b(n) \Big(c_x^\dagger c_{-b}^\dagger \delta(a,y)  -c_x^\dagger c_{a}^\dagger \delta (-b,y)\Big) \nonumber \\
&&+ \frac{1}{2N}\sum_{a,b} e^{i(a-b)l} \Phi_a(n) \Psi_b(n)  c_a^\dagger c_{-b}^\dagger 
-\frac{1}{2} \Big( \frac{1}{N} \sum_{a}|\Psi_a(n)|^2 -\frac{1}{2} \Big) \Big]  |0\rangle, \nonumber
\end{eqnarray}
\begin{eqnarray}
F_x^{l,m}(n)&=& -2^3 \Big[\frac{1}{ N^3} \Big( \sum_{p,a,r,x,y}e^{-i(p-a)l}e^{i(x-y)m}|\Psi_r(n)|^2 \Phi_p^*(n) \Phi_a(n) \delta(p,x) \delta(a,y) \nonumber \\ &&-e^{i(r+a)l}e^{i(x-y)m}  \Psi_r(n)^* \Phi_p^*(n) \Phi_a(n)\Psi_{-p}(n) \delta(-r,x) \delta(a,y) \nonumber \\ 
&& -e^{-i(p+b)l}e^{i(x-y)m} \Psi_{b}(n)  \Psi_{-a}(n)^*  \Phi_{p}(n)^* \Phi_{a}(n)\delta(p,x) \delta(-b,y) \nonumber \\
&& + e^{i(r-b)l}e^{i(x-y)m}\Psi_{b}(n) \Psi_r(n)^* |\Phi_a(n)|^2 \delta(p,a)\delta(-b,y)\Big) \nonumber \\ 
&&- \frac{1}{2} \Big(\frac{1}{N}  \sum_{p}| \Psi_p(n)|^2 -\frac{1}{2}\Big)\Big(\frac{1}{N}  \sum_{a}| \Psi_a(n)|^2-\frac{1}{2}\Big) \Big]   \nonumber  \\
&& -\frac{1}{ 2N^2}  \sum_{p,r} \Big(|\Psi_p(n)|^2 |\Phi_r(n)|^2- \Psi_{-p}(n)  \Psi_r(n)^* \Phi_{p}(n)^* \Phi_{-r}(n) \Big) .
\end{eqnarray}
Since the term 
\begin{eqnarray}
&&  \frac{1}{ 2N^2}  \sum_{p,r} \Big(|\Psi_p(n)|^2 |\Phi_r(n)|^2- \Psi_{-p}(n)  \Psi_r(n)^* \Phi_{p}(n)^* \Phi_{-r}(n) \Big)  \nonumber \\ 
&& +\frac{1}{2} \Big(\frac{1}{N}  \sum_{p}| \Psi_p(n)|^2-\frac{1}{2}\Big)\Big(\frac{1}{N}  \sum_{a}| \Psi_a(n)|^2-\frac{1}{2}\Big) \nonumber 
 \end{eqnarray}
is constant for all number of kicks $(n)$ and system size $(N)$ which comes out to be $\frac{1}{2^3}$. Since $a$ and $b$ are  dummy indices, we replace them with $q$.
Hence, the final formula of TMOTOC is 
\begin{eqnarray}
 F_z^{l,m}(n) &=& 1- \Big(\frac{2}{ N}\Big)^3 \sum_{p,q,r} \Big( e^{i(p-q)(m-l)}|\Psi_r(n)|^2 \Phi_p^{*}(n) \Phi_q(n) \nonumber \\ 
 &&- e^{i(-r-q)(m-l)}  \Psi_r(n)^{*} \Phi_p^{*}(n)  \Phi_q(n) \Psi_{-p}(n)\nonumber \\
 &&-e^{i(p+q)(m-l)} \Psi_{q}(n)   \Phi_{-r}(n)\Psi_{r}(n)^{*}  \Phi_{p}(n)^{*} \nonumber \\
 &&+  e^{i(q-r)(m-l)}\Psi_{q}(n)\Psi_r(n)^* |\Phi_p(n)|^2 \Big).  
\end{eqnarray}
Now, we take a special case in which both the operators are the same local operator i.e.  $W=\sigma_l^z$ and $V=\sigma_l^z$. Then the formula becomes
\begin{eqnarray}
 F_z^{l,l}(n)&=& 1- \Big(\frac{2}{ N}\Big)^3 \sum_{p,q,r}  \Big( |\Psi_r(n)|^2 \Phi_p^*(n) \Phi_q(n)  
-\Psi_{-p}(n) \Psi_r(n)^* \Phi_p^*(n) \Phi_q(n)\nonumber  \\
 && - \Psi_q(n)  \Psi_r(n)^*  \Phi_{p}(n)^* \Phi_{-r}(n)+ \Psi_{q}(n)  \Psi_r(n)^* |\Phi_p(n)|^2 \Big).  
\end{eqnarray}

\section{Calculation of Longitudinal Magnetization OTOC}
\label{AppendixA2}
 Let us attempt to find the analytical expression of LMOTOC so that we calculate the phase structure of LMOTOC for higher system size. After moving some steps in analytical calculation of LMOTOC, we realize that the analytical expression of LMOTOC will take longer time than the numerical calculation. A few initial steps of our calculation of LMOTOCs are given below:
$S_l^x$ in the form of raising and lowering operator:
\begin{eqnarray}
S_l^x=\frac{1}{2}[S_l^{+}+S_l^{-}] 
= \frac{1}{2} \exp\Big [-\pi i \sum_{j=1}^{l-1}c_j^\dagger c_j\Big](c^{l\dagger}+c^l)= - \frac{1}{2} \prod_{j=1}^{l-1}(2 c_j^\dagger c_j-1)(c^{l\dagger}+c^l)  \nonumber  
\end{eqnarray}
The above equation is written by using the relation 
\begin{equation}
S_l^{+}=c^{l\dagger} \exp[\pi i \sum_{j=1}^{l-1}c^{j\dagger} c^j], \quad 
S_l^{-}=  \exp[-\pi i \sum_{j=1}^{l-1}c_j^\dagger c_j]c^l.  \nonumber
\end{equation}
Now, we move in the momentum space by doing the Fourier transform of $c^l$ and $c^{l\dagger}$. Hence $S_l^x$ in momentum space can be written as: 
\begin{eqnarray}
S_l^x &=&-\frac{1}{2} \prod_{j=1}^{l-1}\Big[2 \sum_{q_{j},p_{j}} \frac{1}{N}\exp[i(p_j-q_j)j] c_{q_j}^\dagger c_{p_j}-1\Big] \Big[\sum_{r}\Big(\frac{1}{\sqrt{N}} 
\exp(\frac{-i \pi}{4})\exp(irl) c_r^\dagger \nonumber \\
&&+ \frac{1}{\sqrt{N}} \exp(\frac{i \pi}{4})\exp(-irl) c_r\Big)\Big] 
\end{eqnarray}
For the calculation of  the time evolution of $S_l^x$ {\it i.e.}, $S_l^x(n)$ we have to compute the time evolution of all the operators in the string of length $N$. Time evolution of such a large term will be too much complicated and unfruitful for our purpose because the calculation of LMOTOC involving product of four operators will be too much to handle.

 \renewcommand{\thesection}{B-\Roman{section}}

\renewcommand{\citenumfont}[1]{B#1}
\renewcommand{\bibnumfmt}[1]{[B#1]}

\chapter{Characteristic, dynamic, and near saturation regions of Out-of-time-order correlation in Floquet Ising models}
\label{appendix3}
\section{Calculation of TMOTOC in the non-integrable Floquet system using random state}
\label{Appendix_B1}
 If $\hat V$ and $\hat W$ are two Hermitian operators that are localized on different positions $l$ and $m$, respectively, the OTOC  \cite{larkin1969quasiclassical} is given as:
\begin{equation}
C^{l,m}(n)=-\frac{1}{2}\mbox{Tr} \left([\hat W^l(n), \hat V^m]^2\right),
\label{Ap3_Cn1}
\end{equation}
which is a measure of the noncommutativity of two operators $\hat W^l$ and $\hat V^m$. These are infinite temperature quantities and involve the entire spectrum of $2^{N}$ states. One can use the trick for evaluating OTOC by employing Haar random states of $2^N$ dimensions ($\vert\Psi_R \rangle$) and calculate expectation value over $\vert\Psi_R \rangle$. OTOC will be 
\begin{equation}
C^{l,m}(n)=-2^{N-1} \langle \Psi_R|[\hat W^l(n), \hat V^m]^2\vert\Psi_R\rangle,
\end{equation}
Since, the behaviour of OTOC is similar in both the cases either taking random states or special initial states ($\vert\phi\rangle$ and $\vert\psi\rangle$ accordingly). So we consider special initial states and OTOC will be 
\begin{equation}
C^{l,m}(n)=-2^{N-1} \langle \psi/\phi|[\hat W^l(n), \hat V^m]^2\vert\psi/\phi\rangle,
\end{equation} 

\begin{figure*}[hbt!]
\begin{subfigure}{.490\textwidth}
  \includegraphics[width=.99\linewidth, height=.70\linewidth]{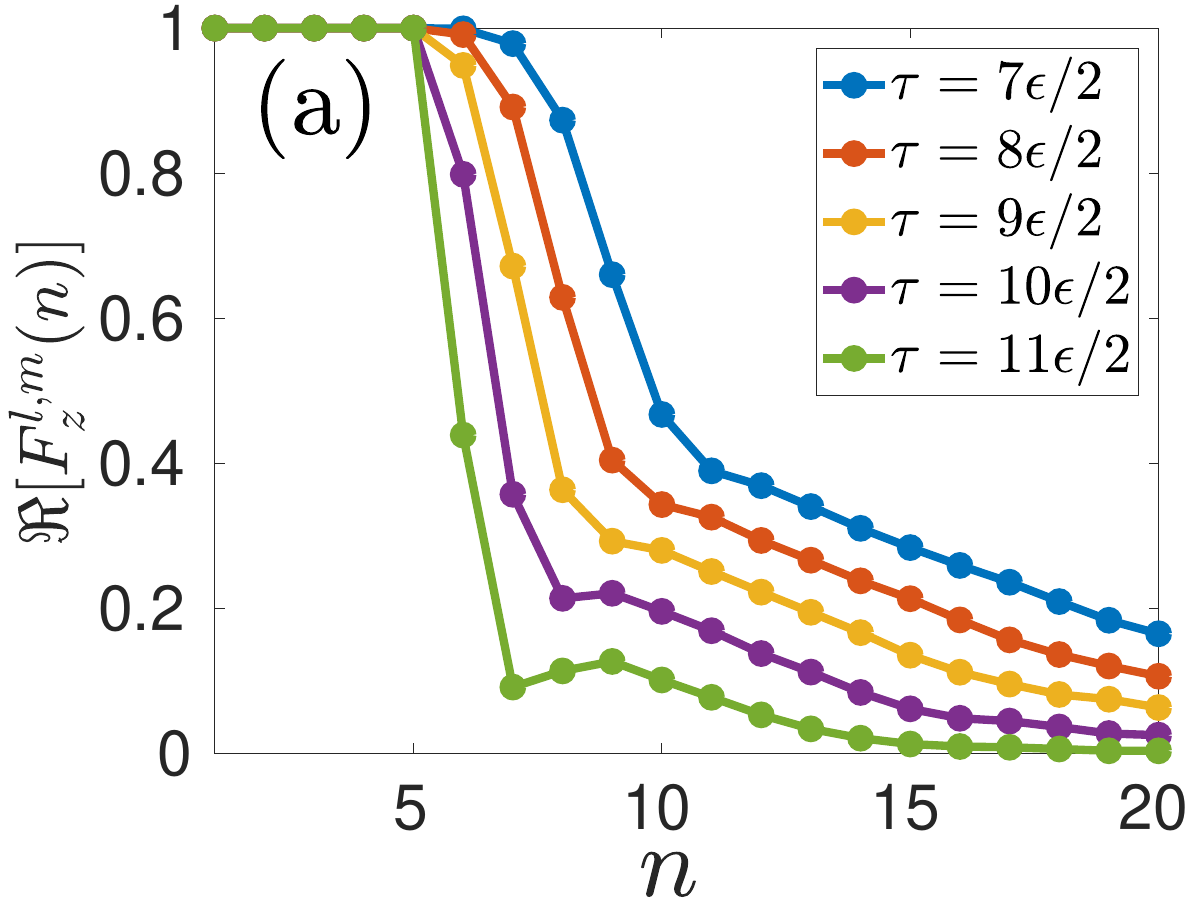}
  \end{subfigure}
  \begin{subfigure}{.490\textwidth}
  \includegraphics[width=.99\linewidth, height=.70\linewidth]{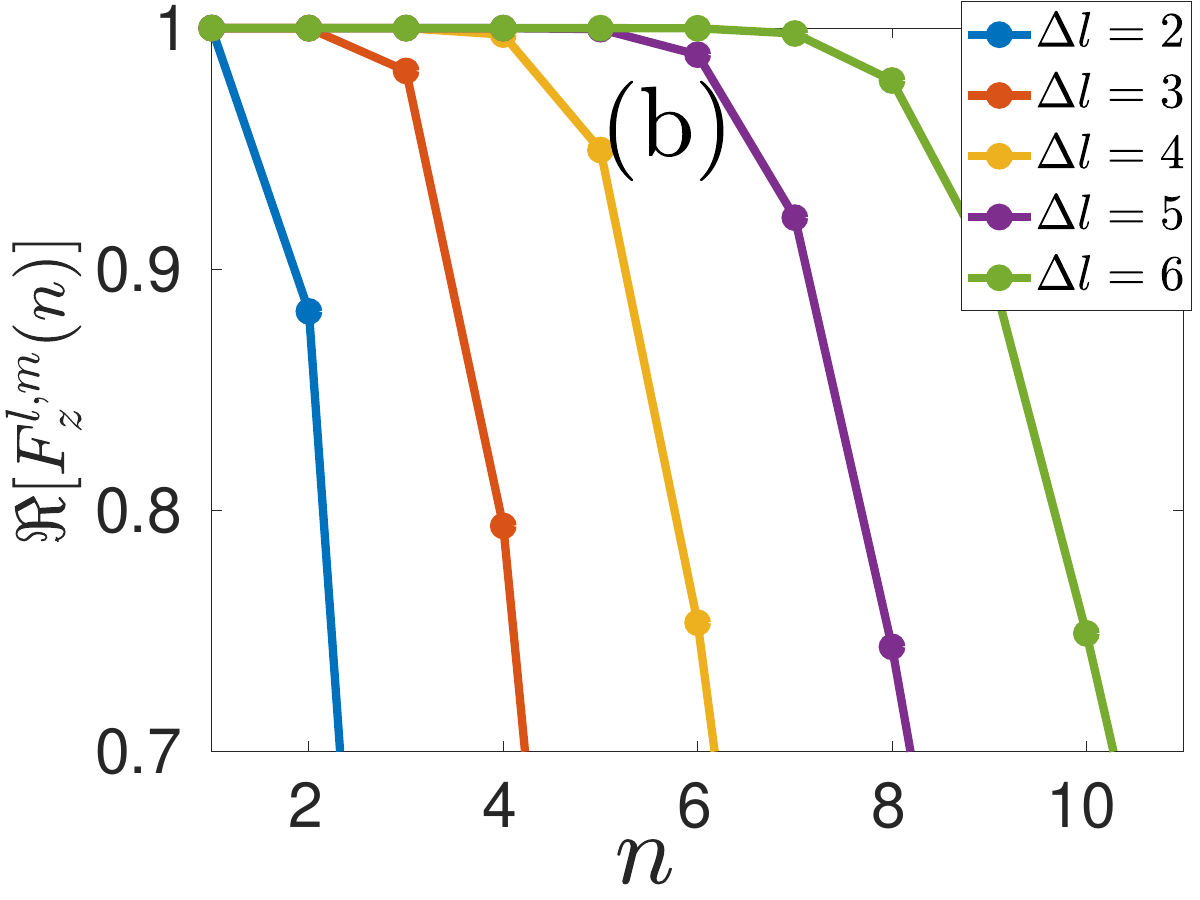}
  \end{subfigure}
\begin{subfigure}{.490\textwidth}
  \includegraphics[width=.99\linewidth, height=.70\linewidth]{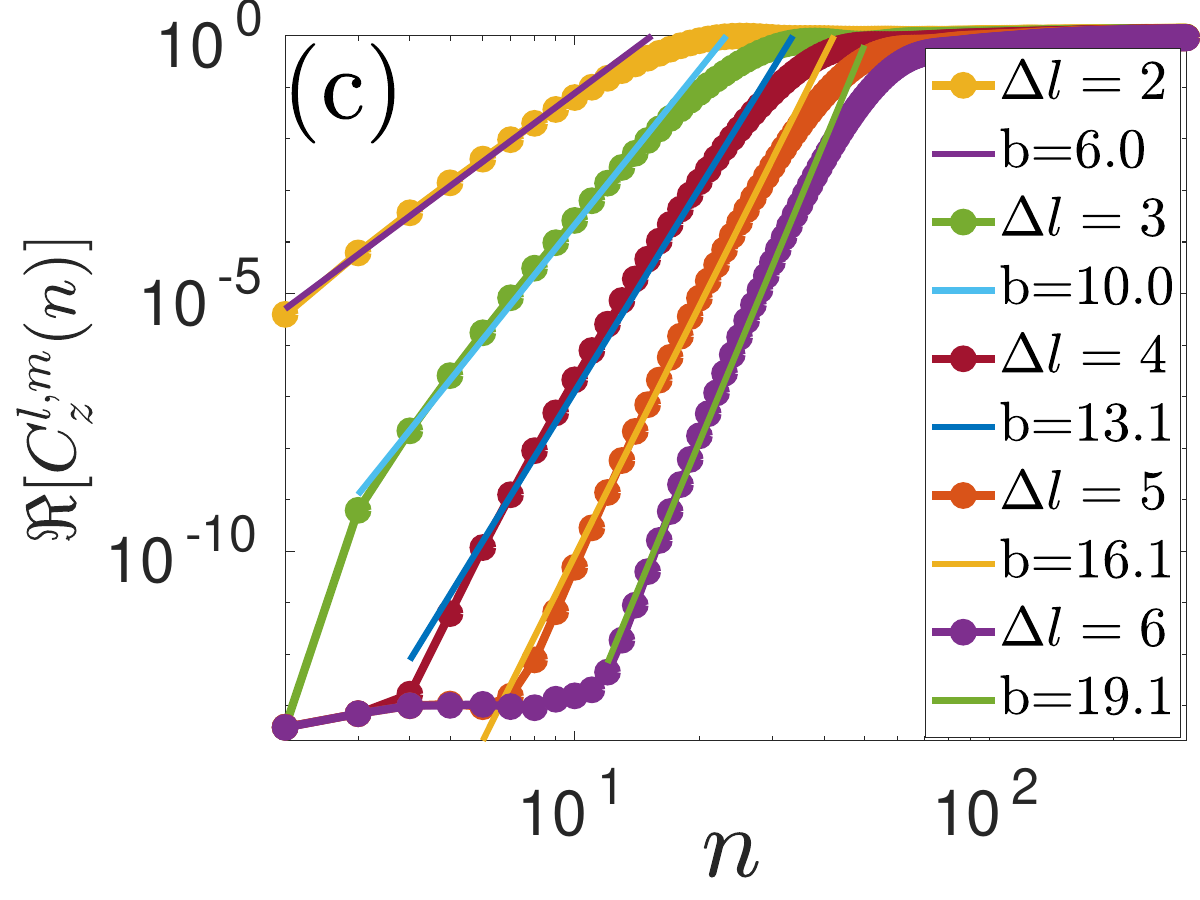}
  \end{subfigure}
  \begin{subfigure}{.490\textwidth}
  \includegraphics[width=.99\linewidth, height=.70\linewidth]{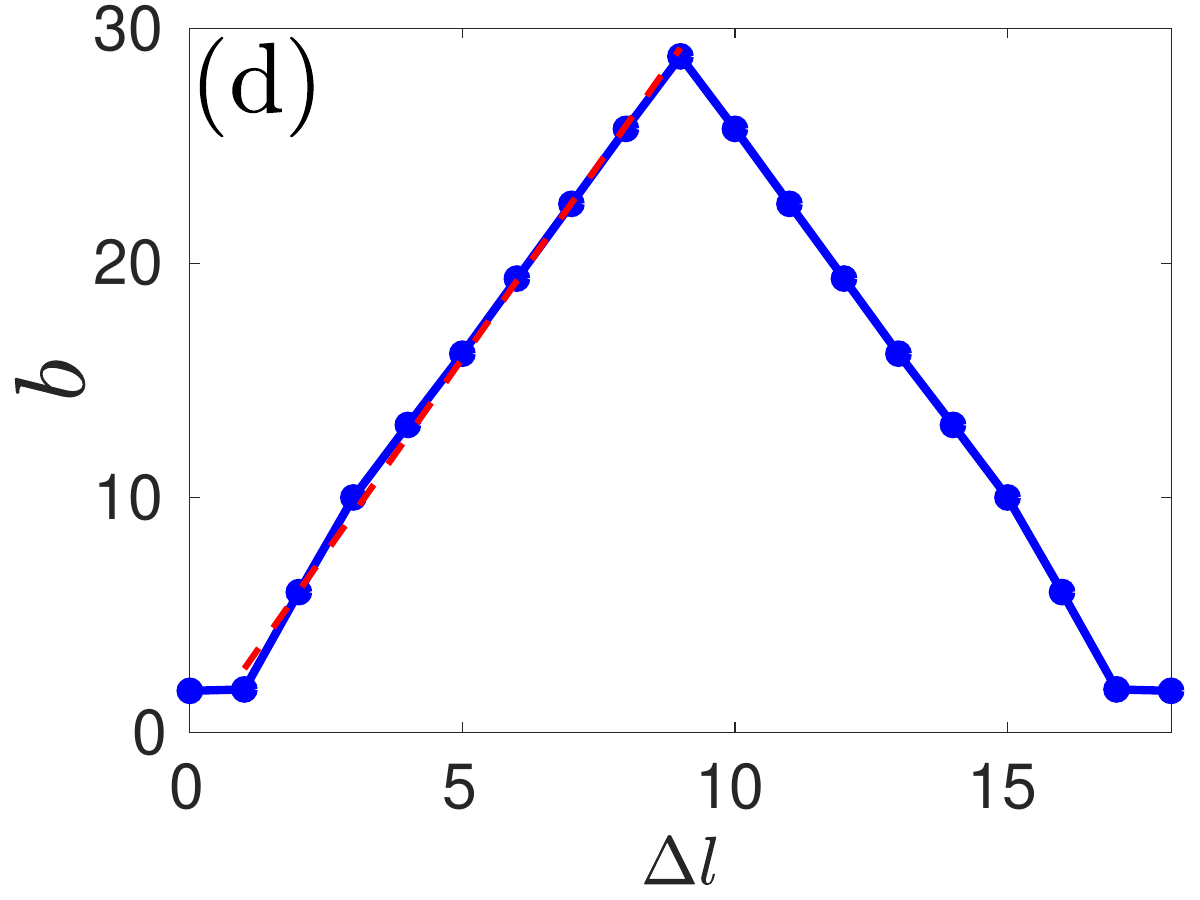}
  \end{subfigure}
  \begin{subfigure}{.490\textwidth}
  \includegraphics[width=.99\linewidth, height=.70\linewidth]{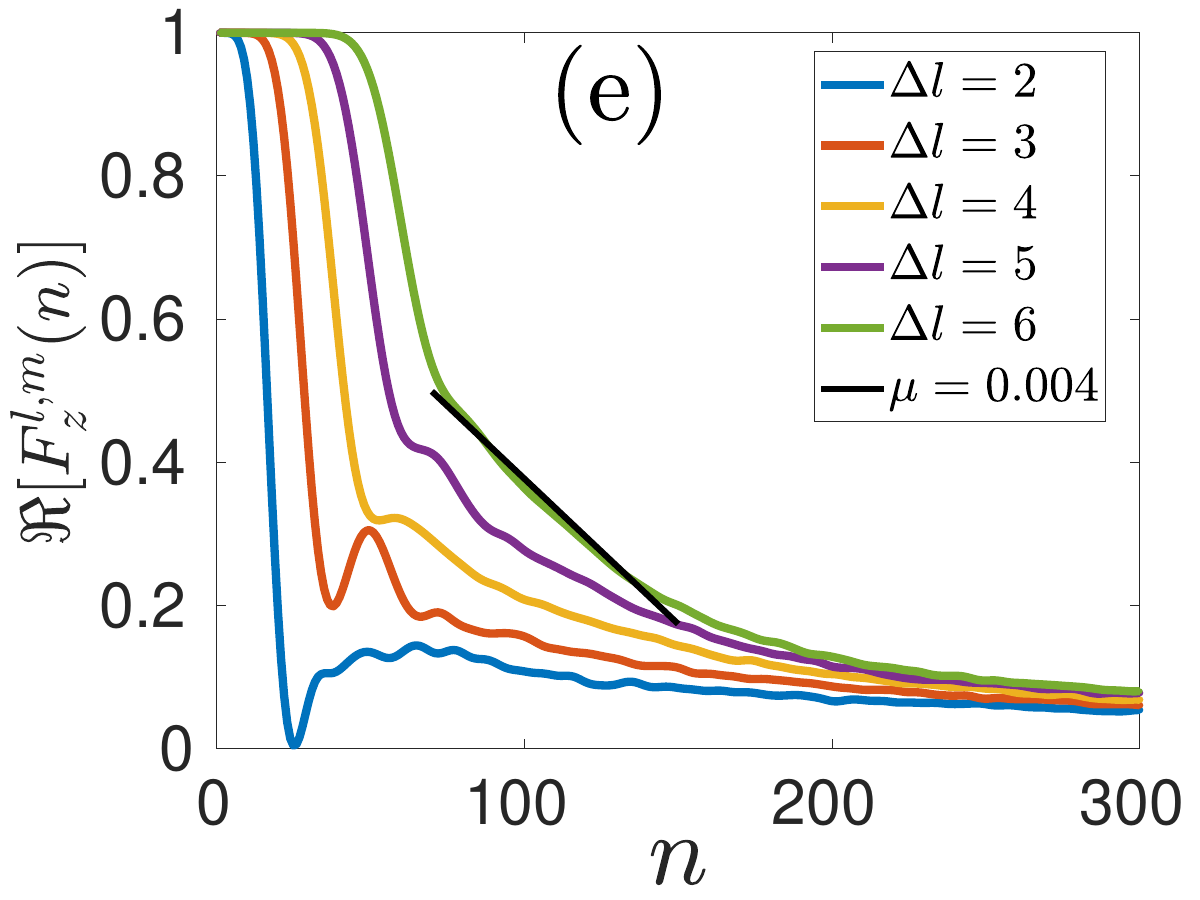}
  \end{subfigure}
 \caption{Nonintegrable closed chain transverse Ising Floquet system  with $J_x=1$, $h_z=1$, and $h_x=1$ of size $N=18$. \textbf{(a)} Behaviour of $TMOTOC$ with number of kicks $(n)$ by increasing Floquet period from  $\frac{7\epsilon}{2}$ to $\frac{11\epsilon}{2}$ differing by  $\frac{\epsilon}{2}$ with fixed $\Delta l=6$ ($\epsilon=\frac{\pi}{28})$.  \textbf{(b)}  Initial region of $F^{l,m}_z$ with number of kicks and increasing distances between the spins ($\Delta l$) with fixed Floquet period  $\tau=6\epsilon/2$. \textbf{(c)}   $C^{l,m}_z$ with number of kicks ($\log-\log$) with increasing ($\Delta l$) at fixed $\tau=\frac{\epsilon}{2}$} .  \textbf{(d)} Changing of exponent of power-law with $\Delta l$. \textbf{(e)} Saturation of $F^{l,m}_z$ with number of kicks.
\label{rand_cf_TMOTOC_nint}
\end{figure*}
Fig.~(\ref{rand_cf_TMOTOC_nint}) is the behaviour of TMOTOC in the nonintegrable $\mathcal{\hat U}_x$ system using random initial state $(\psi_R)$ drawn form the Harr measure. Characteristic time is independent of the Floqeut period [Fig.~\ref{rand_cf_TMOTOC_nint}(a)] and it depends on the separation between the observables. Number of kicks required to depart from unity is equal to separation between the observables [Fig.~\ref{rand_cf_TMOTOC_nint}(b)]. Dynamic region of TMOTOC for the nonintegrable is showing a power-law [Fig.~\ref{rand_cf_TMOTOC_nint}(c)] that is approximately similar to the [Fig.~\ref{cf_TMOTOC_nint}(c)]. The exponent of the power-law $(b)$ depends on $\Delta l$ [Fig.~\ref{rand_cf_TMOTOC_nint}(d)] and its behaviour is approximately similar as Fig.~\ref{cf_TMOTOC_nint}(d). 
Saturation of $\Re[F_z^{l,m}(n)]$ is following a linear decaying behaviour with a very small slope ($0.004$) for all $\Delta l$ [Fig.~\ref{rand_cf_TMOTOC_nint}(e)]. There is very small oscillation in comparison of Fig.~\ref{cf_TMOTOC_nint}(e).
 
 \section{Time evolution of TMOTOC}
 \label{HBC_TMOTOC}
  The Heisenberg evolution of an operator $\hat W(t)$ can be expanded using the Hausdorff-Baker-Campbel (HBC) formula as
\begin{equation}
\label{HBC}
\hat W(t)=\sum_{p=0}^{\infty}
\frac{(i t)^p}{p!}
[\hat H,[ \hat H,\cdots^{{\rm p~times}},[\hat H,\hat W]]]. 
\end{equation}
If $\hat W=\hat \sigma^{z/x}_l$, the HBC formula captures the spread of the operator over the spin sites and how it becomes more complex as time increases. Furthermore, direct replacement of Eq.~\ref{HBC} in Eq.~\ref{gene_OTOC}
highlights the fact that the short-time growth is characterized
by the smallest $p$ on which
\begin{equation}
    [\hat H,[ \hat H,\cdots^{{\rm p~times}},[\hat H,\hat \sigma^{x/z}_l]], \hat \sigma^{x/z}_m] \neq 0,
\end{equation}
due to the time factor $t^n$ that weights the terms in the expansion. We remark that this mechanism points out that the short-time growth is characterized by a general Hamiltonian structure of the system and not by the regular to chaotic regimes observed in the studied spin chains.
\par
We consider Pauli operator in transverse direction of the coupling and $\mathcal{\hat U}_x=\hat U_{xx}\hat U_z$ where, $\hat U_{xx}=\exp\big[-i \tau (J_x \hat H_{xx}+h_{x}\hat H_{x})\big]$ and $ \hat U_{z}=\exp( -i \tau h_{z} \hat H_{z})$.  Using Eq.~\ref{HBC}, the Heisenberg evolution of the spin operator $\hat \sigma_z^l$ is obtained:
\begin{eqnarray}
\hat \sigma_z^l(n)&=&(\hat U_z^{\dagger} \hat U_{xx}^{\dagger})^n\hat \sigma_z^l (\hat U_{xx} \hat U_{z})^n,
\end{eqnarray}
after applying first kick $\hat \sigma_z^l(1)$ is  
\begin{eqnarray}
\hat \sigma_z^l(1)&=& \hat U_z^{\dagger} \hat U_{xx}^{\dagger}\hat \sigma_z^l \hat U_{xx} \hat U_{z}, \nonumber \\ 
&=&\hat U_z^{\dagger}(\hat \sigma_z^l+i\tau[\hat H_{xx}+\hat H_{x}, \hat \sigma_z^l]+\frac{(i \tau)^2}{2!}[\hat H_{xx}+\hat H_{x}, [\hat H_{xx}+\hat H_{x}, \hat \sigma_z^l]+\cdots)\hat U_z, \nonumber \\
&=&\hat \sigma_z^l+ i\tau \Big(\hat U_z^{\dagger} (-2i(\hat \sigma_x^{l-1}\hat \sigma_y^{l} + \hat \sigma_y^l\hat \sigma_x^{l+1}+  \hat \sigma_y^l ) \hat U_{z}\Big)+\cdots, \nonumber \\
&=&\hat \sigma_z^l+2\tau \Big(\hat U_{z}^{\dagger}( \hat \sigma_x^{l-1}\hat \sigma_y^{l}+\hat \sigma_y^l\hat \sigma_x^{l+1}+\hat \sigma_y^l ) \hat U_{z}\Big)+\cdots, \nonumber \\ 
&=& \hat \sigma_z^l+2\tau \Big(\hat \hat \sigma_x^{l-1}\hat \sigma_y^{l}+ \hat \sigma_y^l\hat \sigma_x^{l+1}+\hat \sigma_y^l+i\tau(-2i[\hat \sigma_x^{l-1}\hat \sigma_x^{l}+\hat \sigma_y^{l-1}\hat \sigma_y^{l}+ \hat \sigma^l_x\hat \sigma^{l+1}_x \nonumber \\ 
&&+\hat \sigma^l_y\hat \sigma^{l+1}_y + \hat \sigma_x^l])+\cdots\Big)+\cdots, \nonumber \\ 
&=& \hat \sigma_z^l+ \Big(2\tau(\hat \sigma_x^{l-1}\hat \sigma_y^{l}+ \hat \sigma_y^l\hat \sigma_x^{l+1}+\hat \sigma_y^l)+(2\tau)^2(\hat \sigma_x^{l-1}\hat \sigma_x^{l}+\hat \sigma_y^{l-1}\hat \sigma_y^{l}+ \hat \sigma^l_x\hat \sigma^{l+1}_x \nonumber \\
&&+\hat \sigma^l_y\hat \sigma^{l+1}_y + \hat \sigma_x^l)+\cdots\Big)+\cdots .
\end{eqnarray}
We apply second kick then $\hat \sigma_z^l(2)$ will be
 \begin{eqnarray}
\hat \sigma_z^l(2)&=&\hat U_z^{\dagger} \hat U_{xx}^{\dagger}\Big(\hat \sigma_z^l+ \Big(2\tau(\hat \sigma_x^{l-1}\hat \sigma_y^{l}+ \hat \sigma_y^l\hat \sigma_x^{l+1}+\hat \sigma_y^l)+(2\tau)^2(\hat \sigma_x^{l-1}\hat \sigma_x^{l}+\hat \sigma_y^{l-1}\hat \sigma_y^{l} \nonumber \\
&&+ \hat \sigma^l_x\hat \sigma^{l+1}_x+\hat \sigma^l_y\hat \sigma^{l+1}_y + \hat \sigma_x^l)+\cdots\Big)+\cdots \Big)\hat U_{xx} \hat U_{z},   \nonumber \\  
&=&\hat \sigma_z^l+ \Big(4\tau( \hat \sigma_y^{l-1}\hat \sigma_x^{l}+\hat \sigma_y^l\hat \sigma_x^{l+1}+\hat \sigma_y^l)+(2\tau)^2(\hat \sigma^{l-1}_x\hat \sigma^{l}_x+\hat \sigma^{l-1}_y\hat \sigma^{l}_y + \hat \sigma^l_x\hat \sigma^{l+1}_x \nonumber \\
&&+\hat \sigma^l_y\hat \sigma^{l+1}_y + \hat \sigma_x^l) +(2\tau)^2(\hat U_z^{\dagger} \hat U_{xx}^{\dagger}(\hat \sigma_y^{l-1}\hat \sigma_y^{l}+\hat \sigma_y^l\hat \sigma_y^{l+1})\hat U_{xx}^{\dagger} \hat U_{z}^{\dagger})+\cdots\Big)+\cdots \nonumber\\ \end{eqnarray}
From the above equation, we extract the coefficient of $\tau^2$ which contain $\hat \sigma_y^{l+2}$ term. This is given as
 \begin{equation}
     \begin{split}
(2\tau)^2(\hat U_z^{\dagger} \hat U_{xx}^{\dagger}(\hat \sigma_y^{l-1}\hat \sigma_y^{l}&+\hat \sigma_y^l\hat \sigma_y^{l+1})\hat U_{xx}^{\dagger} \hat U_{z}^{\dagger})\\
&=(2\tau)^2\hat U_z^{\dagger}\Big(\hat \sigma_y^l\hat \sigma_y^{l+1}+i\tau[\hat H_{xx}+\hat H_x, \hat \sigma_y^l\hat \sigma_y^{l+1}]+\cdots\Big)\hat U_z, \\ 
 &=(2\tau)^2\Big(\cdots -2\tau\hat \sigma_y^{l-1}\hat \sigma_z^{l}\hat \sigma_y^{l+1} -2\tau\hat \sigma_y^l\hat \sigma_z^{l+1}\hat \sigma_y^{l+2}+\cdots \Big).
\end{split}
 \end{equation}
 For $m=l+2$, $C_z^{l,m}(2)=64\tau^6$
We apply the third kick then $\hat \sigma_z^l(3)$ will be
 \begin{equation}
     \begin{split}
          \hat \sigma_z^l(3)&=\hat U_z^{\dagger} \hat U_{xx}^{\dagger}\Big(\hat \sigma_z^l+ \Big(4\tau(\hat \sigma_y^{l-1}\hat \sigma_x^{l}+ \hat \sigma_y^l\hat \sigma_x^{l+1}+\hat \sigma_y^l)+(2\tau)^2( \hat \sigma^{l-1}_x\hat \sigma^{l}_x+\hat \sigma^{l-1}_y\hat \sigma^{l}_y  \\
 &+\hat \sigma^l_x\hat \sigma^{l+1}_x+\hat \sigma^l_y\hat \sigma^{l+1}_y + \hat \sigma_x^l)+\cdots\Big)+\cdots\Big)\hat U_{xx} \hat U_{z}, \\ 
&=\Big[\hat \sigma_z^l+ \Big(6\tau( \hat \sigma_y^{l-1}\hat \sigma_x^{l}+\hat \sigma_y^l\hat \sigma_x^{l+1}+\hat \sigma_y^l)+(2\tau)^2(\hat \sigma^{l-1}_x\hat \sigma^{l}_x+\hat \sigma^{l-1}_y\hat \sigma^{l}_y + \hat \sigma^l_x\hat \sigma^{l+1}_x \\
&+\hat \sigma^l_y\hat \sigma^{l+1}_y + \hat \sigma_x^l)+\cdots\Big)+\cdots \Big].
  \end{split}
 \end{equation}

For $\Delta l=1$, $m=l+1$ dominating exponent of the power-law of the OTOC will be 
\begin{eqnarray}
C_z^{l,l+1}(1)&=&4\tau^2\langle \phi_0\vert[( \hat \sigma_y^l\hat \sigma_x^{l+1}+\hat \sigma_y^l), \hat \sigma_z^m]^2\vert\phi_0\rangle, \\ \nonumber &=&4\langle\phi_0\vert(-i\hat \sigma_y^l\hat \sigma_y^{l+1})^2\vert\phi_0\rangle =4 \tau^2.
\end{eqnarray}
Similarly, $C_z^{l,l+1}(2)=16\tau^2$ and $C_z^{l,l+1}(3)=36\tau^2$.

For $\Delta l=2$, $m=l+2$, dominating exponent of the power-law of the OTOC will be 
$C_z^{l,l+2}(1)=0$, $C_z^{l,l+2}(2)=64\tau^6$.
For $\Delta l=2$, $m=l+2$.
This power-law growth approximately matches the dynamic region of the  Eq.~(\ref{C_tmotoc_nint}).  
 \section{Time evolution of LMOTOC}
 \label{HBC_LMOTOC}
We consider Pauli operator in longitudinal direction of the coupling. Using Eq.~\ref{HBC}, the Heisenberg evolution of the spin operator $\hat \sigma_x^l$ is obtained:
\begin{eqnarray}
\hat \sigma_x^l(n)&=&(\hat U_z^{\dagger} \hat U_{xx}^{\dagger})^n\hat \sigma_x^l (\hat U_{xx} \hat U_{z})^n, \nonumber
\end{eqnarray}
after applying first kick $\hat \sigma_x^l(1)$ is  \begin{equation}
\begin{split}
\hat \sigma_x^l(1)&= \hat U_z^{\dagger} \hat U_{xx}^{\dagger}\hat \sigma_x^l \hat U_{xx} \hat U_{z}=\hat U_z^{\dagger}(\hat \sigma_x^l+i\tau[\hat H_{xx}+\hat H_{x}, \hat \sigma_x^l]+\frac{(i \tau)^2}{2!}[\hat H_{xx}+\hat H_{x}, [\hat H_{xx}+\hat H_{x}, \hat \sigma_x^l]+\cdots)\hat U_z. \nonumber
\end{split}
\end{equation}
 Since, $[\hat H_{xx}+\hat H_{x}, \hat \sigma_x^l]=0$, then 
\begin{equation}
    \begin{split}
\hat \sigma_x^l(1)&= \hat U_z^{\dagger} \hat \sigma_x^l  \hat U_{z}=\hat \sigma_x^l+i\tau[\hat H_{z}, \hat \sigma_x^l]+\frac{(i \tau)^2}{2!}[\hat H_{z}, [\hat H_{z}, \hat \sigma_x^l]+\cdots=
\hat \sigma_x^l(1)=\hat \sigma_x^l- 2\tau \hat \sigma_y^l+\cdots.\nonumber
\end{split}
\end{equation}
We apply second kick then $\hat \sigma_x^l(2)$ will be
\begin{equation}
\begin{split}
\hat \sigma_x^l(2)&=\hat U_z^{\dagger} \hat U_{xx}^{\dagger}\hat \sigma_x^l(1) \hat U_{xx} \hat U_{z}=U_z^{\dagger} \hat U_{xx}^{\dagger}(\hat \sigma_x^l- 2\tau \hat \sigma_y^l+\cdots)\hat U_{xx} \hat U_{z},\nonumber \\  
&=\Big(U_z^{\dagger} \hat U_{xx}^{\dagger}\hat \sigma_x^l \hat U_{xx} \hat U_{z} - 2\tau  U_z^{\dagger} \hat U_{xx}^{\dagger}\hat \sigma_y^l\hat U_{xx} \hat U_{z}+\cdots\Big), \nonumber \\
  &=\Big(\hat \sigma_x^l- 2\tau \hat \sigma_y^l-2\tau  U_z^{\dagger} (\hat \sigma_y^l-2\tau (\hat \sigma_y^{l-1} \hat \sigma_z^{l} +\hat \sigma_z^l \hat \sigma_y^{l+1}+  \hat \sigma_z^l)+\cdots) \hat U_{z}+\cdots \Big), \nonumber\\ 
 \end{split}
\end{equation}
\begin{equation}
    \begin{split}
\hat \sigma_x^l(2)&=\Big(\hat \sigma_x^l- 2\tau \hat \sigma_y^l-2\tau  [\hat \sigma_y^l-2\tau \hat \sigma_y^{l-1} \hat \sigma_z^{l}-2\tau \hat \sigma_z^l\hat \sigma_y^{l+1}-2\tau  \hat \sigma_z^l+i\tau(-2i \hat \sigma_x^l \\
&-2\tau  (-2i)\hat \sigma_z^l \hat \sigma_y^{l+1})+\cdots] +\cdots \Big), \\ 
&=\Big(\hat \sigma_x^l- 4\tau \hat \sigma_y^l+(2\tau)^2 (\hat \sigma_z^l\hat \sigma_y^{l+1}+\hat \sigma_x^l+  \hat \sigma_z^l)+(2\tau )^3 \hat \sigma_z^l \hat \sigma_y^{l+1}) +\cdots \Big). \end{split}
\end{equation}
We apply third kick then $\hat \sigma_x^l(3)$ will be
\begin{equation}
    \begin{split}
\hat \sigma_x^l(3)&=\hat U_z^{\dagger} \hat U_{xx}^{\dagger}\Big(\hat \sigma_x^l- 4\tau \hat \sigma_y^l+(2\tau)^2 (\hat \sigma_z^l\hat \sigma_y^{l+1}+ \hat \sigma_x^l+ \hat \sigma_z^l)+(2\tau)^3 \hat \sigma_z^l \hat \sigma_y^{l+1}) +\cdots \Big) \hat U_{xx} \hat U_{z}, \\ 
&=\Big(\hat \sigma_x^l- 6\tau \hat \sigma_y^l+2(2\tau )^2 (\hat \sigma_z^l\hat \sigma_y^{l+1}+\hat \sigma_x^l+ \hat \sigma_z^l)+2(2\tau )^3 \hat \sigma_z^l \hat \sigma_y^{l+1}) +\cdots \Big)+\cdots
\end{split}
\end{equation}
Consider $\Delta l=1,$ $m=l+1$
$C_x^{l,l+1}(1)=0$, $C_x^{l,l+1}(2)=64 \tau^6 $, and $C_x^{l,l+1}(3)=256 \tau^6$. Exponent of the power-law approximately matches the  Eq.~(\ref{bformula_lmotoc}).


\renewcommand{\thesection}{C-\Roman{section}}

\renewcommand{\citenumfont}[1]{C#1}
\renewcommand{\bibnumfmt}[1]{[C#1]}

\chapter{ Out-of-time-order correlators of nonlocal block-spin and random observables in integrable and nonintegrable spin chains}
\section{Calculation of post-scrambling OTOC using random unitary operator}
\label{appendix_C}
We calculate long-time saturation values of OTOC for spin-block operators $\hat V$ and $\hat W$ are calculated by replacing the unitary operator $\hat U$ with random CUE of size $2^N$ and averaging over it. Two- and four-point correlation functions $C_2(n)$ and $C_4(n)$ are calculated as below:  
 \subsection{Calculation of two-point correlation}
 Two point correlation ($C_2(n)$) averaged over random $ U$ drawn from CUE of size $2^N$ is given by
\begin{equation}
\ovl{C_2(n)}^{U}=\frac{1}{d_A d_B}\ovl{\Tr(\hat W(n)^2\hat V^2)}^{U}.
\end{equation}
Since time evolution of $\hat W$ is given by Heisenberg time evolution  as $\hat W(n)=\hat U(n)^{\dagger}\hat W\hat U(n)$. Hence, 
\begin{equation}
\begin{split}
\ovl{C_2(n)}^{U}&=\frac{1}{d_A d_B}\ovl{\Tr(\hat U^{\dagger}\hat W^2\hat U\hat V^2)}^{U}
=\frac{1}{d_A d_B}\sum_{j=1}^{d}\ovl{\langle j|\hat U^{\dagger}\hat W^2\hat U\hat V^2)|j\rangle}^{U}, \\ 
&=\frac{1}{d_A d_B}\sum_{j,k,l,m}\ovl{\langle j|\hat U^{\dagger}|k\rangle \langle k|\hat W^{2}|l\rangle\langle l|\hat U|m\rangle \langle m|\hat V^{2}|j\rangle}^{U}
=\frac{1}{d_A d_B}\sum_{j,k,l,m} \ovl{\hat U_{kj}^{*}\hat U_{lm}}^{U}\hat W^2_{kl}\hat V^2_{mj}.  \nonumber 
\end{split}
\end{equation}
Since, $\ovl{\hat U_{kj}^{*}\hat U_{lm}}^{U}=\sum_{j,k,l,m} \delta_{kl}\delta_{jm} |\hat U_{kj}|^2$ and $|\hat U_{kj}|^2=\frac{1}{d}$
\begin{equation}
    \begin{split}
\ovl{C_2(n)}^{U}=\frac{1}{d_A d_B}\frac{1}{d}\sum_{j,k,l,m} \delta_{kl}\delta_{jm}\hat W^2_{kl}\hat V^2_{mj}
=\frac{1}{d_A d_B}\frac{1}{d}\sum_{k,j} \hat W^2_{kk}\hat V^2_{jj}
=\frac{1}{d_Ad_B}\frac{1}{d} \Tr (\hat W^2) \Tr (\hat V^2). \nonumber
  \end{split}
\end{equation}
Since, $d_Ad_B=2^N$. Hence $C_2(n)$ will be 
$C_2(n)=\frac{1}{2^{2N}} \hat \Tr(\hat W^2)\hat \Tr(\hat V^2).$\\
Since, block observables are localized spin block observables defined by Eq.~(\ref{Block}). Then  calculate $\Tr(\hat W^2)$ will be
\begin{equation}
\label{trw}
\Tr  (\hat W^2)=\frac{4}{N^2}\Tr\Bigg(\sum_{l=1}^{\frac{N}{2}}(\hat \sigma_l^{x})^2+\sum_{l\neq l^{'}}\hat \sigma_l^x\hat \sigma_{l^{'}}^x\Bigg).
\end{equation}
By using the properties of Pauli operator, square of Pauli operators are equal to identity matrix. Hence first term of Eq. \ref{trw} will be equal to $\frac{2}{N}2^N$. And second term, $\sum_{l\neq l^{'}}\hat \sigma_l^x\hat \sigma_{l^{'}}^x$ is equal to zero because Pauli observable follow the anti-commutation relation.
Hence, $C_2(n)$ for the spin block observables is
\begin{equation}
\ovl{C_2(n)}^{U}=\frac{1}{2^{2N}}\frac{4}{N^2}2^{2N}=\frac{4}{N^2}.
\end{equation}
 \subsection{Calculation of four point correlator} 
Four-point correlator [$C_4(n)$] averaged over random $ U$ drawn from CUE of size $2^N$ is given by
\begin{equation}
    \begin{split}
\overline{C_4(n)}^U&= \frac{1}{d_Ad_B}\ovl{\Tr(\hat W(n)\hat V \hat W(n)\hat V)}^{U}
=\frac{1}{d_Ad_B}\ovl{\Tr(\hat U^{\dagger}\hat W\hat U\hat V \hat U^{\dagger}\hat W\hat U\hat V)}^{U},\\ \nonumber
&=\frac{1}{d_Ad_B}\sum_{i_1,i_2,\cdot,i_8} \ovl{\langle i_1|\hat U^{\dagger}|i_2\rangle \langle i_2|\hat W|i_3\rangle\langle i_3|\hat U|i_4\rangle \langle i_4|\hat V|i_5\rangle\langle i_5|\hat U^{\dagger}|i_6\rangle \langle i_6|\hat W|i_7\rangle \langle i_7|\hat U|i_8\rangle \langle i_8|\hat V|i_1\rangle}^{U},\\ \nonumber
&=\frac{1}{d_Ad_B}\sum_{i_1,i_2\cdot i_8} \ovl{\hat U_{i_1,i_2}^{*}\hat U_{i_3,i_4}\hat U_{i_6,i_5}^{*}\hat U_{i_7,i_8}}^{U}\hat W_{i_2,i_3}\hat V_{i_4,i_5}\hat W_{i_6,i_7}\hat V_{i_8,i_1},\\ \nonumber
&=\frac{1}{d_Ad_B}\sum_{i_1,i_2 \cdot i_8}\Bigg(\delta_{i_2,i_3}\delta_{i_1,i_4}\delta_{i_6,i_7}\delta_{i_5,i_8}|\hat U_{i_2,i_1}|^2|\hat U_{i_6,i_5}|^2 \hat W_{i_2,i_3} \hat V_{i_4,i_5} \hat W_{i_6,i_7} \hat V_{i_8,i_1}\\ \nonumber 
&+\delta_{i_2,i_7}\delta_{i_1,i_8}\delta_{i_3,i_6}\delta_{i_4,i_5}|U_{i_2,i_1}|^2|\hat U_{i_3,i_4}|^2 \hat W_{i_2,i_3} \hat V_{i_4,i_5} \hat W_{i_6,i_7} \hat V_{i_8,i_1}\Bigg) \\ \nonumber
&-\frac{1}{d_Ad_B}\sum_{i_1,i_2 \cdot i_8}\Bigg(\delta_{i_2,i_3}\delta_{i_1,i_4}\delta_{i_6,i_7}\delta_{i_5,i_8}\hat U^{*}_{i_2,i_1}\hat U_{i_2,i_4}\hat U^{*}_{i_6,i_5}\hat U_{i_6,i_8}\hat W_{i_2,i_3} \hat V_{i_4,i_5} \hat W_{i_6,i_7} \hat V_{i_8,i_1}\\ \nonumber
&+\delta_{i_2,i_7}\delta_{i_1,i_8}\delta_{i_3,i_6}\delta_{i_4,i_5}U^{*}_{i_2,i_1}\hat U_{i_6,i_5}\hat U^{*}_{i_2,i_1}\hat U_{i_6,i_5}\hat W_{i_2,i_3} \hat V_{i_4,i_5} \hat W_{i_6,i_7} \hat V_{i_8,i_1}\Bigg),\\ \nonumber
&=\frac{1}{d_Ad_B}\frac{1}{d^2-1}\sum_{i_1,i_2 \cdot i_8}\Bigg(\delta_{i_2,i_3}\delta_{i_1,i_4}\delta_{i_6,i_7}\delta_{i_5,i_8} \hat W_{i_2,i_3} \hat V_{i_4,i_5} \hat W_{i_6,i_7} \hat V_{i_8,i_1} \nonumber \\ 
&+\delta_{i_2,i_7}\delta_{i_1,i_8}\delta_{i_3,i_6}\delta_{i_4,i_5} \hat W_{i_2,i_3} \hat V_{i_4,i_5} \hat W_{i_6,i_7} \hat V_{i_8,i_1}\Bigg) \\  \nonumber
&-\frac{1}{d_Ad_B}\frac{1}{d(d^2-1)}\sum_{i_1,i_2 \cdot i_8}\Bigg(\delta_{i_2,i_3}\delta_{i_1,i_4}\delta_{i_6,i_7}\delta_{i_5,i_8} \hat W_{i_2,i_3} \hat V_{i_4,i_5} \hat W_{i_6,i_7} \hat V_{i_8,i_1} \nonumber \\ 
&+\delta_{i_2,i_7}\delta_{i_1,i_8}\delta_{i_3,i_6}\delta_{i_4,i_5} \hat W_{i_2,i_3} \hat V_{i_4,i_5} \hat W_{i_6,i_7} \hat V_{i_8,i_1}\Bigg), \\ \nonumber
&=\frac{1}{d_Ad_B}\frac{1}{d^2-1}\sum_{i_1,i_2 \cdot i_8} \Bigg( \hat W_{i_1,i_2} \hat V_{i_1,i_5} \hat W_{i_6,i_6} \hat V_{i_5,i_1} +\hat W_{i_2,i_3} \hat V_{i_4,i_4} \hat W_{i_3,i_2} \hat V_{i_1,i_1}\Bigg) \\ \nonumber
&-\frac{1}{d_Ad_B}\frac{1}{d(d^2-1)}\sum_{i_1,i_2 \cdot i_8} \Bigg( \hat W_{i_2,i_2}\hat V_{i_4,i_4} \hat W_{i_6,i_6} \hat V_{i_8,i_8} + \hat W_{i_2,i_3} \hat V_{i_4,i_5} \hat W_{i_6,i_7} \hat V_{i_8,i_1} \Bigg), \\ \nonumber
\end{split}
\end{equation}
\begin{equation}
    \begin{split}
\overline{C_4(n)}^U&=\frac{1}{d_Ad_B}\frac{1}{d^2-1}\Bigg((\Tr \hat W)^2 (\Tr \hat V)^2+(\Tr \hat W^2) (\Tr \hat V)^2\Bigg) \\ \nonumber
&-\frac{1}{d_Ad_B}\frac{1}{d(d^2-1)}\Bigg(\Tr (\hat W^2) \Tr(\hat V^2)+(\Tr \hat W)^2 (\Tr \hat V)^2\Bigg)+O\Bigg(\frac{1}{d(d^2-1)}\Bigg).
  \end{split}
\end{equation}
Considering traceless observables such that $\Tr (\hat W)=0$ and $\Tr(\hat V)=0$, and $d_Ad_{B}=d$ we get 
\beqa
\ovl{C_4(n)}^{U}&=&-\frac{1}{d}\frac{1}{d(d^2-1)}(\Tr \hat W^2) (\Tr \hat V^2)=-\frac{1}{d^2(d^2-1)}(\Tr \hat W^2) (\Tr \hat V^2).
\eeqa
For traceless observables $C_2(n)$ will be
\begin{equation}
\ovl{C_2(n)}^{U}=\frac{1}{d^2}\Tr (\hat W^2) \Tr (\hat V^2).
\end{equation}
Hence, OTOC for the traceless observables will be 
\beqa
\ovl{C(n)}^{U}&=&\ovl{C_2(n)}^{U}-\ovl{C_4(n)}^{U}=\frac{1}{d^2}(\Tr \hat W^2) (\Tr \hat V^2)\Bigg(1+\frac{1}{d^2-1}\Bigg),\nonumber \\
&=&\frac{1}{d^2}(\Tr \hat W)^2 (\Tr \hat V)^2\frac{d^2}{d^2-1}=\frac{1}{d^2-1}(\Tr \hat W)^2 (\Tr \hat V)^2=\frac{1}{2^{2N}-1}
\approx \frac{1}{2^{2N}}.\nonumber \\ 
\eeqa
\renewcommand{\thesection}{D-\Roman{section}}

\renewcommand{\citenumfont}[1]{D#1}
\renewcommand{\bibnumfmt}[1]{[D#1]}
\chapter{ Quantum information diode based on a magnonic crystal}
\section{Diagonalization of Hamiltonian of 2D square lattice }
\label{diag_2d_Ham}
2D square-lattice spin system with nearest-neighbor $J_1$ and the next nearest-neighbor $J_2$ coupling constants (taking $\hslash=1$): 
\begin{eqnarray}
\label{Hamiltonian_A4}
\hat{H}&=&J_1\sum\limits_{\langle
 n,m\rangle}\hat\sigma_n\hat\sigma_m+
  J_2\sum\limits_{\langle\langle
  n,m\rangle\rangle}\hat\sigma_n\hat\sigma_m-{\bf
  P}\cdot{\bf E}, \nonumber \\ 
  &=&J_1\sum\limits_{\langle
 n,m\rangle}\hat\sigma_n\hat\sigma_m+
  J_2\sum\limits_{\langle\langle
  n,m\rangle\rangle}\hat\sigma_n\hat\sigma_m-D\sum\limits_{n}(\hat\sigma_{n}\times\hat\sigma_{n+1})_z, \nonumber \\
  &=&\frac{1}{4}\Big[J_1\sum\limits_{\langle
 n,m\rangle}\hat S_n\hat S_m+
  J_2\sum\limits_{\langle\langle
  n,m\rangle\rangle} \hat S_n\hat S_m+\frac{D}{i}\sum_n( \hat S_{n}^+\hat S_{n+1}^{-}-\hat  S_{n}^{-}\hat S_{n+1}^{+})\Big], \nonumber \\
  &=&\frac{1}{4}\Big[J_1 \sum\limits_{\langle
 n,m\rangle} \frac{1}{2}\Big\{\Big(\hat S_n^{-}\hat S_m^{+}+\hat S_n^{+}\hat S_m^{-}\Big)+\hat S_n^{z}\hat S_m^{z}\Big\}
  +J_2\sum\limits_{\langle\langle
  n,m\rangle\rangle} \frac{1}{2}\Big\{\hat S_n^{-}\hat S_m^{+}+\hat S_n^{+}\hat S_m^{-} \Big)+\hat S_n^{z}\hat S_m^{z}\Big\} \nonumber \\
  &+&\frac{D}{i}\sum_n( \hat S_{n}^+\hat S_{n+1}^{-}-\hat  S_{n}^{-}\hat S_{n+1}^{+})\Big].
\end{eqnarray} 

Spin-half systems have two permitted states on each site, {\it i.e.}, $\vert \uparrow\rangle$ and $\vert \downarrow\rangle$. Operation of spin operators on these state are given as
\begin{equation}
\begin{split}
&\hat S^+\vert \downarrow\rangle=\vert \uparrow\rangle,~~ \hat S^+\vert \uparrow\rangle=0,\\
&\hat S^-\vert \uparrow\rangle=\vert \downarrow\rangle,~~ \hat S^-\vert \downarrow\rangle=0, \\
&\hat S^z\vert \uparrow\rangle=\frac{1}{2}\vert \uparrow\rangle,~~ \hat S^z\vert \downarrow\rangle=-\frac{1}{2}\vert \downarrow\rangle,
\end{split}
\end{equation}
Transformation of the spin operators in hard-core bosonic creation and annihilation operators are given as
\begin{equation}
    \begin{split}
&\hat S_{m,n}^+=\hat a_{m,n}, \\
& \hat S_{m,n}^-=\hat a_{m,n}^{\dagger},  \\
&  \hat S_{m,n}^z=1/2-\hat a_{m,n}^{\dagger}\hat a_{m,n}
\end{split}
\end{equation}
Hamiltonian in the bosonic representation is given as 
\begin{eqnarray}
\hat H &=&\frac{1}{4}\Big[J_1 \sum\limits_{\langle
 n,m\rangle} \Big(\hat a_n^{\dagger}\hat a_m+\hat a_n\hat a_m^{\dagger}-\hat a_n^{\dagger}\hat a_n-\hat a_m^{\dagger}\hat a_m\Big)
 \nonumber \\
  &+&J_2\sum\limits_{\langle\langle
  n,m\rangle\rangle} \Big(\hat a_n^{\dagger}\hat a_m+\hat a_n\hat a_m^{\dagger}-\hat a_n^{\dagger}\hat a_n-\hat a_m^{\dagger}\hat a_m\Big) +\frac{D}{i}\sum_n \Big(\hat a_{n}\hat a_{n+1}^{\dagger}-\hat a_{n}^{\dagger}\hat a_{n+1}\Big)\Big].
\end{eqnarray}
Fourier transform of $\hat a_{n}^{\dagger}(\hat a_{n})$ is $\hat a_{\vec{k}}^{\dagger}(\hat a_{\vec{k}})$.
\begin{equation}
\hat a_{\vec{k}}^{\dagger}=\frac{1}{\sqrt{N}}\sum_ne^{i \vec{k}\vec{r}_n}a_n^{\dagger},~~~~~~~~~~~\hat a_{\vec{k}}=\frac{1}{\sqrt{N}}\sum_ne^{i \vec{k}\vec{r}_n}a_n.
\end{equation}
Inverse Fourier transform is given as
\begin{equation}
\hat a_n^{\dagger}=\frac{1}{\sqrt{N}}\sum_ne^{i \vec{k}\vec{r}_n}a_{\vec{k}}^{\dagger},~~~~~~~~~\hat a_n=\frac{1}{\sqrt{N}}\sum_ne^{i \vec{k}\vec{r}_n}a_{\vec{k}}.
\end{equation}

After summing over $n$ we get Hamiltonian  (Eq.~\ref{Hamiltonian_A4}) in $\vec{k}$ space as 
\begin{eqnarray}
\hat H &=& \sum\limits_{\vec{k}}\omega_{\vec{k}} \hat{a}^{\dagger}_{\vec{k}}\hat{a}_{\vec{k}}-D\sum_{\vec{k}} \sin(\vec{k} a)\hat a_{\vec{k}}^{\dagger} \hat a_{\vec{k}}=\sum\limits_{\vec{k}}\omega(\pm D,\textbf{k})\hat{a}^{\dagger}_{\vec{k}}\hat{a}_{\vec{k}}
\end{eqnarray}
where,
 \begin{eqnarray}
 &&\omega(\pm D,\textbf{k})=\big(\omega(\vec{k})\pm\omega_{DM}(\vec{k})\big),  ~~\omega_{DM}(\vec{k})=D\sin(k_xa),\nonumber\\
 &&\omega_k=2J_1(1-\gamma_{1,\textbf{k}})+2J_2(1-\gamma_{2,\textbf{k}}),\,\,\gamma_{1,\textbf{k}}=1/2(\cos k_x+\cos k_y),\nonumber\\
 && \gamma_{2,\textbf{k}}=1/2(\cos (k_x+k_y)+\cos (k_x-k_y)).
  \end{eqnarray}

\section{Calculation of left and right out-of-time ordered correlation functions}
\label{LROTOC_calculation}
We will calculate OTOC exactly for one magnon excitation state given in Eq. (7) as

\begin{equation}
    \begin{split}
        \label{Ap4_occupation number}
  C(t)=\frac{1}{2}\bigg\lbrace\langle\hat \eta_n\hat \eta_m(t)\hat \eta_m(t)\hat \eta_n\rangle
  &+\langle \hat \eta_m(t)\hat \eta_n\hat \eta_n\hat \eta_m(t)\rangle  \\
  &-\langle \hat \eta_m(t)\hat \eta_n\hat \eta_m(t)\hat \eta_n\rangle-\langle
          \hat \eta_n\hat \eta_m(t)\hat \eta_n\hat \eta_m(t)\bigg\rbrace.
          \end{split}
\end{equation}
Here, $\hat \eta_{m/n}= \hat \sigma_{m/n}^z$ is Hermitian and unitary, therefore, Eq.~(\ref{Ap4_occupation number}) transforms in the form given as
\begin{equation}
C(t)=1-\langle \hat \eta_m(t)\hat \eta_n\hat \eta_m(t)\hat \eta_n\rangle=1-F(t),
\end{equation}
where $F(t)$ is given as
\begin{equation}
\label{F_t}
F(t)=\langle \phi \vert \hat a_{n}\hat \eta_m(t)\hat \eta_n\hat \eta_m(t)\hat \eta_na^{\dagger}_
{n}\vert \phi \rangle.
\end{equation}
 In the above equation, the expectation value is taken over one magnon excitation state $\hat a^{\dagger}_n\vert \phi\rangle$, where $\vert \phi\rangle$ is the vacuum state, equivalent to a polarized state. 
First of all we calculate the product four obsevables in $F(t)$ (Eq.~(\ref{F_t}))in bosonic representation as
\begin{eqnarray}
\label{S3}
\hat \eta_m(t)\hat \eta_n\hat \eta_m(t)\hat \eta_n&=&[1-2\hat a^{\dagger}_{
m}\hat a_{
m}(t)][1-2\hat a^{\dagger}_{
n}\hat a_{
n}][1-2\hat a^{\dagger}_{
m}\hat a_{
m}(t)][1-2\hat a^{\dagger}_{
n}\hat a_{
n}], \nonumber \\
&=&\Big[1-2\hat a^{\dagger}_{
m}\hat a_{
m}(t)-2\hat a^{\dagger}_{
n}\hat a_{
n}+4\hat a^{\dagger}_{
m}\hat a_{
m}(t)\hat a^{\dagger}_{
n}\hat a_{
n}\Big] \nonumber \\
&\times&\Big[1-2\hat a^{\dagger}_{
m}\hat a_{
m}(t)-2\hat a^{\dagger}_{
n}\hat a_{
n}+4\hat a^{\dagger}_{
m}\hat a_{
m}(t)\hat a^{\dagger}_{
n}\hat a_{
n}\Big], \nonumber \\
&=&1-4\hat a^{\dagger}_{
m}\hat a_{
m}(t)-4\hat a^{\dagger}_{
n}\hat a_{
n}+4\hat a^{\dagger}_{
m}\hat a_{
m}(t)\hat a^{\dagger}_{
n}\hat a_{
n} +4\hat a^{\dagger}_{
n}\hat a_{
n}\hat a^{\dagger}_{
m}\hat a_{
m}(t)\nonumber \\ 
&+& 4\hat a^{\dagger}_{
m}\hat a_{
m}\hat a^{\dagger}_{
m}\hat a_{
m}(t)+4\hat a^{\dagger}_{
n}\hat a_{
n}\hat a^{\dagger}_{
n}\hat a_{
n}+4\hat a^{\dagger}_{
m}\hat a_{
m}\hat a^{\dagger}_{
n}\hat a_{
n}+4\hat a^{\dagger}_{
n}\hat a_{
n}\hat a^{\dagger}_{
m}\hat a_{
m} \nonumber \\
&-&8\hat a^{\dagger}_{
m}\hat a_{
m}\hat a^{\dagger}_{
m}\hat a_{
m}(t) \hat a^{\dagger}_{
n}\hat a_{
n}-8\hat a^{\dagger}_{
n}\hat a_{
n}\hat a^{\dagger}_{
m}\hat a_{
m}(t) \hat a^{\dagger}_{
n}\hat a_{
n} \nonumber \\
&-&8\hat a^{\dagger}_{
m}\hat a_{
m}(t)\hat a^{\dagger}_{
n}\hat a_{
n}\hat a^{\dagger}_{
m}\hat a_{
m}(t)-8\hat a^{\dagger}_{
m}\hat a_{
m}(t)\hat a^{\dagger}_{
n}\hat a_{
n} \hat a^{\dagger}_{
n}\hat a_{
n} \nonumber \\
&+& 16\hat a^{\dagger}_{
m}\hat a_{
m}(t)\hat a^{\dagger}_{
n}\hat a_{
n}\hat a^{\dagger}_{
m}\hat a_{
m}(t)\hat a^{\dagger}_{
n}\hat a_{
n}.
\end{eqnarray}
Further, we calculate the expectation value of the last term of  Eq. (\ref{S3}) over one magnon excitation state  {\it i. e.},
$\langle \phi \vert \hat a_{n} \hat a^{\dagger}_{
m}\hat a_{
m}(t) \hat a^{\dagger}_{n}\hat a_{n}
\hat a_{n} \hat a^{\dagger}_{
m}\hat a_{
m}(t) \hat a^{\dagger}_{n}\hat a_{n}\hat a^{\dagger}_{n}\vert \phi \rangle, $
using the properties of bosonic operators $[\hat a_i, \hat a_j^{\dagger}]=\delta_{ij}$, $(\hat a_i)^2=0$, and $(\hat a^{\dagger}_i)^2=0$. We get 
\begin{equation}
    \label{S5}
    \begin{split}
        \langle \phi \vert \hat a_{n} \hat a^{\dagger}_{m}\hat a_{m}(t) \hat a^{\dagger}_{n}\hat a_{n}\hat a_{n} \hat a^{\dagger}_{m}\hat a_{m}(t) \hat a^{\dagger}_{n}\hat a_{n}\hat a^{\dagger}_{n}\vert \phi \rangle
&=\langle \phi \vert \hat a_{n} e^{i\hat H t}\hat a^{\dagger}_{
m}\hat a_{
m}e^{-i\hat H t} \hat a^{\dagger}_{n}\hat a_{n}
e^{i\hat H t}\hat a^{\dagger}_{
m}\hat a_{
m}e^{-i\hat H t} \hat a^{\dagger}_{n}\vert \phi \rangle, \\
&=\langle \Psi(t)\vert \Psi(t)  \rangle,
\end{split}
\end{equation}
where $\vert \Psi(t)\rangle=\hat a_{n}
e^{i\hat H t}\hat a^{\dagger}_{
m}\hat a_{
m}e^{-i\hat H t} \hat a^{\dagger}_{n}\vert \phi \rangle.$ Fourier transformation of the $\vert \Psi(t)\rangle$ and  diagonalized Hamiltonian will provide
\begin{eqnarray}
\vert \Psi(t)\rangle &=&\frac{1}{N}\sum_k e^{i(-k(m-n)+\omega_kt/\hslash)} \frac{1}{N}\sum_{k^{'}} e^{i( k^{'} (m-n)-\omega_{k^{'}}t/\hslash)} \vert \phi \rangle  \nonumber \\
&=&\frac{1}{N^2} \Omega_1 \Omega_2 \vert \phi \rangle. \nonumber
\end{eqnarray}
Hence, 
\begin{equation}
\label{S6}
\langle \Psi(t)\vert \Psi(t)  \rangle=\frac{1}{N^4}\Omega_1 \Omega_2\Omega_1 \Omega_2.
\end{equation}
Similarly,
\begin{equation}
\label{S7}
\langle \phi \vert \hat a_{n}
 \hat a_{
m} \hat a_{
m}(t)\hat a^{\dagger}_{n} \vert \phi \rangle=\frac{1}{N^2}\Omega_1 \Omega_2
\end{equation}
After doing some simple bosonic algebra, time dependent terms of Eq.~(\ref{S3}) are converted either in the form of Eq.~(\ref{S5}) or Eq.~(\ref{S7}). By using Eq.~(\ref{S6}) and Eq.~(\ref{S7}), we calculate $F(t)$ as
\begin{eqnarray}
F(t)&=&1-\frac{4}{N^2}\Omega_1\Omega_2-4+\frac{4}{N^2}\Omega_1\Omega_2+\frac{4}{N^2}\Omega_1\Omega_2+\frac{4}{N^2}\Omega_1\Omega_2+\frac{4}{N^2}\Omega_1\Omega_2 \nonumber \\
&&+\frac{4}{N^2}\Omega_1\Omega_2+4-\frac{8}{N^2}\Omega_1\Omega_2-\frac{8}{N^2}\Omega_1\Omega_2 -\frac{8}{N^4}\Omega_1\Omega_2\Omega_1\Omega_2-\frac{8}{N^2}\Omega_1\Omega_2\nonumber \\ &&+\frac{16}{N^4}\Omega_1\Omega_2\Omega_1\Omega_2, \nonumber \\ 
&=&1-\frac{8}{N^2}\Omega_1\Omega_2+\frac{8}{N^4}\Omega_1\Omega_2\Omega_1\Omega_2.
\end{eqnarray}
Then, we get left and right OTOCs’ analytical expression as

\begin{eqnarray}
\label{Ap4_we deduce the analytical formula2}
C_L(t)&=&\frac{8}{N^2}\Omega_1^L\Omega_2^L-\frac{8}{N^4}\Omega_1^L\Omega_2^L\Omega_1^L\Omega_2^L,  \nonumber \\
C_R(t)&=&\zeta^4(D)\Big(\frac{8}{N^2}\Omega_1^R\Omega_2^R-\frac{8}{N^4}\Omega_1^R\Omega_2^R\Omega_1^R\Omega_2^R\Big),  
\end{eqnarray}
where frequencies $\Omega_{1/2}^{L/R}$ are given as
\begin{eqnarray}\label{Ap4_Left and Right}
\Omega_1^R&=&\Omega_2^{R*}=\sum_{m_0}\exp\bigg(-\frac{im_0\pi r_{1,2}} {a_0}\bigg)\exp\bigg(\frac{i \omega_{m_0}t}{ \hslash}\bigg), ~{\rm and}
\nonumber \\
\Omega_1^L&=&\Omega_2^{L*}=\sum_{m_0} \exp(-ik_s^{-}r_{1,2})\exp\bigg(\frac{i \omega_{m_0}t}{ \hslash}\bigg).
\end{eqnarray}

\end{appendices}
\chapter*{\vspace{-3cm} \centering \Large List of Publications}

\begin{enumerate}
\item ``Out-of-time-order correlation and detection of phase structure in  Floquet transverse Ising spin system", {\bf Rohit Kumar Shukla}, Gautam Kamalakar Naik, and Sunil Kumar Mishra, EPL {\bf 132}, 47003 (2021).

\item ``Characteristic, dynamic, and  saturation regions of Out-of-time-order correlation in Floquet Ising models", {\bf Rohit Kumar Shukla} and Sunil Kumar Mishra,  Phys. Rev. A {\bf 132}, 47003 (2022).

\item ``Out-of-time-order correlation of nonlocal block-spin and random observables in integrable and nonintegrable spin chains", {\bf Rohit Kumar Shukla}, Arul Lakshminarayan, and Sunil Kumar Mishra, Phys. Rev. B {\bf 105}, 224307 (2022)

\item ``Quantum information diode based on the  magnonic crystal", {\bf Rohit Kumar Shukla}, L. Chotorlishvili, Vipin Vijayan, Harshit
Verma, A. Ernst, S. S. P. Parkin, and Sunil K. Mishra, under review in npj Quantum Information.
\end{enumerate}

\printthesisindex 

\end{document}